\documentclass[11pt]{article}

\input{epsf}
\usepackage{epsfig}
\usepackage{amssymb}
\usepackage{amsfonts}
\usepackage{amsbsy}
\usepackage[all]{xy}
\usepackage{amsmath}

\usepackage{amssymb,amscd}
\usepackage{mathrsfs}
\usepackage{amsmath,amsthm}
\usepackage{dsfont}

\def\be{ \begin{eqnarray} }
\def\ee{ \end{eqnarray}}

\def\Co0{{\rm Co}_0}

\def\exp{{\rm exp}}

\def\I{{\rm i}}

\def\log{{\rm log}}

\def\p{\partial}

\def\one{{\hbox{ 1\kern-.8mm l}}}

\def\CC {{\cal C}}

\def\CE {{\cal E}}

\def\CH {{\cal H}}
\def\CI {{\cal I}}

\def\CL {{\cal L}}
\def\CM {{\cal M}}
\def\CN {{\cal N}}
\def\CO {{\cal O}}
\def\CP {{\cal P}}
\def\CR {{\cal R}}

\def\CW {{\cal W}}

\def\CO {{\cal O}}
\def\CZ {{\cal Z}}
\def\CE {{\cal E}}

\def\CH {{\cal H}}
\def\CI {{{\cal I}}}

\def\CQ {{\cal Q}}

\def\CT {{\cal T}}

\def\CZ{{\cal Z}}

\def\IC{\mathbb{C}}
\def\ID{\mathbb{D}}

\def\IP{\mathbb{P}}
\def\IR{{\mathbb{R}}}

\def\IV{{\mathbb{V}}}
\def\IW{{\mathbb{W}}}
\def\IZ{{\mathbb{Z}}}

\def\fd{\mathfrak{d}}

\def\fg{\mathfrak{g}}

\def\fs{\mathfrak{s}}

\def\fs{\mathfrak{s}}

\def\fu{\mathfrak{u}}

\def\fI{\mathfrak{I}}

\def\fL{\mathfrak{L}}

\def\fS{\mathfrak{S}}

\def\rmk#1{\bigskip\noindent{\bf Remarks} }

\usepackage{tikz}
\usetikzlibrary{arrows}
\usetikzlibrary{arrows.meta}
\usetikzlibrary{positioning}
\usetikzlibrary{shapes,snakes}
\usetikzlibrary{fit}

\def\lm{\limits}
\def\nn{\nonumber}

\def\S{\draw[ultra thick, purple] (0,0) to[out=90,in=210] (0.5,0.5) to[out=30,in=270] (1,1) (1,0) to[out=90,in=330] (0.6,0.4) (0.4,0.6) to[out=150,in=270] (0,1);}

\def\lm{\limits}

\newtheorem{theorem}{Theorem}
\newtheorem{proposition}{Proposition}
\newtheorem{definition}{Definition}

\hoffset= - 0.9in    

\numberwithin{equation}{section}
\numberwithin{theorem}{section}

\title{Why Is Landau-Ginzburg Link Cohomology Equivalent To Khovanov Homology?}
\author{Dmitry Galakhov\footnote{e-mails: galakhov@berkeley.edu; galakhov@itep.ru}}
\date{}

\begin{document}
\maketitle
\vspace{-1cm}
\begin{center}
	{\it Center for Theoretical Physics, University of California, Berkeley, USA, 94720}\\
	{\it Institute for Information Transmission Problems, Moscow, Russia, 127994}
\end{center}

\begin{abstract}
	In this note we make an attempt to compare a cohomological theory of Hilbert spaces of ground states in the $\CN=(2,2)$ 2d Landau-Ginzburg theory in models describing link embeddings in $\IR^3$ to Khovanov and Khovanov-Rozansky homologies. To confirm the equivalence we exploit the invariance of Hilbert spaces of  ground states for interfaces with respect to homotopy.  In this attempt to study solitons and instantons in the Landau-Giznburg theory we apply asymptotic analysis also known in the literature as exact WKB method, spectral networks method, or resurgence. In particular, we associate instantons in LG model to specific WKB line configurations we call null-webs. 
\end{abstract}


\tableofcontents




\section{Introduction And Discussion}
A quest for a TQFT describing Khovanov invariants \cite{Khovanov}, a categorification of Jones polynomials -- Wilson loop averages in the 3d Chern-Simons theory \cite{Witten:1988hf,Moore:1989vd,Elitzur:1989nr}, -- or more desirably Khovanov-Rozansky invariants \cite{Khovanov_sl_3,KR},  or even more generally superpolynomials \cite{super}, delivering categorification to HOMFLY-PT invariants, has led to a wide and profound development in the both physics and mathematics, see e.g. \cite{Aganagic:2011sg,Anokhina:2014hha,Arthamonov:2015rha,Dolotin:2012sw,Gukov:2004hz,KR,Lauda,Mackaay,Rasmussen,WItten:2011pz,resurgence,Witten:2014xwa,Witten:2011zz,Gukov:2016gkn,Gukov_Hopf,Chun:2015gda}.\footnote{The literature on generalizations and physical constructions of knot/link invariants is extremely rich. Here we try to mention just some of possible directions.} See also a recent nice review \cite{guide}.

In this note we will discuss an approach proposed by Witten \cite{Witten:2010cx, Witten:2011zz} using equivariant Morse theory on a field space emerging as a Kaluza-Klein reduction of six-dimensional (2,0) theory on a cigar. The link is associated to a 'tHooft operator insertion giving specific boundary conditions for fields \cite{Mazzeo:2013zga}. This formulation is enormously complicated for a practical use, say, an explicit calculation of invariants. Then further simplifications were proposed in \cite{Gaiotto:2011nm,GMW}. The theory descends to an effective IR 2d $\CN=(2,2)$ Landau-Ginzburg (LG) theory with the target space given by a universal cover of the monopole moduli space. The link is encoded in a half-supersymmetric interface representing a ``moving'' singular monopole probe.

This note is a companion of \cite{GM} where this idea was checked. It was shown that to construct an analog of Khovanov homology for a link $L$ it is enough to consider a moduli space of a certain fixed number of monopoles. The Hilbert space $\CE(L)$ of perturbative ground states is spanned by vectors associated to critical points of a Morse height functional $h(L)$. It is naturally bi-graded by the fermion number $\bf F$ and by central charge $Z$ appearing in discrete portions of $\pi \I$, so that $Z=\pi \I {\bf P}$, where ${\bf P}\in\IZ$. In the presence of a link interface one nilpotent supercharge $Q$ is preserved. Hilbert space $\CE(L)$ plays the role of Morse-Smale-Witten complex, while $Q$ acts as a differential on this complex \cite{WittenMorse}. The non-perturbative Hilbert space is defined as cohomology:
\be
{\rm\bf LGCoh}(L):=H^{*}(\CE(L),Q)=\bigoplus_{{\bf F},{\bf P}}\CH^{({\bf F},{\bf P})}(L)
\ee
The Euler characteristic - supersymmetric index \cite{CFIV} - of this complex gives Jones polynomials.
In \cite{GM} ${\rm\bf LGCoh}(L)$ was argued to be an invariant of Reidemeister moves and therefore  an invariant of link $L$. As well it was shown in several examples to be isomorphic as a bi-graded vector space to the Khovanov link homology we denote here as ${\rm \bf KHom}(L)$. Generic argument for equivalence between ${\rm\bf LGCoh}(L)$ and ${\rm \bf KHom}(L)$ is still missing:
\be
\label{quasi-isomorphism}
{\rm\bf LGCoh}(L)\mathop{\cong}\lm^{?} {\rm \bf KHom}(L)
\ee
This note is devoted to an attempt to deliver this missing argument.

We use the following arguments. One associates to embedding of link $L$ certain interface path $\hat{\wp}$ on parameter space $\CP$ of the LG theory (see Figure \ref{fig:paths}). The invariance of this construction for different link embeddings follows form the isomorphism of cohomologies:
\be
H^*(\CE(\hat \wp),Q)\cong H^*(\CE(\hat \wp'),Q)
\ee
for homotopic paths $\hat\wp$ and $\hat{\wp}'$. Despite this isomorphism the very complexes $\CE(\hat{\wp})$ and $\CE(\hat{\wp}')$ may be quite \emph{different}. Relations between their generators -- soliton-interface bound states -- are governed by 2d wall-crossing phenomena \cite{CV}. So the cohomology is indeed an invariant of the path homotopy equivalence class $\left[\hat\wp\right]$, while complexes $\CE(\hat\wp)$ may vary from representative $\hat\wp\in \left[\hat \wp\right]$ to another representative. We will find \emph{distinguished} representative $\hat\wp_0$ for each class $\left[\hat\wp\right]$ such that the corresponding complex repeats the combinatorial construction of the Khovanov homology \cite{BarNatan}:
\begin{enumerate}
	\item $\CE(\hat \wp_0)$ is isomorphic to Khovanov complex as a bi-graded vector space. More precisely, we will be able to categorize subspaces of $\CE(\hat \wp_0)$ as vertices of link resolution hypercube.
	\item Supercharge $Q$ acts on $\CE(\hat \wp_0)$ in the same way as the differential in Khovanov's construction by tensor multiplications and co-multiplications.
\end{enumerate}
Equivalence \eqref{quasi-isomorphism} follows tautologically:
\be
{\rm\bf LGCoh}(L)\cong H^*(\CE(\hat \wp_0),Q)\cong {\rm \bf KHom}(L)
\ee

\begin{figure}
\begin{center}
\begin{tikzpicture}
\draw[ultra thick, fill=blue!20!white, opacity=1] (0,2.3) to[out=0,in=90] (3.5,0) to[out=270,in=0] (0,-2.3) to[out=180,in=270] (-3.5,0) to[out=90,in=180] (0,2.3);
\begin{scope}[shift={(1,-1.6)}]
\draw[ultra thick, fill=white] (0,0) to[out=45,in=135] (0.5,0) to[out=225,in=315] (0,0);
\draw[ultra thick] (0,0) -- (-0.15,0.15) (0.5,0) -- (0.65,0.15);
\end{scope}
\begin{scope}[shift={(1.6,-1)}]
\draw[ultra thick, fill=white] (0,0) to[out=45,in=135] (0.5,0) to[out=225,in=315] (0,0);
\draw[ultra thick] (0,0) -- (-0.15,0.15) (0.5,0) -- (0.65,0.15);
\end{scope}
\begin{scope}[shift={(-1.4,1.4)}]
\draw[ultra thick, fill=white] (0,0) to[out=45,in=135] (0.5,0) to[out=225,in=315] (0,0);
\draw[ultra thick] (0,0) -- (-0.15,0.15) (0.5,0) -- (0.65,0.15);
\end{scope}
\draw[dashed, thick]  (-2,-0.5) -- (2,0.5) (-2,0.2) to[out=30,in=170] (2,1.2) (-2,-1.2) to[out=350, in=210] (2,-0.2) (0,-1.2) -- (0,1.2) (1,-0.9) -- (1,1.5) (-1,0.9) -- (-1,-1.5);
\draw[ultra thick, ->, blue] (0,1) to[out=0,in=150] (1.5,0.7) (-1.5,-0.7) to[out=80,in=180] (0,1);
\draw[ultra thick, ->, blue, dashed] (0,0) -- (1.5,0.7) (-1.5,-0.7) -- (0,0);
\draw[ultra thick, ->, blue, dashed] (0,-1) to[out=0,in=260] (1.5,0.7) (-1.5,-0.7) to[out=330,in=180] (0,-1);
\draw[fill=blue] (-1.5,-0.7) circle (0.1) (1.5,0.7) circle (0.1);
\begin{scope}[shift={(-2,1)}]
\draw[ultra thick, violet] (-0.1,-0.1) -- (0.1,0.1) (0.1,-0.1) -- (-0.1,0.1);
\end{scope}
\begin{scope}[shift={(0,-1.8)}]
\draw[ultra thick, violet] (-0.1,-0.1) -- (0.1,0.1) (0.1,-0.1) -- (-0.1,0.1);
\end{scope}
\begin{scope}[shift={(2,1)}]
\draw[ultra thick, violet] (-0.1,-0.1) -- (0.1,0.1) (0.1,-0.1) -- (-0.1,0.1);
\end{scope}
\node[left] at (-1.6,-0.7) {$A$}; \node[right] at (1.6,0.7) {$B$};
\node[above] at (0,1.1) {$\color{blue}\hat\wp$};
\end{tikzpicture}
\end{center}
\caption{Paths on parameter space $\CP$.}\label{fig:paths}
\end{figure}

This note is divided in four sections.

An important role in the LG model is played by Lefschetz thimbles used to construct critical points of the height functional $h$ and the action of the supercharge $Q$. In Section \ref{sec:SAILGM} we review and extend asymptotic counting of Lefschetz thimbles by their contribution to the Picard-Lefschetz monodromy of integrals also known in the literature as spectral network technique \cite{SN,GLM}, exact  Wentzel-Kramers-Brillouin (WKB) method \cite{Elyutin}, or resurgence \cite{resurgence,res_review}. This analysis has been proven to be useful in estimates of eta-invariants of Dirac operators in the soliton background (see \cite[Appendix A]{GM}) and we use it to count $h$-critical points in our setup. Moreover we conjecture that instantons in the LG model correspond to a specific critical WKB web configurations we call {\bf null-webs}.

In Section \ref{sec:su_2} we review the construction of ${\rm\bf LGCoh}(L)$ under the prism of spectral analysis, consider wall-crossings acting by morphisms on ${\rm\bf LGCoh}(L)$, and finally construct a wall crossing morphims mapping ${\rm\bf LGCoh}(L)$ to a complex isomorphic to ${\rm \bf KHom}(L)$. The vacua in the LG model on the $SU(n)$ monopole moduli space are in one-to-one correspondence with tensor powers of $SU(n)$ irreducible representations. However this set has two descriptions - two ``nice" basis choices: tensor product basis and isotypical basis. $R$-matrix acquires a simple diagonal form in the isotypical basis. And to establish the morphism between ${\rm\bf LGCoh}(L)$ and ${\rm\bf KHom}(L)$ we exploit an ``isotypical" interface performing isotypical decomposition of LG vacua.

Afterwards in Section \ref{sec:su_n} we construct $\mathfrak{su}_n$ Landau-Ginzburg link cohomology ${}_n{\rm\bf LGCoh}(L)$ starting from the monopole moduli space arising in a (2,0) six-dimensional theory with a gauge algebra $\mathfrak{su}_n$ \cite{Braverman}. We argue that ${}_n{\bf LGCoh}(L)$ is an invariant of link $L$.

And, finally, in Section \ref{sec:compare} we compare ${}_n{\rm\bf LGCoh}(L)$ to Khovanov-Rozansky link homology \cite{KR}, we denote it as ${}_n{\rm \bf KRHom}(L)$. In particular, we argue that LG cohomology for a link and its resolutions should satisfy an exact triangle relation:
$$
\begin{array}{c}
	\begin{tikzpicture}
	\node(A) {${}_n {\bf LGCoh}\left[\begin{array}{c}
		\begin{tikzpicture}[scale=0.7]
		\draw[ultra thick, ->] (-0.5,-0.5) -- (0.5,0.5);
		\draw[ultra thick] (0.5,-0.5) -- (0.1,-0.1);
		\draw[ultra thick, ->] (-0.1,0.1) -- (-0.5,0.5);
		\end{tikzpicture}
		\end{array}\right]$};
	\node(B) at (-4.5,0) {${}_n {\bf LGCoh}\left[\begin{array}{c}
		\begin{tikzpicture}[scale=0.7]
		\draw [ultra thick,->] (-0.5,-0.5) to [out= 45, in =315] (-0.5,0.5); \draw [ultra thick,->] (0.5,-0.5) to [out=135, in=225] (0.5,0.5);
		\end{tikzpicture}
		\end{array}\right]$};
	\node(C) at (4.5,0) {${}_n {\bf LGCoh}\left[\begin{array}{c}
		\begin{tikzpicture}[scale=0.7]
		\draw [ultra thick] (-0.5,-0.5) -- (0,-0.2) -- (0.5,-0.5);
		\draw [ultra thick, ->] (0,0.2) -- (-0.5,0.5); 
		\draw [ultra thick, ->] (0,0.2) -- (0.5,0.5);
		\draw[fill=white] (0.05,0.2) -- (0.05,-0.2) -- (-0.05,-0.2) -- (-0.05,0.2) -- cycle;
		\end{tikzpicture}
		\end{array}\right]$};
	\path (B) edge[->] (A) (A) edge[->] (C) (C) edge[bend left, ->] node[above] {$t$} (B);
	\end{tikzpicture}
\end{array}
$$
This exact triangle relation is a categorification of skein relations for the HOMFLY link diagrams, a starting point for Khovanov-Rozansky construction.

In Appendix we put some examples of explicit applications of methods discussed along this note, in particular, we calculate explicitly ${}_3 {\bf LGCoh}(\CH)$ for Hopf link $\CH$ and observe it coincides with ${}_3 {\bf KRHom}(\CH)$.

\paragraph{Remarks:}
\begin{enumerate}
	\item It is well-known that the link invariants are actually invariants of knotted and linked ribbons rather than thin strands due to framing anomaly. Ribbon twists affect an overall monomial multiplier of the link polynomial. Stripping off this monomial factor one gets link invariants. Therefore we consider all the polynomial equivalences modulo overall monomial factor. Alternatively, we claim bi-graded vector space isomorphisms modulo overall degree shifts $({\bf P},{\bf F})\to ({\bf P}+a,{\bf F}+b)$.
	\item We should mention that asymptotic expansion we use differs slightly from one used in the spectral networks literature \cite{SN,ADE,Hollands:2016kgm,GLM}, see footnote \ref{foot} on page \pageref{foot}. Therefore despite the methodology does not change we trade the term ``spectral network" for ``WKB web".
	\item In \cite{GMW} rich algebraic $A_{\infty}$- and $L_{\infty}$-structures were discovered hidden in the instanton processes governing LG models. Our construction of null-webs is motivated by a search for a simpler LG instanton construction. So it would be interesting to observe  $A_{\infty}$- and $L_{\infty}$-structures hidden in null-webs if there are ones.
	\item A relation between a notion of Chern-Simons $(q,t)$-refinement introduced in \cite{Aganagic:2011sg} and LG cohomology in the formulation used in this note is still missing. One may hope to present such a relation in terms of spectral analysis. In this spectral analysis differentiable connections should be substituted by lattice-like difference operators, see e.g. \cite{Maulik-Okunkov,Okuonkov_Lect,Zenkevich:2015rua}, and recent development \cite{Eager:2016yxd} in exponential deformation of spectral networks may turn out to be a suitable description on this route.
	\item We argue that one can choose a distinguished path $\hat \wp_0$ on $\CP$ such that the LG complex becomes isomorphic to Khovanov's one, however there is an intriguing question if one can distinguish path $\hat \wp_0$ from other representatives in its homotopy class giving different complexes \emph{a priori}.
\end{enumerate}

\section*{Acknowledgements}
I would like to thank Mina Aganagic, Joel Clingempeel, Pietro Longhi, Alexei Morozov, Gregory Moore, Andrey Smirnov for fruitful discussions and valuable suggestions and remarks in different stages of this project. My research is supported in part by the Berkeley Center for Theoretical Physics, by the Simons Foundation, and by NSF grant \#1521446 and RFBR grant 16-01-00291. 

\section{Spectral Analysis Of Instantons In The Landau-Ginzburg Model}
\label{sec:SAILGM}
\subsection{Spectral analysis of Lefschetz thimbles}\label{subsec:SpecAn}
\subsubsection{Lefschetz thimbles and Picard-Lefschetz monodromy}
Consider a K\"ahler manifold $X$ of complex dimension $N$, we index its coordinates with capital Latin letters $I=1,\ldots,N$. This manifold is endowed with the K\"ahler metric $g_{I\bar J}$. Also consider a holomorphic function $W:\; X\to \IC$, in physical setup we refer to it as a superpotential.

We will refer to critical points of $W$ as vacua and denote them by small Latin letter indices from a vacuum set: $i\in\IV$.

Define a left (right) Lefschetz thimble $\CL_i(\zeta)$ of phase $\zeta$ for a critical point $\phi_i$ of $W$ as a $N$-dimensional \emph{real} surface in $X$, a union of paths satisfying:
\be\label{thimble}
\p_s \phi^I(s)=-\zeta g^{I\bar J}\overline{\p_J W(\phi^I(s))}
\ee
approaching a critical point $\phi_i$ in the limit $s\to-\infty$ ($s\to+\infty$).

Lefschetz thimbles are nice integration cycles for the integrals of type:
\be\label{integral}
B_i=\int\lm_{\CL_{i}(\zeta)} \prod\lm_I d\phi^I e^{\zeta^{-1} W(\phi^I)}
\ee
Notice in the $W$-plane Lefschetz thimbles are represented by lines of constant slope $\pi\I+{\rm Arg}\;\zeta$, consequently integral \eqref{integral} taken along a Lefschetz thimble converges:
\begin{center}
	\begin{tikzpicture}
	\begin{scope}
	\draw[thick, ->] (0,0) -- (3,1);
	\draw (0,0) -- (1,0) ([shift=(0:0.8)]0,0) arc (0:19:0.8);
	\node[right] at (1.2,0.2) {$\vartheta$};
	\node[left] at (0,-0.1) {$i_1$};
	\draw[purple, ultra thick] (-0.1,-0.1) -- (0.1,0.1) (0.1,-0.1) -- (-0.1,0.1);
	\end{scope}
	\begin{scope}[shift={(0.2,-0.7)}]
	\draw[thick, ->] (0,0) -- (3,1);
	\draw (0,0) -- (1,0) ([shift=(0:0.8)]0,0) arc (0:19:0.8);
	\node[right] at (1.2,0.2) {$\vartheta$};
	\node[left] at (0,-0.1) {$i_2$};
	\draw[purple, ultra thick] (-0.1,-0.1) -- (0.1,0.1) (0.1,-0.1) -- (-0.1,0.1);
	\end{scope}
	\begin{scope}[shift={(1.6,-1.1)}]
	\draw[thick, ->] (0,0) -- (3,1);
	\draw (0,0) -- (1,0) ([shift=(0:0.8)]0,0) arc (0:19:0.8);
	\node[right] at (1.2,0.2) {$\vartheta$};
	\node[left] at (0,-0.1) {$i_3$};
	\draw[purple, ultra thick] (-0.1,-0.1) -- (0.1,0.1) (0.1,-0.1) -- (-0.1,0.1);
	\end{scope}
	\draw (-1.7,0) -- (-1,0) -- (-1,0.7);
	\node[above left] at (-1,0) {$W$};
	\node at (7,0) {$-\zeta=e^{\I\vartheta}$};
	\end{tikzpicture}
\end{center}

Now we consider an $n$-dimensional complex space $\CP$ spanned by coordinates $z_a$ and a family of functions $W:\; X\times\CP\to \IC$ holomorphic in both types of variables. A bundle of integrals over $\CP$:
\be\label{intgrl}
B_i(z_a)=\int\lm_{\CL_i(\zeta)} \prod\lm_I d\phi^I e^{\zeta^{-1} W(\phi^I|z_a)}
\ee
has monodromies called Picard-Lefschetz monodromies corresponding to intersecting Lefschetz thimbles \cite{Hori:2000ck}.

If a critical point $j$ intersects a Lefschetz thimble $\CL_i(\zeta)$ the integral \eqref{intgrl} acquires a monodromy:
\be\label{jumps}
B_i(z_a)\mapsto B_i(z_a)+\mu_{ij} B_j
\ee
Here $\mu_{ij}$ is up to a sign an intersection number of thimbles $\CL_{i}(\zeta)$ and $\CL_{j}(\zeta)$ and is also referred to as a Maslov index \cite{Maslov}. From the physical point of view $\mu_{ij}$ has an interpretation of an index \cite{CV,Mirror,Hori:2000ck} of $ij$-solitons.

\subsubsection{Asymptotic Berry monodromy}
An alternative way to construct this monodromy and therefore to count contribution of intersections of Lefschetz thimbles is to consider a natural Berry connection on this bundle of integrals. This connection constuction is somewhat analogous to the construction of $tt^*$-connection \cite{CV}\footnote{More elaborated relation between $G$-opers and Gauss-Manin connection constructed from solution to a $G$-Hitchin system can be found in \cite{DFKMMN}.}. In particular, in the quasi-classical limit $\zeta\to 0$ the major contribution to the integral \eqref{intgrl} is accumulated near critical points of $W$, therefore any insertion in the integrand $\CO_a$ having a multiplier of $dW$ along any direction in the $X$-space is quasi-classically 0. Therefore insertions $\CO_a$ form a ring isomorphic to $\IC[\phi^I]/dW$, usually called a chiral ring of operators in the 2-dimensional $\CN=(2,2)$ Landau-Ginzburg model \cite{Mirror}.

Furthermore we notice that the derivatives $\p_{z_a}$ along the base $\CP$ act quasi-classically by multiplication by $\p_{z_a}W(\phi^I|z_a)$ in this ring, this operation induces a natural connection. To be more specific suppose we choose a set of $\hat\CO_a:=\p_{z_a}W(\phi^I|z_a)$ in the chiral ring, then the vector $\Psi:=(\langle \CO_1\rangle,\ldots, \langle \CO_n\rangle)$, where $\langle\star\rangle$ denotes averaging with measure \eqref{intgrl}, is a section of a connection\footnote{Notice that the real $tt^*$-connection has an anti-holomorphic part in contrast inducing in this way a non-trivial Zamolodchikov metric on the parameter space $\CP$ \cite{tt^*}.}:
\be
\nabla_a \Psi:=\left(\p_{z_a}-\zeta^{-1}C_a+A_a\right)\Psi=0
\ee
where $C_a$ are ring structure constants:
\be
\hat\CO_a\CO_b=(C_a)_b^c\CO_c+\sum\lm_I P_I \p_{\phi^I}W
\ee
and $A_a$ are connection components of order $O(\zeta^0)$.
Further we consider a complex line $z_a(x): \IC\to \CP$ and pull back our connection $\nabla_a$ to $\nabla:=\nabla_a\frac{dz_a}{dx}$ on $\IC$. Then $\Psi$ is a solution of a linear system of ordinary differential equations:
\be\label{connection}
\frac{d}{dx}\Psi-\zeta^{-1} C_a \frac{dz_a}{dx}\Psi +A_a\frac{dz_a}{dx}\Psi=0
\ee
Then the analysis is quite similar to the study of differential equation solution asymptotics due to G.~Stokes, also known as a Wentzel-Kramers-Brillouin (WKB) method in physics (see e.g.\cite{Elyutin}). We search for a solution in the following form:
\be\label{asympt_form}
\Psi(x)\sim e^{\zeta^{-1}\int\lm^x p(t)dt+O(\zeta^0)}
\ee
The zero order approximation reduces to an algebraic equation for the eikonal $p dx$, this algebraic equation defines a branched cover $\Sigma$ over $\IC$:
\be\label{spec_curve}
\mathop{\rm Det}\left(\mathds{1}\cdot p(x)dx+C_a dz_a\right)=0
\ee
This is an equation defining a complex curve $\Sigma$ with a meromorphic differential $\lambda:=p\; dx$.

\begin{figure}[h!]
	\begin{center}
		\begin{tikzpicture}
		\begin{scope}[shift={(-1,0)}]
		\node[above]{$(ij)$};
		\begin{scope}[rotate=270]
		\draw [orange, domain=0:1] plot (\x, {0.1*sin(20*\x r)});
		\draw [purple, ultra thick] (-0.1,-0.1) -- (0.1,0.1);
		\draw [purple, ultra thick] (-0.1,0.1) -- (0.1,-0.1);
		\end{scope};
		\end{scope};
		
		\begin{scope}[shift={(0,1)}]
		\node[above]{$(kl)$};
		\begin{scope}[rotate=270]
		\draw [orange, domain=0:1] plot (\x, {0.1*sin(20*\x r)});
		\draw [purple, ultra thick] (-0.1,-0.1) -- (0.1,0.1);
		\draw [purple, ultra thick] (-0.1,0.1) -- (0.1,-0.1);
		\end{scope};
		\end{scope};	
		
		\begin{scope}[shift={(1,0)}]
		\node[above]{$(im)$};
		\begin{scope}[rotate=270]
		\draw [orange, domain=0:1] plot (\x, {0.1*sin(20*\x r)});
		\draw [purple, ultra thick] (-0.1,-0.1) -- (0.1,0.1);
		\draw [purple, ultra thick] (-0.1,0.1) -- (0.1,-0.1);
		\end{scope};
		\end{scope};		
		\end{tikzpicture}
	\end{center}
	\caption{Branching points and cuts}\label{fig:cuts_gen}
\end{figure}
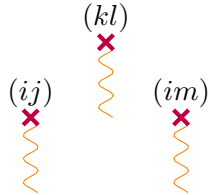

We choose some trivialization of the branched cover $\pi:\Sigma\to{\IC}$: a collection of branching points -- zeroes of the discriminant of \eqref{spec_curve} with respect to $p$ -- and a collection of cuts (see fig.\ref{fig:cuts_gen}). On a set ${\cal C}^c={\IC}\setminus\{{\rm branch\; cuts} \}$ a global choice of the order of the roots $p^{(i)}$ can be made. Thus we have a well defined map $s: \;\Sigma\setminus\pi^{-1}\{ {\rm cuts} \}\to \IZ$ that associates to each point on $\Sigma$ not lying on the cut an order number of the cover sheet it is lying on. We continue the definition of the root ordering to the branching points and associate to each branch point and cut an element of the Galois group of the spectral cover (\ref{spec_curve}) as a polynomial in $p$. 
Different roots $p^{(i)}$ are naturally identified with the choice of the Lefschetz thimble, or vacuum, in \eqref{intgrl}. In theories whith some extra symmetry where the vacua can be labelled by weights of some group the Galois group of the cover can be naturally identified with the Weyl group \cite{SN,ADE,Hollands:2016kgm}.

Now one has a linear space of asymptotic solutions to \eqref{connection}, however, it turns out, an expansion over the asymptotic solutions with constant coefficients can not be a good global solution to equation \eqref{connection}. An easy way to see this is to notice that in branching points the asympotic solutions get Galois monodromies as we mentioned, however connection $\nabla$ is flat in projections of the branching points down to $\IC$. And it is clear that in the neighbourhood of the branching point of type $(ij)$ the asymptotic form \eqref{asympt_form} is invalid since $p^{(i)}-p^{(j)}$ is close to zero. Let us have here a recipe to construct a corrected version of the asymptotic expansion, details and derivation of it can be found in various literature sources \cite{Elyutin,SN,GLM}.

\paragraph{WKB recipe.}

\begin{figure}[h!]
\begin{center}
\begin{tikzpicture}
\node at (-5,0) {
$\begin{array}{c}
\begin{tikzpicture}
\draw[->] (0,0) -- (0,1.5);
\draw[->] (0,0) -- (1.3,-0.75);
\draw[->] (0,0) -- (-1.3,-0.75);
\node[above] at (0,1.5) {$ij$};
\node[below left] at (-1.3,-0.75) {$ji$};
\node[below right] at (1.3,-0.75) {$ji$};
\node[above left] at (0,0) {$(ij)$};
\begin{scope}[shift={(0,0)}]
\begin{scope}[rotate=270]
\draw [orange, domain=0:1] plot (\x, {0.1*sin(20*\x r)});
\draw [purple, ultra thick] (-0.1,-0.1) -- (0.1,0.1);
\draw [purple, ultra thick] (-0.1,0.1) -- (0.1,-0.1);
\end{scope};
\end{scope};
\end{tikzpicture}
\end{array}$
};
\node at (0,0) {
	$\begin{array}{c}
	\begin{tikzpicture}
	\draw[->] (0,0) -- (0,1.5);
	\draw[->] (0,0) -- (1.3,-0.75);
	\draw[->] (0,0) -- (-1.3,-0.75);
	\node[above] at (0,1.5) {$ji$};
	\node[below left] at (-1.3,-0.75) {$ij$};
	\node[below right] at (1.3,-0.75) {$ij$};
	\node[above left] at (0,0) {$(ij)$};
	\begin{scope}[shift={(0,0)}]
	\begin{scope}[rotate=270]
	\draw [orange, domain=0:1] plot (\x, {0.1*sin(20*\x r)});
	\draw [purple, ultra thick] (-0.1,-0.1) -- (0.1,0.1);
	\draw [purple, ultra thick] (-0.1,0.1) -- (0.1,-0.1);
	\end{scope};
	\end{scope};
	\end{tikzpicture}
	\end{array}$
};
\node at (5,0) {
	$\begin{array}{c}
	\begin{tikzpicture}
	\draw[->] (-1,1) -- (-0.5,0.5);
	\draw[->] (-0.5,0.5) -- (0.5,-0.5);
	\draw[->] (-1,-1) -- (-0.5,-0.5);
	\draw[->] (-0.5,-0.5) -- (0.5,0.5);
	\draw[->] (0,0) -- (0.5,0);
	\draw (0.5,0.5) -- (1,1) (0.5,-0.5) -- (1,-1) (0.5,0) -- (1.5,0);
	\node[below left] at (-1,-1) {$ij$};
	\node[above right] at (1,1) {$ij$};
	\node[above left] at (-1,1) {$jk$};
	\node[below right] at (1,-1) {$jk$};
	\node[right] at (1.5,0) {$ik$};
	\end{tikzpicture}
	\end{array}$
};
\node at (-5,-2.3) {(a)};
\node at (0,-2.3) {(b)};
\node at (5,-2.3) {(c)};
\end{tikzpicture}
\end{center}
\caption{Examples of WKB lines.}\label{WKB_lines}
\end{figure}
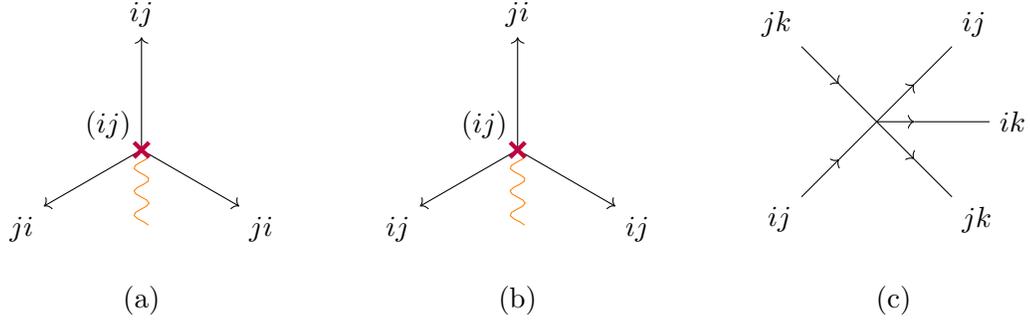

We cut the base $\IC$ by {\bf real} WKB lines (also called Stokes lines, or $S$-walls). The set of lines consists of two subsets: primary lines and descendant ones. A primary line of type $ij$ starts from the projection of the branching point of type $(ij)$ and satisfies:
\be\label{WKB}
{\mathfrak{i}}_{\p_t}(\lambda^{(i)}-\lambda^{(j)})\in \zeta\IR_{>0}
\ee
where $\p_t$ is a tangent vector to the line (see Figure \ref{WKB_lines}(a,b)). Notice that $ij$-line and $ji$-line are not equivalent, and there are two possible generic patterns of WKB lines near a projection of a simple branching point. 

 In joints of primary lines new descendant lines are born, in particular, if $ij$- and $jk$-lines intersect each other, a new $ik$ line is born in the joint, it satisfies the same equation \eqref{WKB} (see Figure \ref{WKB_lines}(c)).
 
Then any new intersection of primary lines with descendant and of descendant ones among themselves gives a birth to new descendant lines.
We call a collection of all WKB lines for given $\zeta$ a WKB web $\CW_{\zeta}$\footnote{\label{foot} 
The rules for constructing WKB webs and calculating their contributions are the same as ones for spectral networks \cite{SN}. However spectral networks appear in a limit of large conformal dimensions in CFT formulation \cite{LargeD}, or in the limit $\epsilon_{1,2}\to 0$ of the $\Omega_{\epsilon_1,\epsilon_2}$-background in the class S theory formulation. From the point of view of connection like \eqref{KZ_2_conn} WKB web limit corresponds to $\zeta\to 0$ limit while spectral network limit corresponds to large representation spin limit \cite{Galakhov:2014aha}.
}.

\begin{figure}[h]
	\begin{center}
		\begin{tikzpicture}
		\draw (-2,-1) -- (2,1) (-2,0) -- (2,0) (-2,1) -- (2,-1);
		\draw [->] (-2,-1) -- (-1,-0.5);
		\draw [->] (-2,0) -- (-1,0);
		\draw [->] (-2,1) -- (-1,0.5);
		\draw [->] (0,0) -- (1,0);
		\draw [->] (0,0) -- (1,0.5);
		\draw [->] (0,0) -- (1,-0.5);
		\filldraw (0,0) circle (0.07);
		\node[left] at (-2,-1) {$(ij)$};
		\node[left] at (-2,0) {$(ik)$};
		\node[left] at (-2,1) {$(jk)$};
		\node[right] at (2,-1) {$(jk)'$};
		\node[right] at (2,0) {$(ik)'$};
		\node[right] at (2,1) {$(ij)'$};
		\node[below] at (0,-0.1) {$p_{\rm jun}$};
		\end{tikzpicture}
	\end{center}
	\caption{Junction of WKB lines}\label{fig:joint}
\end{figure}
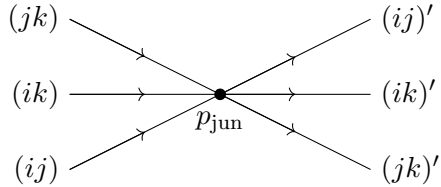

Consider a set $\fL$ formed by formal sums of oriented real paths on $\Sigma$ with coefficients being non-negative integers. We can endow this set with associative unital multiplication by  imposing a multiplication law for two arbitrary paths $\wp_1$ and $\wp_2$:
\be
\wp_1\circ \wp_2=\left\{\begin{array}{l}
	0,\;{\rm if\;the\; beggining\;of}\;\wp_1{\rm\; does\; not\; coincide\; with
		\; the\; end\; of}\; \wp_2 \\
	{\rm concatenation\; of\;} \wp_1 {\rm \; and \;} \wp_2, {\rm \; otherwise} 
\end{array}\right.
\ee
We construct detours $\ID(\wp,\zeta)\in\fL$ depending on oriented path $\wp$ on $\IC$ and phase $\zeta$ following two rules:
\begin{enumerate}
	\item If $\wp$ does not intersect $\CW_{\zeta}$ then
	\be\label{eq:rule1a}
	\ID(\wp,\zeta)=\pi^{-1}(\wp)=\sum\lm_i \pi^{-1}_{(i)}(\wp)
	\ee
	Where $\pi^{-1}$ implies a lift of path $\wp$ to all the sheets of the cover $\Sigma$.
	\item If the path $\wp$ intersects $\CW_{\zeta}$ along a line $\ell$ in a point $p$, the point $p$ splits path $\wp$ into two parts $\wp_-$ and $\wp_+$. Then new detours are
	\be\label{sandwitch}
	\ID(\wp,\zeta)=\ID(\wp_-,\zeta)(1+\eta\;\fd_{p,\ell})\ID(\wp_+,\zeta)
	\ee 
\end{enumerate}
Here $\eta$ is a formal parameter.
The elementary detour $\fd_{p,\ell}$ for a $ij$-WKB line $\ell$ is calculated as follows. One considers a line segment $\bar \ell$ going from the source of the line $\ell$ to the point $p$, then
\be
\fd_{p,\ell}=\pi^{-1}_{(i)}(\bar{\ell}^{-1})\;\fd_{\rm source}\;\pi^{-1}_{(j)}(\bar{\ell})
\ee
Where $\bar{\ell}^{-1}$ implies that we consider the opposite orientation of the segment $\bar \ell$, and $\fd_{\rm source}$ is simple for a primary WKB line, it is just a gluing of lifts to $i^{\rm th}$ and $j^{\rm th}$ sheets across the cut, and for a new-born $ik$ line in a generic junction $p_{\rm jun}$ (see Figure \ref{fig:joint}) we have:
\be
\fd_{\rm source}=\fd_{p_{\rm jun},\ell_{ik}}+\fd_{p_{\rm jun},\ell_{ij}}\fd_{p_{\rm jun},\ell_{jk}}
\ee
The system of detours has a natural grading - ``detour degree" - weighted by a formal parameter $\eta$, generically we imply $\eta=1$ however sometimes we will use this grading denoting it as $\ID^{(k)}$:
\be
\ID(\wp,\zeta)=\sum\lm_k \eta^k \ID^{(k)}(\wp,\zeta)
\ee

Having constructed this system one can define an asymptotic solution of \eqref{connection} being a path-oriented exponential as a matrix in coordinates associated to vacua $\IV$ having the following expansion:
\be\label{conn_pb}
\left({\rm Pexp}\;\zeta^{-1}\int\lm_{\wp}\nabla\right)_{ij}=\sum\lm_{\gamma\in \ID(\wp,\zeta)}e^{\zeta^{-1}\int\lm_{\gamma}\lambda+\pi \I\int\lm_{\gamma} \omega+O(\zeta)}E_{\p\gamma}
\ee
Here $E_{a-b}:=\delta_{i,s(a)}\delta_{j,s(b)}$, where $s$ is a map from points on $\Sigma$ enumerating cover sheets defined after \eqref{spec_curve}. And the form $\omega$ is defined in the following way (see \cite[Appendix A]{GM}). For each root $p^{(i)}$ we have a one-dimensional null space $\psi^{(i)}$:
\be
\left(\mathds{1}\cdot p^{(i)}(x)dx+C_a dz_a\right)\psi^{(i)}=0
\ee
Then on the $i^{\rm th}$ sheet of the cover the form $\omega^{(i)}$ is defined as
\be
\omega^{(i)}=-\frac{1}{\pi \I}\frac{{\rm Det}\left[\psi^{(1)},\ldots,(d+A)\psi^{(i)},\ldots,\psi^{(n)}\right]}{{\rm Det}\left[\psi^{(1)},\ldots,\psi^{(n)}\right]}
\ee

We should stress that the path ordered exponent and similarly detour decomposition satisfy traditional evolution composition property :
\be\label{composability1}
\sum\lm_j\left({\rm Pexp}\;\zeta^{-1}\int\lm_{\wp}\nabla\right)_{ij} \left({\rm Pexp}\;\zeta^{-1}\int\lm_{\wp'}\nabla\right)_{jk}=\left({\rm Pexp}\;\zeta^{-1}\int\lm_{\wp\circ\wp'}\nabla\right)_{ik}
\ee

\subsubsection{Thimbles from WKB lines}\label{th_WKB}
We have considered a complex line $z_a:\; \IC\to \CP$, now we consider a one-dimensional path $\wp\subset\IC$, map $z_a$ lifts it to some path $\hat{\wp}$ in $\CP$. We can calculate parallel transport in the bundle of the integrals $B_i$ along the path $\hat{\wp}$ using jumps \eqref{jumps} and pull it back to $\wp$, or alternatively we can calculate the same parallel transport using the pulled back connection as in \eqref{conn_pb}, the results should coincide. Using this equivalence one can associate detours to intersections of Lefschetz thimbles. Advantage of this approach is two-fold. First of all, an explicit form of the thimbles, a solution to \eqref{thimble} with prescribed boundary conditions is hard to calculate even numerically since this is not a canonical Cauchy problem, in the WKB web formalism this problem is mapped to a one-dimensional one that can be easily solved at least numerically\footnote{Powerful software for constructing spectral networks is available at \cite{loom}.}. Then one has a simple expression as an integral of form $\omega$ for the Maslov index identified with the eta-invariant of the Dirac operator in the soliton background \cite{GM, GMW} that is also hard to calculate from the first principles.

Let us be more specific. Suppose our path $\wp$ intersects the WKB web $\CW_{\zeta}$ in  a collection of points $p_{\alpha}$:
\be
\wp \cap \CW_{\zeta}=\{p_{\alpha} \}_{\alpha=1}^r
\ee
Suppose to a point $p_{\alpha}$ we associate an elementary detour $\fd_{p_{\alpha}}$ in the sandwiching formula \eqref{sandwitch} from intersection with a $ij$-WKB line. Therefore we claim that at the lift $\hat p_{\alpha}$ of $p_{\alpha}$ to $\hat \wp$ the $i^{\rm th}$ Lefschetz thimble intersects the $j^{\rm th}$ critical value. The corresponding difference of values of $W$ in the critical points and the index read
\be\label{WKB_solit}
W_i-W_j=\int\lm_{\fd_{p_{\alpha}}}\lambda\\
\mu_{ij}\sim e^{\pi \I \int\lm_{\fd_{p_{\alpha}}}\omega}
\ee
We have not put an equality for the index since generically the integral gives a complex number. However we can rescale an overall phase of the integrals $B_i$, similarly it is reasonable to compare only  relative indices for the same boundary vacua values $i$ and $j$ from physical reasons \cite{GMW}. In this case the ratio of indices is an integer.

\subsection{Half-supersymmetric interfaces in the Landau-Ginzburg theory}

Those data we have used to define Lefshetz thimbles in Subsection \ref{subsec:SpecAn} can be applied to define 2d $\CN=(2,2)$ Landau-Ginzburg theory. We consider chiral superfields $\Phi^I=\phi^I+\ldots$, then the action reads:
\be
S=\int d^2 x \left[\int d^4\theta\; K(\Phi,\bar{\Phi})+\frac{1}{2}\left(\int d^2\theta\; W(\Phi)+\int d^2\bar\theta\; \bar W(\bar\Phi)\right)\right]
\ee

A detailed description of this theory can be found in \cite{Mirror,GMW}.

Here we just remind some basic properties:
\begin{enumerate}
	\item The vacua of the theory on the real line are in one-to-one correspondence with the critical points of $W$ we denoted $\phi_i$, $i\in\IV$.
	\item Classical finite energy field configurations saturating BPS bound, \emph{solitons}, are solutions to \eqref{thimble} interpolating from vacuum $i$ to vacuum $j$ and with the fixed phase $\zeta_{ij}=\frac{Z_{ij}}{|Z_{ij}|}$. Here the central charge reads $Z_{ij}=W_i-W_j$.
\end{enumerate}

We introduce half-supersymmetric interfaces following \cite{GMW} by considering a Morse height functional $h: \; Y\to \IR$ on the space $Y={\rm Map}(\IR\to X)$ depending on some path $\hat \wp\subset\CP$:
\be
h=-\int\left(\sigma-{\rm Im}\;\left[\zeta^{-1}W\left(\phi^I(x)|z_a(x)\right)\right]dx\right)
\ee
Where $\sigma$ is a 1-form defined locally as $d\sigma=\frac{\I}{2}g_{I\bar J}\; d\phi^I\wedge d\bar\phi^{\bar J}$.
Then the field theory is formulated as supersymmetric quantum mechanics (SQM) on the target space $Y$. We will discuss the SQM formulation in the next subsection. Here we just review the construction of critical field configurations to the functional $h$. The interface preserves supercharges
$$
\CQ_{\zeta}=Q_- -\zeta^{-1}\bar Q_+,\quad \bar \CQ_{\zeta}=\bar Q_- -\zeta Q_+
$$
in standard notations of \cite{Mirror}.

Critical field configurations of the functional $h$ are described as solutions of the following equation:
\be\label{forced_soliton}
\p_x \phi^I(x)=-\zeta g^{I\bar J}\overline{\p_J W\left(\phi^I(x)|z_a(x)\right)}
\ee

In an adiabatic limit where the change  rate for parameters is very slow  $|dz_a/dx|\ll \Lambda_W^{-1}$, where $\Lambda_W$ is the maximal soliton width, solution descends to two classes. 

The first class of solutions - the
\emph{hovering solutions} - are simply reminiscent of the local vacuum solution. They are described by fields which at each $x$ are critical points for $W(\phi;z(x))$ for the same value of $x$.
In equations:  if $\phi^I_i(x)$ is a critical point:
\be
dW(\phi^I|z_a(x))\Big|_{\phi_i^I(x)}=0
\ee
(where the exterior derivative only takes derivatives with respect to $\phi^I$)
then, since the $z_a(x)$ vary continuously along some path $\hat{\wp}\subset\CP$ once we have a chosen critical point $i$ at some $x_0$ the field
$\phi^I_i(x)$ will evolve continuously. (Here we use the assumption that we have a family of Morse superpotentials
so that critical points never collide.)
We denote this solution ${\rm Hov}_{\hat \wp,\zeta}(i,i)$.

In addition to the hovering solutions there are also solutions where solitons are localized near \emph{binding points}. Consider an effective $x$-dependent central charge for a pair of hovering solutions:
\be\label{eff_cc}
{\CZ}_{ij}(x):=W(\phi_i^I(x)|z_a(x))-W(\phi_j^I(x)|z_a(x))
\ee
Binding points $x_c$ are defined by an equation:
\be
\zeta^{-1}{\CZ}_{ij}(x_c)>0
\ee
In the adiabatic limit we may think parameters $z_a(x_c)$ are approximately constant.
In the vicinity of binding point $x_c$ the soliton solution to \eqref{thimble} having small width of order $\Lambda_W$ satisfies approximately \eqref{forced_soliton} up to corrections of order $O\left(\frac{dz_a}{dx}\Lambda_W\right)$. This solution interpolates from vacuum $i$ to vacuum $j$ where it may be approximated by the hovering solution. We denote this solution as ${\rm Slt}_{\hat \wp,\zeta}(i,j)$.

Then we construct the space of adiabatic solutions $\fS(\wp)$ depending on the path $\hat{\wp}$ of parameters $z_a$ in $\CP$ to equation \eqref{forced_soliton} following simple rules:
\begin{enumerate}
	\item If the path $\wp$ does not have  binding points then:
	\be
	\fS(\hat\wp,\zeta)=\bigoplus\lm_i {\rm Hov}_{\hat \wp,\zeta}(i,i)
	\ee
	\item If the path has a binding point $x_c$ it splits the path $\hat\wp$ into two parts  $\hat\wp_-$ and $\hat\wp_+$, then the solution has a sandwich structure:
	\be\label{rule1}
	\fS(\hat\wp,\zeta)=\fS(\hat\wp_-,\zeta)\left(\mathds{1}\oplus {\rm Slt}_{\hat \wp,\zeta}(i,j)\right)\fS(\hat\wp_+,\zeta)
	\ee
	We decompose this expression by a simple distributive law, and in this decomposition one should match vacua at all the non-loose ends, for example,
	\be
	{\rm Solution}_{\hat \wp,\zeta}(i,j)\times {\rm Solution}_{\hat \wp',\zeta}(k,l)\sim \delta_{j,k}\;	{\rm Solution}_{\hat \wp\circ \hat \wp',\zeta}(i,l)
	\ee
	\item If the path intersects multiple binding points one applies the rule \eqref{rule1} iteratively.
\end{enumerate}

Compare these rules to the rules for constructing detours \eqref{eq:rule1a} and \eqref{sandwitch}. Applying identification between detours and contributions of intersecting Lefschetz thimbles discussed in \ref{th_WKB} - solitons in the binding points - we conclude there is an isomorphism $\xi$:
\be\label{equivalence}
\fS(z_a(I),\zeta)\mathop{\rightleftarrows}_{\xi^{-1}}^{\xi} \ID(I,\zeta)
\ee
where $I$ is a real interval $I\subset \IC$, and it is implied that the map $z_a:\;\IR\to \CP$ can be continued analytically to a complex line $z_a:\;\IC\to \CP$.

One of the crucial properties of interfaces is their \emph{composability}.

If $\wp$ denotes a particular path in $\CP$, or equivalently a path of theories, then
reference \cite{GMW} explains that one can define an object $\fI[\wp]$ in an
$A_\infty$-category of interfaces between the theories associated with the initial and final
points of the path. For composable paths $\wp_1, \wp_2$ one can moreover define a product $\boxtimes$ of
interfaces so that
\be\label{eq:ComposeInterfaces}
\fI[\wp_1 \circ \wp_2 ] \sim  \fI[\wp_1] \boxtimes \fI[\wp_2]
\ee
where $\sim$ denotes homotopy equivalence of interfaces.

For adiabatic solutions to the critical field configuration equation we have the same property of composability:
\be\label{composability}
\fS_{ik}[\wp_1\circ \wp_2]=\bigoplus\lm_j \fS_{ij}[\wp_1]\fS_{jk}[\wp_2]
\ee  
following naturally from composability of parallel transports \eqref{composability1}.

\subsection{Spectral analysis of instantons in supersymmetric quantum mechanics}

Let us briefly review Witten's setup \cite{WittenMorse} of the supersymmetric quantum mechanics for the needs of the Morse theory.

Consider a \emph{real} $n$-dimensional Riemannian manifold $Y_{\IR}$ with coordinates $u^I$, $I=1,\ldots,N$, metric $g_{IJ}(u)$ and Morse height function $h: \; Y_{\IR}\to \IR$. 

One promotes coordinates $u^I$ to superfields $U^I=u^I+\theta \psi^I+\bar{\theta}\bar \psi^I+\theta\bar{\theta} D^I$ then the action of the supersymetric quantum mechanics (SQM) reads\cite{GMW}:
\be\label{SQM_action}
S=\int d\tau d^2\theta\left(\frac{1}{2}g_{IJ}(U)DU^I\bar D U^J-h(U)\right)
\ee

The corresponding Morse-Smale-Witten (MSW) complex is defined as a vector space spanned by perturbative ground state wave functions with the differential defined by the action of supercharge $Q$.\footnote{In LG cohomology construction the role of $Q$ is played by $\CQ_\zeta$. In principle, one may be interested in considering the whole family of $\CQ_{\zeta}$ parameterized by $\zeta$. In our setup we are interested in one fixed value of $\zeta$, thus we do not distinguish $\CQ_{\zeta}$ and $Q$ from now on.}

If we integrate over auxiliary fields in the SQM action \eqref{SQM_action}, the resulting field potential is proportional to $(\p h)^2$, therefore perturbative vacua in the theory are labelled by critical points of the height function. Thus we define the MSW complex as a vector space in the following way\footnote{Naively, the complex should inherit a complex structure form the quantum mechanics Hilbert space, therefore from this point of view the complex should be a module over $\IC$. \emph{However} we should stress that both the Morse theory and the Khovanov theory imply naturally an integer structure, so we trade our underlying ring for $\IZ$, and all our rules will make sense as well.}:
\be
\CE:=\bigoplus\lm_{p:\; dh_p=0}\IZ|p\rangle
\ee

This complex has a natural grading by the fermion number ${\bf F}_p=\frac{1}{2}(n_-(p)-n_+(p))$, a half-difference of positive and negative eigen values of the Hessian of $h$ in the critical point $p$.

To define non-perturbative vacua we should take into account non-perturbative corrections due to instantons. Instantons are defined as \emph{real} action \eqref{SQM_action} saddle points, solutions to the steepest ascend flow (function $h$ increases along the flow) equation:
\be\label{instanton}
\frac{du^I}{d\tau}=g^{IJ}(u)\frac{\p h(u)}{\p u^J}
\ee
interpolating between critical point of $h$. Let us denote the space of solutions to \eqref{instanton} interpolating from $a$ to $b$ as $\CI_{a,b}^{\IR}$, then the action of the supercharge $Q$ on the complex $\CE$ reads:
\be\label{Q-me}
\langle p'|Q|p\rangle=\delta_{{\bf F}_{p'}+1,{\bf F}_p}\sum\lm_{\mathfrak{i}\in \CI_{p,p'}}\frac{ {\rm Det}'{\nabla\!\!\!\!/}_{\mathfrak{i}} }{|{\rm Det}'{\nabla\!\!\!\!/}_{\mathfrak{i}} |}
\ee
where ${\nabla\!\!\!\!/}_{\mathfrak{i}}$ is the corresponding Dirac operator on the instanton background \cite{Mirror}.

The Hilbert space of ground states is defined non-perturbatively as cohomology:
\be
\CH:=H^{*}(\CE,Q)
\ee
Obviously, the supercharge shifts the fermion degree by one, so the fermion grading coincides with the cohomological grading of the complex.

To study instantons with the use of proposed spectral analysis techniques we should identify somehow this problem to the construction of Picard-Lefschetz monodromies. We are going to propose a complexified version of the height function, then use it as a new superpotenial. In this setup instanton transitions would appear similar to soliton transitions in the interface background when we perturb our system a little bit like induced boiling of overheated liquid.

Steepest ascend equation \eqref{instanton} is the special case of the thimble equation \eqref{thimble}, however the crucial difference is that the quantum mechanics target space $Y_{\IR}$ in comparison to $X$ does not have a complex structure. A rather generic complexification continuation of the quantum mechanics was exploited in \cite{NewLook} to relate integration cycles with branes in a two-dimensional A-model. We will follow the easiest way. Define an $n$-dimensional complex manifold $Y_{\IC}=Y_{\IR}\times \IR^n$, with coordinates $v^I$ on $\IR^n$. We introduce complex coordinates $Z^I=u^I+\I v^I$ on $Y_{\IC}$. We pull back the metric and extend analytically the Morse height function to $Y_{\IC}$:
\be
G_{I\bar J}dZ^I\overline{dZ^J}=g_{IJ}(du^I+\I dv^I)(du^J-\I dv^J)=g_{IJ}du^Idu^J+g_{IJ}dv^Idv^J\\
H(Z)=h(u+\I v)
\ee
And consider new integrals:
\be
B=\int \prod\lm_I dZ^I \;  e^{\zeta^{-1}H(Z)}
\ee
Behavior of these integrals is identical to discussed in Subsection \ref{subsec:SpecAn}.

We can consider Lefschetz thimbles in this setup, in real coordinates it is given by the following equation, we call ``complex" instantons $\CI^{\IC}$:
\be\label{compl_inst}
\frac{d}{ds}(u^I+\I v^I)=-\zeta\; g^{IJ}(u)\; \p_J h(u-\I v)
\ee
The space of instanton solutions $\CI^{\IR}$  is a subspace of complex instanton solutions $\CI^{\IC}$ specified to $\zeta=-1$, $v^I=0$. To filter out real instantons from generically complex ones we notice that in the case $\zeta=-1$ equation \eqref{compl_inst} admits a complex conjugation mapping a solution to a new solution. Therefore the complex conjugation induces a $\IZ_2$-\emph{action} on $\CI^{\IC}$, and truly real instantons are the only fixed points of this action. Hence we use the proposed spectral analysis technique to construct the space $\CI^{\IC}$ and define $\CI^{\IR}$ as its conjugation-invariant subspace.

\begin{figure}[h!]
	\begin{center}
		\begin{tikzpicture}
		\draw[thick, red] (-2,0) -- (0,0)  (1.5,0) -- (0,0) (2,0.5) -- (1.5,0) -- (2,-0.5) (1.5,0) -- (2.5,0);
		\draw  (-1,1) -- (0,0) (-1,-1) -- (0,0) (1,1.2) to[out=270,in=45] (0,0) (1,-1.2) to[out=90,in=315] (0,0);
		\draw[ultra thick] (0,-0.2) -- (0,0.2);
		\node[below] at (0,-0.2) {$0$};
		\begin{scope}[shift={(-2,0)}]
		\draw[ultra thick, purple] (-0.1,-0.1) -- (0.1,0.1) (0.1,-0.1) -- (-0.1,0.1);
		\end{scope}
		\begin{scope}[shift={(2,0.5)}]
		\draw[ultra thick, purple] (-0.1,-0.1) -- (0.1,0.1) (0.1,-0.1) -- (-0.1,0.1);
		\end{scope}
		\begin{scope}[shift={(2,-0.5)}]
		\draw[ultra thick, purple] (-0.1,-0.1) -- (0.1,0.1) (0.1,-0.1) -- (-0.1,0.1);
		\end{scope}
		\begin{scope}[shift={(-1,1)}]
		\draw[ultra thick, purple] (-0.1,-0.1) -- (0.1,0.1) (0.1,-0.1) -- (-0.1,0.1);
		\end{scope}
		\begin{scope}[shift={(-1,-1)}]
		\draw[ultra thick, purple] (-0.1,-0.1) -- (0.1,0.1) (0.1,-0.1) -- (-0.1,0.1);
		\end{scope}
		\begin{scope}[shift={(1,1.2)}]
		\draw[ultra thick, purple] (-0.1,-0.1) -- (0.1,0.1) (0.1,-0.1) -- (-0.1,0.1);
		\end{scope}
		\begin{scope}[shift={(1,-1.2)}]
		\draw[ultra thick, purple] (-0.1,-0.1) -- (0.1,0.1) (0.1,-0.1) -- (-0.1,0.1);
		\end{scope}
		\begin{scope}[shift={(2.5,0)}]
		\draw[ultra thick, purple] (-0.1,-0.1) -- (0.1,0.1) (0.1,-0.1) -- (-0.1,0.1);
		\end{scope}
		\end{tikzpicture}
	\end{center}
	\caption{Example of a spectral network for instanton calculus}\label{fig:Inst_ex}
\end{figure}
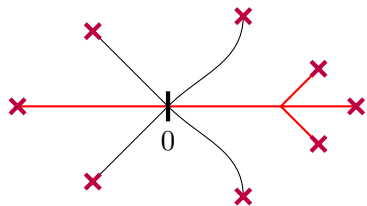

Suppose we consider a family of models with height functions $h(u^I|t_a)$, by analogy we construct a family of complex models with superpotentials $H(Z^I|z_a)$ depending now on generically complex parameters $z_a$ covering some parameter space $\CP$.
We are interested in a  study of instantons in some specific real point $z^{(0)}$ of the parameter space. Let us choose some holomorphic map $\IC\to \CP$ mapping $\{ 0 \}$ to $z^{(0)}$ and $\IR\subset \IC$ to a subspace of $\CP$. Then we can construct a spectral cover for this map using $H(Z^I|z_a)$ as a superpotential. Having done so, we consider a small path $\wp_0$ going from $-0\I$ to $+0\I$. As we have discussed in the previous subsection the space of solutions to \eqref{compl_inst} for $\zeta=-1$ is counted by detours, we conclude:
\be
\CI^{\IC}\cong \ID^{(1)}(\wp_0,-1)
\ee
where we used $\eta$-grading to stress only a subset of single detours.
As we have discussed the space of ``complex" instantons admits an action of complex conjugation, under this action the spectrla cover is conjugated as well $(p,x)\mapsto(p^*,x^*)$. Some of the detours are fixed points of this action we denote them $\ID_{\IR}$, therefore we conclude:
\be
\CI^{\IR}\cong \ID^{(1)}_{\IR}(\wp_0,-1)
\ee
Let us propose an example of such WKB web in Figure \ref{fig:Inst_ex}. We have depicted only those parts of the spectral network giving detours. Those WKB lines
that give rise to detours fixed under conjugation are marked by red. For a simple example of this calculation see Appendix \ref{sec:Airy}.

A tricky problem we may attempt to solve in this approach is the \emph{sign} of the instanton contribution to the supercharge $Q$ matrix element \eqref{Q-me}. Naively, one has to calculate either the determinant (actually determinant bundle) of the Dirac operator in the instanton background or parallel transport of the differential form along the instanton path in the target space (see \cite{Mirror} and discussion in \cite[Appendix F]{GMW}). In our case the corresponding sign can be easily derived from integral $\int\omega$. \emph{However} in our formulation we substitute a free problem by an interface. Due to the  interface all the eigenvalues of the Dirac operator are shifted by some small amount. In particular, a zero eigen mode of the Dirac operator is assigned to each solution modulus by the supersymmetry. However, in the interface background a BPS solution is bound to certain binding point $x_c$ and has no moduli. Consequently, the zero mode of the Dirac operator gets shifted to either positive or negative bulk of eigenvalues. We can estimate this shift using the conventional perturbation theory \cite{Landau}: the first order correction to the eigen value reads:
\be
\Delta \kappa=\frac{1}{||\p_x \phi||^2}{\rm Re}\;\int dx\left[\p^2_{IJ}\p_{z_a}W\right]\p_x\phi^I\p_x\phi^J \frac{dz_a}{dx} (x-x_c)+O(|x-x_c|^2)
\ee
where we have expanded the dependence of the parameters $z_a$ in orders of $|x-x_c|$ since beyond the strip of soliton width $\Lambda_W$ the solution is exponentially close to the vacuum one. From this formula it is obvious that if we change $dz_a/dx$ to the opposite, i.e. if we cross by $\wp_0$ the WKB line in the opposite direction, the correction to the Dirac operator zero eigen value also changes sign.

For example for these two paths $\wp_+$ and $\wp_-$
\begin{center}
	\begin{tikzpicture}
	\draw[thick] (0,0) -- (2,0);
	\draw[ultra thick, purple] (-0.1,-0.1) -- (0.1,0.1) (0.1,-0.1) -- (-0.1,0.1);
	\draw[ultra thick, ->] (1,-0.5) -- (1,0.5);
	\draw[ultra thick, ->] (1.8,0.5) -- (1.8,-0.5);
	\node[below] at (1,-0.5) {$\wp_+$}; \node[below] at (1.8,-0.5) {$\wp_-$};
	\end{tikzpicture}
\end{center}
we have the fermion numbers in the presence of the interface differing by 1:
\be
F_{\wp_-}=F_{\wp_+}\pm 1
\ee
To get the contribution of only ${\rm Det}'$ where the zero mode contribution is stripped off rather than ${\rm Det}$ we subtract contribution of the zero mode, or alternatively, multiply ${\rm Det}$ by the difference $F_{\wp_+}-F_{\wp_-}$:
\be
\frac{{\rm Det}\;{\nabla\!\!\!\!/}_{\mathfrak{i}}}{|{\rm Det}'\;{\nabla\!\!\!\!/}_{\mathfrak{i}}|}\sim e^{\pi \I \int\lm_{\xi({\mathfrak{i}})}\omega}\left[\int\lm_{\xi({\mathfrak{i}})-\xi({\mathfrak{i}})^{-1}}\omega\right]\theta
\ee
where $\xi({\mathfrak{i}})$ is the detour on the spectral cover corresponding the given instanton solution intersecting the WKB line in either transverse direction and $\theta$ is a formal Grassman variable corresponding to the instanton zero mode. For a simple example of this calculation see Appendix \ref{sec:Airy-2}.

Finally, in this  subsection let us discuss another aspect of the spectral analysis of the instantons. When choosing the map $z:\; \IC\to \CP$ one sees there is too much freedom in this choice. Apparently there may be ``good" and ``bad" choices. To discuss them and a possible way out let us consider branching points of would-be spectral cover for some choice of map $z$.

The roots of the spectral curve are in one-to-one correspondence with vacua $\IV$ of the theory. Therefore branching points correspond to singularities where some vacua become degenerate. 

Consider a space of real parameters $\CP_{\IR}$ of a family of height functions $h$. $\CP_{\IR}$ is just a real slice of $\CP$ when we promote height function $h$ to superpotential $H$. Let us choose an open chart $U\subset\CP_{\IR}$, such that one can define an ordering of vacua $\IV$ on the whole $U$. The boundary of closure $\bar U$ can contain singular loci where $i^{\rm th}$ and $j^{\rm th}$ vacua collide. Denote these loci $\sigma_U(i,j)$
 
We call map $z:\;\IC\to\CP$ ``good'' if in addition to this map we can choose such a neighbourhood $U_0$ of $0$, and a real segment $I_{\IR}$ containing $0$ such that
\be
z(0)\in U_0,\quad z(I_{\IR})\cap \sigma_{U_0}(i,j)\neq \emptyset
\ee

For a ``good" map we have following properties:
\begin{enumerate}
	\item There is an $(ij)$-branching point on the spectral curve given by the map $z$. The branching point is projected to $\IR$.
	\item There is an $ij$-instanton at $z_a(0)$.
\end{enumerate}
The first statement just follows from  the fact that branching points appear when vacua collide.  The second statement was proposed in \cite[Section 12.4]{GMW}. Indeed in the neighbourhood of a branching point one can approximate the theory by the Airy cubic model (see Appendix \ref{sec:Airy}), there is an instanton between two real vacua, and we assume it is not spoiled when we go far away from the branching point. In terms of WKB webs this situation looks like a contribution from the branching point in the left in Figure \ref{fig:Inst_ex}, the WKB line from the branching point goes along the real line. 

We call map $z$ ``bad" when for any choice of $U_0$ $z(I_{\IR})\cap \sigma_{U_0}(i,j)= \emptyset$. In this case even if $ij$-instanton exists we can not connect $0$ and the branching point by a simple WKB line, since there is no $(ij)$-branching point projected to the real line. The situation is more cumbersome, and the instanton WKB line is ``glued" from several descendant WKB lines from other vacua degeneration loci like a contribution from 3 branching points on the right in Figure \ref{fig:Inst_ex}. We propose an illustrative example of ``good" and ``bad" choices of $z$ in Appendix \ref{sec:good-bad}.
 
Now let us suppose that the ``good" choice of the map $z$ can always be made. Then instead of searching for a solution to a gradient flow \eqref{instanton} between vacua $i$ and $j$ we can search for a point $s_{ij}$ on the real parameter space $\CP_{\IR}$ where these two vacua collide. We make the following
\begin{proposition}\label{prop:1}
The $Q$-matrix element between $i^{\rm th}$ and $j^{\rm th}$ vacua is counted by possible homotopy classes of real paths from $z(0)$ to $s_{ij}$.
\end{proposition}

\subsection{Landau-Ginzburg cohomology and WKB web degeneration}\label{sec:LG-coh-SN}
Form the point of view of the spectral analysis the Landau-Ginzburg model is two-fold:
\begin{enumerate}
	\item Critical points of the height functional are counted by Lefschetz thimble intersections along a path $\hat\wp$ in the parameter space.
	\item Instantons are counted by intersecting real Lefschetz thimbles, generated by real gradient flows, on the space ${\rm Map}(\IR \to X)$ with respect to the height functional $h$.
\end{enumerate}

To construct the MSW complex we should take the space of solutions to \eqref{forced_soliton} and promote it to a complex. We can use equivalence map $\xi$ \eqref{equivalence} and construct the following functor:
\be
\ID(\wp,\zeta)=\sum\lm_i n_i \cdot \fd_i\quad \mapsto \quad \CE(\hat{\wp},\zeta)=\bigoplus\lm_i \IZ^{n_i}|\fd_i\rangle
\ee
The only purpose of this functor: it promotes the set of $h$-critical points to the Hilbert space of perturbative ground states.

The second part is much more subtle since one should construct complexification of the space ${\rm Map}(\IR \to X)$ and define corresponding chiral ring, etc. Instead we use the idea presented in Proposition \ref{prop:1}. We consider a family of interfaces given by functions $z_a$ and search for a map $z_a^{(0)}$ when two different critical points given by, say, $\fd_i$ and $\fd_j$ coincide. This would give us a special degenerate kind of a WKB web we call {\bf null-web} $\IW_{\zeta}(\fd_j,\fd_i)$, so that for the $Q$-matrix element one has:
\be\label{null-web}
\langle \fd_j|Q|\fd_i\rangle=\delta_{F_j,F_i+1}\#\left[\IW_{\zeta}(\fd_j,\fd_i)\right]
\ee

\subsubsection{Instantons and null-webs}
Generically instantons solve the differential equation:
\be\label{LG_instanton}
\left(\p_x+\I \p_{\tau}\right)\phi^I(x,\tau)=-\zeta g^{I\bar J}\overline{\p_J W\left(\phi^I(x,\tau)|z_a(x)\right)}
\ee
In the limit $\tau\to\pm\infty$ the solution should approach either solution of \eqref{forced_soliton}.

A method to solve the instanton equation is  proposed in \cite{GMW}. One should consider so called boosted solitons. Indeed under the boost - a rotation in the Euclidean space-time to angle $\varphi$ - equation \eqref{LG_instanton} transforms covariantly with a shift $\zeta\mapsto \zeta e^{-\I\varphi}$. One considers the solitons bound to the interface as 2d quasi-particles, then they can migrate with the boost velocity so that supersymmetry is preserved. The world-line of such quasi-particle is given by the following equation:
\be\label{eff_particle}
\frac{dx}{d\tau}=\varphi=-\frac{{\rm Im}\; \left[\zeta^{-1}\CZ_{ij}(x) \right]}{{\rm Re}\; \left[\zeta^{-1}\CZ_{ij}(x) \right]}
\ee
where we used the effective central charge definition \eqref{eff_cc}.

This approach gives a rise to a natural definition of the \emph{future} and \emph{past} stable binding points. In the neighborhood of the binding point real and imaginary parts of the central charge have the following behavior:
\be
\begin{split}
{\rm Im}\; \left[\zeta^{-1}\CZ_{ij}(x) \right]={\rm Im}\; \left[\zeta^{-1}\CZ_{ij}'(x_c) \right](x-x_c)+O((x-x_c)^2)\\
{\rm Re}\; \left[\zeta^{-1}\CZ_{ij}(x) \right]={\rm Re}\; \left[\zeta^{-1}\CZ_{ij}(x_c) \right]+O((x-x_c))\\
\end{split}
\ee
Then asymptotic solution to \eqref{eff_particle} has the following form in the neighbourhood of the binding point:
\be
x(\tau)\sim x_c+C \; \exp\left[-\frac{{\rm Im}\; \left[\zeta^{-1}\CZ_{ij}'(x_c) \right]}{{\rm Re}\; \left[\zeta^{-1}\CZ_{ij}(x_c) \right]}\tau\right]
\ee
Since ${\rm Re}\; \left[\zeta^{-1}\CZ_{ij}(x_c) \right]$ is always positive one has two following possibilities:
\begin{itemize}
	\item ${\rm Im}\; \left[\zeta^{-1}\CZ_{ij}'(x_c) \right]>0$: the solution converges to the critical point $x_c$ in the far future $\tau\to +\infty$, the binding point $x_c$ is called \emph{future} stable binding point
	\item ${\rm Im}\; \left[\zeta^{-1}\CZ_{ij}'(x_c) \right]<0$: the solution converges to the critical point $x_c$ in the far past $\tau\to-\infty$, the binding point $x_c$ is called \emph{past} stable binding point
\end{itemize}

It is simple to define if the binding point is future or past from the form of the WKB web. Notice that 
\be
\CZ_{ij}'(x_c)=p^{(i)}-p^{(j)}
\ee
in the intersection point $p_0$ of path $\wp$ and $ij$-WKB line $\ell$. Moreover equation \eqref{WKB} implies
\be
{\rm Im}\; \left[\zeta^{-1}\CZ_{ij}'(x_c) \right]={\rm Im}\; dz^{-1}|_{p_0}
\ee
So the angle between tangent vectors $\p_t$ to $\ell$ and to $\wp$ define the type of the binding point:
\be
\begin{split}
	\p_t|_{\ell}\wedge \p_t|_{\wp}>0 \quad\to\quad{\it future}\\
	\p_t|_{\ell}\wedge \p_t|_{\wp}<0 \quad\to\quad{\it past}\\
\end{split}
\ee
For example,
\begin{center}
\begin{tikzpicture}
\draw[ultra thick,->] (0,0) -- (3,0);
\node[right] at (3,0) {$\wp$};
\draw[thick, ->](0.5,0.5) -- (0.5,-0.5);
\draw[thick, ->](2.5,-0.5) -- (2.5,0.5);
\begin{scope}[shift={(0.5,0.5)}]
\draw[ultra thick, purple] (-0.1,-0.1) -- (0.1,0.1) (0.1,-0.1) -- (-0.1,0.1);
\end{scope}
\begin{scope}[shift={(2.5,-0.5)}]
\draw[ultra thick, purple] (-0.1,-0.1) -- (0.1,0.1) (0.1,-0.1) -- (-0.1,0.1);
\end{scope}
\node[below] at (0.5,-0.6) {future};
\node[below] at (2.5,-0.6) {past};
\end{tikzpicture}
\end{center}

Now let us construct a simple example of a null-web. Suppose there is an instanton connecting future and past binding points:
\begin{center}
	\begin{tikzpicture}
	\draw[<->] (0,2) -- (0,0) -- (3,0);
	\node[right] at (3,0) {$x$};
	\node[above] at (0,2) {$\tau$};
	\draw[dashed] (0,2) -- (3,2);
	\node[below left] at (0,0) {$-\infty$};
	\node[below left] at (0,2) {$+\infty$};
	\node[below] at (3,0) {$+\infty$};
	\draw[ultra thick] (1,0) to[out=90,in=180] (1.5,1) to[out=0,in=270] (2,2);
	\begin{scope}[shift={(1,0)}]
	\filldraw[red] (-0.1,0.1) -- (0.1,0.1) -- (0.1,-0.1) -- (-0.1,-0.1) -- (-0.1,0.1);
	\end{scope}
	\begin{scope}[shift={(2,2)}]
	\filldraw[red] (-0.1,0.1) -- (0.1,0.1) -- (0.1,-0.1) -- (-0.1,-0.1) -- (-0.1,0.1);
	\end{scope}
	\node[below] at (1,-0.1) {$\fd_1$};
	\node[above] at (2,2.1) {$\fd_2$};
	\node at (0.5,1) {$i$};
	\node at (2.5,1) {$j$};
	\end{tikzpicture}
\end{center}
Notice that finiteness of the instanton action implies that the field configuration tends to some vacua at spatial infinities $x\to\pm\infty$ common for all values of $\tau$. In other words there are no boundary instanton emission vertices at spatial infinities. It is simple to construct corresponding spectral cover and corresponding detours:
\be\label{diagram}
\begin{array}{c}
	\begin{tikzpicture}
	\begin{scope}[shift={(0.5,-0.5)}]
	\begin{scope}[rotate=270]
	\draw [orange, domain=0:1] plot (\x, {0.1*sin(20*\x r)});
	\end{scope};
	\end{scope};
	\begin{scope}[shift={(2.5,0.5)}]
	\begin{scope}[rotate=90]
	\draw [orange, domain=0:1] plot (\x, {0.1*sin(20*\x r)});
	\end{scope};
	\end{scope};
	\draw[ultra thick,->] (0,0) -- (3,0);
	\node[right] at (3,0) {$\wp$};
	\draw[thick, ->](2.5,0.5) -- (2.5,-0.5);
	\draw[thick, ->](0.5,-0.5) -- (0.5,0.5);
	\begin{scope}[shift={(2.5,0.5)}]
	\draw[ultra thick, purple] (-0.1,-0.1) -- (0.1,0.1) (0.1,-0.1) -- (-0.1,0.1);
	\end{scope}
	\begin{scope}[shift={(0.5,-0.5)}]
	\draw[ultra thick, purple] (-0.1,-0.1) -- (0.1,0.1) (0.1,-0.1) -- (-0.1,0.1);
	\end{scope}
	\node[above] at (0.5,0.5) {$ij$};
	\node[below] at (2.5,-0.5) {$ij$};
	\draw (0,-0.1) -- (0.2,-0.1) to[out=0,in=90] (0.3,-0.6) to[out=270,in=270] (0.7,-0.6) to[out=90,in=180] (0.8,-0.1) -- (3,-0.1);
	\begin{scope}[shift={(3,0)}]
	\begin{scope}[scale=-1]
	\draw (0,-0.1) -- (0.2,-0.1) to[out=0,in=90] (0.3,-0.6) to[out=270,in=270] (0.7,-0.6) to[out=90,in=180] (0.8,-0.1) -- (3,-0.1);
	\end{scope}
	\end{scope}
	\node[above left] at (0,-0.1) {$i$};
	\node[below left] at (0,0.1) {$i$};
	\node[above right] at (3,-0.1) {$j$};
	\node[below right] at (3,0.1) {$j$};
	\node[above] at (1.5,0.1) {$\fd_2$};
	\node[below] at (1.5,-0.1) {$\fd_1$};
	\end{tikzpicture}
\end{array}\mapsto
\begin{array}{c}
\begin{tikzpicture}
	\begin{scope}[shift={(0.5,-0.5)}]
	\begin{scope}[rotate=270]
	\draw [orange, domain=0:1] plot (\x, {0.1*sin(20*\x r)});
	\end{scope};
	\end{scope};
	\begin{scope}[shift={(2.5,0.5)}]
	\begin{scope}[rotate=90]
	\draw [orange, domain=0:1] plot (\x, {0.1*sin(20*\x r)});
	\end{scope};
	\end{scope};
	\draw[ultra thick,->] (0,0) -- (3,0);
	\node[right] at (3,0) {$\wp$};
	\draw[thick, ->](2.5,0.5) -- (1.5,0);
	\draw[thick, ->](0.5,-0.5) -- (1.5,0);
	\begin{scope}[shift={(2.5,0.5)}]
	\draw[ultra thick, purple] (-0.1,-0.1) -- (0.1,0.1) (0.1,-0.1) -- (-0.1,0.1);
	\end{scope}
	\begin{scope}[shift={(0.5,-0.5)}]
	\draw[ultra thick, purple] (-0.1,-0.1) -- (0.1,0.1) (0.1,-0.1) -- (-0.1,0.1);
	\end{scope}
	\begin{scope}[shift={(1.5,0)}]
	\begin{scope}[rotate=-64]
	\draw[blue,thick, <-] (-0.2,0) -- (-0.2,1.3) to[out=90,in=90] (0.2,1.3) -- (0.2,-1.3) to[out=270,in=270] (-0.2,-1.3) -- (-0.2,0);
	\end{scope}
	\end{scope}
	\node[above left] at (1.3,0.2) {$\color{blue} \gamma$};
\end{tikzpicture}
\end{array}
\ee

Both detours $\fd_1$ and $\fd_2$ start on the $i^{\rm th}$ sheet and end on the the $j^{\rm th}$ sheet. Following our paradigm we put in Proposition \ref{prop:1} there should be a map $z$ such that solitons corresponding to detours $\fd_1$ and $\fd_2$ coincide, i.e.  intersection points of WKB lines with the path $\wp$ should coincide as in the right part of diagram \eqref{diagram}. The path $\fd_1\circ \fd_2^{-1}$ is a closed cycle on $\Sigma$ we denoted $\gamma$. Since according to \eqref{WKB_solit}:
\be
\int\lm_{\fd_1}\lambda= \int\lm_{\fd_2}\lambda=Z_{ij}=W_i-W_j
\ee 
one sees easily that
\be\label{nw_cond}
\oint\lm_{\gamma} \lambda=0
\ee
For this value of the contour integral we call this asymptotic state of the WKB web {\bf null-web} $\IW_{\zeta}(\fd_j,\fd_i)$.
 
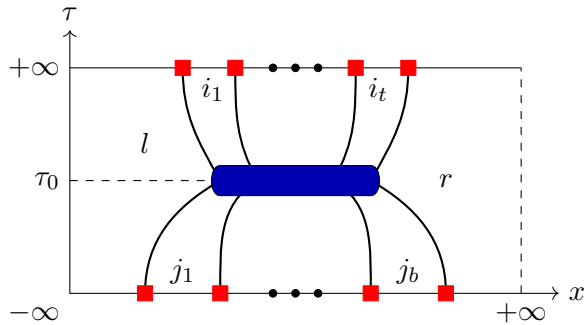
\begin{figure}[h]
	\begin{center}
	\begin{tikzpicture}
		\draw[<->] (0,3.5) -- (0,0) -- (6.5,0);
		\node[above] at (0,3.5) {$\tau$}; \node[right] at (6.5,0) {$x$};
		\draw (0,3) -- (6,3);
		\draw[dashed] (6,0) -- (6,3) (0,1.5) -- (2,1.5);
		\draw[thick] (1,0) to[out=90,in=210] (2,1.5) to[out=120,in=270] (1.5,3) (5,0) to[out=90,in=330] (4,1.5) to[out=60,in=270] (4.5,3) (2,0) to[out=90,in=210] (2.5,1.5) to[out=120,in=270] (2.2,3) (4,0) to[out=90,in=330] (3.5,1.5) to[out=60,in=270] (3.8,3);
		\draw[fill=blue!70!black] (2,1.7) -- (4,1.7) to[out=0,in=0] (4,1.3) -- (2,1.3) to[out=180,in=180] cycle;
		\begin{scope}[shift={(1,0)}]
		\filldraw[red] (-0.1,0.1) -- (0.1,0.1) -- (0.1,-0.1) -- (-0.1,-0.1) -- (-0.1,0.1);
		\end{scope}
		\begin{scope}[shift={(5,0)}]
		\filldraw[red] (-0.1,0.1) -- (0.1,0.1) -- (0.1,-0.1) -- (-0.1,-0.1) -- (-0.1,0.1);
		\end{scope}
		\begin{scope}[shift={(2,0)}]
		\filldraw[red] (-0.1,0.1) -- (0.1,0.1) -- (0.1,-0.1) -- (-0.1,-0.1) -- (-0.1,0.1);
		\end{scope}
		\begin{scope}[shift={(4,0)}]
		\filldraw[red] (-0.1,0.1) -- (0.1,0.1) -- (0.1,-0.1) -- (-0.1,-0.1) -- (-0.1,0.1);
		\end{scope}
		\begin{scope}[shift={(1.5,3)}]
		\filldraw[red] (-0.1,0.1) -- (0.1,0.1) -- (0.1,-0.1) -- (-0.1,-0.1) -- (-0.1,0.1);
		\end{scope}
		\begin{scope}[shift={(4.5,3)}]
		\filldraw[red] (-0.1,0.1) -- (0.1,0.1) -- (0.1,-0.1) -- (-0.1,-0.1) -- (-0.1,0.1);
		\end{scope}
		\begin{scope}[shift={(2.2,3)}]
		\filldraw[red] (-0.1,0.1) -- (0.1,0.1) -- (0.1,-0.1) -- (-0.1,-0.1) -- (-0.1,0.1);
		\end{scope}
		\begin{scope}[shift={(3.8,3)}]
		\filldraw[red] (-0.1,0.1) -- (0.1,0.1) -- (0.1,-0.1) -- (-0.1,-0.1) -- (-0.1,0.1);
		\end{scope}
		\node[left] at (0,1.5) {$\tau_0$};
		\node[below left] at (0,0) {$-\infty$};
		\node[left] at (0,3) {$+\infty$};
		\node[below] at (6,0) {$+\infty$};
		\node at (1,2) {$l$};
		\node at (5,1.5) {$r$};
		\node[above] at (1.5,0) {$j_1$}; \node[above] at (4.5,0) {$j_b$};
		\node[below] at (1.9,3) {$i_1$}; \node[below] at (4.1,3) {$i_t$};
		\filldraw (3,0) circle (0.05) (2.7,0) circle (0.05) (3.3,0) circle (0.05) (3,3) circle (0.05) (2.7,3) circle (0.05) (3.3,3) circle (0.05);
	\end{tikzpicture}
    \end{center}
	\caption{Typical form of an instanton in 2d LG model.}\label{fig:caterpillar}
\end{figure}

Let us argue that condition \eqref{nw_cond} is common for all instanton configurations. A common form of an instanton (a curved web in the language of \cite{GMW}) is a horizontal ``caterpillar" (see Figure \ref{fig:caterpillar}). The ``body" consists of short curved domain walls and bulk joint vertices we will discuss in the next subsection. Legs are domain walls approaching asymptotically binding points. The collection of vacua at the top (infinite future) and at the bottom (infinite past) are denoted correspondingly as $i_1,\ldots, i_t$ and $j_1,\ldots,j_b$. Vacua $l$ and $r$ at the most left and at the most right are common for past and future states. Indeed, the instanton action is proportional \cite{WittenMorse} to the difference $\Delta h$ of the values of the height function for two critical field configurations. The major contribution to this difference from spatial infinities is given by $\int {\rm Im}\;\left[\zeta^{-1} \Delta W\right] dx$. For the instanton contribution to be finite difference $\Delta W$ at spatial infinities should be zero, therefore in a generic case corresponding vacua should be common for top and bottom critical field configurations. Consequently detours $\fd_t$ and $\fd_b$ corresponding to top and bottom critical field configurations start on the sheet corresponding to vacuum $l$ and go to the sheet corresponding to vacuum $r$, moreover
$$
\int\lm_{\fd_t}\lambda=\int\lm_{\fd_b}\lambda=W_l-W_r
$$
We can form a cycle $\gamma=\fd_t\circ\fd_b^{-1}$ closed on cover $\Sigma$ and satisfying \eqref{nw_cond}.

\begin{definition}
	Generically, we define a {\bf null-web} $\IW_{\zeta}(\fd_j,\fd_i)$ for a pair of detours $\fd_i$ and $\fd_j$ as an asymptotic state of the WKB web, such that a part of the web giving contribution to these detours lifts to a finite closed cycle $\gamma$ on $\Sigma$, moreover $\oint\lm_{\gamma}\lambda=0$.
\end{definition}

This definition might seem somewhat \emph{paradoxical}. It is different form a usual triangulation flip used to discuss cluster transformations in the spectral network literature \cite{SN,GLM,ADE,Hollands:2016kgm}. Flips appear when $ij$-WKB line $\ell$ starting from a ramification point $p_1$ hits another ramification point $p_2$, at the same parameter setting there is $ji$-WKB line solution $\ell'$ going from $p_2$ to $p_1$. Unlike flips null-webs appear from junctions of two $ij$-WKB lines. In the junction point they form a cross-like structure (see an example in Figure \ref{fig:nw-example} on page \pageref{fig:nw-example}). Actually, the cross-like junction corresponds to a ``fake" branching point. The fake branching point is a point $p_0$ where the discriminant of the spectral cover $\Sigma$ has a double zero, so the sheets of the cover only touch each other without forming a cut. The contour $\gamma$ is pinched in the fake branching point and can be divided in two closed cycles $\gamma_1$ and $\gamma_2$ projected to paths connecting $p_1$ to $p_0$ and $p_0$ to $p_2$ correspondingly:
\be
\oint\lm_{\gamma}\lambda=\int\lm_{p_1}^{p_0}\lambda^{(j)}+\int\lm_{p_0}^{p_2}\lambda^{(j)}+\int\lm_{p_2}^{p_0}\lambda^{(i)}+\int\lm_{p_0}^{p_1}\lambda^{(i)}\mathop{\scalebox{3}[1]{=}}^{\eqref{WKB}}\zeta\left[\left|\oint\lm_{\gamma_1}\lambda\right|-\left|\oint\lm_{\gamma_2}\lambda\right|\right]
\ee


We will consider more illustrative examples of null-webs in what follows and will comment more on this matter.

Null-web $\IW_{\zeta}(\fd_j,\fd_i)$ corresponds to an instanton connecting two critical values of the height functional corresponding  to detours $\fd_i$ and $\fd_j$, so we propose to count instanton contributions as null-webs, see \eqref{null-web}.

Actually, the condition of a null-web is \emph{a priori} stronger than needed for existence of an instanton. In practice, the null-web guarantees the existence of a hovering solution to a ``forced" instanton equation:
\be\label{forced_instanton}
\left(\p_x+\I \p_{\sigma}\right)\phi^I(x,\sigma)=-\zeta g^{I\bar J}\overline{\p_J W\left(\phi^I(x,\sigma)|z_a(x,\sigma)\right)}
\ee
Here $\sigma$ is a continuous homotopy parameter interpolating between two interfaces corresponding to paths $\hat{\wp}^{(1)}$ and $\hat{\wp}^{(2)}$.

As explained in \cite[Section 10.7]{GMW} (and was heavily exploited in \cite{GM} to argue interface homotopy invariance) counting rigid ``forced" $\sigma\zeta$-instantons
produces a chain map
\be
U :\; \CE(\wp^{(1)},\zeta) \rightarrow \CE(\wp^{(2)},\zeta)
\ee
that preserves bi-gradings up to an overall shift and moreover (anti)-commutes with the differentials
\be
U \CQ_{\zeta}^{(1)} = \pm \CQ_{\zeta}^{(2)} U
\ee
To establish an isomorphism of cohomologies $H^*(\CE^{(1)},\CQ_{\zeta}^{(1)}) \cong H^*(\CE^{(2)},\CQ_{\zeta}^{(2)})$ one must find a chain map $U'$ in the opposite direction:
\be
U' :\; \CE(\wp^{(2)},\zeta) \rightarrow \CE(\wp^{(1)},\zeta)
\ee
that graded commutes with the differentials and moreover satisfies
\be
UU' = {\rm \bf Id} + T_1 Q^{(2)} + Q^{(2)} T_2  \qquad \quad U'U = {\rm \bf Id} + T_3 Q^{(1)} + Q^{(1)} T_4
\ee
for some maps $T_i$. Apparently, inverted interface homotopy gives $U'$.

It is enough to have this map to establish an isomorphism:
\be\label{fake_arg}
H^*\left(\CE(\wp^{(1)},\zeta),Q^{(1)}\right)\cong H^*\left(\CE(\wp^{(2)},\zeta),Q^{(2)}\right)
\ee

And the presence of null-web $\IW_{\zeta}(\fd_1,\fd_2)$ implies that critical field configurations corresponding to $\fd_1$ and $\fd_2$ emerge or merge by pairs like in the famous illustration (see, for example, \cite[Example 10.5.1]{Mirror}) of the MSW complex invariance on a two-sphere under homotopy. This pair represents a ``fake" pair of BPS ground states lifted non-perturbatively. Statement \eqref{fake_arg} implies that we can form a subcomplex containing these two states and cancel it as cohomologous to zero. In what follows we will refer to this argument as a {\bf fake pair argument}.

Let us complete this subsection by the simplest example of a null-web: just a single branching point. Suppose we have a path $\wp$ intersecting two of WKB lines emanated from a single branching point as it is depicted in the following diagram (left):
\be
\begin{array}{c}
\begin{tikzpicture}
\begin{scope}[rotate=220]
\draw [orange, domain=0:0.94] plot (\x, {0.1*sin(20*\x r)});
\begin{scope}[shift={(0.94,0)}]
\draw [orange, domain=0:0.94] plot (\x, {0.1*sin(20*\x r)});
\end{scope}
\begin{scope}[shift={(1.88,0)}]
\draw [orange, domain=0:0.94] plot (\x, {0.1*sin(20*\x r)});
\end{scope}
\end{scope}
\draw[thick] (0,0) -- (0,-2) (0,0) -- (3,1) (0,0) -- (-3,1);
\begin{scope}[shift={(0,0)}]
\draw[purple, ultra thick] (-0.1,-0.1) -- (0.1,0.1) (0.1,-0.1) -- (-0.1,0.1);
\end{scope}
\draw[ultra thick, ->] ([shift=(180:1.5)]0,0) arc (180:0:1.5);
\node[below] at (1.5,0) {$\wp$};
\begin{scope}[rotate=71.56]
\draw[thick] (0,1.5) to[out=240,in=180] (0,-0.5) to[out=0,in=300] (0,1.5);
\end{scope}
\begin{scope}[rotate=-71.56]
\draw[thick] (0,1.5) to[out=240,in=180] (0,-0.5) to[out=0,in=300] (0,1.5);
\end{scope}
\node at (-0.5,0.8) {$\fd_1$};
\node at (0.5,0.8) {$\fd_2$};
\node at (-2.5,1.1) {$ij$};
\node at (2.5,1.1) {$ji$};
\end{tikzpicture}
\end{array}\quad
\begin{array}{c}
\begin{tikzpicture}
\draw[<->] (0,2) -- (0,0) -- (3,0);
\node[right] at (3,0) {$x$};
\node[above] at (0,2) {$\tau$};
\draw[dashed] (0,2) -- (3,2);
\node[below left] at (0,0) {$-\infty$};
\node[below left] at (0,2) {$+\infty$};
\node[below] at (3,0) {$+\infty$};
\draw[ultra thick] (1,0) to[out=90,in=180] (1.5,1) to[out=0,in=90] (2,0);
\begin{scope}[shift={(1,0)}]
\filldraw[red] (-0.1,0.1) -- (0.1,0.1) -- (0.1,-0.1) -- (-0.1,-0.1) -- (-0.1,0.1);
\end{scope}
\begin{scope}[shift={(2,0)}]
\filldraw[red] (-0.1,0.1) -- (0.1,0.1) -- (0.1,-0.1) -- (-0.1,-0.1) -- (-0.1,0.1);
\end{scope}
\node[below] at (1,-0.1) {$\fd_1$};
\node[below] at (2,-0.1) {$\fd_2$};
\node at (1.5,0.5) {$j$};
\node at (1.5,1.5) {$i$};
\end{tikzpicture}
\end{array}
\label{diag2}
\ee

Here we compare a detour constructed from two consequent detours $\fd_1$ and $\fd_2$ and an ``empty'' detour corresponding to lift of $\wp$ to $i^{\rm th}$ sheet. When we shrink the arc of the path $\wp$, so that path $\wp$ crosses the ramification point, cycle $\gamma=\fd_1\circ\fd_2^{-1}$ can be projects to a loop encircling the ramification point twice, obviously, this gives a null-web $\IW_{\zeta}(\fd_1\fd_2,\emptyset)$, the corresponding instanton representing merging of an instanton-anti-instanton pair is depicted at the right part of diagram \eqref{diag2}.

\subsubsection{Bulk vertices and wall-crossing}
The boosted soliton solutions can scatter during migration through bulk vertices \cite{GMW}:
\be\label{bulk vertex}
\beta_{\{i_1,\ldots,i_p\}}[z_a]\rightsquigarrow \quad\left.
\begin{array}{c}
	\begin{tikzpicture}
		\draw[thick, ->] (0,0) -- (0,1.5);
		\draw[thick, ->] (0,0) -- (-1.5,0);
		\draw[thick, ->] (0,0) -- (1.3,1);
		\draw[thick, ->] (0,0) -- (1.3,-1);
		\draw[thick, ->] (0,0) -- (-1.3,-1);
		\begin{scope}[rotate=90]
		\draw[fill=black] (-1,0) circle (0.05);
		\end{scope}
		\begin{scope}[rotate=105]
		\draw[fill=black] (-1,0) circle (0.05);
		\end{scope}
		\begin{scope}[rotate=75]
		\draw[fill=black] (-1,0) circle (0.05);
		\end{scope}
		\draw[fill = blue] (0,0) circle (0.1);
		\node at (-1,1) {$i_1$};
		\node at (0.5,1.2) {$i_2$};
		\node at (1.3,0) {$i_3$};
		\node at (-1.3,-0.4) {$i_p$};
		\begin{scope}[shift={(-1.5,1.5)}]
		\draw (0,0.5) -- (0,0) -- (-0.5,0);
		\node[above left] at (0,0) {$\IR^2$};
		\end{scope}
	\end{tikzpicture}
\end{array}\quad \right|\quad
\begin{array}{c}
\begin{tikzpicture}
\draw[fill=blue, opacity=0.7] (0,0) circle (1.2);
\begin{scope}[rotate=135]
\draw[fill=orange] (1.2,0) circle (0.1);
\end{scope}
\draw[fill=orange] (1.2,0) circle (0.1);
\begin{scope}[rotate=60]
\draw[fill=orange] (1.2,0) circle (0.1);
\end{scope}
\begin{scope}[rotate=200]
\draw[fill=orange] (1.2,0) circle (0.1);
\end{scope}
\begin{scope}[rotate=90]
\draw[fill=black] (-1.2,0) circle (0.05);
\end{scope}
\begin{scope}[rotate=105]
\draw[fill=black] (-1.2,0) circle (0.05);
\end{scope}
\begin{scope}[rotate=75]
\draw[fill=black] (-1.2,0) circle (0.05);
\end{scope}
\node[above left] at (-0.8,1) {$i_1$};
\node[above right] at (0.6,1.1) {$i_2$};
\node[right] at (1.3,0) {$i_3$};
\node[left] at (-1.2,-0.5) {$i_p$};
\begin{scope}[shift={(-1.5,1.5)}]
\draw (0,0.5) -- (0,0) -- (-0.5,0);
\node[above left] at (0,0) {$X$};
\end{scope}
\end{tikzpicture}
\end{array}
\ee
The bulk vertex $\beta_{\{i_1,\ldots,i_p\}}[z_a]$ is given by the instanton solution at certain point $z_a$ of parameter space $\CP$ and with boundary conditions given at infinity by a fan of vacua $\{i_1,\ldots,i_p\}$.

In target space $X$ the situation is somewhat dual. Vacua $i_k$ from the boundary condition fan are points in the target space, boosted solitons are paths in $X$ connecting vacua and forming a closed curve in $X$, and the instanton (a map $\IR^2\to X$) is a disk stretched on this curve as it is depicted in the right part of diagram \eqref{bulk vertex}. The instanton solution is differentiable, hence the disk should be a differential contractible manifold.

In \cite{GMW} rich algebraic $A_{\infty}$- and $L_{\infty}$-structures hidden in the soliton scattering processes were discovered. In particular, a generating function of bulk vertices satisfies the Maurer-Cartan equation. This generates a set of mutual relations between bulk vertices, however these relations are not defining, so we are not able to fix all the bulk vertices from these constraints.

However an explicit value of the bulk vertex is irrelevant in our consideration since the action of the whole instanton is given by the change of the height function. Nevertheless we are interested if a certain bulk vertex is zero or not. In the case the vertex is zero there is no solution to the instanton equation with prescribed boundary conditions, for example if $X$ has a non-trivial homology $H_1(X,\IZ)$ the curve formed by boundary solitons in $X$ may have a non-zero homology class, then a smooth instanton disk is unable to be stretched to this boundary.

We will try to solve the problem of defining necessary bulk vertices from a different perspective: we use consistency conditions with wall-crossing, or, equivalently, homotopy invariance of interfaces. Indeed, the same bulk vertices contribute to the hovering solution of ``forced" $\sigma\zeta$-instanton equation \eqref{forced_instanton} counting chain maps for interface homotopy. A hovering solution including vertex $\beta_{\{i_1,\ldots,i_p\}}[z_a]$ implies that this vertex is non-zero. The interface homotopy is naturally implemented in the WKB web formalism, hence it will be not difficult to implement bulk vertex definition in the null-web formalism.

Let us illustrate this idea in the simplest example when $ij$ and $jk$ solitons scatter into a $ik$ soliton. We construct corresponding diagrams following approaches we mentioned: a contribution of bulk vertices to the ``forced" instanton equation, descendant WKB line creation, disk in the target space:
\be
\begin{array}{c|c|c}
\begin{array}{c}
\begin{tikzpicture}
\draw (-2,-1.5) -- (2,-1.5) (-2,1.5) -- (2,1.5);
\draw[thick] (-1,-1.5) to[out=90,in=225] (0,0) to[out=315,in=90] (1,-1.5) (0,0) -- (0,1.5);
\draw[thick, dashed] (0,0) to[out=45,in=270] (1.5,1.5) (0,0) to[out=135,in=270] (-1.5,1.5);
\draw[fill=blue] (0,0) circle (0.1);
\begin{scope}[shift={(-1,-1.5)}]
\draw[fill=red] (-0.1,0.1) -- (0.1,0.1) -- (0.1,-0.1) -- (-0.1,-0.1) -- cycle;
\end{scope}
\begin{scope}[shift={(1,-1.5)}]
\draw[fill=red] (-0.1,0.1) -- (0.1,0.1) -- (0.1,-0.1) -- (-0.1,-0.1) -- cycle;
\end{scope}
\begin{scope}[shift={(0,1.5)}]
\draw[fill=red] (-0.1,0.1) -- (0.1,0.1) -- (0.1,-0.1) -- (-0.1,-0.1) -- cycle;
\end{scope}
\node at (0,-1) {$j$};
\node at (-1,0) {$i$};
\node at (1,0) {$k$};
\node[left] at (-2,-1.5) {$\CE(\hat\wp_1,\zeta)$};
\node[left] at (-2,1.5) {$\CE(\hat\wp_2,\zeta)$};
\begin{scope}[shift={(-3,2)}]
\draw (-0.5,0) -- (0,0) -- (0,0.5);
\node[above left] {$\IR^2$};
\end{scope}
\end{tikzpicture} 
\end{array}
&
\begin{array}{c}
\begin{tikzpicture}
\begin{scope}[shift={(-1,2)}]
\draw (-0.5,0) -- (0,0) -- (0,0.5);
\node[above left] {$\Sigma$};
\end{scope}
\draw[thick] (-1,-1) -- (1,1) (1,-1) -- (-1,1) (0,0) -- (0,1.5);
\begin{scope}[shift={(-1,-1)}]
\draw[purple, ultra thick] (-0.1,-0.1) -- (0.1,0.1) (0.1,-0.1) -- (-0.1,0.1);
\end{scope}
\begin{scope}[shift={(1,-1)}]
\draw[purple, ultra thick] (-0.1,-0.1) -- (0.1,0.1) (0.1,-0.1) -- (-0.1,0.1);
\end{scope}
\node[below] at (-1,-1.1) {$(ij)$};
\node[below] at (1,-1.1) {$(jk)$};
\node[above left] at (-1,1) {$jk$};
\node[above right] at (1,1) {$ij$};
\node[above] at (0,1.5) {$ik$};
\draw[ultra thick,->] (-1.2,-0.2) to[out=0,in=180] (0,-0.5) to[out=0,in=180] (1.2,-0.2);
\draw[ultra thick,->] (-1.2,0.2) to[out=0,in=180] (0,0.5) to[out=0,in=180] (1.2,0.2);
\node[right] at (1.2,-0.2) {$\wp_1$};
\node[right] at (1.2,0.2) {$\wp_2$};
\end{tikzpicture}
\end{array}
&
\begin{array}{c}
\begin{tikzpicture}
\begin{scope}[shift={(0,1)}]
\draw (-0.5,0) -- (0,0) -- (0,0.5);
\node[above left] {$X$};
\end{scope}
\draw[fill=blue, opacity=0.7] (0,0) to[bend left] (2,0) to[bend left] (1,-1.41) to[bend left] cycle;
\draw[fill=orange] (0,0) circle (0.1) (2,0) circle (0.1) (1,-1.41) circle (0.1);
\node[above left] at (0,0) {$i$};
\node[above right] at (2,0) {$k$};
\node[below] at (1,-1.42) {$j$};
\end{tikzpicture}
\end{array}
\\
{\rm (a)} & {\rm (b)} & {\rm (c)}\\
\end{array}\label{vertex}
\ee

The wall-crossing formulas in this formulation  explicitly correspond to a statement of invariance of the Euler characteristic of interfaces under homotopy, this statement is replicated by equivalence of detours:
\be
\begin{split}
	\chi\left[\CE(\hat\wp_1,\zeta)\right]=\chi\left[\CE(\hat\wp_2,\zeta)\right]\\
	\ID(\wp_1,\zeta)=\ID(\wp_2,\zeta)
\end{split}
\ee
However if we restrict ourselves to $ik$ sector then the complexes are one-dimensional and equivalence of Euler characteristics implies an isomorphism, therefore there is a corresponding invertible chain map $U$ represented by a hovering ``forced" instanton solution containing bulk vertex $\beta_{ijk}$ as it is depicted in diagram (\ref{vertex}(a)). Surely, this reasoning can be generalized to more complicated bulk vertices.

Notice that counting bulk vertices through descendant WKB lines is implemented intrinsically in the formulation of the null-webs. The peculiarity of the null-web in this case is that it includes descendant WKB lines rather than only primary ones.

{\bf However} the reader should be {\bf warned} that the proposed reasoning fails in the case when the target space has a non-trivial fundamental group. In this case the boundary of implied instanton disk solution depicted in diagram (\ref{vertex}(c)) may encircle some defects in the target space, and hence there is no smooth disk instanton solution. And this is exactly a discrepancy between so called Yang-Yang and monopole LG models proposed in \cite{GM}, we will also mention this situation in what follows. So while applying methods from this section we always imply $\pi_1(X)=\{1\}$. An accompanying interpretation in terms of wall-crossing formulas is the following: punctures of $X$ give rise to purely ``flavor" solitons (see \cite[Section 7.1]{GM}) that spoils the wall-crossing formulas.

\section{Landau-Ginzburg $\fs\fu_2$ Link Cohomology}\label{sec:su_2}

\subsection{Tangle interfaces in the Landau-Ginzburg model}

\begin{figure}[h]
	\begin{center}
		\begin{tikzpicture}
		\begin{scope}[scale=1.5]
		\begin{scope}[shift={(0,-1)}]
		\draw [ultra thick] (-0.25,-0.5) to [out=90,in=210] (0,0) to [out=30, in=270] (0.25,0.5) (-0.25,0.5) to [out=270, in=150] (-0.1,0.1) (0.1,-0.1) to [out=330, in=90] (0.25,-0.5);
		\end{scope}
		\begin{scope}[shift={(0.25,0)}]
		\draw [ultra thick] (-1,-0.5)-- (-1,-1.5) to [out=270, in=270] (-0.5,-1.5);
		\end{scope}
		\begin{scope}[shift={(-0.25,0)}]
		\begin{scope}[xscale=-1]
		\draw [ultra thick] (-1,-0.5)-- (-1,-1.5) to [out=270, in=270] (-0.5,-1.5);
		\end{scope}
		\end{scope}
		\begin{scope}[shift={(-1.75,-0.5)}]
		\draw[ultra thick, fill=blue, opacity=0.4] (0,0) to[out=90,in=180] (1.5,0.5) to[out=0,in=180] (2.75,1) to[out=0,in=90] (3.5,0) to[out=270,in=0] (2,-0.5) to [out=180,in=0] (0.75,-1) to[out=180,in=270] (0,0);
		\begin{scope}[shift={(0.5,-0.5)}]
		\draw[ultra thick, fill=white] (0,0) to[out=45,in=135] (0.5,0) to[out=225,in=315] (0,0);
		\draw[ultra thick] (0,0) -- (-0.15,0.15) (0.5,0) -- (0.65,0.15);
		\end{scope}
		\begin{scope}[shift={(2.5,0.5)}]
		\draw[ultra thick, fill=white] (0,0) to[out=45,in=135] (0.5,0) to[out=225,in=315] (0,0);
		\draw[ultra thick] (0,0) -- (-0.15,0.15) (0.5,0) -- (0.65,0.15);
		\end{scope}
		\begin{scope}[shift={(1,0)}]\draw[ultra thick, purple] (-0.1,-0.1)--(0.1,0.1) (0.1,-0.1)--(-0.1,0.1); \end{scope}
		\begin{scope}[shift={(1.5,0)}]\draw[ultra thick, purple] (-0.1,-0.1)--(0.1,0.1) (0.1,-0.1)--(-0.1,0.1); \end{scope}
		\begin{scope}[shift={(2,0)}]\draw[ultra thick, purple] (-0.1,-0.1)--(0.1,0.1) (0.1,-0.1)--(-0.1,0.1); \end{scope}
		\begin{scope}[shift={(2.5,0)}]\draw[ultra thick, purple] (-0.1,-0.1)--(0.1,0.1) (0.1,-0.1)--(-0.1,0.1); \end{scope}
		\node[below left] at (1,0) {$z_1$};
		\node[below left] at (1.5,0) {$z_2$};
		\node[below left] at (2,0) {$z_3$};
		\node[below left] at (2.5,0) {$z_4$};
		\end{scope}
		\draw [ultra thick] (-0.25,-0.5) to [out=90,in=210] (0,0) to [out=30, in=270] (0.25,0.5) (-0.25,0.5) to [out=270, in=150] (-0.1,0.1) (0.1,-0.1) to [out=330, in=90] (0.25,-0.5);
		\begin{scope}[shift={(0.25,0)}]
		\draw [ultra thick] (-0.5,0.5) to [out=90, in=90] (-1,0.5) --(-1,-0.5);
		\end{scope}
		\begin{scope}[shift={(-0.25,0)}]
		\begin{scope}[xscale=-1]
		\draw [ultra thick] (-0.5,0.5) to [out=90, in=90] (-1,0.5) --(-1,-0.5);
		\end{scope}
		\end{scope}
		\draw[ultra thick](-1.75,-0.5) to[out=270,in=180] (0,-2.25) to[out=0,in=270] (1.75,-0.5) to[out=90,in=0] (0,1.25) to[out=180,in=90] (-1.75,-0.5);
		\node [right] at (-1.7,-0.5) {$\cal C$};
		\node at (-1.75,1.25){${\cal M}_3$};
		\draw[->] (2,-2.25) -- (2,1.25);
		\node[above right] at (2,1.25) {$x$};
		\end{scope}
		\end{tikzpicture}
	\end{center}
	\caption{A tangle embedded in a three-manifold ${\cal M}_3$. Transverse slices to the $x$-plane are a Riemann surface $\CC$ so the
		tangle is described by an evolving set of points $z_a(x)$ in $\CC$. In this paper we will take $\CC$ to be the complex plane so that
		$\CM_3$ is just $\IC \times \IR = \IR^3$.  } \label{fig:tangle}
\end{figure}
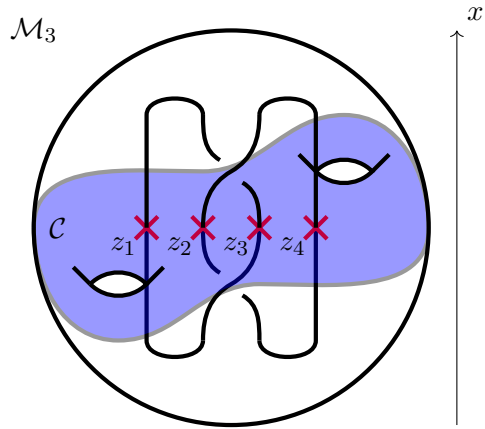

Let us briefly recall a few points from Section 18.4 of  \cite{GMW}.
The link $L$ whose link-homology we wish to formulate will be presented as a tangle, that is, as
set of distinct points   $z_a$ in the complex plane continuously evolving with a parameter $x$. The points are allowed to be
created or annihilated in pairs at critical values of $x$ but otherwise cannot collide\footnote{In principle one could consider punctures on a general Riemann surface $\CC$, and hence discuss links in
	$\CC\times \IR$. However, in this paper we will strictly limit ourselves to the case where $\CC$ is the complex plane.}. See Figure \ref{fig:tangle}.

In link-homology the strands of the tangle, and hence the points $z_a$ are labeled by
irreducible representations $R_a$ of some compact simple Lie group $G$. And indeed the Chern-Simons
invariant associated to the Euler character of the link homology will be that for gauge
group $G$ with Wilson lines in irreducible representations $R_a$. Actually this formulation allows one to calculate invariants of an oriented link. So if the orientation of the strand in the tangle coincides with the evolution direction $x$ orientation we mark this strand by the original representation $R_a$, however if it is opposite the strand should be marked by the conjugate representation $\bar R_a$.

In the Landau-Ginzburg approach to link homology collection  $\{ z_a, R_a\} $ parameterizes a family of
massive $1+1$ dimensional quantum field  theories with $(2,2)$ supersymmetry, $\CT(z_a, R_a)$. The spatial
axis of the ``worldsheet'' on which the quantum field theory is defined is to be identified
with the $x$ axis in Figure\ref{fig:tangle}, and the time axis is not shown in that figure.
Parameters $z_a\in \IC$ span parameter space $\CP$ of the quantum field theory.
In a tangle the $z_a$'s vary as a function of $x$, and therefore the parameters of the quantum field
theory vary as functions of $x$. This defines what is known as a ``Janus'' or an ``interface.''

As shown in \cite{GMW}, interfaces define a notion of parallel transport of the category
of branes over the parameter space of the quantum field theory. Moreover, suitably interpreted, the
parallel transport is given by a ``flat connection.'' Therefore we can hope that homotopic deformations
of the path of parameters $\{ z_a(x) \}$ will lead to chain homotopic complexes. Therefore, this is a
natural setting for the formulation of link homology.

\subsection{Review of LG $\fs\fu_2$ link cohomology construction}

We follow \cite{GMW} and consider Landau-Ginzburg model of \cite{Gaiotto:2011nm} on the monopole moduli space. The target space of this
model is the universal cover of the moduli space of $m$ magnetic monopoles in $SU(2)$
Yang-Mills-Higgs theory on $\IR^3$. The moduli space has the form
$\IR^4 \times \CM_0$ where $\CM_0$ is the simply-connected strongly-centered moduli space.
The most natural metric would be the hyperk\"ahler metric, although fortunately we will
not need to know any details of this metric.

In order to describe the superpotential we use the identification of the moduli space
with a space of rational maps $\IC \IP^1 \to \IC \IP^1$ of degree $m$ and preserving the point at infinity
\cite{Donaldson,Atiyah:1988jp,Hurtubise}.
Thus the rational map can be taken to be $ u \mapsto P(u)/Q(u)$ where $Q$ is a monic polynomial of degree $m$
and $P$ has degree $m-1$ with nonvanishing leading coefficient. The polynomials $P,Q$ are relatively prime
so $P$ never vanishes at the roots of $Q$.  We will
denote those roots (which need not be distinct)  by $w_i$:
\be
Q(u) = \prod_{i=1}^m(u-w_i).
\ee
The superpotential is
\be\label{eq:MLG-SUPERPOT}
W_{\rm monopole}(P,Q|z_a)=\sum\lm_{i=1}^m \Biggl\{
\mathop{\rm Res}\lm_{u=w_i} \left[K(u) \frac{ P(u)}{Q(u)} du \right]
-     \log  P(w_i) +c  w_i \Biggr\}
\ee
where $K(u):=\prod\lm_{a=1}^n (u-z_a)^{k_a}$. Here $k_a$ ia a charge of a singular monopole located at $z_a$.
This superpotential is multivalued on the monopole moduli space but single valued on a cyclic cover.
This cyclic cover coincides with the simply connected cover of monopole moduli space.

If monopole centers are widely separated  we can define   $e^{Y_i}:= P(w_i)$
then evaluation of the residue in \eqref{eq:MLG-SUPERPOT} gives a new superpotential:
\be\label{q-c_mono}
W_{\rm q-c\; mono}(w_i,Y_i|z_a)=c\sum\lm_i w_i+\sum\lm_i Y_i+\sum\lm_i \frac{\prod\lm_a (w_i-z_a)^{k_a}}{\prod\lm_{j\neq i}(w_i-w_j)}e^{-Y_i}
\ee
Now coordinates $(w_i,Y_i)$ on the target space have a nice quasi-classical interpretation: for a single monopole we could write: $w = x^2 + \I x^3$ and $Y= \mu y + \I \vartheta$ where $(y,x^2,x^3)$ denote
the center of mass of a monopole in $\IR_+\times \IR^2$ and $\vartheta$ denotes the electric phase. All the monopoles are placed in a half-space $y>0$ and singular monopoles are sitting at the plane $y=0$ (a tip of a ``cigar" in the M5-brane formulation of \cite{Witten:2011zz}).

We can go further and integrate over fields $Y_i$, the resulting superpotential - called Yang-Yang superpotential - reads:
\be\label{eq:YY-superpotential}
W_{\rm YY}(w_i|z_a)=\sum\lm_{a=1}^n\sum\lm_{i=1}^m k_a\log(w_i-z_a)-2\sum\lm_{1\leq i<j\leq m}\log(w_i-w_j)+c\sum\lm_i w_i
\ee

From this point of view $W_{\rm YY}$ can be considered as a low energy effective superpotential for the monopole Landau-Ginzburg model, from the point of view of manipulations with the chiral ring we have assigned to some of the operators their classical expectation values. This procedure does not change asymptotic behavior of the connection \eqref{connection}, and therefore it  does not affect the spectral curve. \emph{However} the resulting model is different in a crucial way as it was demonstrated in \cite{GM}. As the result of integration the target space acquires singularities and a non-trivial fundamental group, this breaks flatness of the parallel transport and invariance of the Hilbert spaces of ground states under link regular homotopic transforms. In particular, the wall-crossing arguments for constructing instanton bulk amplitudes fail due to extra flavour wall-crossing that is not captured by the usual wall-crossing formula (see \cite{GM}). 

Notice that the Yang-Yang superpotential \eqref{eq:YY-superpotential} substituted in the integral \eqref{intgrl} gives a description of conformal blocks in the free field fomalism \cite{GMOMS}. And it is well-known that monodromies of the conformal blocks are braid invariants and after braid closure become link invariants. The asymptotic monodromy of the integral is given by the Euler characteristic
 of the corresponding MSW complex, hence the Euler characteristic of the MSW complex is a link invariant by construction, it is not complicated to check that it coincides with the link Jones polynomial.

In the remaining part of this section we are going to apply methods of spectral analysis from the previous section, for this purpose the superpotential $W_{YY}$ is suitable, however as we mentioned we imply that the target space has a trivial fundamental group, there are no wall-crossings involving purely flavour solitons. Thus secretly we work  with $W_{\rm monopole}$.

\subsubsection{Spectral curve and classification of vacua}

Consider the set of vacua $\IV(\{z_a\};\{k_a\};m)$ of the superpotential  \eqref{eq:YY-superpotential} for $m$ monopoles. If $z_a$ are widely separated ($|c\Delta z|\gg 1$) then the space of vacua can be decomposed as a space of individual vacua for each puncture $z_a$:
\be\label{vac_dec}
\IV(\{z_a\};\{k_a\};m)=\bigoplus\lm_{\sum\lm_{a=1}^{n}\mu_a=m}\IV(z_1;k_1;\mu_1)\times\ldots\times \IV(z_n;k_n;\mu_n)
\ee
Therefore it is enough to study the vacua for a single puncture. For a single puncture we have (see Appendix \ref{sec:vac_su_2})
\be
\IV(z;k;m)\cong v_{k-2m},\quad m=0,\ldots,k
\ee
where $v_{\alpha}$ is a one-dimensional invariant subspace of weight $\alpha$. And we have generically
\be
\bigoplus\lm_{m=0}^{k}\IV(z;k;m)\cong \left.\begin{array}{c}
 \begin{tikzpicture}
 \begin{scope}[scale=0.4]
 \draw (0,0) --(0,4) -- (1,4) -- (1,0) -- cycle (0,1)--(1,1) (0,2)--(1,2) (0,3)--(1,3);
 \end{scope}
 \end{tikzpicture}
 \end{array}\right\}k\;{\rm boxes}
\ee

We will consider the simplest case of the fundamental representations, in the $\fs\fu_2$ case representations are isomorphic to complex conjugate ones, then it is enough to consider just the case of $k_a=1$. There are just two possible solutions for a single puncture - $\IV(z;1;0)$ and $\IV(z;1;1)$ - we denote as $(-)$ and $(+)$ spin projections correspondingly.

The next basic example is two punctures, correspondingly we have:
\be
\begin{array}{c}\label{3subspaces}
\IV(\{z_a,z_b\};\{1,1\},0)=(-,-)\\
\IV(\{z_a,z_b\};\{1,1\},1)=(+,-)\oplus (-,+)\\
\IV(\{z_a,z_b\};\{1,1\},2)=(+,+)
\end{array}
\ee

The spectral curve in this case for two punctures $z_a$ and $z_b$ can be easily constructed (see \cite[Appendix A]{GM}). However in a more generic situation this construction is technically involved. The major difficulty on this route is to find a suitably ``nice" basis in the chiral ring. Such a basis was found in \cite{Maulik-Okunkov} (see also \cite{Aganagic:2017smx,EFK,Aganagic:2016jmx,Matsuo,Okuonkov_Lect,Mironov:2016yue,Zenkevich:2015rua}) in a more generic case -- it is called a \emph{stable envelope} basis -- of ``quantized'' or Jackson integrals and it does not seem to have a single line formulation in our framework. However we get used to the quantum $q$-Langlands correspondence
\begin{theorem} (\cite[Equation (1.7)]{Aganagic:2017smx}) 
	Spaces of conformal blocks of $U_{\hbar}\left(\widehat{{}^L\fg}\right)$ and $\CW_{q,t}\left(\fg\right)$ are isomorphic for certain representations. 
\end{theorem}
This theorem is important for us since conformal blocks of $\CW_{q,t}\left(\fg\right)$ have a nice representation in terms of Jackson type integrals generalizing \eqref{intgrl}, on the other hand conformal blocks of $U_{\hbar}\left(\widehat{{}^L\fg}\right)$ satisfy $q$-Knizhnik-Zamolodchikov difference equations.
The only thing we should do now is to adopt this result to our non-quantized case, in this case we get a 
\begin{theorem}
	There exists a basis in the chiral ring such that connection \eqref{connection} has the following form:
	\be\label{KZ_2_conn}
	\nabla=d-\zeta^{-1}\sum\lm_a dz_a\left[\sum\lm_{b\neq a} \frac{h^{(a)}\otimes h^{(b)} +e^{(a)}\otimes f^{(b)}+f^{(a)}\otimes e^{(b)}}{z_a-z_b}+ c\; m^{(a)}\right]
	\ee
\end{theorem}
where $h$, $e$, $f$ are standard generators of $\fs\fu_2$:
\be
\left[h,e\right]=2e,\quad\left[h,f\right]=-2f,\quad\left[e,f\right]=h
\ee
Apparently, connection \eqref{KZ_2_conn} acts diagonally in all three spaces \eqref{3subspaces}. For spaces with $m=0$ and $m=2$ the spectral curve becomes trivially one-fold covering, this implies there are no branching points, therefore any interface gives only hovering solutions.

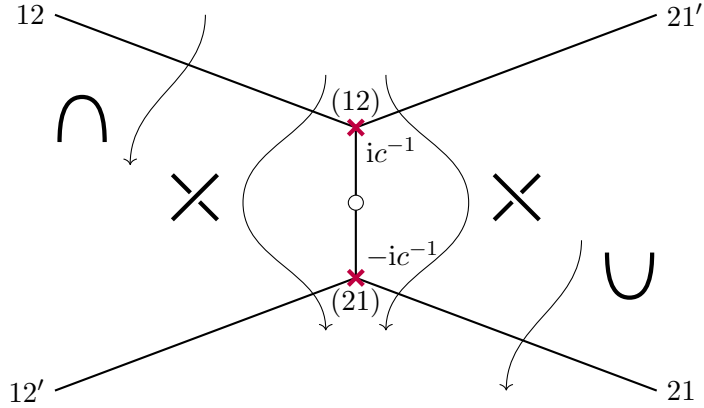
\begin{figure}
	\begin{center}
		\begin{tikzpicture}
		\draw[thick] (0,-1) -- (0,1) (-4,-2.5) to (0,-1) to (4,-2.5) (-4,2.5) to (0,1) to (4,2.5);
		\node[left] at (-4,2.5) {$12$};
		\node[left] at (-4,-2.5) {$12'$};
		\node[right] at (4,2.5) {$21'$};
		\node[right] at (4,-2.5) {$21$};
		\node[above] at (0,1) {$(12)$};
		\node[below] at (0,-1) {$(21)$};
		\draw[fill=white] (0,0) circle (0.1);
		\begin{scope}[shift={(0,1)}]
		 \draw[purple, ultra thick] (-0.1,-0.1) -- (0.1,0.1) (0.1,-0.1) -- (-0.1,0.1);
		\end{scope}
		\begin{scope}[shift={(0,-1)}]
		\draw[purple, ultra thick] (-0.1,-0.1) -- (0.1,0.1) (0.1,-0.1) -- (-0.1,0.1);
		\end{scope}
		\node[above right] at (0,-1) {$-\I c^{-1}$};
		\node[below right] at (0,1) {$\I c^{-1}$};
		\draw[->] (-0.4,1.7) to[out=270,in=90] (-1.5,0) to[out=270,in=90] (-0.4,-1.7);
		\draw[->] (0.4,1.7) to[out=270,in=90] (1.5,0) to[out=270,in=90] (0.4,-1.7);
		\node[left] at (-1.5,0){$
			\begin{array}{c}
			\begin{tikzpicture}
			\begin{scope}[scale=0.6]
			\draw[ultra thick] (-0.5,-0.5) -- (0.5,0.5) (0.5,-0.5) -- (0.1,-0.1) (-0.1,0.1) -- (-0.5,0.5);
			\end{scope}
			\end{tikzpicture}
			\end{array}$
			};
		\node[right] at (1.5,0){$
			\begin{array}{c}
			\begin{tikzpicture}
			\begin{scope}[scale=0.6,xscale=-1]
			\draw[ultra thick] (-0.5,-0.5) -- (0.5,0.5) (0.5,-0.5) -- (0.1,-0.1) (-0.1,0.1) -- (-0.5,0.5);
			\end{scope}
			\end{tikzpicture}
			\end{array}$
		};
		\draw[->] (-2,2.5) to[out=270,in=90] (-3,0.5);
		\node[above left] at (-3,0.5) {$
			\begin{array}{c}
			\begin{tikzpicture}
			\begin{scope}[scale=0.6]
			\draw[ultra thick] (-0.5,0) to[out=90,in=180] (0,1) to[out=0,in=90] (0.5,0);
			\end{scope}
			\end{tikzpicture}
			\end{array}
			$};
		\draw[->] (3,-0.5) to[out=270,in=90] (2,-2.5);
		\node[below right] at (3,-0.5) {$
				\begin{array}{c}
				\begin{tikzpicture}
				\begin{scope}[scale=0.6,yscale=-1]
				\draw[ultra thick] (-0.5,0) to[out=90,in=180] (0,1) to[out=0,in=90] (0.5,0);
				\end{scope}
				\end{tikzpicture}
				\end{array}
				$};
		\end{tikzpicture}
	\end{center}
\caption{Spectral curve for two punctures in $\fs\fu_2$ and interface paths.}\label{fig:sc_su_2}
\end{figure}

Only a mutual positioning of the punctures matters, so we take $z_a=z$, $z_b=-z$. Then the space of parameters $\CP\cong\IC$ and we choose a trivial map $z:\;\IC\to \IC$, then the spectral curve reads:
\be
p^2+ \frac{2p}{x}-c^2=0,\quad \lambda=p\; dx
\ee
The two roots coincide with two vacua $(+,-)$ and $(-,+)$ and we construct easily the WKB web (see Figure \ref{fig:sc_su_2}).

\subsubsection{Braiding, fusion/defusion interfaces}

Composability property of interfaces \eqref{composability} allows one not to construct a new interface for a new link each time, instead we can decompose any interface into a product of elementary interfaces similarly to Kirillov-Reshetikhin-Turaev (see e.g. \cite{KRT,RT}) $U_q({\fs\fu}_2)$ formulation of link invariants. Such basic interfaces are braiding interfaces analogous to generators of the braid group - we call them $\CR$ and $\CR^{-1}$-interfaces by analogy to $R$-matrices - and caps/cups, or fusion/defusion interfaces joining/creating a pair of strands into/from a closure. The latter are analogs of $q$-Clebsh-Gordan coefficients.

Without loss of generality we can choose parameters to be real $c>0$, $\zeta<0$. Then the dominant term in \eqref{eq:YY-superpotential} is $c\sum\lm_i w_i$. The superpotential along the Lefschetz thimble grows as $W(t)=-\zeta t$, $t\in\IR_+$, therefore asymptotically the Lefschetz thimble can be approximated by a line $w(t)=-\zeta c^{-1} t$, if the choice of parameters is made as we proposed, these are lines parallel to the real axis flowing to the positive infinity. In the case of a single monopole the Lefschetz thimble looks like this:
\begin{center}
\begin{tikzpicture}
\draw (0,0) circle (0.1) (0,1) circle (0.1);
\begin{scope}[shift={(0,1)}]
\draw[thick] (2,-0.25) -- (0,-0.25) to[out=180,in=270] (-0.5,0) to[out=90,in=180] (0,0.25) -- (2,0.25);
\end{scope}
\begin{scope}[shift={(-0.5,0)}]
\draw[purple,ultra thick] (-0.1,-0.1) -- (0.1,0.1) (0.1,-0.1) -- (-0.1,0.1);
\end{scope}
\begin{scope}[shift={(-0.5,1)}]
\draw[purple,ultra thick] (-0.1,-0.1) -- (0.1,0.1) (0.1,-0.1) -- (-0.1,0.1);
\end{scope}
\end{tikzpicture}
\end{center}
It is natural to associate a collection of constant braids in the diagram to a collection of constant trajectories $z_a$ lying on the imaginary axis:
\be
\begin{array}{c}
\begin{tikzpicture}
\draw[ultra thick] (0,0) -- (0,2) (0.5,0) -- (0.5,2) (1,0) -- (1,2) (1.5,0) -- (1.5,2);
\node[below] at (0,0) {$z_1$};
\node[below] at (0.5,0) {$z_2$};
\node[below] at (1,0) {$z_3$};
\node[below] at (1.5,0) {$z_4$};
\end{tikzpicture}
\end{array}\mapsto
\begin{array}{c}
\begin{tikzpicture}
\draw (0,1.5) circle (0.1) (0,1) circle (0.1) (0,0.5) circle (0.1) (0,0) circle (0.1);
\node[right] at (0.1,1.5) {$z_1$};
\node[right] at (0.1,1) {$z_2$};
\node[right] at (0.1,0.5) {$z_3$};
\node[right] at (0.1,0) {$z_4$};
\end{tikzpicture}
\end{array}
\ee
Then two braiding elements are given by the following plate diagrams:
\be
\CR:=\begin{array}{c}
	\begin{tikzpicture}
	\begin{scope}[scale=0.6]
	\draw[ultra thick] (-0.5,-0.5) -- (0.5,0.5) (0.5,-0.5) -- (0.1,-0.1) (-0.1,0.1) -- (-0.5,0.5);
	\end{scope}
	\end{tikzpicture}
\end{array},\quad 
\CR^{-1}:=\begin{array}{c}
	\begin{tikzpicture}
	\begin{scope}[scale=0.6,xscale=-1]
	\draw[ultra thick] (-0.5,-0.5) -- (0.5,0.5) (0.5,-0.5) -- (0.1,-0.1) (-0.1,0.1) -- (-0.5,0.5);
	\end{scope}
	\end{tikzpicture}
\end{array}
\ee
and they correspond to paths on $\CP=\IC$ depicted in Figure \ref{fig:sc_su_2} correspondingly.

Calculating detours and taking into account the hovering solutions in vacua $(-,-)$ and $(+,+)$ we derive the following expressions for the interfaces (to compare to \cite[eq. (4.5), (4.7)]{GM} one should switch $(q,t)$-gradings to opposite ones):
\begin{subequations}
\be\label{rules_a}
\begin{split}
	\CE\left(\begin{array}{c}
		\begin{tikzpicture}
		\begin{scope}[scale=0.8]
		\draw[ultra thick] (-0.5,-0.5) -- (0.5,0.5) (0.5,-0.5)-- (0.1,-0.1) (-0.1,0.1) -- (-0.5,0.5); 
		\end{scope}
		\end{tikzpicture}
	\end{array}\right)= q^{-1} \begin{array}{c}
	\begin{tikzpicture}
	\begin{scope}[scale=0.8]
	\node[above] at (0,2) {$+$};
	\node[above] at (1,2) {$+$};
	\node[below] at (0,0) {$+$};
	\node[below] at (1,0) {$+$};
	\draw[ultra thick, purple] (0,0) -- (0,0.5) to[out=90,in=210] (0.5,1) to[out=30,in=270] (1,1.5) -- (1,2) (1,0) -- (1,0.5) to[out=90,in=330] (0.6,0.9) (0.4,1.1) to[out=150,in=270] (0,1.5) -- (0,2);
	\end{scope}
	\end{tikzpicture}
\end{array}\oplus q^{-1}\begin{array}{c}
\begin{tikzpicture}
\begin{scope}[scale=0.8]
\node[above] at (0,2) {$-$};
\node[above] at (1,2) {$-$};
\node[below] at (0,0) {$-$};
\node[below] at (1,0) {$-$};
\draw[ultra thick, purple] (0,0) -- (0,0.5) to[out=90,in=210] (0.5,1) to[out=30,in=270] (1,1.5) -- (1,2) (1,0) -- (1,0.5) to[out=90,in=330] (0.6,0.9) (0.4,1.1) to[out=150,in=270] (0,1.5) -- (0,2);
\end{scope}
\end{tikzpicture}
\end{array}\oplus \begin{array}{c}
\begin{tikzpicture}
\begin{scope}[scale=0.8]
\node[above] at (0,2) {$-$};
\node[above] at (1,2) {$+$};
\node[below] at (0,0) {$+$};
\node[below] at (1,0) {$-$};
\draw[ultra thick, purple] (0,0) -- (0,0.5) to[out=90,in=210] (0.5,1) to[out=30,in=270] (1,1.5) -- (1,2) (1,0) -- (1,0.5) to[out=90,in=330] (0.6,0.9) (0.4,1.1) to[out=150,in=270] (0,1.5) -- (0,2);
\end{scope}
\end{tikzpicture}
\end{array}\oplus\\ \oplus \begin{array}{c}
\begin{tikzpicture}
\begin{scope}[scale=0.8]
\node[above] at (0,2) {$+$};
\node[above] at (1,2) {$-$};
\node[below] at (0,0) {$-$};
\node[below] at (1,0) {$+$};
\draw[ultra thick, purple] (0,0) -- (0,0.5) to[out=90,in=210] (0.5,1) to[out=30,in=270] (1,1.5) -- (1,2) (1,0) -- (1,0.5) to[out=90,in=330] (0.6,0.9) (0.4,1.1) to[out=150,in=270] (0,1.5) -- (0,2);
\end{scope}
\end{tikzpicture}
\end{array}\oplus q t^{-1}\begin{array}{c}
\begin{tikzpicture}
\begin{scope}[scale=0.8]
\node[above] at (0,2) {$+$};
\node[above] at (1,2) {$-$};
\node[below] at (0,0) {$+$};
\node[below] at (1,0) {$-$};
\draw[->] (0.6,0.5) -- (1,0.5) (0,0.5) -- (0.6,0.5);
\node[below] at (0.5,0.5) {$12$};
\draw[ultra thick, purple] (0,0) -- (0,0.5) to[out=90,in=210] (0.5,1) to[out=30,in=270] (1,1.5) -- (1,2) (1,0) -- (1,0.5) to[out=90,in=330] (0.6,0.9) (0.4,1.1) to[out=150,in=270] (0,1.5) -- (0,2);
\end{scope}
\end{tikzpicture}
\end{array}\oplus q^{-1}\begin{array}{c}
\begin{tikzpicture}
\begin{scope}[scale=0.8]
\node[above] at (0,2) {$+$};
\node[above] at (1,2) {$-$};
\node[below] at (0,0) {$+$};
\node[below] at (1,0) {$-$};
\draw[->] (0.6,0.5) -- (1,0.5) (0,0.5) -- (0.6,0.5);
\node[below] at (0.5,0.5) {$12'$};
\draw[ultra thick, purple] (0,0) -- (0,0.5) to[out=90,in=210] (0.5,1) to[out=30,in=270] (1,1.5) -- (1,2) (1,0) -- (1,0.5) to[out=90,in=330] (0.6,0.9) (0.4,1.1) to[out=150,in=270] (0,1.5) -- (0,2);
\end{scope}
\end{tikzpicture}
\end{array}
\end{split}
\ee

\be
\label{rules_b}
\begin{split}
	\CE\left(\begin{array}{c}
		\begin{tikzpicture}
		\begin{scope}[scale=0.8, xscale=-1]
		\draw[ultra thick] (-0.5,-0.5) -- (0.5,0.5) (0.5,-0.5)-- (0.1,-0.1) (-0.1,0.1) -- (-0.5,0.5); 
		\end{scope}
		\end{tikzpicture}
	\end{array}\right)= q \begin{array}{c}
	\begin{tikzpicture}
	\begin{scope}[scale=0.8, xscale=-1]
	\node[above] at (0,2) {$+$};
	\node[above] at (1,2) {$+$};
	\node[below] at (0,0) {$+$};
	\node[below] at (1,0) {$+$};
	\draw[ultra thick, purple] (0,0) -- (0,0.5) to[out=90,in=210] (0.5,1) to[out=30,in=270] (1,1.5) -- (1,2) (1,0) -- (1,0.5) to[out=90,in=330] (0.6,0.9) (0.4,1.1) to[out=150,in=270] (0,1.5) -- (0,2);
	\end{scope}
	\end{tikzpicture}
\end{array}\oplus q\begin{array}{c}
\begin{tikzpicture}
\begin{scope}[scale=0.8, xscale=-1]
\node[above] at (0,2) {$-$};
\node[above] at (1,2) {$-$};
\node[below] at (0,0) {$-$};
\node[below] at (1,0) {$-$};
\draw[ultra thick, purple] (0,0) -- (0,0.5) to[out=90,in=210] (0.5,1) to[out=30,in=270] (1,1.5) -- (1,2) (1,0) -- (1,0.5) to[out=90,in=330] (0.6,0.9) (0.4,1.1) to[out=150,in=270] (0,1.5) -- (0,2);
\end{scope}
\end{tikzpicture}
\end{array}\oplus \begin{array}{c}
\begin{tikzpicture}
\begin{scope}[scale=0.8, xscale=-1]
\node[above] at (0,2) {$-$};
\node[above] at (1,2) {$+$};
\node[below] at (0,0) {$+$};
\node[below] at (1,0) {$-$};
\draw[ultra thick, purple] (0,0) -- (0,0.5) to[out=90,in=210] (0.5,1) to[out=30,in=270] (1,1.5) -- (1,2) (1,0) -- (1,0.5) to[out=90,in=330] (0.6,0.9) (0.4,1.1) to[out=150,in=270] (0,1.5) -- (0,2);
\end{scope}
\end{tikzpicture}
\end{array}\oplus\\ \oplus \begin{array}{c}
\begin{tikzpicture}
\begin{scope}[scale=0.8, xscale=-1]
\node[above] at (0,2) {$+$};
\node[above] at (1,2) {$-$};
\node[below] at (0,0) {$-$};
\node[below] at (1,0) {$+$};
\draw[ultra thick, purple] (0,0) -- (0,0.5) to[out=90,in=210] (0.5,1) to[out=30,in=270] (1,1.5) -- (1,2) (1,0) -- (1,0.5) to[out=90,in=330] (0.6,0.9) (0.4,1.1) to[out=150,in=270] (0,1.5) -- (0,2);
\end{scope}
\end{tikzpicture}
\end{array}\oplus q^{-1}t\begin{array}{c}
\begin{tikzpicture}
\begin{scope}[scale=0.8, xscale=-1]
\node[above] at (0,2) {$+$};
\node[above] at (1,2) {$-$};
\node[below] at (0,0) {$+$};
\node[below] at (1,0) {$-$};
\draw[->] (0,1.5) -- (0.4,1.5) (1,1.5) -- (0.4,1.5);
\node[above] at (0.5,1.5) {$21$};
\draw[ultra thick, purple] (0,0) -- (0,0.5) to[out=90,in=210] (0.5,1) to[out=30,in=270] (1,1.5) -- (1,2) (1,0) -- (1,0.5) to[out=90,in=330] (0.6,0.9) (0.4,1.1) to[out=150,in=270] (0,1.5) -- (0,2);
\end{scope}
\end{tikzpicture}
\end{array}\oplus q\begin{array}{c}
\begin{tikzpicture}
\begin{scope}[scale=0.8, xscale=-1]
\node[above] at (0,2) {$+$};
\node[above] at (1,2) {$-$};
\node[below] at (0,0) {$+$};
\node[below] at (1,0) {$-$};
\draw[->] (0,1.5) -- (0.4,1.5) (1,1.5) -- (0.4,1.5);
\node[above] at (0.5,1.5) {$21'$};
\draw[ultra thick, purple] (0,0) -- (0,0.5) to[out=90,in=210] (0.5,1) to[out=30,in=270] (1,1.5) -- (1,2) (1,0) -- (1,0.5) to[out=90,in=330] (0.6,0.9) (0.4,1.1) to[out=150,in=270] (0,1.5) -- (0,2);
\end{scope}
\end{tikzpicture}
\end{array}
\end{split}
\ee

In \cite{GM} it was argued that the fusion/defusion interfaces can be thought of as certain half-twist also depicted in Figure \ref{fig:sc_su_2}. The resulting expressions are:
\be
\CE\left(\begin{array}{c}
	\begin{tikzpicture}
	\begin{scope}[scale=0.8,yscale=-1]
	\draw[ultra thick] (0,1) to[out=270,in=180] (0.5,0) to[out=0,in=270] (1,1);
	\end{scope}
	\end{tikzpicture}
\end{array}\right)=\begin{array}{c}
\begin{tikzpicture}
\begin{scope}[scale=0.8]
\draw[ultra thick, purple] (-0.25,1) -- (1.25,1) (0,0) -- (0,1) (1,0) -- (1,1);
\node[below] at (0,0) {$-$};
\node[above] at (0,1) {$-$};
\node[below] at (1,0) {$+$};
\node[above] at (1,1) {$+$};
\end{scope}
\end{tikzpicture}
\end{array}\oplus q t^{-1}\begin{array}{c}
\begin{tikzpicture}
\begin{scope}[scale=0.8]
\draw[->] (0.6,0.5) -- (1,0.5) (0,0.5) -- (0.6,0.5);
\node[below] at (0.5,0.5) {$12$};
\draw[ultra thick, purple] (-0.25,1) -- (1.25,1) (0,0) -- (0,1) (1,0) -- (1,1);
\node[below] at (0,0) {$+$};
\node[above] at (0,1) {$-$};
\node[below] at (1,0) {$-$};
\node[above] at (1,1) {$+$};
\end{scope}
\end{tikzpicture}
\end{array}\label{rules_c}\ee
\be\label{rules_d}
\CE\left(\begin{array}{c}
	\begin{tikzpicture}
	\begin{scope}[scale=0.8]
	\draw[ultra thick] (0,1) to[out=270,in=180] (0.5,0) to[out=0,in=270] (1,1);
	\end{scope}
	\end{tikzpicture}
\end{array}\right)= q^{-1}t\begin{array}{c}
\begin{tikzpicture}
\begin{scope}[scale=0.8]
\draw[->] (0.6,0.5) -- (1,0.5) (0,0.5) -- (0.6,0.5);
\node[above] at (0.5,0.5) {$21$};
\draw[ultra thick, purple] (-0.25,0) -- (1.25,0) (0,0) -- (0,1) (1,0) -- (1,1);
\node[below] at (0,0) {$+$};
\node[above] at (0,1) {$-$};
\node[below] at (1,0) {$-$};
\node[above] at (1,1) {$+$};
\end{scope}
\end{tikzpicture}
\end{array}\oplus \begin{array}{c}
\begin{tikzpicture}
\begin{scope}[scale=0.8]
\draw[ultra thick, purple] (-0.25,0) -- (1.25,0) (0,0) -- (0,1) (1,0) -- (1,1);
\node[below] at (0,0) {$+$};
\node[above] at (0,1) {$+$};
\node[below] at (1,0) {$-$};
\node[above] at (1,1) {$-$};
\end{scope}
\end{tikzpicture}
\end{array}
\ee
\end{subequations}

Here we used grading $q^{\bf P} t^{\bf F}$ for all the solutions:
\be
{\bf P}_{\fs}=\frac{1}{\pi\I}\oint\lm_{\xi(\fs)}\lambda,\quad {\bf F}_{\fs}=-\oint\lm_{\xi(\fs)}\omega
\ee

By horizontal lines we have depicted soliton solutions bound to the interface, they are labelled by indices of corresponding WKB lines. We splitted these indices into ones without a prime and ones with a prime. The reason for that is that soliton paths in the target space $X$ have different topology for prime and non-prime binding points. We will not comment here on this point, however this fact can be easily seen from comparing $\bf P$ degrees of solutions, also one can think of a path in $\CP$ corresponding to $\CR^{-1}$-twist as a path corresponding to $\CR$-twist flowing in the opposite direction. If we reverse the path corresponding to $\CR$ we see that it first intersects the primed WKB line then non-primed one.

We define the Landau-Ginzburg cohomology as
\be
{\bf LGCoh}(L):=H^{*}(\CE(L),Q)
\ee
where the action of supercharge $Q$ is given by \eqref{null-web}.

\subsubsection{Null-web examples}
We will not re-cite all the arguments of link invariance under the Reidemeister moves listed in \cite[Section 6]{GM}, let us consider here just the Reidemeister move II and show how the corresponding complex contributions cancel due to presence of null-webs.

The Reidemeister move II implies that two following links are equivalent:
\be
\begin{array}{c}
	\begin{tikzpicture}
	\begin{scope}[scale=0.8]
	\draw[ultra thick] (0,0) to[out=90,in=210] (0.5,0.5) to[out=30,in=270] (1,1) (1,0) to [out=90,in=330] (0.6,0.4) (0.4,0.6) to[out=150,in=270] (0,1);
	\begin{scope}[xscale=-1,shift={(-1,1)}]
	\draw[ultra thick] (0,0) to[out=90,in=210] (0.5,0.5) to[out=30,in=270] (1,1) (1,0) to [out=90,in=330] (0.6,0.4) (0.4,0.6) to[out=150,in=270] (0,1);
	\end{scope}
	\end{scope}
	\end{tikzpicture}
\end{array}\sim\begin{array}{c}
\begin{tikzpicture}
\begin{scope}[scale=0.8]
\draw[ultra thick] (0,0) to[out=70,in=290] (0,2) (1,0) to[out=110,in=250] (1,2);
\end{scope}
\end{tikzpicture}
\end{array}
\ee
However expanding the complex in the left hand side one derives the following contributions:
\be\label{eq:RII-complex}
\begin{split}
	\CE\left(
	\begin{array}{c}
		\begin{tikzpicture}
		\begin{scope}[scale=0.7]
		\draw[ultra thick] (0,0) to[out=90,in=210] (0.5,0.5) to[out=30,in=270] (1,1) (1,0) to [out=90,in=330] (0.6,0.4) (0.4,0.6) to[out=150,in=270] (0,1);
		\begin{scope}[xscale=-1,shift={(-1,1)}]
		\draw[ultra thick] (0,0) to[out=90,in=210] (0.5,0.5) to[out=30,in=270] (1,1) (1,0) to [out=90,in=330] (0.6,0.4) (0.4,0.6) to[out=150,in=270] (0,1);
		\end{scope}
		\end{scope}
		\end{tikzpicture}
	\end{array}\right)=
	\begin{array}{c}
		\begin{tikzpicture}
		\begin{scope}[scale=0.7]
		\draw[ultra thick,purple] (0,-0.5) -- (0,0) to[out=90,in=210] (0.5,0.5) to[out=30,in=270] (1,1) (1,-0.5) -- (1,0) to [out=90,in=330] (0.6,0.4) (0.4,0.6) to[out=150,in=270] (0,1);
		\begin{scope}[yscale=-1,shift={(0,-2.5)}]
		\draw[ultra thick,purple] (0,-0.5) -- (0,0) to[out=90,in=210] (0.5,0.5) to[out=30,in=270] (1,1) (1,-0.5) -- (1,0) to [out=90,in=330] (0.6,0.4) (0.4,0.6) to[out=150,in=270] (0,1);
		\end{scope}
		\node[below] at (0,-0.5) {$-$};\node[below] at (1,-0.5) {$-$};
		\node at (0,1.25) {$-$};\node at (1,1.25) {$-$};
		\node[above] at (0,3) {$-$};\node[above] at (1,3) {$-$};
		\end{scope}
		\end{tikzpicture}
	\end{array}\oplus
	\begin{array}{c}
		\begin{tikzpicture}
		\begin{scope}[scale=0.7]
		\draw[ultra thick,purple] (0,-0.5) -- (0,0) to[out=90,in=210] (0.5,0.5) to[out=30,in=270] (1,1) (1,-0.5) -- (1,0) to [out=90,in=330] (0.6,0.4) (0.4,0.6) to[out=150,in=270] (0,1);
		\begin{scope}[yscale=-1,shift={(0,-2.5)}]
		\draw[ultra thick,purple] (0,-0.5) -- (0,0) to[out=90,in=210] (0.5,0.5) to[out=30,in=270] (1,1) (1,-0.5) -- (1,0) to [out=90,in=330] (0.6,0.4) (0.4,0.6) to[out=150,in=270] (0,1);
		\end{scope}
		\node[below] at (0,-0.5) {$+$};\node[below] at (1,-0.5) {$+$};
		\node at (0,1.25) {$+$};\node at (1,1.25) {$+$};
		\node[above] at (0,3) {$+$};\node[above] at (1,3) {$+$};
		\end{scope}
		\end{tikzpicture}
	\end{array}\oplus
	\begin{array}{c}
		\begin{tikzpicture}
		\begin{scope}[scale=0.7]
		\draw[ultra thick,purple] (0,-0.5) -- (0,0) to[out=90,in=210] (0.5,0.5) to[out=30,in=270] (1,1) (1,-0.5) -- (1,0) to [out=90,in=330] (0.6,0.4) (0.4,0.6) to[out=150,in=270] (0,1);
		\begin{scope}[yscale=-1,shift={(0,-2.5)}]
		\draw[ultra thick,purple] (0,-0.5) -- (0,0) to[out=90,in=210] (0.5,0.5) to[out=30,in=270] (1,1) (1,-0.5) -- (1,0) to [out=90,in=330] (0.6,0.4) (0.4,0.6) to[out=150,in=270] (0,1);
		\end{scope}
		\node[below] at (0,-0.5) {$-$};\node[below] at (1,-0.5) {$+$};
		\node at (0,1.25) {$+$};\node at (1,1.25) {$-$};
		\node[above] at (0,3) {$-$};\node[above] at (1,3) {$+$};
		\end{scope}
		\end{tikzpicture}
	\end{array}\oplus
	\begin{array}{c}
		\begin{tikzpicture}
		\begin{scope}[scale=0.7]
		\draw[ultra thick,purple] (0,-0.5) -- (0,0) to[out=90,in=210] (0.5,0.5) to[out=30,in=270] (1,1) (1,-0.5) -- (1,0) to [out=90,in=330] (0.6,0.4) (0.4,0.6) to[out=150,in=270] (0,1);
		\begin{scope}[yscale=-1,shift={(0,-2.5)}]
		\draw[ultra thick,purple] (0,-0.5) -- (0,0) to[out=90,in=210] (0.5,0.5) to[out=30,in=270] (1,1) (1,-0.5) -- (1,0) to [out=90,in=330] (0.6,0.4) (0.4,0.6) to[out=150,in=270] (0,1);
		\end{scope}
		\node[below] at (0,-0.5) {$+$};\node[below] at (1,-0.5) {$-$};
		\node at (0,1.25) {$-$};\node at (1,1.25) {$+$};
		\node[above] at (0,3) {$+$};\node[above] at (1,3) {$-$};
		\end{scope}
		\end{tikzpicture}
	\end{array}\oplus\\
	\oplus q^{-1}\left(t\begin{array}{c}
		\begin{tikzpicture}
		\begin{scope}[scale=0.7]
		\draw[->] (0.6,2.5) -- (1,2.5) (0,2.5) -- (0.6,2.5);
		\node[above] at (0.5,2.5) {$21$};
		\draw[ultra thick,purple] (0,-0.5) -- (0,0) to[out=90,in=210] (0.5,0.5) to[out=30,in=270] (1,1) (1,-0.5) -- (1,0) to [out=90,in=330] (0.6,0.4) (0.4,0.6) to[out=150,in=270] (0,1);
		\begin{scope}[yscale=-1,shift={(0,-2.5)}]
		\draw[ultra thick,purple] (0,-0.5) -- (0,0) to[out=90,in=210] (0.5,0.5) to[out=30,in=270] (1,1) (1,-0.5) -- (1,0) to [out=90,in=330] (0.6,0.4) (0.4,0.6) to[out=150,in=270] (0,1);
		\end{scope}
		\node[below] at (0,-0.5) {$+$};\node[below] at (1,-0.5) {$-$};
		\node at (0,1.25) {$-$};\node at (1,1.25) {$+$};
		\node[above] at (0,3) {$+$};\node[above] at (1,3) {$-$};
		\end{scope}
		\end{tikzpicture}
	\end{array}\oplus \begin{array}{c}
	\begin{tikzpicture}
	\begin{scope}[scale=0.7]
	\draw[->] (0.6,0) -- (1,0) (0,0) -- (0.6,0);
	\node[below] at (0.5,0) {$12'$};
	\draw[ultra thick,purple] (0,-0.5) -- (0,0) to[out=90,in=210] (0.5,0.5) to[out=30,in=270] (1,1) (1,-0.5) -- (1,0) to [out=90,in=330] (0.6,0.4) (0.4,0.6) to[out=150,in=270] (0,1);
	\begin{scope}[yscale=-1,shift={(0,-2.5)}]
	\draw[ultra thick,purple] (0,-0.5) -- (0,0) to[out=90,in=210] (0.5,0.5) to[out=30,in=270] (1,1) (1,-0.5) -- (1,0) to [out=90,in=330] (0.6,0.4) (0.4,0.6) to[out=150,in=270] (0,1);
	\end{scope}
	\node[below] at (0,-0.5) {$+$};\node[below] at (1,-0.5) {$-$};
	\node at (0,1.25) {$+$};\node at (1,1.25) {$-$};
	\node[above] at (0,3) {$+$};\node[above] at (1,3) {$-$};
	\end{scope}
	\end{tikzpicture}
\end{array}\right) \oplus q\left(\begin{array}{c}
\begin{tikzpicture}
\begin{scope}[scale=0.7]
\draw[->] (0.6,2.5) -- (1,2.5) (0,2.5) -- (0.6,2.5);
\node[above] at (0.5,2.5) {$21'$};
\draw[ultra thick,purple] (0,-0.5) -- (0,0) to[out=90,in=210] (0.5,0.5) to[out=30,in=270] (1,1) (1,-0.5) -- (1,0) to [out=90,in=330] (0.6,0.4) (0.4,0.6) to[out=150,in=270] (0,1);
\begin{scope}[yscale=-1,shift={(0,-2.5)}]
\draw[ultra thick,purple] (0,-0.5) -- (0,0) to[out=90,in=210] (0.5,0.5) to[out=30,in=270] (1,1) (1,-0.5) -- (1,0) to [out=90,in=330] (0.6,0.4) (0.4,0.6) to[out=150,in=270] (0,1);
\end{scope}
\node[below] at (0,-0.5) {$+$};\node[below] at (1,-0.5) {$-$};
\node at (0,1.25) {$-$};\node at (1,1.25) {$+$};
\node[above] at (0,3) {$+$};\node[above] at (1,3) {$-$};
\end{scope}
\end{tikzpicture}
\end{array}\oplus t^{-1}\begin{array}{c}
\begin{tikzpicture}
\begin{scope}[scale=0.7]
\draw[->] (0.6,0) -- (1,0) (0,0) -- (0.6,0);
\node[below] at (0.5,0) {$12$};
\draw[ultra thick,purple] (0,-0.5) -- (0,0) to[out=90,in=210] (0.5,0.5) to[out=30,in=270] (1,1) (1,-0.5) -- (1,0) to [out=90,in=330] (0.6,0.4) (0.4,0.6) to[out=150,in=270] (0,1);
\begin{scope}[yscale=-1,shift={(0,-2.5)}]
\draw[ultra thick,purple] (0,-0.5) -- (0,0) to[out=90,in=210] (0.5,0.5) to[out=30,in=270] (1,1) (1,-0.5) -- (1,0) to [out=90,in=330] (0.6,0.4) (0.4,0.6) to[out=150,in=270] (0,1);
\end{scope}
\node[below] at (0,-0.5) {$+$};\node[below] at (1,-0.5) {$-$};
\node at (0,1.25) {$+$};\node at (1,1.25) {$-$};
\node[above] at (0,3) {$+$};\node[above] at (1,3) {$-$};
\end{scope}
\end{tikzpicture}
\end{array}\right)
\end{split}
\ee

\begin{figure}[h]
\begin{center}
	\begin{tikzpicture}
	\node at (-7,0) {$\begin{array}{c}
		\includegraphics[scale=0.25]{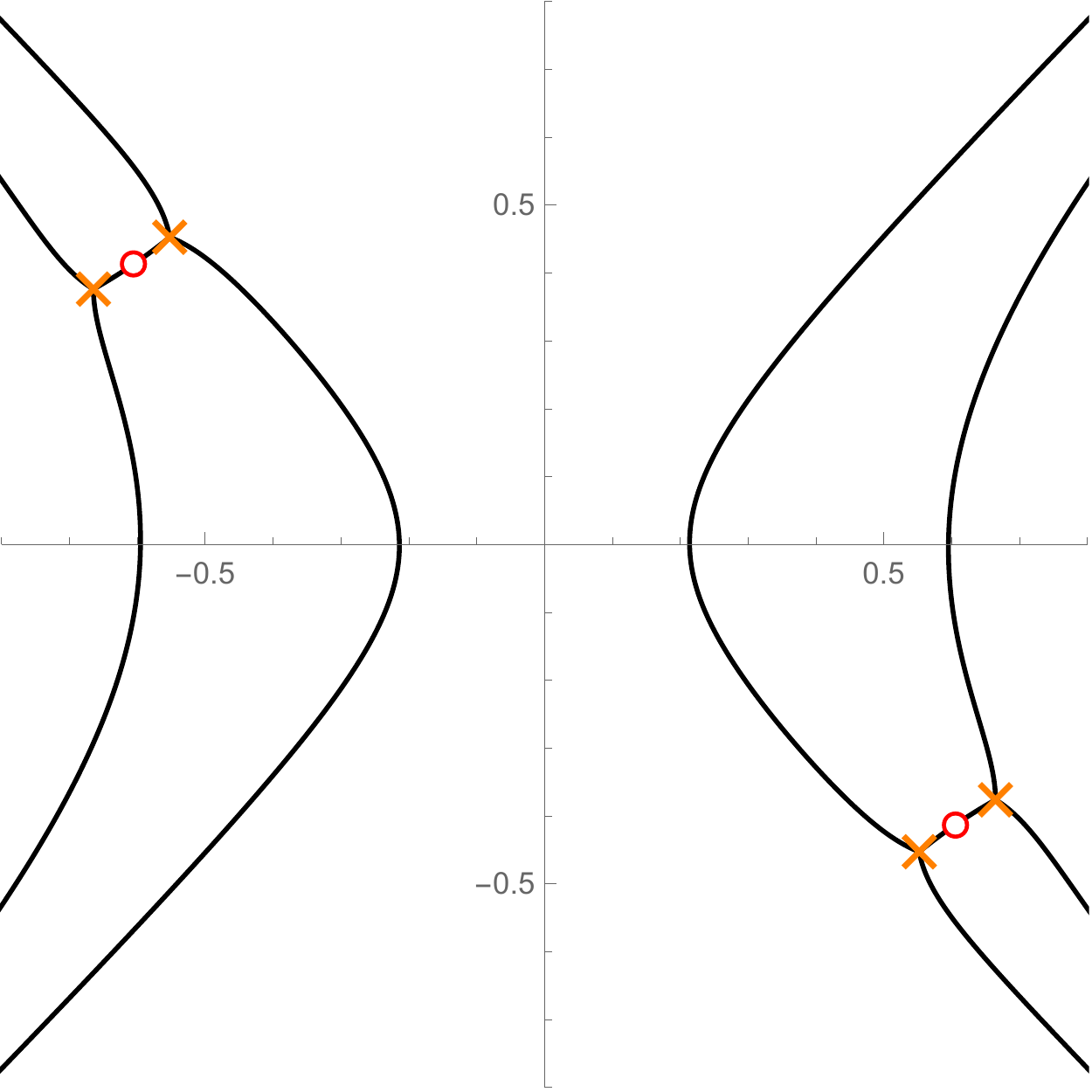}
		\end{array}$};
	\node at (-3.5,0) {$\begin{array}{c}
		\includegraphics[scale=0.25]{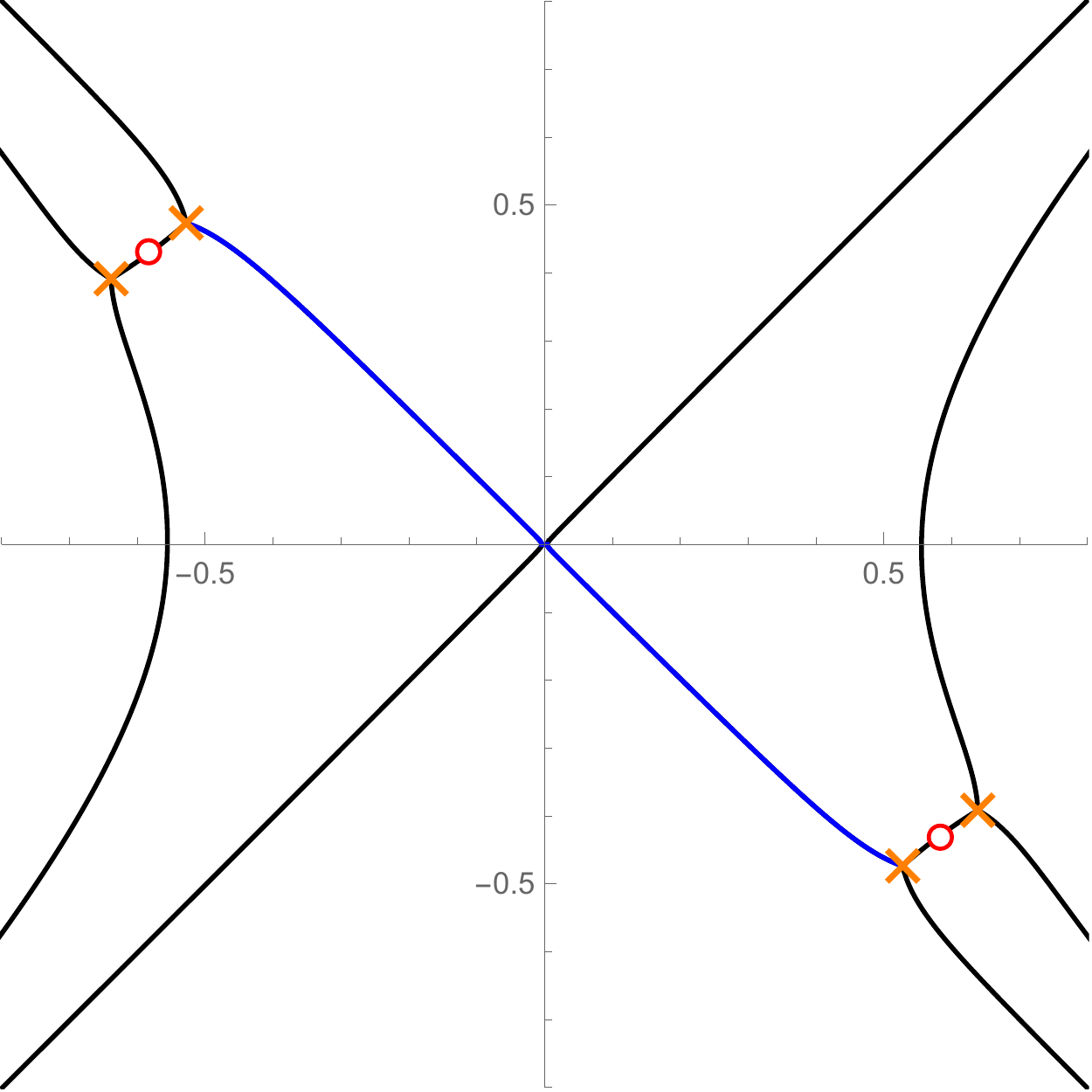}
		\end{array}$};
	\node at (0,0) {$\begin{array}{c}
		\includegraphics[scale=0.25]{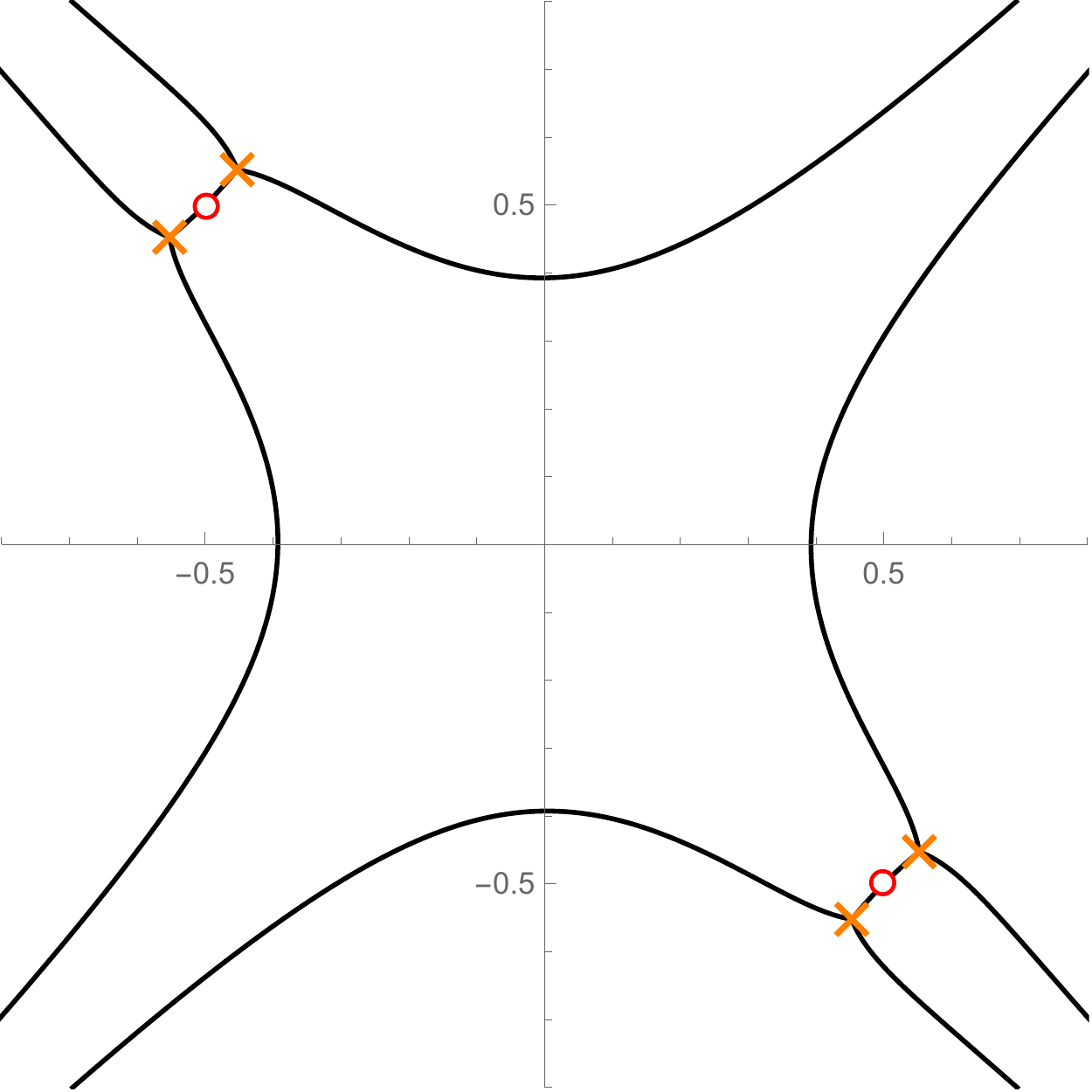}
		\end{array}$};
	\node at (3.5,0) {$\begin{array}{c}
		\includegraphics[scale=0.25]{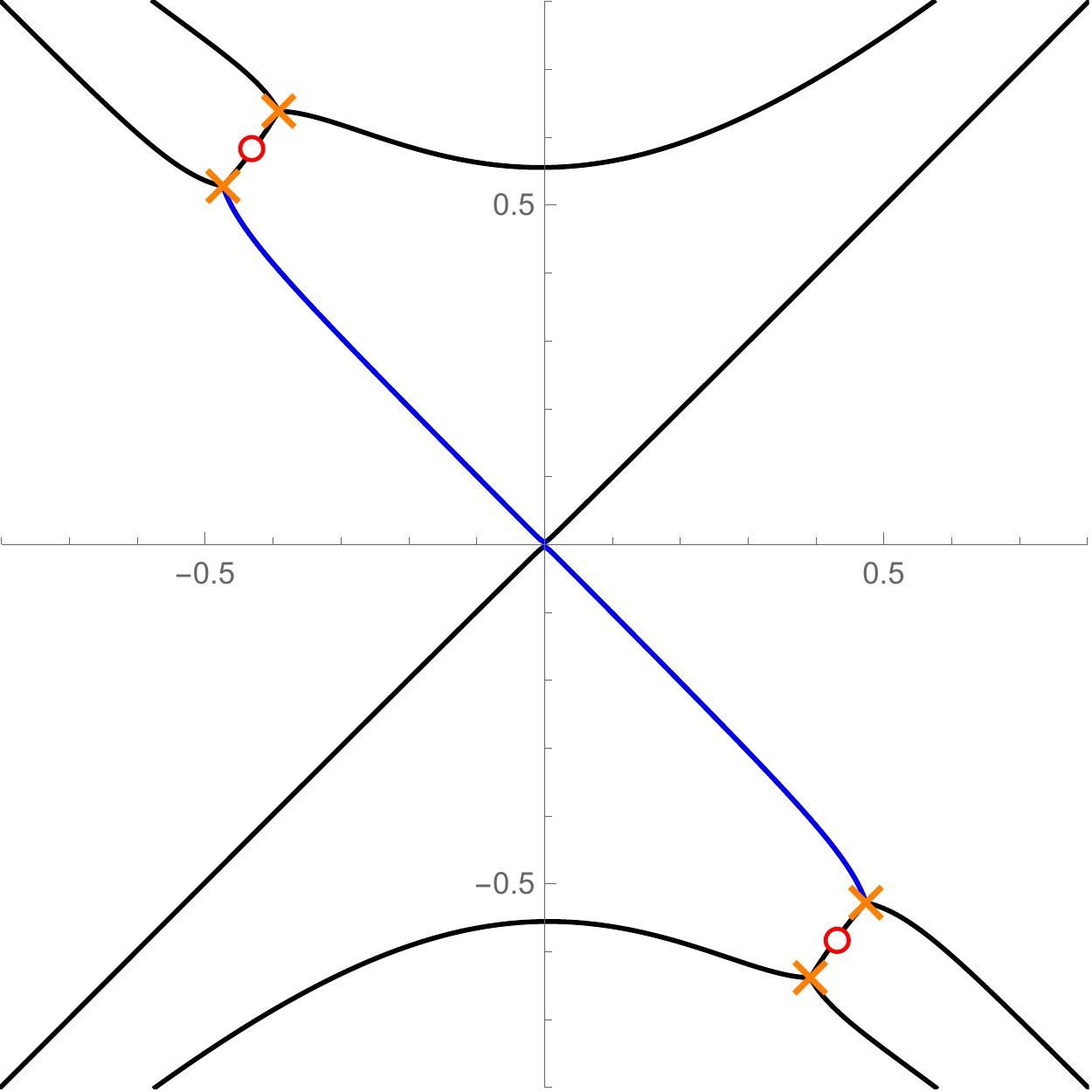}
		\end{array}$};
	\node at (7,0) {$\begin{array}{c}
		\includegraphics[scale=0.25]{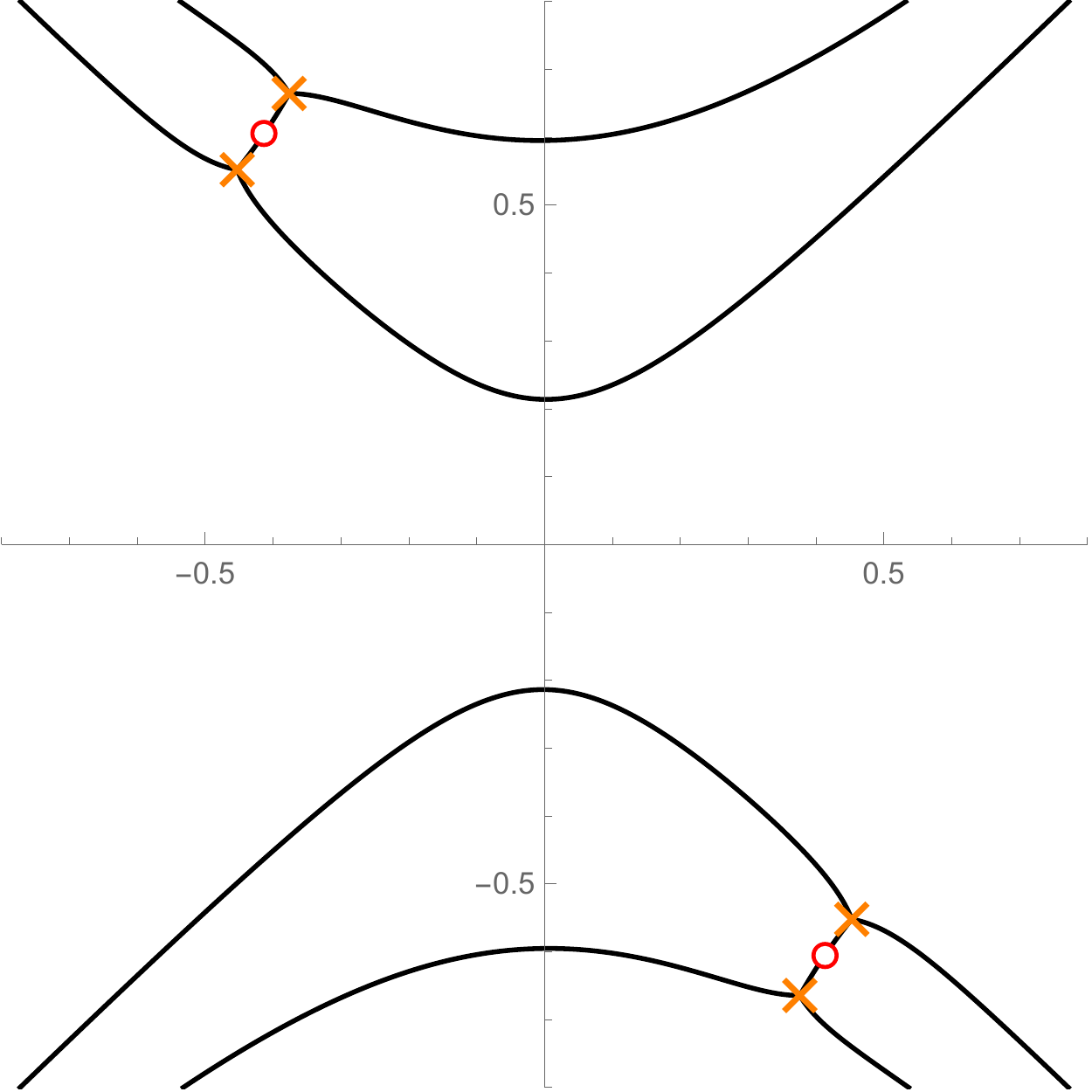}
		\end{array}$};
	\draw[dashed] (-5.25,-1.75) -- (-5.25,1.75) (-1.75,-1.75) -- (-1.75,1.75) (1.75,-1.75) -- (1.75,1.75) (5.25,-1.75) -- (5.25,1.75);
	\draw[->] (-8,-1.9) -- (8,-1.9);
	\draw[ultra thick] (-7,-1.8) -- (-7,-2) (0,-1.8) -- (0,-2) (7,-1.8) -- (7,-2);
	\begin{scope}[shift={(-7,-2.6)}]
	\begin{scope}[scale=0.5]
	\draw[thick] (-0.5,-1) to[out=90,in=210] (0,-0.5) to[out=30,in=270] (0.5,0) to[out=90,in=330] (0,0.5) to[out=150,in=270] (-0.5,1) (0.5,-1) to[out=90,in=330] (0.1,-0.6) (-0.1,-0.4) to[out=150,in=270] (-0.5,0) to[out=90,in=210] (-0.1,0.4) (0.1,0.6) to[out=30,in=270] (0.5,1);
	\end{scope}
	\end{scope}
	\begin{scope}[shift={(0,-2.6)}]
	\begin{scope}[scale=0.5]
	\draw[thick] (-0.5,-1) to[out=90,in=270] (0,0) to[out=90,in=270] (-0.5,1) (0.5,-1) to[out=90,in=270] (0,0) to[out=90,in=270] (0.5,1);
	\end{scope}
	\end{scope}
	\begin{scope}[shift={(7,-2.6)}]
	\begin{scope}[scale=0.5]
	\draw[thick] (-0.5,-1) to[out=90,in=270] (-0.2,0) to[out=90,in=270] (-0.5,1) (0.5,-1) to[out=90,in=270] (0.2,0) to[out=90,in=270] (0.5,1);
	\end{scope}
	\end{scope}
	\node at (-3.5,-2.6) {$\color{blue} \IW^{(-1)}$};
	\node at (3.5,-2.6) {$\color{blue} \IW^{(+1)}$};
	\end{tikzpicture}
\end{center}
\caption{Family of link diagrams and corresponding spectral networks.}\label{fig:RRinv}
\end{figure}
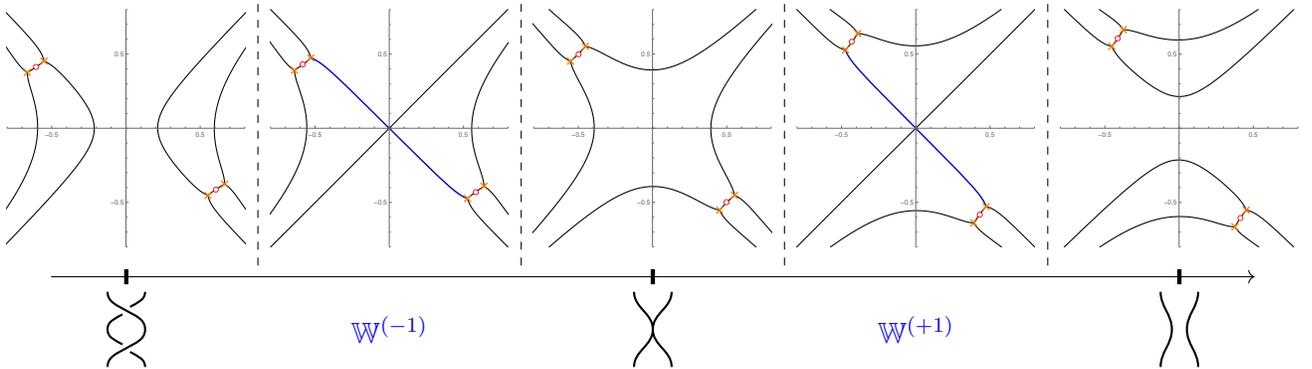

Terms in brackets cancel due to instantons and we present corresponding null-webs, it is simple to construct them since as one moves apart the strands the intersection points on the link diagram move towards each other, and soliton binding points move with them. So we can choose the following path $\hat\wp$ family:
\be
z_a(x)=\I(r-x^2)-\frac{1}{2}, \quad z_b(x)=-\left(\I(r-x^2)-\frac{1}{2}\right)
\ee
In a diagram drawn in Figure \ref{fig:RRinv} we present corresponding link diagrams and WKB webs.

We have marked null-webs corresponding to instantons cancelling contributions of $\bf P$-degrees $-1$ and $+1$ as $\IW^{(-1)}$ and $\IW^{(+1)}$ and by a blue marker correspondingly. 
At this point let us stress again that a null-web \emph{does not} represent a cluster mutation  mapping an ideal triangulation to an ideal triangulation, and the situation in question is rather illustrative. In this case the spectral cover is two-fold, usually one assigns  to a WKB web an ideal triangulation of the base curve $\IC$ according to three rules \cite{SN}: WKB lines play the role of ``medians" of triangles, branching points correspond to ``centers'' of faces of triangles, punctures correspond to vertices of triangles. And triangulation is \emph{ideal} if all the edges of triangulation have triangle faces on both sides. We have depicted a what-would-it-be trangulation according to this rules at $r=0$ by dashed red lines. Obviously, in the middle we get some rhombus rather than an ideal triangulation. 

A generic strategy to construct a null-web is the following. Solitons are usually bound to certain elements like intersections, or caps/cups. We can move them homotopically in the link diagram so solitons (intersection points of WKB webs with path $\wp$) move as well. Null-webs appear near configurations corresponding to shrinking diagram resolutions as in the previous example. However, this strategy may fail when applied to multiple twists, like in the following diagram:
$$
\begin{array}{c}
\begin{tikzpicture}
\node(A) at (0,0) {$\begin{array}{c}
	\begin{tikzpicture}
	\begin{scope}[scale=0.8]
		\draw[ultra thick] (0,0) to[out=90,in=210] (0.5,0.5) to[out=30,in=270] (1,1) (1,0) to[out=90,in=330] (0.6,0.4) (0.4,0.6) to[out=150,in=270] (0,1);
		\begin{scope}[shift={(0,1)}]
		\draw[ultra thick] (0,0) to[out=90,in=210] (0.5,0.5) to[out=30,in=270] (1,1) (1,0) to[out=90,in=330] (0.6,0.4) (0.4,0.6) to[out=150,in=270] (0,1);
		\end{scope}
	\end{scope}
	\end{tikzpicture}
	\end{array}$};
\node(B) at (-3,0) {$\begin{array}{c}
	\begin{tikzpicture}
	\begin{scope}[scale=0.8]
	\draw[ultra thick] (0,0) to[out=90,in=270] (0.5,1) to[out=90,in=270] (0,2) (1,0) to[out=90,in=270] (0.5,1) to[out=90,in=270] (1,2);
	\filldraw[black] (0.5,1) circle (0.1);
	\end{scope}
	\end{tikzpicture}
	\end{array}$};
\node(C) at (3,0) {$\begin{array}{c}
	\begin{tikzpicture}
	\begin{scope}[scale=0.8]
	\draw[ultra thick] (0,0) to[out=90,in=210] (0.5,0.5) to[out=30,in=270] (1,1) (1,0) to[out=90,in=330] (0.6,0.4) (0.4,0.6) to[out=150,in=315] (0.1,0.9);
	\begin{scope}[shift={(0,1)}]
	\draw[ultra thick] (-0.1,0.1) -- (-0.5,0.5) to[out=135,in=225] (-0.5,-0.5) -- (0,0) to[out=45,in=210] (0.5,0.5) to[out=30,in=270] (1,1) (1,0) to[out=90,in=330] (0.6,0.4) (0.4,0.6) to[out=150,in=270] (0,1);
	\end{scope}
	\end{scope}
	\end{tikzpicture}
	\end{array}$};
\node(D) at (6,0) {$\begin{array}{c}
	\begin{tikzpicture}
	\begin{scope}[scale=0.8]
	\draw[ultra thick] (1,0) to[out=90,in=315] (0.6,0.9) (0.4,1.1) -- (0,1.5) to[out=135,in=225] (0,0.5) -- (0.5,1) to[out=45,in=270] (1,2) (0,0) to[out=45,in=270] (0.5,0.85) (0.5,1.15) to[out=90,in=315] (0,2);
	\end{scope}
	\end{tikzpicture}
	\end{array}$};
\path (A) edge[->] (B) (A) edge[->] (C) (C) edge[->] (D) (D) edge[->, bend right] (B);
\begin{scope}[shift={(-1.5,0)}]
\draw (-0.3,-0.3) -- (0.3,0.3) (0.3,-0.3) -- (-0.3,0.3);
\end{scope}
\end{tikzpicture}
\end{array}
$$
If one naively shrinks a double twist (moving to the left in the diagram) to a point the WKB web becomes too singular. To resolve this singularity we use Reidemeister move I (move to the right in the diagram), then in this spider-like diagram the intersection can be shrinked painlessly. For example, let us show that indeed $12$ and $12'$ solitons form an soliton-anti-soliton pair, therefore they can be canceled. This is hardly seen from the simple double twist diagrams. We should squeeze branching points so we denote corresponding null-web as merge of two branching points:
\be
\begin{array}{c}
\begin{tikzpicture}
\begin{scope}[scale=0.7]
\draw[->] (0.6,0.5) -- (1,0.5) (0,0.5) -- (0.6,0.5);
\node[below] at (0.5,0.5) {$12$};
\draw[ultra thick, purple] (0,0) -- (0,0.5) to[out=90,in=210] (0.5,1) to[out=30,in=270] (1,1.5) (1,0) -- (1,0.5) to[out=90,in=330] (0.6,0.9) (0.4,1.1) to[out=150,in=270] (0,1.5);
\begin{scope}[shift={(0,2)}]
\draw[->] (0.6,0.5) -- (1,0.5) (0,0.5) -- (0.6,0.5);
\node[below] at (0.5,0.5) {$12'$};
\draw[ultra thick, purple] (0,0) -- (0,0.5) to[out=90,in=210] (0.5,1) to[out=30,in=270] (1,1.5) (1,0) -- (1,0.5) to[out=90,in=330] (0.6,0.9) (0.4,1.1) to[out=150,in=270] (0,1.5);
\end{scope}
\node[below] at (0,0) {$+$}; \node[below] at (1,0) {$-$};
\node[above] at (0,3.5) {$+$}; \node[above] at (1,3.5) {$-$};
\node at (0,1.75) {$+$}; \node at (1,1.75) {$-$};
\end{scope}
\end{tikzpicture}
\end{array}\oplus t\begin{array}{c}
\begin{tikzpicture}
\begin{scope}[scale=0.7]
\draw[ultra thick, purple] (0,0) -- (0,0.5) to[out=90,in=210] (0.5,1) to[out=30,in=270] (1,1.5) (1,0) -- (1,0.5) to[out=90,in=330] (0.6,0.9) (0.4,1.1) to[out=150,in=270] (0,1.5);
\begin{scope}[shift={(0,2)}]
\draw[ultra thick, purple] (0,0) -- (0,0.5) to[out=90,in=210] (0.5,1) to[out=30,in=270] (1,1.5) (1,0) -- (1,0.5) to[out=90,in=330] (0.6,0.9) (0.4,1.1) to[out=150,in=270] (0,1.5);
\end{scope}
\node[below] at (0,0) {$+$}; \node[below] at (1,0) {$-$};
\node[above] at (0,3.5) {$+$}; \node[above] at (1,3.5) {$-$};
\node at (0,1.75) {$-$}; \node at (1,1.75) {$+$};
\end{scope}
\end{tikzpicture}
\end{array},\qquad \IW=\begin{array}{c}
\begin{tikzpicture}
\draw[thick, ->] (0,0.8) -- (0,1.5) (0,0) -- (0,0.8);
\draw[thick, ->] (0.2,0.8) -- (0.2,1.5) (0.2,0) -- (0.2,0.8);
\begin{scope}[shift={(0,0)}]
\draw[purple, ultra thick] (-0.1,-0.1) -- (0.1,0.1) (0.1,-0.1) -- (-0.1,0.1);
\end{scope}
\begin{scope}[shift={(0.2,0)}]
\draw[purple, ultra thick] (-0.1,-0.1) -- (0.1,0.1) (0.1,-0.1) -- (-0.1,0.1);
\end{scope}
\draw (-0.1,0.2) -- (0.3,0.2) to[out=0,in=0] (0.3,-0.2) -- (-0.1,-0.2) to[out=180,in=180] (-0.1,0.2);
\node[left] at (0,1.5) {$12$};
\node[right] at (0.2,1.5) {$12'$};
\end{tikzpicture}
\end{array}
\ee
By this null-web we imply a resolution by a spider like move:
\be
\begin{array}{c}
	\begin{tikzpicture}
	\begin{scope}[scale=0.7]
	\draw[->] (0.6,0.5) -- (1,0.5) (0,0.5) -- (0.6,0.5);
	\node[below] at (0.5,0.5) {$12$};
	\draw[ultra thick, purple] (0,0) -- (0,0.5) to[out=90,in=210] (0.5,1) to[out=30,in=270] (1,1.5) (1,0) -- (1,0.5) to[out=90,in=330] (0.6,0.9) (0.4,1.1) to[out=150,in=270] (0,1.5);
	\begin{scope}[shift={(0,4)}]
	\draw[->] (0.6,0.5) -- (1,0.5) (0,0.5) -- (0.6,0.5);
	\node[below] at (0.5,0.5) {$12'$};
	\draw[ultra thick, purple] (0,0) -- (0,0.5) to[out=90,in=210] (0.5,1) to[out=30,in=270] (1,1.5) (1,0) -- (1,0.5) to[out=90,in=330] (0.6,0.9) (0.4,1.1) to[out=150,in=270] (0,1.5);
	\end{scope}
	\begin{scope}[shift={(-1,2)}]
	\draw[ultra thick, purple] (0,0) -- (0,0.5) to[out=90,in=210] (0.5,1) to[out=30,in=270] (1,1.5) (1,0) -- (1,0.5) to[out=90,in=330] (0.6,0.9) (0.4,1.1) to[out=150,in=270] (0,1.5);
	\end{scope}
	\begin{scope}[shift={(-2,0.5)}]
	\draw[->] (0.6,0.5) -- (1,0.5) (0,0.5) -- (0.6,0.5);
	\node[above] at (0.5,0.5) {$21$};
	\draw[purple, ultra thick] (-0.25,0) -- (1.25,0) (0,0) -- (0,4.5) (-0.25,4.5) -- (1.25,4.5) (1,4.5) -- (1,3.5) (1,0) -- (1,1) (3,1.5) -- (3,3);
	\end{scope}
	\node[below] at (0,0) {$+$}; \node[below] at (1,0) {$-$};
	\node at (-1,1.75) {$+$}; \node at (0,1.75) {$+$}; \node at (1,1.75) {$-$};
	\node at (-1,3.75) {$+$}; \node at (0,3.75) {$+$}; \node at (1,3.75) {$-$};
	\node[above] at (0,5.5) {$+$}; \node[above] at (1,5.5) {$-$};
	\node[below] at (-2,0.5) {$+$}; \node[below] at (-1,0.5) {$-$};
	\node[above] at (-2,5) {$-$}; \node[above] at (-1,5) {$+$};
	\end{scope}
	\end{tikzpicture}
\end{array}\oplus
t\begin{array}{c}
	\begin{tikzpicture}
	\begin{scope}[scale=0.7]
	\draw[ultra thick, purple] (0,0) -- (0,0.5) to[out=90,in=210] (0.5,1) to[out=30,in=270] (1,1.5) (1,0) -- (1,0.5) to[out=90,in=330] (0.6,0.9) (0.4,1.1) to[out=150,in=270] (0,1.5);
	\begin{scope}[shift={(0,4)}]
	\draw[ultra thick, purple] (0,0) -- (0,0.5) to[out=90,in=210] (0.5,1) to[out=30,in=270] (1,1.5) (1,0) -- (1,0.5) to[out=90,in=330] (0.6,0.9) (0.4,1.1) to[out=150,in=270] (0,1.5);
	\end{scope}
	\begin{scope}[shift={(-1,2)}]
	\draw[->] (0.6,0.5) -- (1,0.5) (0,0.5) -- (0.6,0.5);
	\node[below] at (0.5,0.5) {$\tilde{12}'$};
	\draw[ultra thick, purple] (0,0) -- (0,0.5) to[out=90,in=210] (0.5,1) to[out=30,in=270] (1,1.5) (1,0) -- (1,0.5) to[out=90,in=330] (0.6,0.9) (0.4,1.1) to[out=150,in=270] (0,1.5);
	\end{scope}
	\begin{scope}[shift={(-2,0.5)}]
	\draw[->] (0.6,0.5) -- (1,0.5) (0,0.5) -- (0.6,0.5);
	\node[above] at (0.5,0.5) {$21$};
	\draw[purple, ultra thick] (-0.25,0) -- (1.25,0) (0,0) -- (0,4.5) (-0.25,4.5) -- (1.25,4.5) (1,4.5) -- (1,3.5) (1,0) -- (1,1) (3,1.5) -- (3,3);
	\end{scope}
	\node[below] at (0,0) {$+$}; \node[below] at (1,0) {$-$};
	\node at (-1,1.75) {$+$}; \node at (0,1.75) {$-$}; \node at (1,1.75) {$+$};
	\node at (-1,3.75) {$+$}; \node at (0,3.75) {$-$}; \node at (1,3.75) {$+$};
	\node[above] at (0,5.5) {$+$}; \node[above] at (1,5.5) {$-$};
	\node[below] at (-2,0.5) {$+$}; \node[below] at (-1,0.5) {$-$};
	\node[above] at (-2,5) {$-$}; \node[above] at (-1,5) {$+$};
	\end{scope}
	\end{tikzpicture}
\end{array}
,\qquad \IW=\begin{array}{c}
	\begin{tikzpicture}
	\draw[thick] (0,1) -- (0,0) (-0.7,-0.5) -- (0,0) (0.7,-0.5) -- (0,0);
	\begin{scope}[shift={(0,1)}]
	\draw[purple, ultra thick] (-0.1,-0.1) -- (0.1,0.1) (0.1,-0.1) -- (-0.1,0.1);
	\end{scope}
	\begin{scope}[shift={(-0.7,-0.5)}]
	\draw[purple, ultra thick] (-0.1,-0.1) -- (0.1,0.1) (0.1,-0.1) -- (-0.1,0.1);
	\end{scope}
	\begin{scope}[shift={(0.7,-0.5)}]
	\draw[purple, ultra thick] (-0.1,-0.1) -- (0.1,0.1) (0.1,-0.1) -- (-0.1,0.1);
	\end{scope}
	\node[above] at (0,1.1) {$\tilde{12}'$};
	\node[right] at (0.8,-0.5) {$12'$};
	\node[left] at (-0.8,-0.5) {$12$};
	\end{tikzpicture}
\end{array}
\ee

\begin{figure}
\begin{center}
\begin{tikzpicture}
\begin{scope}[scale=0.8]
\node {\includegraphics[scale=0.4]{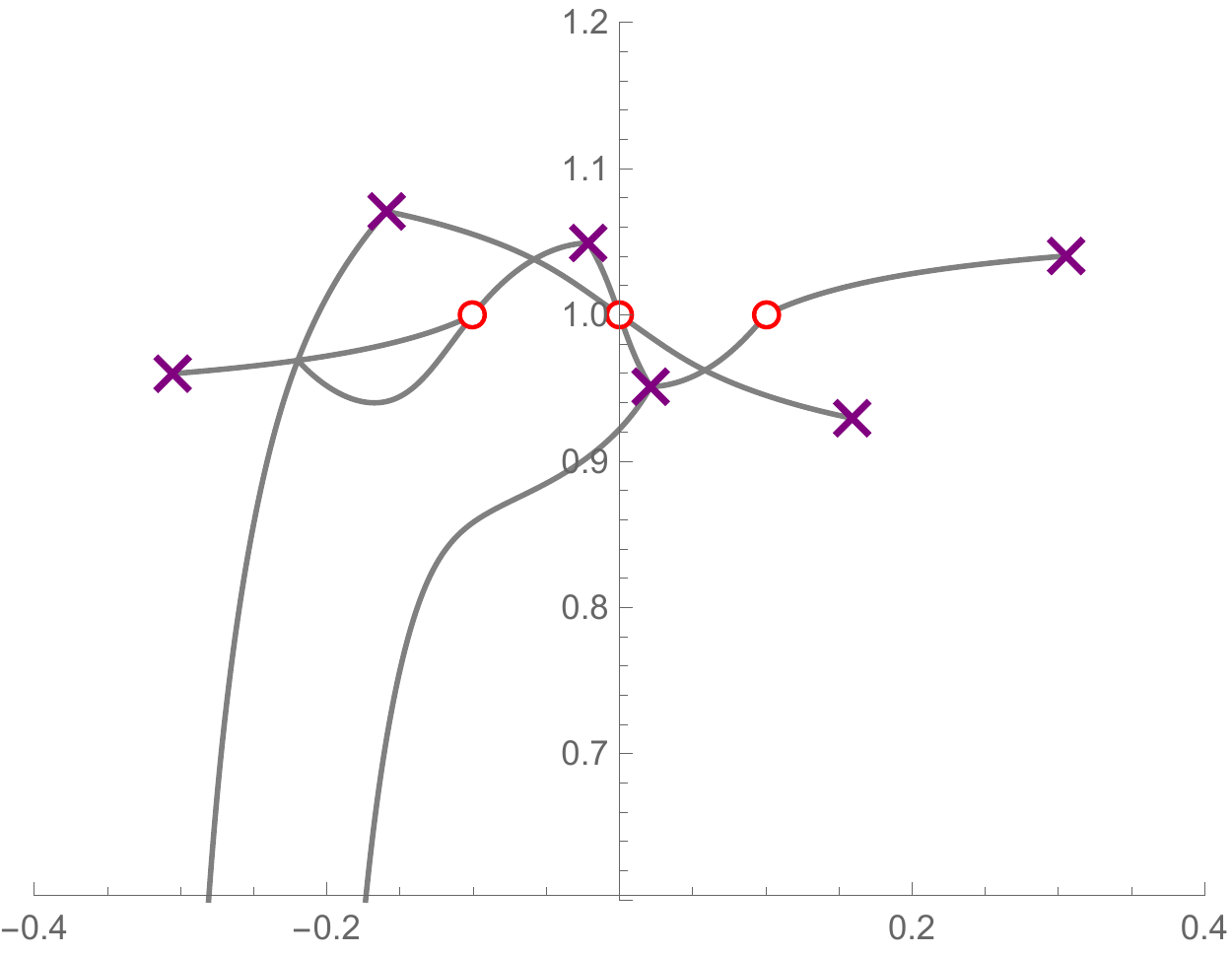}};
\draw[ultra thick, ->] (-3,-1.2) -- (3,-1.2); 
\node[right] at (3,-1.2) {$\wp$};
\node at (-2.7,0.6) {$\tilde{12}$};
\node at (-1.2,1.7) {$12$};
\node at (0.4,0.2) {$\tilde{\tilde{12}}$};
\node at (0,-2.7) {$\rho=0.1$};
\end{scope}
\end{tikzpicture}\quad
\begin{tikzpicture}
\begin{scope}[scale=0.8]
\node {\includegraphics[scale=0.4]{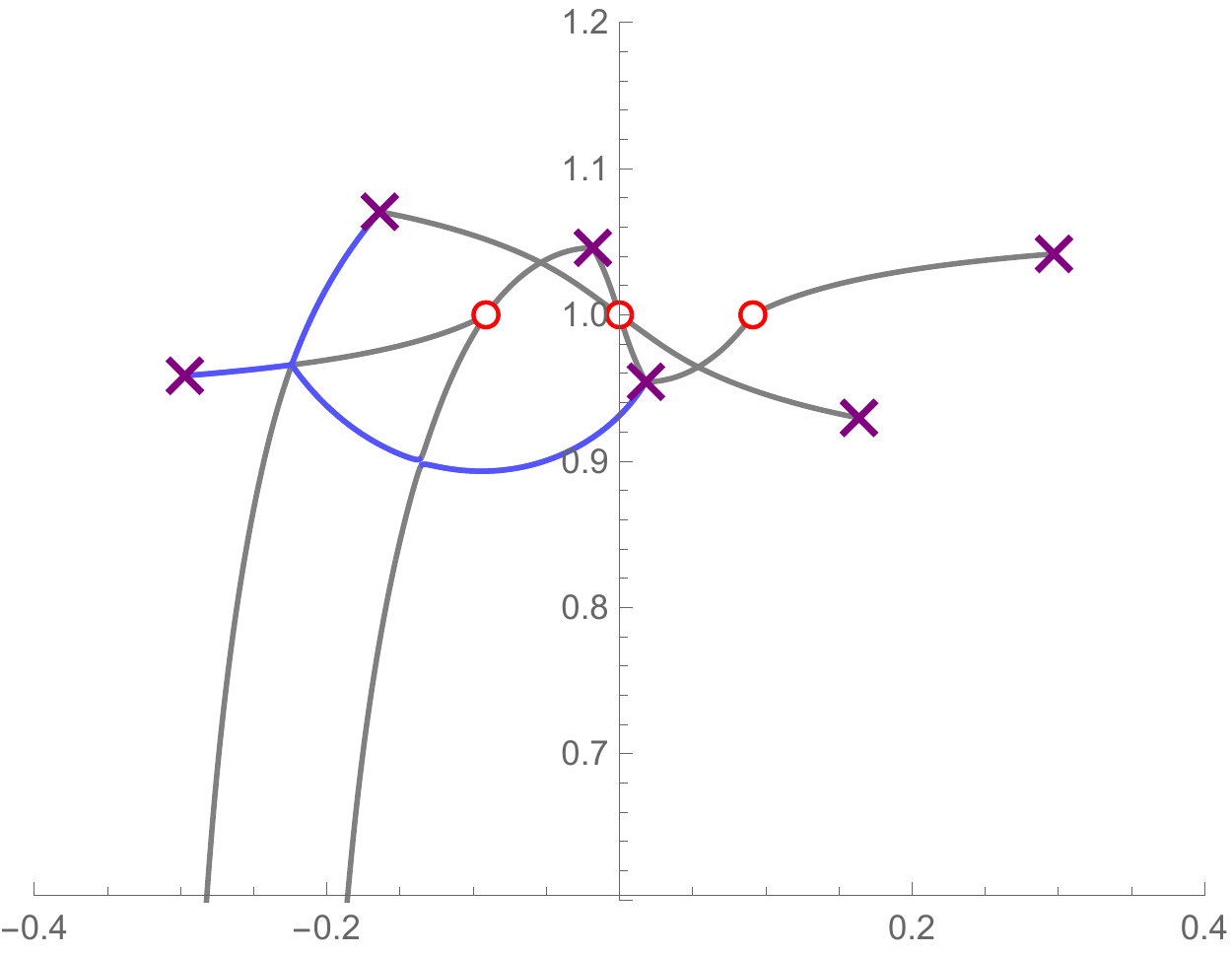}};
\draw[ultra thick, ->] (-3,-1.2) -- (3,-1.2); 
\node[right] at (3,-1.2) {$\wp$};
\node at (-2.7,0.6) {$\tilde{12}$};
\node at (-1.2,1.7) {$12$};
\node at (0.4,0.2) {$\tilde{\tilde{12}}$};
\node at (0,-2.7) {$\rho=0.091$};
\end{scope}
\end{tikzpicture}\quad
\begin{tikzpicture}
\begin{scope}[scale=0.8]
\node {\includegraphics[scale=0.4]{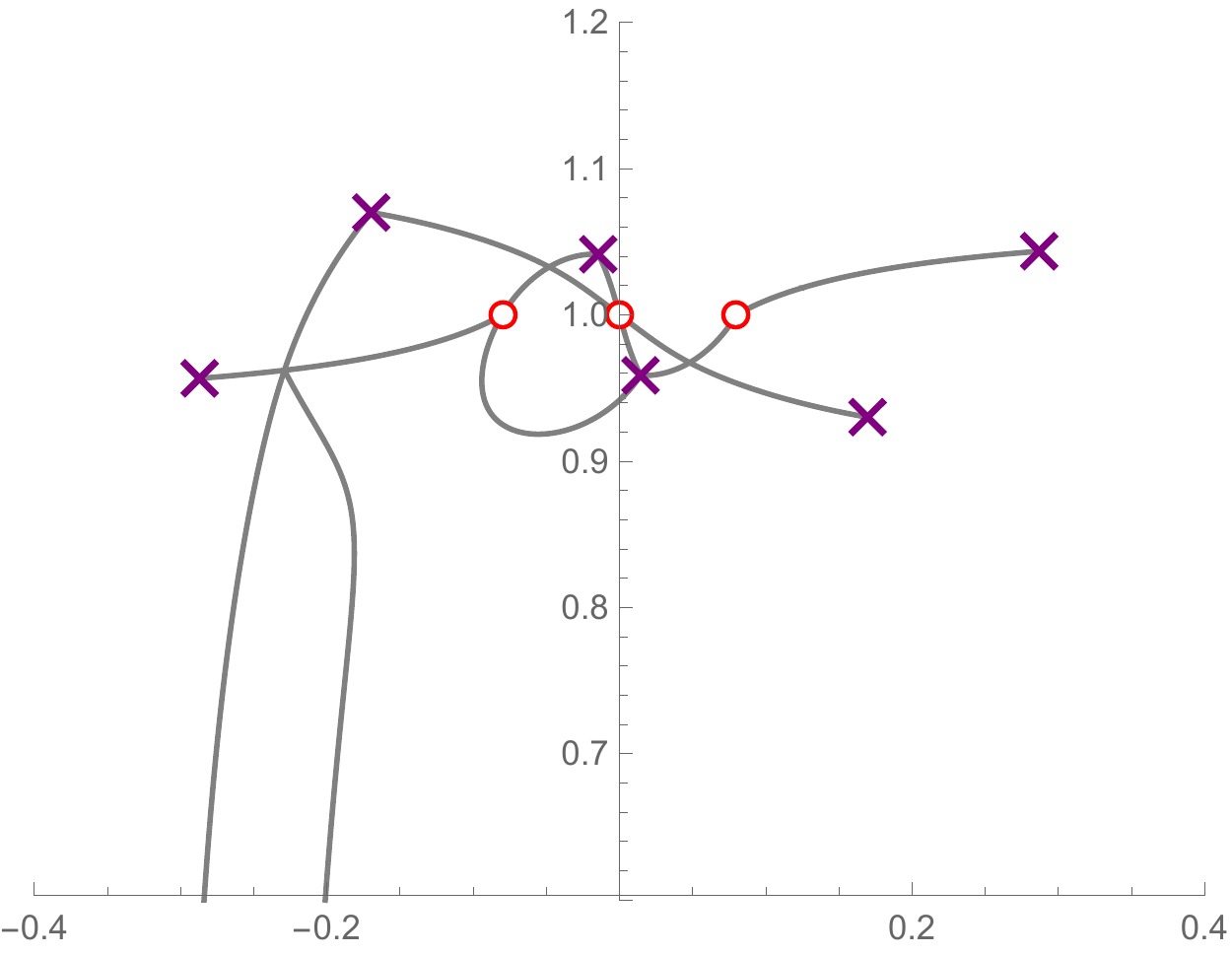}};
\draw[ultra thick, ->] (-3,-1.2) -- (3,-1.2); 
\node[right] at (3,-1.2) {$\wp$};
\node at (-2.7,0.6) {$\tilde{12}$};
\node at (-1.2,1.7) {$12$};
\node at (0.4,0.2) {$\tilde{\tilde{12}}$};
\node at (0,-2.7) {$\rho=0.08$};
\end{scope}
\end{tikzpicture}
\end{center}
\caption{Emerging null-web example for Reidemeister III for specification $z_1=-1-\I x$, $z_2=\I \rho$, $z_3=1+\I x$, $c=10$.}\label{fig:nw-example}
\end{figure}
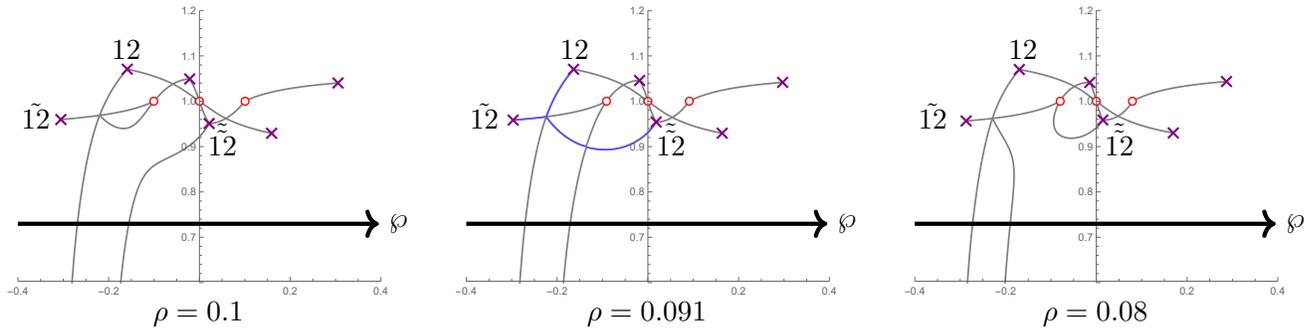

Let us put here an explicit example of a null-web governing an instanton canceling two following terms required for validity of Reidemeister III move:
\def\S{\draw[ultra thick, purple] (0,0) to[out=90,in=210] (0.5,0.5) to[out=30,in=270] (1,1) (1,0) to[out=90,in=330] (0.6,0.4) (0.4,0.6) to[out=150,in=270] (0,1);}
\be
\begin{array}{c}
	\begin{tikzpicture}
	\begin{scope}[scale=0.7]
	\draw[->] (1.6,1.7) -- (1.9,1.7) (1.1,1.7) -- (1.6,1.7);
	\node[right] at (2,1.7) {$12$};
	\S;
	\begin{scope}[shift={(1,1.5)}]
	\S;
	\end{scope}
	\begin{scope}[shift={(0,3)}]
	\S;
	\end{scope}
	\draw[ultra thick, purple] (0,1.5) -- (0,2.5);
	\node[below] at (0,0) {$+$};\node[below] at (1,0) {$-$};
	\node at (0,1.25) {$-$}; \node at (1,1.25) {$+$};\node at (2,1.25) {$-$};
	\node at (0,2.75) {$-$}; \node at (1,2.75) {$+$};\node at (2,2.75) {$-$};
	\node[above] at (0,4) {$+$}; \node[above] at (1,4) {$-$};
	\end{scope}
	\end{tikzpicture}
\end{array}\oplus t\begin{array}{c}
	\begin{tikzpicture}
	\begin{scope}[scale=0.7]
	\draw[->] (0.6,0.2) -- (0.9,0.2) (0.1,0.2) -- (0.6,0.2);
	\draw[->] (0.6,3.2) -- (0.9,3.2) (0.1,3.2) -- (0.6,3.2);
	\node[left] at (0,0.2) {$\tilde{12}$};
	\node[left] at (0,3.2) {$\tilde{\tilde{12}}$};
	\S;
	\begin{scope}[shift={(1,1.5)}]
	\S;
	\end{scope}
	\begin{scope}[shift={(0,3)}]
	\S;
	\end{scope}
	\draw[ultra thick, purple] (0,1.5) -- (0,2.5);
	\node[below] at (0,0) {$+$};\node[below] at (1,0) {$-$};
	\node at (0,1.25) {$+$}; \node at (1,1.25) {$-$};\node at (2,1.25) {$-$};
	\node at (0,2.75) {$+$}; \node at (1,2.75) {$-$};\node at (2,2.75) {$-$};
	\node[above] at (0,4) {$+$}; \node[above] at (1,4) {$-$};
	\end{scope}
	\end{tikzpicture}
\end{array}
\ee
We have chosen a configuration with a single smooth monopole since it is simpler to analyze the spectral curve in this case. In particular, the locus of branching points is given by the discriminant of vacuum equation $\p_{\phi^I}W=0$ that is a cubic polynomial in this case. The corresponding null-web is depicted in Figure \ref{fig:nw-example}.

\subsection{Sketchy review of Khovanov homology construction}
As it was indicated in the introduction our final aim is to compare cohomological theory emerging in the Landau-Ginzburg model description of tangles to Khovanov homology. Let us put here a brief review of the Khovanov homology construction following \cite{BarNatan,Dolotin:2012sw}.

Let us denote by $E(L)$ a complex corresponding to a link $L$  as a vector space. One constructs the complex following simple rules.
One puts a resolution in each vertex according to the rule:
\be\label{Khovan}
E\left[
\begin{array}{c}
	\begin{tikzpicture}[scale=0.5,xscale=-1]
	\draw[ultra thick] (-0.5,-0.5) -- (0.5,0.5);
	\draw[ultra thick] (0.5,-0.5) -- (0.1,-0.1);
	\draw[ultra thick] (-0.1,0.1) -- (-0.5,0.5);
	\end{tikzpicture}
\end{array}\right]=E\left[
\begin{array}{c}
	\begin{tikzpicture}[scale=0.5]
	\draw [ultra thick] (-0.5,-0.5) to [out= 45, in =315] (-0.5,0.5) (0.5,-0.5) to [out=135, in=225] (0.5,0.5);
	\end{tikzpicture}
\end{array}\right]\oplus \left[q^{-1}t^{-1}\right]\;E\left[
\begin{array}{c}
	\begin{tikzpicture}[scale=0.5]
	\draw [ultra thick] (-0.5,-0.5) to [out= 45, in =135] (0.5,-0.5) (-0.5,0.5) to [out=315, in=225] (0.5,0.5);
	\end{tikzpicture}
\end{array}
\right]
\ee
The complex is bi-graded by $\bf P$-degree and $\bf F$-degree (homological degree), we denote their shifts by $q^{\bf P}t^{\bf F}$.
Then the complex corresponding to link $L$ decomposes eventually into a direct sum of complexes corresponding to disjoint sets of cycles with appropriate degree shifts.
One maps them to a bi-graded linear spaces according to a rule:
\be
E\left[
\begin{array}{c}
	\begin{tikzpicture}[scale=0.5]
	\draw [ultra thick] (0,0) circle (0.5);
	\end{tikzpicture}
\end{array} L
\right]=V\otimes E[L]
\ee
Where $V$ is a two-dimensional space spanned by vectors $v_+$ and $v_-$ with $\bf P$-degrees +1 and -1 correspondingly.

\def\mul{\draw[thick] (0,0.7) to[out=0,in=180] (2,0.2) (0,-0.7) to[out=0,in=180] (2,-0.2) (0,0.3) .. controls (0.9,0.2) and (0.9,-0.2) .. (0,-0.3) (2,0.2) to[out=0,in=0] (2,-0.2) to[out=180,in=180] (2,0.2) (0,0.7) to[out=180,in=180] (0,0.3) (0,-0.7) to[out=180,in=180] (0,-0.3); \draw[dashed] (0,0.7) to[out=0,in=0] (0,0.3) (0,-0.7) to[out=0,in=0] (0,-0.3);}

\def\comul{\begin{scope}[xscale=-1,shift={(-2,0)}] \draw[thick] (0,0.7) to[out=0,in=180] (2,0.2) (0,-0.7) to[out=0,in=180] (2,-0.2) (0,0.3) .. controls (0.9,0.2) and (0.9,-0.2) .. (0,-0.3) (2,0.2) to[out=0,in=0] (2,-0.2) (0,0.7) to[out=180,in=180] (0,0.3) (0,-0.7) to[out=180,in=180] (0,-0.3) (0,0.7) to[out=0,in=0] (0,0.3) (0,-0.7) to[out=0,in=0] (0,-0.3); 
	\end{scope} \draw[dashed] (0,-0.2) to[out=0,in=0] (0,0.2);}

\def\cilind{\draw[thick] (0,0.2) -- (2,0.2) (0,-0.2) -- (2,-0.2) (2,0.2) to[out=0,in=0] (2,-0.2) to[out=180,in=180] (2,0.2) (0,0.2) to[out=180,in=180] (0,-0.2); \draw[thick, dashed] (0,0.2) to[out=0,in=0] (0,-0.2);}

Then one considers a set of differentials $d_{\chi}^2=0$ having homological degree 1 and acting in each intersection $\chi$ by either:
\begin{itemize}
	\item Joining cycles (multiplication): \be\label{mul}
	\begin{split}
		\begin{array}{c}
			\begin{tikzpicture}
			\begin{scope}[scale=0.7]
			\draw [ultra thick] (-1,0) to [out=90, in=180] (-0.5,0.5) to [out=0, in=90] (-0.2,0) to[out=270, in=0] (-0.5, -0.5) to [out=180, in=270] (-1,0);
			\begin{scope}[scale=-1]
			\draw [ultra thick] (-1,0) to [out=90, in=180] (-0.5,0.5) to [out=0, in=90] (-0.2,0) to[out=270, in=0] (-0.5, -0.5) to [out=180, in=270] (-1,0);
			\end{scope}
			\end{scope}
			\end{tikzpicture}
		\end{array}\to \begin{array}{c}
		\begin{tikzpicture}
		\begin{scope}[scale=0.7]
		\draw [ultra thick] (-1,0) to [out=90, in=180] (-0.5,0.5) to [out=0, in=180] (0,0.2) (0,-0.2) to[out=180, in=0] (-0.5, -0.5) to [out=180, in=270] (-1,0);
		\begin{scope}[scale=-1]
		\draw [ultra thick] (-1,0) to [out=90, in=180] (-0.5,0.5) to [out=0, in=180] (0,0.2) (0,-0.2) to[out=180, in=0] (-0.5, -0.5) to [out=180, in=270] (-1,0);
		\end{scope}
		\end{scope}
		\end{tikzpicture}
	\end{array},\\
	V\otimes V\mathop{\longrightarrow}\lm^{m} V,\quad m:\left\{\begin{array}{c}
		v_+\otimes v_-\mapsto v_+\\
		v_-\otimes v_+\mapsto v_+
	\end{array}\right. \left\{\begin{array}{c}
	v_+\otimes v_+\mapsto 0\\
	v_-\otimes v_-\mapsto v_-
\end{array}\right.\\
\begin{array}{c}
\begin{tikzpicture}
\mul
\end{tikzpicture}
\end{array}
\end{split}
\ee
	\item Cutting cycles (co-multiplication):\be\label{co-mul}
	\begin{split}
		\begin{array}{c}
			\begin{tikzpicture}
			\begin{scope}[scale=0.7]
			\draw [ultra thick] (-1,0) to [out=90, in=180] (-0.5,0.5) to [out=0, in=180] (0,0.2) (0,-0.2) to[out=180, in=0] (-0.5, -0.5) to [out=180, in=270] (-1,0);
			\begin{scope}[scale=-1]
			\draw [ultra thick] (-1,0) to [out=90, in=180] (-0.5,0.5) to [out=0, in=180] (0,0.2) (0,-0.2) to[out=180, in=0] (-0.5, -0.5) to [out=180, in=270] (-1,0);
			\end{scope}
			\end{scope}
			\end{tikzpicture}
		\end{array}\to\begin{array}{c}
		\begin{tikzpicture}
		\begin{scope}[scale=0.7]
		\draw [ultra thick] (-1,0) to [out=90, in=180] (-0.5,0.5) to [out=0, in=90] (-0.2,0) to[out=270, in=0] (-0.5, -0.5) to [out=180, in=270] (-1,0);
		\begin{scope}[scale=-1]
		\draw [ultra thick] (-1,0) to [out=90, in=180] (-0.5,0.5) to [out=0, in=90] (-0.2,0) to[out=270, in=0] (-0.5, -0.5) to [out=180, in=270] (-1,0);
		\end{scope}
		\end{scope}
		\end{tikzpicture}
	\end{array},\\
	V\mathop{\longrightarrow}\lm^{\Delta} V\otimes V,\quad \Delta:\left\{\begin{array}{c}
		v_+\mapsto v_+\otimes v_+\\
		v_-\mapsto v_+\otimes v_- + v_-\otimes v_+
	\end{array}\right. \\
\begin{array}{c}
	\begin{tikzpicture}
	\comul
	\end{tikzpicture}
\end{array}
\end{split}
\ee
\end{itemize}

The total differential is defined as 
\be
d=\sum\lm_{{\rm all}\; \chi} d_{\chi}
\ee
The multiplication and co-multiplication make from complex elements a Frobenious algebra mimicked by a TQFT, corresponding string diagrams are depicted below  equations \eqref{mul} and \eqref{co-mul}. Homotopic equivalence of string diagrams produces corresponding commutativity diagrams for multiplication and co-multiplication.
The differential is local in the sense that all individual differentials $d_{\chi}$ anti-commute with each other. This follows from homotopy equivalence of the string diagrams:
\be\label{commutatitivity}
\begin{split}
\begin{array}{c}
\begin{tikzpicture}
\begin{scope}[scale=0.7]
\mul
\begin{scope}[shift={(2,0)}]
\comul
\end{scope}
\end{scope}
\end{tikzpicture}
\end{array}=\begin{array}{c}
\begin{tikzpicture}
\begin{scope}[scale=0.7]
\mul
\begin{scope}[shift={(-2,-1)}]
\comul
\end{scope}
\begin{scope}[shift={(0,-1.5)}]
\cilind
\end{scope}
\begin{scope}[shift={(-2,0.5)}]
\cilind
\end{scope}
\end{scope}
\end{tikzpicture}
\end{array}=
\begin{array}{c}
	\begin{tikzpicture}
	\begin{scope}[scale=0.7]
	\mul
	\begin{scope}[shift={(-2,1)}]
	\comul
	\end{scope}
	\begin{scope}[shift={(0,1.5)}]
	\cilind
	\end{scope}
	\begin{scope}[shift={(-2,-0.5)}]
	\cilind
	\end{scope}
	\end{scope}
	\end{tikzpicture}
\end{array}\\
\begin{array}{c}
\begin{tikzpicture}
\begin{scope}[scale=0.7]
\mul
\begin{scope}[shift={(2,-0.5)}]
\mul
\end{scope}
\begin{scope}[shift={(0,-1)}]
\cilind
\end{scope}
\end{scope}
\end{tikzpicture}
\end{array}=
\begin{array}{c}
\begin{tikzpicture}
\begin{scope}[scale=0.7]
\mul
\begin{scope}[shift={(2,0.5)}]
\mul
\end{scope}
\begin{scope}[shift={(0,1)}]
\cilind
\end{scope}
\end{scope}
\end{tikzpicture}
\end{array}; \quad
\begin{array}{c}
	\begin{tikzpicture}
	\begin{scope}[scale=0.7]
	\comul
	\begin{scope}[shift={(2,-0.5)}]
	\comul
	\end{scope}
	\begin{scope}[shift={(2,0.5)}]
	\cilind
	\end{scope}
	\end{scope}
	\end{tikzpicture}
\end{array}=
\begin{array}{c}
	\begin{tikzpicture}
	\begin{scope}[scale=0.7]
	\comul
	\begin{scope}[shift={(2,0.5)}]
	\comul
	\end{scope}
	\begin{scope}[shift={(2,-0.5)}]
	\cilind
	\end{scope}
	\end{scope}
	\end{tikzpicture}
\end{array}
\end{split}
\ee
The Khovanov homology is defined as
\be
{\bf KHom}(L):=H_{*} (E(L),d)
\ee

By definition, the Khovanov homology satisfies a skein exact triangle categorifying the skein relations for the Jones polynomials:
\be
\ldots \to {\bf KHom}\left[\begin{array}{c}
	\begin{tikzpicture}[scale=0.5]
	\draw [ultra thick] (-0.5,-0.5) to [out= 45, in =135] (0.5,-0.5) (-0.5,0.5) to [out=315, in=225] (0.5,0.5);
	\end{tikzpicture}
\end{array}\right] \to
{\bf KHom}\left[\begin{array}{c}
	\begin{tikzpicture}[scale=0.5,xscale=-1]
	\draw[ultra thick] (-0.5,-0.5) -- (0.5,0.5);
	\draw[ultra thick] (0.5,-0.5) -- (0.1,-0.1);
	\draw[ultra thick] (-0.1,0.1) -- (-0.5,0.5);
	\end{tikzpicture}
\end{array}\right]\to
{\bf KHom}\left[\begin{array}{c}
	\begin{tikzpicture}[scale=0.5]
	\draw [ultra thick] (-0.5,-0.5) to [out= 45, in =315] (-0.5,0.5) (0.5,-0.5) to [out=135, in=225] (0.5,0.5);
	\end{tikzpicture}
\end{array}\right]\to
t\;{\bf KHom}\left[\begin{array}{c}
	\begin{tikzpicture}[scale=0.5]
	\draw [ultra thick] (-0.5,-0.5) to [out= 45, in =135] (0.5,-0.5) (-0.5,0.5) to [out=315, in=225] (0.5,0.5);
	\end{tikzpicture}
\end{array}\right] \to\ldots
\ee

\subsection{``Isotypical" interface}

The structure like in \eqref{Khovan} is inherited naturally from the isotypical decomposition. One can re-write the $R$-matrix acting in $\Box\otimes\Box$ as
\be
R=I-q P_{\emptyset} 
\ee
where $P_{\emptyset}$ is a projector to the scalar in the isotypical decomposition. 

On the other hand the structure like in \eqref{rules_b} appears in the natural basis in the tensor product, the Euler characteristic of that complex is just ordinary representation of the Drinfeld universal $R$-matrix \cite{Drinfeld}:
\be
R=\sigma\; q^{h\otimes h}\left(1+(q-q^{-1})E\otimes F\right)
\ee
where $\sigma:\; a\otimes b\mapsto b\otimes a$ is a permutation in the tensor square. 

The isotypical decomposition of tensor products of irreducible representations reads:
\be\label{isotypical}
r\otimes r'=\bigoplus\lm_{Q\vdash |r|+|r'|}D_Q\otimes Q
\ee
where $Q$ are again irreducible representations and $D_Q$ are invariant modules, numbers ${\rm dim}\;D_Q$ are called multiplicities. In this basis the $R$-matrix has a very simple form since it acts by intertwiners:
\be
R=\bigoplus\lm_Q q^{c_2(Q)-c_2(r)-c_2(r')}I_Q\otimes 1
\ee

From this point of view Euler characteristics of twist $R$-complexes derived from Landau-Ginzburg model and Khovanov's approach differ by a conjugation corresponding to a transition between canonical and isotypical bases in the tensor power of representations. To perform this transition on the level of complexes one should perform an action of a specific interface we call an \emph{isotypical} interface $\CI$. In what follows we will consider a conjugated $\CR$-interface:
\be
\tilde\CR=\CI^{-1}\boxtimes\CR\boxtimes \CI
\ee
where $\boxtimes$ is a consequent interface $\zeta$-gluing defined in \cite{GMW}, however we will not need the explicit formulation of this operation, instead we will construct direclty the conjugated interface $\tilde{\CR}$ and compare it to Khovanov's construction.

However before following this route one should make a minor change of variables. So far for an unknot we have:
\be
{\bf LGCoh}\left(\begin{array}{c}
	\begin{tikzpicture}
	\draw[ultra thick] (0,0) circle (0.3);
	\end{tikzpicture}
\end{array} \right)=q t^{-1}\; \IZ\oplus q^{-1}t\; \IZ,\quad {\bf KHom}\left(\begin{array}{c}
\begin{tikzpicture}
\draw[ultra thick] (0,0) circle (0.3);
\end{tikzpicture}
\end{array} \right)=q\; \IZ\oplus q^{-1}\; \IZ
\ee
Actually, these homologies are isomorphic to each other; the only thing we should do is to shift $\bf F$-degree to ${\bf F}+{\bf P}$, one can achieve the same result if applies a gauge transformation to the form $\omega\mapsto \omega + dW$, these transforms do not change the complex since the differential (supercharge $Q$) acts only between subspaces of the same $\bf P$-degree and $\bf F$-degrees differing by one. So we change variables $(q,t)\mapsto (qt,t)$ and call this construction $\widetilde{\bf LGCoh}$. By construction ${\bf LGCoh}$ and $\widetilde{\bf LGCoh}$ are isomorphic.

A construction for an isotypical interface is naturally implemented in the 2d CFT limit of the LG model: integrals \eqref{intgrl} give a free field representiation of conformal blocks. In the CFT there is a natural operation of vertex operator fusion mimicking isotypical decomposition \eqref{isotypical}. Fusion implies that we break initial assumption $|c \Delta z|\gg 1$ when consider vacuum classification \eqref{vac_dec}.

\begin{figure}[h!]
	\begin{center}
		\includegraphics[scale=0.5]{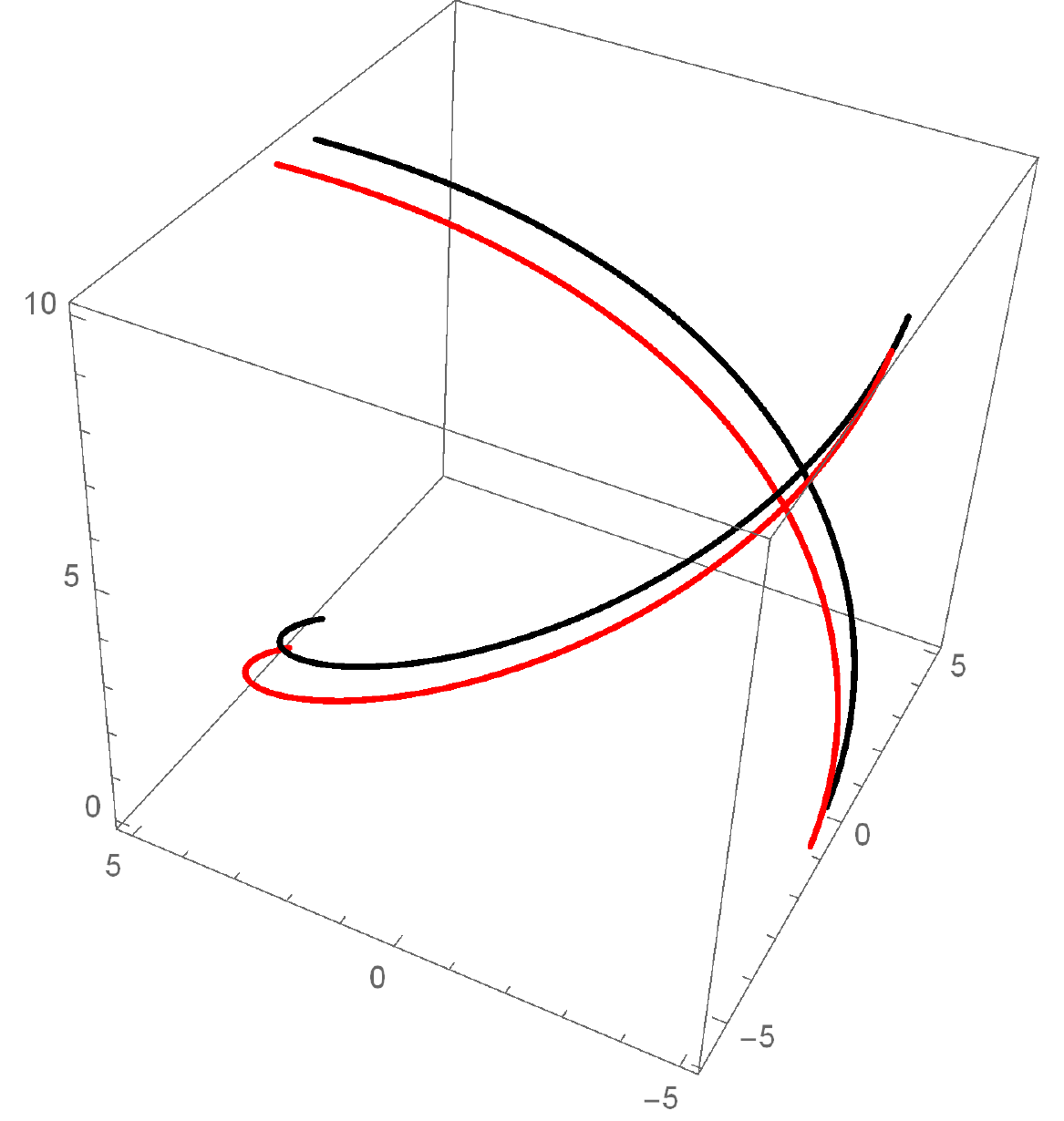}\quad \includegraphics[scale=0.5]{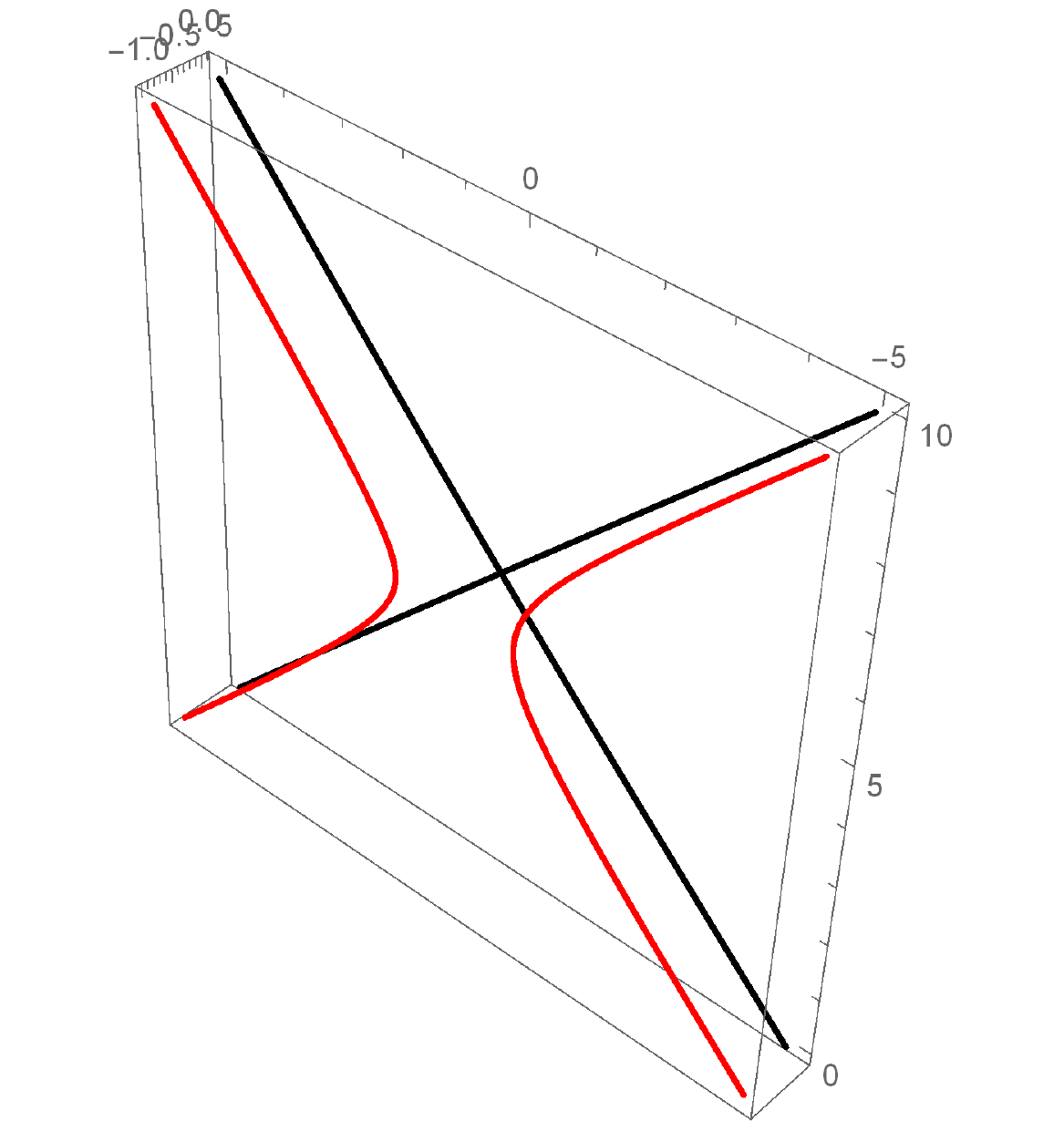}
		\caption{Hovering solution for $++$ vacuum for two interfaces.}\label{Two_int}
	\end{center}
\end{figure}

One can start following the change of the model behavior from the hovering solution in the case of $(+,+)$ vacua (see Figure \ref{Two_int}). In this case the only solution is the hovering one.

In Figure \ref{Two_int} the red trajectories correspond to hovering solutions for $w_i$ while the black trajectories correspond to punctures. It is clear that if one stays above the critical value $|c\Delta z|\sim 1$ the hovering trajectories twist as in \eqref{rules_a}, and if one moves punctures closer the solution behaves as in identity interface in \eqref{Khovan}.

In the sector for $+-$ and $-+$ vacua we should deform the path $\wp$ in such a way that it comes to the puncture closer (notice that during this deformation one slides the path through branching points $\Delta z=\pm \I c^{-1}$, so again a fusion transition of the vacua occurs).

Consider the following diagram:
\be\label{isotyp}
\begin{array}{c}
		\begin{tikzpicture}
		\draw[orange, ultra thick, domain=-0.72:0.72, samples=200]  plot ({1.51*(-3/4 + cos(\x r))*(1+0.05*sin(50*\x r))}, {1.51*(sin(\x r))*(1+0.05*sin(50*\x r))});
		\draw[thick, blue] (0,-1) -- (0,1) (0,-1) to (4,-2.5) (0,1) to (4,2.5);
		\draw[thick, red](-4,-2.5) to (0,-1) (-4,2.5) to (0,1);
		\draw[fill=white] (0,0) circle (0.1);
		\begin{scope}[shift={(0,1)}]
		\draw[purple, ultra thick] (-0.1,-0.1) -- (0.1,0.1) (0.1,-0.1) -- (-0.1,0.1);
		\end{scope}
		\begin{scope}[shift={(0,-1)}]
		\draw[purple, ultra thick] (-0.1,-0.1) -- (0.1,0.1) (0.1,-0.1) -- (-0.1,0.1);
		\end{scope}
		\draw[ultra thick,->] (-1,2.5) to[out=270,in=90] (-2,0) to[out=270,in=90] (-1,-2.5);
		\node[below] at (-1,-2.5) {$\wp_1$};
		\draw[ultra thick,->] (-0.5,2.5) to[out=270,in=135] (-0.5,0) to[out=315,in=90] (0.5,-1) to[out=270,in=90] (0,-1.8) -- (0,-2.5);
		\node[below] at (0,-2.5) {$\wp_3$};
		\draw[ultra thick,->] (0,2.5) -- (0,1.8) to[out=270,in=90] (0.5,1) to [out=270,in=45] (-0.4,0.1) (-0.6,-0.1) to[out=225,in=90] (-0.5,-2.5) ; 
		\node[below] at (-0.5,-2.5) {$\wp_2$};
		\node[left] at (-4,2.5) {$\fd_1^{(12)}$};
		\node[left] at (-4,-2.5) {$\fd_2^{(12)}$};
		\node[right] at (4,2.5) {$\fd_3^{(21)}$};
		\node[right] at (4,-2.5) {$\fd_4^{(21)}$};
		\node[above right] at (0.5,-1) {$\fd_6^{(21)}$};
		\node[below right] at (0.5,1) {$\fd_5^{(21)}$};
		\end{tikzpicture}
\end{array}
\ee
We have just redrawn the diagram depicted in Figure \ref{fig:sc_su_2}: now the cut is shown explicitly, 12-WKB lines are marked by the red color, 21-WKB lines are marked by the blue one. We consider three homotopic paths $\wp_k$. Let us schematically write all the detours for those paths. We denote by $\fd_i^{(ab)}$ a detour due to intersection with the corresponding WKB line starting on the $a^{\rm th}$ sheet and ending on the $b^{\rm th}$ sheet. A path on $\Sigma$ without detours we will denote as $\ell^{(kk)}$, and the sheet permutation due to going through the cut we will denote as $\sigma^{(kl)}$.

Corresponding detours are:
\be\label{su_2_paths}
\begin{split}
	\ID(\wp_1,\zeta)={\color{red} \ell^{(11)}}+{\color{green!30!black}\ell^{(22)}}+{\color{blue}\fd_1^{(12)}}+{\color{purple}\fd_2^{(12)}}\\
	\ID(\wp_2,\zeta)={\color{red}\sigma^{(12)}\fd_5^{(21)}}+ {\color{green!30!black}\fd_3^{(21)}\sigma^{(12)}}+{\color{blue}\sigma^{(12)}}+{\color{purple}\sigma^{(12)}\fd_5^{(21)}\fd_2^{(12)}}+\\ +\left(\sigma^{(21)}+\fd_3^{(21)}\sigma^{(12)}\fd_5^{(21)}\right)+\left(\sigma^{(21)}\fd_2^{(12)}+\fd_3^{(21)}\sigma^{(12)}\fd_5^{(21)}\fd_2^{(12)}\right)\\
	\ID(\wp_3,\zeta)={\color{red}\tilde\sigma^{(12)}\fd_4^{(21)}}+{\color{green!30!black}\fd_6^{(21)}\tilde\sigma^{(12)}}+{\color{blue}\fd_1^{(12)}\fd_6^{(21)}\tilde\sigma^{(12)}}+{\color{purple}\tilde\sigma^{(12)}}+\\+\left(\tilde\sigma^{(21)}+\fd_6^{(21)}\tilde\sigma^{(12)}\fd_4^{(21)}\right)+\left(\fd_1^{(12)}\tilde\sigma^{(21)}+\fd_1^{(12)}\fd_6^{(21)}\tilde\sigma^{(12)}\fd_4^{(21)}\right)
\end{split}
\ee
Here we introduced a color code to emphasise what detours are equivalent paths on $\Sigma$, terms in brackets cancel with each other according to the fake pair argument.

Therefore we see that all the three paths give equivalent interfaces:
\be
H^*(\CE(\wp_1,\zeta),Q)\cong H^*(\CE(\wp_2,\zeta),Q)\cong H^*(\CE(\wp_3,\zeta),Q)
\ee

Neither of paths \eqref{su_2_paths} gives an interface that coincides with the desired one, so we combine some contributions from paths $\wp_2$ and $\wp_3$. We assume that if one starts with the $(+,-)$ vacuum (sheet $1$) then one takes detours form $\wp_3$, otherwise from $\wp_2$, and we take one of terms in the brackets:
\be
\ID(\wp_0,\zeta)={\color{red}\tilde\sigma^{(12)}\fd_4^{(21)}}+{\color{green!30!black}\fd_3^{(21)}\sigma^{(12)}}+{\color{blue}\fd_1^{(12)}\fd_6^{(21)}\tilde\sigma^{(12)}}+{\color{purple}\tilde\sigma^{(12)}}+\left(\sigma^{(21)}+\fd_3^{(21)}\sigma^{(12)}\fd_5^{(21)}\right)
\ee
For these manipulations to be valid we should switch to an equivalence class of complexes, so we say that this interface corresponds to an abstract path $\wp_0$ about what we know only it is homologous to $\wp_1$, $\wp_2$, $\wp_3$.
We should notice that here again there are four contributions and two terms in brackets that can be cancelled by the fake pair argument, hence we have not changed cohomology:
\be
H^*(\CE(\wp_0,\zeta),Q)\cong H^*(\CE(\wp_1,\zeta),Q)
\ee
The definition of the $Q$-charge action does not change, however the complex does change, and it may happen that for some path choice $\wp_0$ the supercharge act locally on the link digram.

All the terms contain permutation $\sigma^{(ij)}$ due to intersection of the cut, this permutation is important since we would like to get an untwisted identity interface like in \eqref{Khovan}, and this permutation exactly undoes the twist for the trajectory of $w$-field. Taking into account all the contributions, rearranging terms, making a substitution $(q,t)\mapsto(qt,t)$ and suppressing an overall $q,t$-monomial factor we arrive to the following interfaces:
\begin{subequations}
\be\label{new_compl}
\begin{split}
	\tilde\CE\left(\begin{array}{c}
		\begin{tikzpicture}
		\begin{scope}[scale=0.8]
		\draw[ultra thick] (-0.5,-0.5) -- (0.5,0.5) (0.5,-0.5)-- (0.1,-0.1) (-0.1,0.1) -- (-0.5,0.5); 
		\end{scope}
		\end{tikzpicture}
	\end{array}\right)=q^{-1}t^{-1}\left(\begin{array}{c}
	\begin{tikzpicture}
	\begin{scope}[scale=0.8]
		\draw[purple,ultra thick] (0,0) -- (0,2) (1,0) -- (1,2);
		\node[below] at (0,0) {$+$}; \node[below] at (1,0) {$+$};
		\node[above] at (0,2) {$+$}; \node[above] at (1,2) {$+$};
	\end{scope}
	\end{tikzpicture}
\end{array}\oplus \begin{array}{c}
\begin{tikzpicture}
\begin{scope}[scale=0.8]
\draw[purple,ultra thick] (0,0) -- (0,2) (1,0) -- (1,2);
\node[below] at (0,0) {$+$}; \node[below] at (1,0) {$-$};
\node[above] at (0,2) {$+$}; \node[above] at (1,2) {$-$};
\end{scope}
\end{tikzpicture}
\end{array}\oplus \begin{array}{c}
\begin{tikzpicture}
\begin{scope}[scale=0.8]
\draw[purple,ultra thick] (0,0) -- (0,2) (1,0) -- (1,2);
\node[below] at (0,0) {$-$}; \node[below] at (1,0) {$+$};
\node[above] at (0,2) {$-$}; \node[above] at (1,2) {$+$};
\end{scope}
\end{tikzpicture}
\end{array}\oplus \begin{array}{c}
\begin{tikzpicture}
\begin{scope}[scale=0.8]
\draw[purple,ultra thick] (0,0) -- (0,2) (1,0) -- (1,2);
\node[below] at (0,0) {$-$}; \node[below] at (1,0) {$-$};
\node[above] at (0,2) {$-$}; \node[above] at (1,2) {$-$};
\end{scope}
\end{tikzpicture}
\end{array}\right)\oplus\\
\oplus\left(
\begin{array}{c}
	\begin{tikzpicture}
	\begin{scope}[scale=0.8]
	\draw[purple,ultra thick] (0,0) -- (0,2) (1,0) -- (1,2);
	\draw[->] (0.6,1.5) -- (1,1.5) (0,1.5) -- (0.6,1.5);
	\node[below] at (0.5,1.5) {$\fd_4$};
	\node[below] at (0,0) {$+$}; \node[below] at (1,0) {$-$};
	\node[above] at (0,2) {$-$}; \node[above] at (1,2) {$+$};
	\end{scope}
	\end{tikzpicture}
\end{array}\oplus 
\begin{array}{c}
	\begin{tikzpicture}
	\begin{scope}[scale=0.8]
	\draw[purple,ultra thick] (0,0) -- (0,2) (1,0) -- (1,2);
	\draw[->] (0.4,0.5) -- (0,0.5) (1,0.5) -- (0.4,0.5);
	\node[above] at (0.5,0.5) {$\fd_3$};
	\node[below] at (0,0) {$-$}; \node[below] at (1,0) {$+$};
	\node[above] at (0,2) {$+$}; \node[above] at (1,2) {$-$};
	\end{scope}
	\end{tikzpicture}
\end{array}\oplus q
\begin{array}{c}
	\begin{tikzpicture}
	\begin{scope}[scale=0.8]
	\draw[purple,ultra thick] (0,0) -- (0,2) (1,0) -- (1,2);
	\draw[->] (0.6,0.5) -- (1,0.5) (0,0.5) -- (0.6,0.5);
	\draw[->] (0.4,1) -- (0,1) (1,1) -- (0.4,1); 
	\node[above] at (0.5,1) {$\fd_6$};
	\node[below] at (0.5,0.5) {$\fd_1$};
	\node[below] at (0,0) {$+$}; \node[below] at (1,0) {$-$};
	\node[above] at (0,2) {$+$}; \node[above] at (1,2) {$-$};
	\end{scope}
	\end{tikzpicture}
\end{array}
\oplus q^{-1}
\begin{array}{c}
	\begin{tikzpicture}
	\begin{scope}[scale=0.8]
	\draw[purple,ultra thick] (0,0) -- (0,2) (1,0) -- (1,2);
	\draw[->] (0.4,0.5) -- (0,0.5) (1,0.5) -- (0.4,0.5);
	\draw[->] (0.6,1) -- (1,1) (0,1) -- (0.6,1); 
	\node[above] at (0.5,1) {$\fd_5$};
	\node[below] at (0.5,0.5) {$\fd_3$};
	\node[below] at (0,0) {$-$}; \node[below] at (1,0) {$+$};
	\node[above] at (0,2) {$-$}; \node[above] at (1,2) {$+$};
	\end{scope}
	\end{tikzpicture}
\end{array}
\right)
\end{split}
\\
\tilde\CE\left(\begin{array}{c}
	\begin{tikzpicture}
	\begin{scope}[scale=0.8,yscale=-1]
	\draw[ultra thick] (0,1) to[out=270,in=180] (0.5,0) to[out=0,in=270] (1,1);
	\end{scope}
	\end{tikzpicture}
\end{array}\right)=q^{-\frac{1}{2}}\begin{array}{c}
\begin{tikzpicture}
\begin{scope}[scale=0.8]
\draw[ultra thick, purple] (-0.25,1) -- (1.25,1) (0,0) -- (0,1) (1,0) -- (1,1);
\node[below] at (0,0) {$-$};
\node[above] at (0,1) {$-$};
\node[below] at (1,0) {$+$};
\node[above] at (1,1) {$+$};
\end{scope}
\end{tikzpicture}
\end{array}\oplus q^{\frac{1}{2}}\begin{array}{c}
\begin{tikzpicture}
\begin{scope}[scale=0.8]
\draw[->] (0.6,0.5) -- (1,0.5) (0,0.5) -- (0.6,0.5);
\node[below] at (0.5,0.5) {$12$};
\draw[ultra thick, purple] (-0.25,1) -- (1.25,1) (0,0) -- (0,1) (1,0) -- (1,1);
\node[below] at (0,0) {$+$};
\node[above] at (0,1) {$-$};
\node[below] at (1,0) {$-$};
\node[above] at (1,1) {$+$};
\end{scope}
\end{tikzpicture}
\end{array}, \quad
 \CE\left(\begin{array}{c}
\begin{tikzpicture}
\begin{scope}[scale=0.8]
\draw[ultra thick] (0,1) to[out=270,in=180] (0.5,0) to[out=0,in=270] (1,1);
\end{scope}
\end{tikzpicture}
\end{array}\right)= q^{-\frac{1}{2}}\begin{array}{c}
\begin{tikzpicture}
\begin{scope}[scale=0.8]
\draw[->] (0.6,0.5) -- (1,0.5) (0,0.5) -- (0.6,0.5);
\node[above] at (0.5,0.5) {$21$};
\draw[ultra thick, purple] (-0.25,0) -- (1.25,0) (0,0) -- (0,1) (1,0) -- (1,1);
\node[below] at (0,0) {$+$};
\node[above] at (0,1) {$-$};
\node[below] at (1,0) {$-$};
\node[above] at (1,1) {$+$};
\end{scope}
\end{tikzpicture}
\end{array}\oplus q^{\frac{1}{2}}\begin{array}{c}
\begin{tikzpicture}
\begin{scope}[scale=0.8]
\draw[ultra thick, purple] (-0.25,0) -- (1.25,0) (0,0) -- (0,1) (1,0) -- (1,1);
\node[below] at (0,0) {$+$};
\node[above] at (0,1) {$+$};
\node[below] at (1,0) {$-$};
\node[above] at (1,1) {$-$};
\end{scope}
\end{tikzpicture}
\end{array}\label{new_compl_cc}
\ee
\end{subequations}

\subsection{Comparison of Landau-Ginzburg cohomology and Khovanov homology}

Now we would like to compare complexes following from \eqref{Khovan} and \eqref{new_compl}. We will argue that \eqref{new_compl} is equivalent to \eqref{Khovan} in two steps:
\begin{enumerate}
	\item We show that 
	\be\label{item1}
	\tilde\CE\left[
	\begin{array}{c}
		\begin{tikzpicture}[scale=0.5]
		\draw[ultra thick] (-0.5,-0.5) -- (0.5,0.5);
		\draw[ultra thick] (0.5,-0.5) -- (0.1,-0.1);
		\draw[ultra thick] (-0.1,0.1) -- (-0.5,0.5);
		\end{tikzpicture}
	\end{array}\right]=[q^{-1}t^{-1}] \; \tilde\CE\left[
	\begin{array}{c}
		\begin{tikzpicture}[scale=0.5]
		\draw [ultra thick] (-0.5,-0.5) to [out= 45, in =315] (-0.5,0.5) (0.5,-0.5) to [out=135, in=225] (0.5,0.5);
		\end{tikzpicture}
	\end{array}\right]\oplus \tilde\CE\left[
	\begin{array}{c}
		\begin{tikzpicture}[scale=0.5]
		\draw [ultra thick] (-0.5,-0.5) to [out= 45, in =135] (0.5,-0.5) (-0.5,0.5) to [out=315, in=225] (0.5,0.5);
		\end{tikzpicture}
	\end{array}
	\right]
	\ee
	as in \eqref{Khovan}
	\item We show that the differential $Q$ can be represented as
	\be\label{item2}
	Q=\sum\lm_{{\rm all}\;\chi} Q_{\chi}
	\ee
	where $\chi$ are intersections in the link diagram, moreover $Q_{\chi}$ acts on intersection $\chi$ by gluing or cutting cycles in the resolution as in \eqref{mul} and \eqref{co-mul}
\end{enumerate}

\subsubsection{Checking item 1.}

Obviously, the first four terms in brackets in \eqref{new_compl} form an identity interface, so the question is why the next four term in the brackets form a pair of a cup and a cap. Let us for brevity denote these terms as $\begin{array}{c}
\begin{tikzpicture}[scale=0.5]
\draw[->, thick] (0,-0.2) -- (0,0.3); 
\draw[fill=white] (-0.1,-0.3) -- (-0.1,0.3) -- (0.1,0.3) -- (0.1,-0.3) -- cycle;
\draw [ultra thick] (-0.5,-0.5) to [out= 45, in =135] (0.5,-0.5) (-0.5,0.5) to [out=315, in=225] (0.5,0.5);
\end{tikzpicture}
\end{array}$

Then one can use a ``hump" move:
\be \label{hump}
\begin{array}{c}
\begin{tikzpicture}
\begin{scope}[xscale=-1]
\draw[ultra thick] (0,1) -- (0,0.5) to[out=270, in=270] (0.5,0.5) to[out=90,in=90] (1,0.5) -- (1,0);
\end{scope}
\end{tikzpicture}
\end{array}=
\begin{array}{c}
\begin{tikzpicture}
\draw[ultra thick] (0,0) -- (0,1);
\end{tikzpicture}
\end{array}= \begin{array}{c}
\begin{tikzpicture}
\draw[ultra thick] (0,1) -- (0,0.5) to[out=270, in=270] (0.5,0.5) to[out=90,in=90] (1,0.5) -- (1,0);
\end{tikzpicture}
\end{array}
\ee
and consider the following diagram identities:
\be
\begin{array}{c}
	\begin{tikzpicture}[scale=0.8]
	\draw[->, thick] (0,-0.2) -- (0,0.3); 
	\draw[fill=white] (-0.065,-0.3) -- (-0.065,0.3) -- (0.065,0.3) -- (0.065,-0.3) -- cycle;
	\draw [ultra thick] (-0.8,-1) to[out=90,in=45] (-0.5,-0.5) to [out= 45, in =135] (0.5,-0.5) to[out=315,in=225] (1,-0.5) to[out=45,in=270] (1.4,1) (-1.4,-1) to[out=90,in=225] (-1,0.5) to[out=45,in=135] (-0.5,0.5) to [out=315, in=225] (0.5,0.5) to[out=45,in=270] (0.8,1);
	\end{tikzpicture}
\end{array}=\begin{array}{c}
\begin{tikzpicture}
\begin{scope}[scale=0.8]
\draw[thick, ->] (0.5,0) -- (-0.5,0);
\draw[fill=white] (-0.5,0.065) -- (0.5,0.065) -- (0.5,-0.065) -- (-0.5,-0.065) -- cycle;
\draw[ultra thick] (-1,-1) to[out=45,in=270] (-0.5,0) to[out=90,in=315] (-1,1) (1,-1) to[out=135,in=270] (0.5,0) to[out=90,in=225] (1,1);
\end{scope}
\end{tikzpicture}
\end{array}\mathop{=}\lm^{?}	\begin{array}{c}
\begin{tikzpicture}[scale=0.5]
\draw [ultra thick] (-0.5,-0.5) to [out= 45, in =315] (-0.5,0.5) (0.5,-0.5) to [out=135, in=225] (0.5,0.5);
\end{tikzpicture}
\end{array}\label{chain}
\ee
Thus to confirm item 1 in the list it is enough to confirm equivalence between initial and final terms in chain \eqref{chain}.

Using expressions \eqref{new_compl}, \eqref{new_compl_cc} one derives
\be
\tilde{\CE}\left[\begin{array}{c}
	\begin{tikzpicture}[scale=0.5]
	\draw[->, thick] (0,-0.2) -- (0,0.3); 
	\draw[fill=white] (-0.1,-0.3) -- (-0.1,0.3) -- (0.1,0.3) -- (0.1,-0.3) -- cycle;
	\draw [ultra thick] (-0.8,-1) to[out=90,in=45] (-0.5,-0.5) to [out= 45, in =135] (0.5,-0.5) to[out=315,in=225] (1,-0.5) to[out=45,in=270] (1.4,1) (-1.4,-1) to[out=90,in=225] (-1,0.5) to[out=45,in=135] (-0.5,0.5) to [out=315, in=225] (0.5,0.5) to[out=45,in=270] (0.8,1);
	\end{tikzpicture}
\end{array}\right]=
\begin{array}{c}
\begin{tikzpicture}
\begin{scope}[scale=0.8]
\begin{scope}[shift={(2,0)}]
\draw[->] (0.6,0.5) -- (1,0.5) (0,0.5) -- (0.6,0.5);
\node[above] at (0.5,0.5) {$\tilde{21}$};
\end{scope}
\begin{scope}[shift={(0,3)}]
\draw[->] (0.6,0.5) -- (1,0.5) (0,0.5) -- (0.6,0.5);
\node[below] at (0.5,0.5) {$12$};
\end{scope}
\begin{scope}[shift={(1,1)}]
\draw[->] (0.6,1.5) -- (1,1.5) (0,1.5) -- (0.6,1.5);
\node[below] at (0.5,1.5) {$\fd_4$};
\end{scope}
\draw[ultra thick, purple] (0,0) -- (0,4) (1,0) -- (1,4) (-0.25,4) -- (1.25,4) (2,0) -- (2,4) (3,0) -- (3,4) (1.75,0) -- (3.25,0);
\node[below] at (2,0) {$+$}; \node[below] at (3,0) {$-$};
\node[above] at (0,4) {$-$}; \node[above] at (1,4) {$+$};
\node[below] at (0,0) {$+$}; \node[below] at (1,0) {$+$};
\node[above] at (2,4) {$+$}; \node[above] at (3,4) {$+$};
\end{scope}
\end{tikzpicture}
\end{array}
\oplus
\begin{array}{c}
	\begin{tikzpicture}
	\begin{scope}[scale=0.8]
	\begin{scope}[shift={(1,1)}]
	\draw[->] (0.4,0.5) -- (0,0.5) (1,0.5) -- (0.4,0.5);
	\node[above] at (0.5,0.5) {$\fd_3$};
	\end{scope}
	\draw[ultra thick, purple] (0,0) -- (0,4) (1,0) -- (1,4) (-0.25,4) -- (1.25,4) (2,0) -- (2,4) (3,0) -- (3,4) (1.75,0) -- (3.25,0);
	\node[below] at (2,0) {$+$}; \node[below] at (3,0) {$-$};
	\node[above] at (0,4) {$-$}; \node[above] at (1,4) {$+$};
	\node[below] at (0,0) {$-$}; \node[below] at (1,0) {$-$};
	\node[above] at (2,4) {$-$}; \node[above] at (3,4) {$-$};
	\end{scope}
	\end{tikzpicture}
\end{array}
\oplus
\begin{array}{c}
	\begin{tikzpicture}
	\begin{scope}[scale=0.8]
	\begin{scope}[shift={(2,0)}]
	\draw[->] (0.6,0.5) -- (1,0.5) (0,0.5) -- (0.6,0.5);
	\node[above] at (0.5,0.5) {$\tilde{21}$};
	\end{scope}
	\begin{scope}[shift={(1,1)}]
	\draw[->] (0.6,0.5) -- (1,0.5) (0,0.5) -- (0.6,0.5);
	\draw[->] (0.4,1) -- (0,1) (1,1) -- (0.4,1); 
	\node[above] at (0.5,1) {$\fd_6$};
	\node[below] at (0.5,0.5) {$\fd_1$};
	\end{scope}
	\draw[ultra thick, purple] (0,0) -- (0,4) (1,0) -- (1,4) (-0.25,4) -- (1.25,4) (2,0) -- (2,4) (3,0) -- (3,4) (1.75,0) -- (3.25,0);
	\node[below] at (2,0) {$+$}; \node[below] at (3,0) {$-$};
	\node[above] at (0,4) {$-$}; \node[above] at (1,4) {$+$};
	\node[below] at (0,0) {$-$}; \node[below] at (1,0) {$+$};
	\node[above] at (2,4) {$-$}; \node[above] at (3,4) {$+$};
	\end{scope}
	\end{tikzpicture}
\end{array}
\oplus
\begin{array}{c}
	\begin{tikzpicture}
	\begin{scope}[scale=0.8]
	\begin{scope}[shift={(0,3)}]
	\draw[->] (0.6,0.5) -- (1,0.5) (0,0.5) -- (0.6,0.5);
	\node[below] at (0.5,0.5) {$12$};
	\end{scope}
	\begin{scope}[shift={(1,1)}]
	\draw[->] (0.4,0.5) -- (0,0.5) (1,0.5) -- (0.4,0.5);
	\draw[->] (0.6,1) -- (1,1) (0,1) -- (0.6,1); 
	\node[above] at (0.5,1) {$\fd_5$};
	\node[below] at (0.5,0.5) {$\fd_3$};
	\end{scope}
	\draw[ultra thick, purple] (0,0) -- (0,4) (1,0) -- (1,4) (-0.25,4) -- (1.25,4) (2,0) -- (2,4) (3,0) -- (3,4) (1.75,0) -- (3.25,0);
	\node[below] at (2,0) {$+$}; \node[below] at (3,0) {$-$};
	\node[above] at (0,4) {$-$}; \node[above] at (1,4) {$+$};
	\node[below] at (0,0) {$+$}; \node[below] at (1,0) {$-$};
	\node[above] at (2,4) {$+$}; \node[above] at (3,4) {$-$};
	\end{scope}
	\end{tikzpicture}
\end{array}
\ee
If one considers these contributions closer one observes these are exactly four contributions of an identity interface for $(+,+)$, $(-,-)$, $(-,+)$ and $(+,-)$ vacua correspondingly. One can modify the set of detours in each case to a simple ``empty" detour on surface $\Sigma$. Let us demonstrate this satement in the case of $(+,+)$ vacuum. We modify the curve and corresponding detours as follows:
\be
\begin{array}{c}
\begin{tikzpicture}
\begin{scope}[shift={(0.5,-0.5)}]
\begin{scope}[rotate=-90]
\draw [orange, domain=0:1] plot (\x, {0.1*sin(20*\x r)});
\end{scope};
\end{scope};
\begin{scope}[shift={(2,-0.5)}]
\begin{scope}[rotate=90]
\draw [orange, domain=0:1] plot (\x, {0.1*sin(20*\x r)});
\end{scope};
\end{scope};
\begin{scope}[shift={(3.5,0.5)}]
\begin{scope}[rotate=90]
\draw [orange, domain=0:1] plot (\x, {0.1*sin(20*\x r)});
\end{scope};
\end{scope};
\draw[thick] (0.5,-0.5) -- (0.5,1);
\draw[thick] (3.5,0.5) -- (3.5,-1);
\draw[thick] (1,0.5) -- (2,-0.5) -- (3,0.5);
\begin{scope}[shift={(0.5,-0.5)}]
\draw[purple, ultra thick] (-0.1,-0.1) -- (0.1,0.1) (0.1,-0.1) -- (-0.1,0.1);
\end{scope}
\begin{scope}[shift={(2,-0.5)}]
\draw[purple, ultra thick] (-0.1,-0.1) -- (0.1,0.1) (0.1,-0.1) -- (-0.1,0.1);
\end{scope}
\begin{scope}[shift={(3.5,0.5)}]
\draw[purple, ultra thick] (-0.1,-0.1) -- (0.1,0.1) (0.1,-0.1) -- (-0.1,0.1);
\end{scope}
\draw[ultra thick, ->] (0,0) -- (4,0); \node[right] at (4,0) {$\wp$};
\draw[thick] (0.5,0) to[out=240,in=90] (0.2,-0.5) to[out=270,in=270] (0.8,-0.5) to[out=90,in=315] (0.5,0) (2.5,0) to[out=202.5,in=90] (1.8,-0.5) to[out=270,in=270] (2.3,-0.5) to[out=90,in=247.5] (2.5,0) (3.5,0) to[out=135,in=270] (3.2,0.5) to[out=90,in=90] (3.8,0.5) to[out=270,in=45] (3.5,0);
\node[above] at (0.5,1) {$\tilde{21}$};
\node[above] at (1,0.5) {$\fd_6$};
\node[above] at (3,0.5) {$\fd_4$};
\node[below] at (3.5,-1) {$12$};
\end{tikzpicture}
\end{array}\rightsquigarrow
\begin{array}{c}
\begin{tikzpicture}
\begin{scope}[shift={(0.5,0.5)}]
\begin{scope}[rotate=-90]
\draw [orange, domain=0:1] plot (\x, {0.1*sin(20*\x r)});
\end{scope};
\end{scope};
\begin{scope}[shift={(2,0.5)}]
\begin{scope}[rotate=90]
\draw [orange, domain=0:1] plot (\x, {0.1*sin(20*\x r)});
\end{scope};
\end{scope};
\begin{scope}[shift={(3.5,-0.5)}]
\begin{scope}[rotate=90]
\draw [orange, domain=0:1] plot (\x, {0.1*sin(20*\x r)});
\end{scope};
\end{scope};
\draw[ultra thick, ->] (0,0) -- (4,0); \node[right] at (4,0) {$\wp$};
\begin{scope}[shift={(0.5,0.5)}]
\draw[purple, ultra thick] (-0.1,-0.1) -- (0.1,0.1) (0.1,-0.1) -- (-0.1,0.1);
\end{scope}
\begin{scope}[shift={(2,0.5)}]
\draw[purple, ultra thick] (-0.1,-0.1) -- (0.1,0.1) (0.1,-0.1) -- (-0.1,0.1);
\end{scope}
\begin{scope}[shift={(3.5,-0.5)}]
\draw[purple, ultra thick] (-0.1,-0.1) -- (0.1,0.1) (0.1,-0.1) -- (-0.1,0.1);
\end{scope}
\end{tikzpicture}
\end{array}
\ee 
So the path $\wp$ gives a single ``empty" detour - a lift of $\wp$ to corresponding sheet. $\wp$ intersects two cuts. The first cut gives a permutation $(+,+,+,-)\to(+,+,-,+)$, then the hovering solution corresponds to a twist $(+,+,-,+)\to(+,-,+,+)$, and, finally, another cut maps $(+,-,+,+)\to(-,+,+,+)$. This detour is equivalent to a hovering solution corresponding to $(+,+)$ vacuum. Similarly, one may check that any instanton for an identity interface corresponds to an instanton solution for this modified interface using an implant argument from \cite{GM}.

\subsubsection{Checking item 2}

To check the second item in our list we will just construct the action of the local differentials $Q_{\chi}$ and compare it with rules \eqref{mul} and \eqref{co-mul}.

The first step one should take is to assign vectors $v_+$ and $v_-$ to two types of cycles. This can be easily done by consideration of the unknot, it is given exactly by two contributions with $\bf P$-degrees $\pm 1$, using \eqref{new_compl} one has
\be
v_+=\begin{array}{c}
\begin{tikzpicture}[scale=0.7]
\draw [ultra thick] (0,0) circle (0.5);
\node[left] at (-0.5,0) {$+$};
\node[right] at (0.5,0) {$-$};
\end{tikzpicture}
\end{array},\quad
v_-=\begin{array}{c}
	\begin{tikzpicture}[scale=0.7]
	\draw [ultra thick] (0,0) circle (0.5);
	\node[left] at (-0.5,0) {$-$};
	\node[right] at (0.5,0) {$+$};
	\end{tikzpicture}
\end{array}
\ee

Internal relations between between link diagrams preserving cohomology: a ``hump" move, and a ``tri-dent" move:
\be\label{trident}
\begin{array}{c}
\begin{tikzpicture}
\begin{scope}[scale=0.5]
\draw[ultra thick] (1,0) -- (1,2) (0,2) to[out=270,in=135] (0.8,1.2) (1.2,0.8) to[out=315,in=180] (1.75,0) to[out=0,in=270] (2,2);
\end{scope}
\end{tikzpicture}
\end{array}= \begin{array}{c}
\begin{tikzpicture}
\begin{scope}[xscale=-1]
\begin{scope}[scale=0.5]
\draw[ultra thick] (1,0) -- (1,2) (0,2) to[out=270,in=135] (0.8,1.2) (1.2,0.8) to[out=315,in=180] (1.75,0) to[out=0,in=270] (2,2);
\end{scope}
\end{scope}
\end{tikzpicture}
\end{array}
\ee
allow one to check \eqref{mul} and \eqref{co-mul} only for a twist (an inverse twist check follows) and in the simplest setting when cycles are represented by circles. Now we just present a form of the maps in terms of soliton link diagrams and depict null-webs responsible for these maps:
\begin{itemize}
	\item Multiplication: 
	\begin{itemize}
		\item  $v_+\otimes v_-\mapsto v_-$: in this case the null-web appears in descendant WKB lines (we denote a new line as $\ell$):
		$$
		\begin{array}{c}
		\begin{tikzpicture}
		\begin{scope}[scale=0.8]
		\node at (1,1.25) {-}; \node at (2,1.25) {-};
		\node at (1,3.75) {-}; \node at (2,3.75) {-};
		\begin{scope}[shift={(0,4)}]
		\draw[->] (0.6,0.5) -- (1,0.5) (0,0.5) -- (0.6,0.5);
		\node[below] at (0.5,0.5) {$12$};
		\end{scope}
		\begin{scope}[shift={(2,0)}]
		\draw[->] (0.6,0.5) -- (1,0.5) (0,0.5) -- (0.6,0.5);
		\node[above] at (0.5,0.5) {$\tilde{21}$};
		\end{scope}
		\draw[purple, ultra thick] (0,0) -- (0,5) (3,0) -- (3,5) (1,0) -- (1,1) (2,0) -- (2,1) (1,5) -- (1,4) (2,5) -- (2,4) (1,1.5) -- (1,3.5) (2,1.5) -- (2,3.5) (-0.25,0) -- (1.25,0) (1.75,0) -- (3.25,0) (-0.25,5) -- (1.25,5) (1.75,5) -- (3.25,5); 
		\node[below] at (0,0) {$+$}; \node[below] at (1,0) {$-$};
		\node[below] at (2,0) {$+$}; \node[below] at (3,0) {$-$};
		\node[above] at (0,5) {$-$}; \node[above] at (1,5) {$+$};
		\node[above] at (2,5) {$-$}; \node[above] at (3,5) {$+$};
		\end{scope}
		\end{tikzpicture}
		\end{array}\mapsto
		%
		%
		\begin{array}{c}
		\begin{tikzpicture}
		\begin{scope}[scale=0.8]
		\node at (1,1.25) {+}; \node at (2,1.25) {-};
		\node at (1,3.75) {+}; \node at (2,3.75) {-};
		\begin{scope}[shift={(1,1.5)}]
		\draw[->] (0.6,0.5) -- (1,0.5) (0,0.5) -- (0.6,0.5);
		\draw[->] (0.4,1) -- (0,1) (1,1) -- (0.4,1); 
		\node[above] at (0.5,1) {$\fd_6$};
		\node[below] at (0.5,0.5) {$\fd_1$};
		\end{scope}
		\begin{scope}[shift={(0,4)}]
		\end{scope}
		\begin{scope}[shift={(0,0)}]
		\draw[->] (0.6,0.5) -- (1,0.5) (0,0.5) -- (0.6,0.5);
		\node[above] at (0.5,0.5) {$21$};
		\end{scope}
		\begin{scope}[shift={(2,0)}]
		\draw[->] (0.6,0.5) -- (1,0.5) (0,0.5) -- (0.6,0.5);
		\node[above] at (0.5,0.5) {$\tilde{21}$};
		\end{scope}
		\draw[purple, ultra thick] (0,0) -- (0,5) (3,0) -- (3,5) (1,0) -- (1,1) (2,0) -- (2,1) (1,5) -- (1,4) (2,5) -- (2,4) (1,1.5) -- (1,3.5) (2,1.5) -- (2,3.5) (-0.25,0) -- (1.25,0) (1.75,0) -- (3.25,0) (-0.25,5) -- (1.25,5) (1.75,5) -- (3.25,5); 
		\node[below] at (0,0) {$+$}; \node[below] at (1,0) {$-$};
		\node[below] at (2,0) {$+$}; \node[below] at (3,0) {$-$};
		\node[above] at (0,5) {$-$}; \node[above] at (1,5) {$+$};
		\node[above] at (2,5) {$-$}; \node[above] at (3,5) {$+$};
		\end{scope}
		\end{tikzpicture}
		\end{array}\quad\quad
		%
		%
		\IW=\begin{array}{c}
		\begin{tikzpicture}
		\draw[thick] (0,0) -- (1,0) (0,0) -- (-0.5,1.5) (0,0) -- (-0.5,-1.5) (1,0) -- (2,-1) (1,0) -- (1.5,1.5) (-0.5,1.5) -- (1,0);
		\draw[fill=white] (0.5,0.5) circle (0.07);
		\begin{scope}[shift={(-0.5,1.5)}]
		\draw[purple, ultra thick] (-0.1,-0.1) -- (0.1,0.1) (0.1,-0.1) -- (-0.1,0.1);
		\end{scope}
		\begin{scope}[shift={(-0.5,-1.5)}]
		\draw[purple, ultra thick] (-0.1,-0.1) -- (0.1,0.1) (0.1,-0.1) -- (-0.1,0.1);
		\end{scope}
		\begin{scope}[shift={(1.5,1.5)}]
		\draw[purple, ultra thick] (-0.1,-0.1) -- (0.1,0.1) (0.1,-0.1) -- (-0.1,0.1);
		\end{scope}
		\begin{scope}[shift={(2,-1)}]
		\draw[purple, ultra thick] (-0.1,-0.1) -- (0.1,0.1) (0.1,-0.1) -- (-0.1,0.1);
		\end{scope}
		\node[left] at (-0.25,-0.75) {$21$};
		\node[left] at (-0.25,0.75) {$\fd_1$};
		\node[right] at (1.25,0.75) {$12$};
		\node[right] at (1.5,-0.5) {$\fd_6$};
		\node[below] at (0.5,0) {$\ell$};
		\end{tikzpicture}
		\end{array}
		$$
		\item $v_-\otimes v_+\mapsto v_-$:  in this case the null-web appears in descendant WKB lines (we denote a new line as $\ell$)
		$$
		\begin{array}{c}
		\begin{tikzpicture}
		\begin{scope}[scale=0.8]
		\node at (1,1.25) {+}; \node at (2,1.25) {+};
		\node at (1,3.75) {+}; \node at (2,3.75) {+};
		\begin{scope}[shift={(2,4)}]
		\draw[->] (0.6,0.5) -- (1,0.5) (0,0.5) -- (0.6,0.5);
		\node[below] at (0.5,0.5) {$\tilde{12}$};
		\end{scope}
		\begin{scope}[shift={(0,0)}]
		\draw[->] (0.6,0.5) -- (1,0.5) (0,0.5) -- (0.6,0.5);
		\node[above] at (0.5,0.5) {$21$};
		\end{scope}
		\draw[purple, ultra thick] (0,0) -- (0,5) (3,0) -- (3,5) (1,0) -- (1,1) (2,0) -- (2,1) (1,5) -- (1,4) (2,5) -- (2,4) (1,1.5) -- (1,3.5) (2,1.5) -- (2,3.5) (-0.25,0) -- (1.25,0) (1.75,0) -- (3.25,0) (-0.25,5) -- (1.25,5) (1.75,5) -- (3.25,5); 
		\node[below] at (0,0) {$+$}; \node[below] at (1,0) {$-$};
		\node[below] at (2,0) {$+$}; \node[below] at (3,0) {$-$};
		\node[above] at (0,5) {$-$}; \node[above] at (1,5) {$+$};
		\node[above] at (2,5) {$-$}; \node[above] at (3,5) {$+$};
		\end{scope}
		\end{tikzpicture}
		\end{array}\mapsto
		%
		%
		\begin{array}{c}
		\begin{tikzpicture}
		\begin{scope}[scale=0.8]
		\node at (1,1.25) {+}; \node at (2,1.25) {-};
		\node at (1,3.75) {+}; \node at (2,3.75) {-};
		\begin{scope}[shift={(1,1.5)}]
		\draw[->] (0.6,0.5) -- (1,0.5) (0,0.5) -- (0.6,0.5);
		\draw[->] (0.4,1) -- (0,1) (1,1) -- (0.4,1); 
		\node[above] at (0.5,1) {$\fd_6$};
		\node[below] at (0.5,0.5) {$\fd_1$};
		\end{scope}
		\begin{scope}[shift={(0,4)}]
		\end{scope}
		\begin{scope}[shift={(0,0)}]
		\draw[->] (0.6,0.5) -- (1,0.5) (0,0.5) -- (0.6,0.5);
		\node[above] at (0.5,0.5) {$21$};
		\end{scope}
		\begin{scope}[shift={(2,0)}]
		\draw[->] (0.6,0.5) -- (1,0.5) (0,0.5) -- (0.6,0.5);
		\node[above] at (0.5,0.5) {$\tilde{21}$};
		\end{scope}
		\draw[purple, ultra thick] (0,0) -- (0,5) (3,0) -- (3,5) (1,0) -- (1,1) (2,0) -- (2,1) (1,5) -- (1,4) (2,5) -- (2,4) (1,1.5) -- (1,3.5) (2,1.5) -- (2,3.5) (-0.25,0) -- (1.25,0) (1.75,0) -- (3.25,0) (-0.25,5) -- (1.25,5) (1.75,5) -- (3.25,5); 
		\node[below] at (0,0) {$+$}; \node[below] at (1,0) {$-$};
		\node[below] at (2,0) {$+$}; \node[below] at (3,0) {$-$};
		\node[above] at (0,5) {$-$}; \node[above] at (1,5) {$+$};
		\node[above] at (2,5) {$-$}; \node[above] at (3,5) {$+$};
		\end{scope}
		\end{tikzpicture}
		\end{array}\quad\quad
		%
		%
		\IW=\begin{array}{c}
		\begin{tikzpicture}
		\draw[thick] (0,0) -- (1,0) (0,0) -- (-0.5,1.5) (0,0) -- (-0.5,-1.5) (1,0) -- (2,-1) (1,0) -- (1.5,1.5) (-0.5,1.5) -- (1,0);
		\draw[fill=white] (0.5,0.5) circle (0.07);
		\begin{scope}[shift={(-0.5,1.5)}]
		\draw[purple, ultra thick] (-0.1,-0.1) -- (0.1,0.1) (0.1,-0.1) -- (-0.1,0.1);
		\end{scope}
		\begin{scope}[shift={(-0.5,-1.5)}]
		\draw[purple, ultra thick] (-0.1,-0.1) -- (0.1,0.1) (0.1,-0.1) -- (-0.1,0.1);
		\end{scope}
		\begin{scope}[shift={(1.5,1.5)}]
		\draw[purple, ultra thick] (-0.1,-0.1) -- (0.1,0.1) (0.1,-0.1) -- (-0.1,0.1);
		\end{scope}
		\begin{scope}[shift={(2,-1)}]
		\draw[purple, ultra thick] (-0.1,-0.1) -- (0.1,0.1) (0.1,-0.1) -- (-0.1,0.1);
		\end{scope}
		\node[left] at (-0.25,-0.75) {$\tilde{21}$};
		\node[left] at (-0.25,0.75) {$\fd_1$};
		\node[right] at (1.25,0.75) {$\tilde{12}$};
		\node[right] at (1.5,-0.5) {$\fd_6$};
		\node[below] at (0.5,0) {$\ell$};
		\end{tikzpicture}
		\end{array}
		$$
		\item $v_+\otimes v_+\mapsto v_+$: in this case the null-web is the simplest one described in the end of section \ref{sec:LG-coh-SN}:
		$$
		\begin{array}{c}
		\begin{tikzpicture}
		\begin{scope}[scale=0.8]
		\node at (1,1.25) {-}; \node at (2,1.25) {+};
		\node at (1,3.75) {-}; \node at (2,3.75) {+};
		\begin{scope}[shift={(2,4)}]
		\draw[->] (0.6,0.5) -- (1,0.5) (0,0.5) -- (0.6,0.5);
		\node[below] at (0.5,0.5) {$\tilde{12}$};
		\end{scope}
		\begin{scope}[shift={(0,4)}]
		\draw[->] (0.6,0.5) -- (1,0.5) (0,0.5) -- (0.6,0.5);
		\node[below] at (0.5,0.5) {$12$};
		\end{scope}
		\draw[purple, ultra thick] (0,0) -- (0,5) (3,0) -- (3,5) (1,0) -- (1,1) (2,0) -- (2,1) (1,5) -- (1,4) (2,5) -- (2,4) (1,1.5) -- (1,3.5) (2,1.5) -- (2,3.5) (-0.25,0) -- (1.25,0) (1.75,0) -- (3.25,0) (-0.25,5) -- (1.25,5) (1.75,5) -- (3.25,5); 
		\node[below] at (0,0) {$+$}; \node[below] at (1,0) {$-$};
		\node[below] at (2,0) {$+$}; \node[below] at (3,0) {$-$};
		\node[above] at (0,5) {$-$}; \node[above] at (1,5) {$+$};
		\node[above] at (2,5) {$-$}; \node[above] at (3,5) {$+$};
		\end{scope}
		\end{tikzpicture}
		\end{array}\mapsto
		%
		%
		\begin{array}{c}
		\begin{tikzpicture}
		\begin{scope}[scale=0.8]
		\node at (1,1.25) {-}; \node at (2,1.25) {+};
		\node at (1,3.75) {-}; \node at (2,3.75) {+};
		\begin{scope}[shift={(1,1.5)}]
		\draw[->] (0.4,0.5) -- (0,0.5) (1,0.5) -- (0.4,0.5);
		\draw[->] (0.6,1) -- (1,1) (0,1) -- (0.6,1); 
		\node[above] at (0.5,1) {$\fd_5$};
		\node[below] at (0.5,0.5) {$\fd_3$};
		\end{scope}
		\begin{scope}[shift={(2,4)}]
		\draw[->] (0.6,0.5) -- (1,0.5) (0,0.5) -- (0.6,0.5);
		\node[below] at (0.5,0.5) {$\tilde{12}$};
		\end{scope}
		\begin{scope}[shift={(0,4)}]
		\draw[->] (0.6,0.5) -- (1,0.5) (0,0.5) -- (0.6,0.5);
		\node[below] at (0.5,0.5) {$12$};
		\end{scope}
		\draw[purple, ultra thick] (0,0) -- (0,5) (3,0) -- (3,5) (1,0) -- (1,1) (2,0) -- (2,1) (1,5) -- (1,4) (2,5) -- (2,4) (1,1.5) -- (1,3.5) (2,1.5) -- (2,3.5) (-0.25,0) -- (1.25,0) (1.75,0) -- (3.25,0) (-0.25,5) -- (1.25,5) (1.75,5) -- (3.25,5); 
		\node[below] at (0,0) {$+$}; \node[below] at (1,0) {$-$};
		\node[below] at (2,0) {$+$}; \node[below] at (3,0) {$-$};
		\node[above] at (0,5) {$-$}; \node[above] at (1,5) {$+$};
		\node[above] at (2,5) {$-$}; \node[above] at (3,5) {$+$};
		\end{scope}
		\end{tikzpicture}
		\end{array}\quad\quad
		%
		%
		\IW=\begin{array}{c}
		\begin{tikzpicture}
		\draw[thick] (-1.5,0) -- (0,0) (0,0) -- (0.8,1.5) (0,0) -- (0.8,-1.5);
		\begin{scope}[shift={(0,0)}]
		\draw[purple, ultra thick] (-0.1,-0.1) -- (0.1,0.1) (0.1,-0.1) -- (-0.1,0.1);
		\end{scope}
		\node[below right] at (0.8,1.5) {$\fd_3$};
		\node[above right] at (0.8,-1.5) {$\fd_5$};
		\draw (0,0) circle (0.2);
		\end{tikzpicture}
		\end{array}
		$$
		\item  $v_-\otimes v_-\mapsto 0$: The term $v_-\otimes v_-$ is of lowest $\bf P$-degree $-3$, there is no term on the other side of same degree therefore $Q=0$ in this situation. 
	\end{itemize}
	\item Co-multiplication: 
	\begin{itemize}
		\item $v_-\mapsto v_-\otimes v_-$: in this case the null-web is the simplest one described in the end of section \ref{sec:LG-coh-SN}:
		$$
		\begin{array}{c}
		\begin{tikzpicture}
		\begin{scope}[scale=0.8]
		\draw[->] (0.6,0.5) -- (1,0.5) (0,0.5) -- (0.6,0.5);
		\node[above] at (0.5,0.5) {$21$};
		\draw[ultra thick, purple] (-0.25,0) -- (1.25,0) (0,0) -- (0,1) (1,0) -- (1,1);
		\node[below] at (0,0) {$+$};
		\node at (0,1.25) {$-$};
		\node[below] at (1,0) {$-$};
		\node at (1,1.25) {$+$};
		\begin{scope}[shift={(0,1.5)}]
		\draw[purple,ultra thick] (0,0) -- (0,2) (1,0) -- (1,2);
		\end{scope}
		\begin{scope}[shift={(0,4)}]
		\draw[ultra thick, purple] (-0.25,1) -- (1.25,1) (0,0) -- (0,1) (1,0) -- (1,1);
		\node at (0,-0.25) {$-$};
		\node[above] at (0,1) {$-$};
		\node at (1,-0.25) {$+$};
		\node[above] at (1,1) {$+$};
		\end{scope}
		\end{scope}
		\end{tikzpicture}
		\end{array}\mapsto
		\begin{array}{c}
		\begin{tikzpicture}
		\begin{scope}[scale=0.8]
		\draw[->] (0.6,0.5) -- (1,0.5) (0,0.5) -- (0.6,0.5);
		\node[above] at (0.5,0.5) {$21$};
		\draw[ultra thick, purple] (-0.25,0) -- (1.25,0) (0,0) -- (0,1) (1,0) -- (1,1);
		\node[below] at (0,0) {$+$};
		\node at (0,1.25) {$-$};
		\node[below] at (1,0) {$-$};
		\node at (1,1.25) {$+$};
		\begin{scope}[shift={(0,1.5)}]
		\draw[->] (0.4,0.5) -- (0,0.5) (1,0.5) -- (0.4,0.5);
		\draw[->] (0.6,1) -- (1,1) (0,1) -- (0.6,1); 
		\node[above] at (0.5,1) {$\fd_5$};
		\node[below] at (0.5,0.5) {$\fd_3$};
		\draw[purple,ultra thick] (0,0) -- (0,2) (1,0) -- (1,2);
		\end{scope}
		\begin{scope}[shift={(0,4)}]
		\draw[ultra thick, purple] (-0.25,1) -- (1.25,1) (0,0) -- (0,1) (1,0) -- (1,1);
		\node at (0,-0.25) {$-$};
		\node[above] at (0,1) {$-$};
		\node at (1,-0.25) {$+$};
		\node[above] at (1,1) {$+$};
		\end{scope}
		\end{scope}
		\end{tikzpicture}
		\end{array}\quad\quad
		%
		%
		\IW=\begin{array}{c}
		\begin{tikzpicture}
			\draw[thick] (-1.5,0) -- (0,0) (0,0) -- (0.8,1.5) (0,0) -- (0.8,-1.5);
			\begin{scope}[shift={(0,0)}]
			\draw[purple, ultra thick] (-0.1,-0.1) -- (0.1,0.1) (0.1,-0.1) -- (-0.1,0.1);
			\end{scope}
			\node[below right] at (0.8,1.5) {$\fd_3$};
			\node[above right] at (0.8,-1.5) {$\fd_5$};
			\draw (0,0) circle (0.2);
		\end{tikzpicture}
		\end{array}
		$$
		\item $v_+\mapsto v_+\otimes v_- + v_-\otimes v_+$:
		$$
		\begin{array}{c}
		\begin{tikzpicture}
		\begin{scope}[scale=0.8]
		\draw[ultra thick, purple] (-0.25,0) -- (1.25,0) (0,0) -- (0,1) (1,0) -- (1,1);
		\node[below] at (0,0) {$+$};
		\node at (0,1.25) {$+$};
		\node[below] at (1,0) {$-$};
		\node at (1,1.25) {$-$};
		\begin{scope}[shift={(0,1.5)}]
		\draw[purple,ultra thick] (0,0) -- (0,2) (1,0) -- (1,2);
		\end{scope}
		\begin{scope}[shift={(0,4)}]
		\draw[->] (0.6,0.5) -- (1,0.5) (0,0.5) -- (0.6,0.5);
		\node[below] at (0.5,0.5) {$12$};
		\draw[ultra thick, purple] (-0.25,1) -- (1.25,1) (0,0) -- (0,1) (1,0) -- (1,1);
		\node at (0,-0.25) {$+$};
		\node[above] at (0,1) {$-$};
		\node at (1,-0.25) {$-$};
		\node[above] at (1,1) {$+$};
		\end{scope}
		\end{scope}
		\end{tikzpicture}
		\end{array}
		\mapsto
		%
		%
		\begin{array}{c}
		\begin{tikzpicture}
		\begin{scope}[scale=0.8]
		\draw[->] (0.6,0.5) -- (1,0.5) (0,0.5) -- (0.6,0.5);
		\node[above] at (0.5,0.5) {$21$};
		\draw[ultra thick, purple] (-0.25,0) -- (1.25,0) (0,0) -- (0,1) (1,0) -- (1,1);
		\node[below] at (0,0) {$+$};
		\node at (0,1.25) {$-$};
		\node[below] at (1,0) {$-$};
		\node at (1,1.25) {$+$};
		\begin{scope}[shift={(0,1.5)}]
		\draw[->] (0.4,0.5) -- (0,0.5) (1,0.5) -- (0.4,0.5);
		\node[below] at (0.5,0.5) {$\fd_3$};
		\draw[purple,ultra thick] (0,0) -- (0,2) (1,0) -- (1,2);
		\end{scope}
		\begin{scope}[shift={(0,4)}]
		\draw[->] (0.6,0.5) -- (1,0.5) (0,0.5) -- (0.6,0.5);
		\node[below] at (0.5,0.5) {$12$};
		\draw[ultra thick, purple] (-0.25,1) -- (1.25,1) (0,0) -- (0,1) (1,0) -- (1,1);
		\node at (0,-0.25) {$+$};
		\node[above] at (0,1) {$-$};
		\node at (1,-0.25) {$-$};
		\node[above] at (1,1) {$+$};
		\end{scope}
		\end{scope}
		\end{tikzpicture}
		\end{array}
		\oplus
		\begin{array}{c}
		\begin{tikzpicture}
		\begin{scope}[scale=0.8]
		\draw[ultra thick, purple] (-0.25,0) -- (1.25,0) (0,0) -- (0,1) (1,0) -- (1,1);
		\node[below] at (0,0) {$+$};
		\node at (0,1.25) {$+$};
		\node[below] at (1,0) {$-$};
		\node at (1,1.25) {$-$};
		\begin{scope}[shift={(0,1.5)}]
		\draw[->] (0.6,1.5) -- (1,1.5) (0,1.5) -- (0.6,1.5);
		\node[below] at (0.5,1.5) {$\fd_4$};
		\draw[purple,ultra thick] (0,0) -- (0,2) (1,0) -- (1,2);
		\end{scope}
		\begin{scope}[shift={(0,4)}]
		\draw[ultra thick, purple] (-0.25,1) -- (1.25,1) (0,0) -- (0,1) (1,0) -- (1,1);
		\node at (0,-0.25) {$-$};
		\node[above] at (0,1) {$-$};
		\node at (1,-0.25) {$+$};
		\node[above] at (1,1) {$+$};
		\end{scope}
		\end{scope}
		\end{tikzpicture}
		\end{array}\quad\quad
		%
		%
		\IW=\begin{array}{c}
		\begin{tikzpicture}
		\draw[thick, ->] (0,0.8) -- (0,1.5) (0,0) -- (0,0.8);
		\draw[thick, ->] (0.2,0.8) -- (0.2,1.5) (0.2,0) -- (0.2,0.8);
		\begin{scope}[shift={(0,0)}]
		\draw[purple, ultra thick] (-0.1,-0.1) -- (0.1,0.1) (0.1,-0.1) -- (-0.1,0.1);
		\end{scope}
		\begin{scope}[shift={(0.2,0)}]
		\draw[purple, ultra thick] (-0.1,-0.1) -- (0.1,0.1) (0.1,-0.1) -- (-0.1,0.1);
		\end{scope}
		\draw (-0.1,0.2) -- (0.3,0.2) to[out=0,in=0] (0.3,-0.2) -- (-0.1,-0.2) to[out=180,in=180] (-0.1,0.2);
		\node[left] at (0,1.5) {$21$};
		\node[right] at (0.2,1.5) {$\fd_3$};
		\end{tikzpicture}
		\end{array}\oplus
		\begin{array}{c}
		\begin{tikzpicture}
		\draw[thick] (-1,-1) -- (1,1);
		\begin{scope}[shift={(-1,-1)}]
		\draw[purple, ultra thick] (-0.1,-0.1) -- (0.1,0.1) (0.1,-0.1) -- (-0.1,0.1);
		\end{scope}
		\begin{scope}[shift={(1,1)}]
		\draw[purple, ultra thick] (-0.1,-0.1) -- (0.1,0.1) (0.1,-0.1) -- (-0.1,0.1);
		\end{scope}
		\node[above left] at (-0.5,-0.5) {$\fd_4$};
		\node[below right] at (0.5,0.5) {$12$};
		\end{tikzpicture}
		\end{array}
		$$
	\end{itemize}
\end{itemize}

So far we have constructed a system of differentials $Q_{\chi}$ acting in diagram intersections. Complex $\tilde\CE$ can be seen as a hypercube of resolutions in the complete analogy to the Khovanov complex: vertices are represented by different resolutions of link diagram intersections $\chi$ in \eqref{new_compl}, edges are $Q_{\chi}$. The action of the supercharge form a Frobenious algebra given by relations \eqref{mul} and \eqref{co-mul} and satisfying \eqref{commutatitivity}, therefore for edges supercharges $Q_{\chi}$ anti-commute. However to claim \eqref{commutatitivity} one should make one more step and argue that there are no differentials other than $Q_{\chi}$.

Suppose there are other types of differentials. Consider a cube of resolutions, for simplicity, let us consider a visualisable 3d cube:
\begin{center}
\begin{tikzpicture}
	\begin{scope}[shift={(0,-2.2)}, scale=2]
		\draw[rounded corners=15mm, fill=gray, opacity=0.3] (-2,0) -- (2,0) -- (0,0.5) -- cycle;
	\end{scope}
	\begin{scope}[shift={(0,-3.4)}, scale=2, yscale=-1]
	\draw[rounded corners=15mm, fill=green!60!black, opacity=0.3] (-2,0) -- (2,0) -- (0,0.5) -- cycle;
	\end{scope}
	\node(111) at (0,0) {$\Psi_{111}$};
	\node(011) at (-2,-2) {$\Psi_{011}$};
	\node(101) at (0,-1.5) {$\Psi_{101}$};
	\node(110) at (2,-2) {$\Psi_{110}$};
	\node(010) at (0,-4) {$\Psi_{010}$};
	\node(001) at (-2,-3.5) {$\Psi_{001}$};
	\node(100) at (2,-3.5) {$\Psi_{100}$};
	\node(000) at (0,-5.5) {$\Psi_{000}$};
	\path (111) edge[->] (011) (111) edge[->] (101) (111) edge[->] (110) (011) edge[->] (001) (110) edge[->] (100) (011) edge[->] (010) (110) edge[->] (010) (101) edge[->] (001) (101) edge[->] (100) (010) edge[->] (000) (001) edge[->] (000) (100) edge[->] (000);
	\node[right] at (3.1,0) {${\bf F}=0$};
	\node[right] at (3.1,-1.83) {${\bf F}=1$};
	\node[right] at (3.1,-3.67) {${\bf F}=2$};
	\node[right] at (3.1,-5.5) {${\bf F}=3$};
	\path (110) edge[thick, red, ->] (001) (110) edge[thick, blue, ->, bend right] (111) (111) edge[thick, blue, ->, bend right] (011) (011) edge[thick, blue, ->, bend right] (001) (110) edge[thick, orange, ->, bend left] (111) (111) edge[thick, orange, ->, bend right] (101) (101) edge[thick, orange, ->, bend right] (001);
\end{tikzpicture}
\end{center}
The vertices of the cube are marked according to the resolution choice. There are three types of differentials each shifting one of the resolutions:
\be
Q_1: \; \Psi_{1 **}\to \Psi_{0 **},\quad Q_2: \; \Psi_{*1*}\to \Psi_{*0*},\quad Q_3: \; \Psi_{**1}\to \Psi_{**0}
\ee
These differentials form edges of the cube.
Suppose there is some new type of differential, it gives a new map $\tilde Q:\; \Psi_{110} \to \Psi_{001}$ marked by the red arrow in the diagram. One can represent it as more elementary moves consequently shifting all the resolutions $\tilde Q\sim Q_2Q_1Q_3^{-1}$. This combination can not be considered as a combination of instantons since one of them is an anti-instanton $Q_3^{-1}$, however there may be present a singe modulus BPS instanton somewhere on the boundary of the moduli space of this configuration. Product $\tilde Q\sim Q_2Q_1Q_3^{-1}$ is depicted by a blue path in the diagram. There is another path $Q_1Q_2Q_3^{-1}$ connecting $\Psi_{110}$ and $\Psi_{001}$ marked by the orange color. Then map $\tilde Q$ should get contributions from both 
$$\tilde Q\sim Q_2Q_1Q_3^{-1}+Q_1Q_2Q_3^{-1}\sim 0$$
In a generic situation all such terms contain numbers of $Q$'s and $Q^{-1}$'s of opposite parity, therefore they are summed up to a zero. Alternatively, we can consider a matrix element of $Q^2$ that should be zero, it is given by a sum over all possible paths:
\be
0=\langle \Psi_{001} |Q^2|\Psi_{111}\rangle=\langle \Psi_{001} |Q_1Q_2+Q_2Q_1+\tilde Q Q_3|\Psi_{111}\rangle=\langle \Psi_{001} |\tilde Q Q_3|\Psi_{111}\rangle= \langle \Psi_{001} |\tilde Q|\Psi_{110}\rangle
\ee

\subsubsection{Discussion}

So far we have argued that the $\fs\fu_2$ Landau-Giunzburg cohomology is equivalent  to the Khovanov homology. This statement was not obvious in the initial formulation \eqref{rules_a} - \eqref{rules_d} since the resulting complexes and differentials acting on them are quite different. However, the Landau-Ginzburg theory proposes not just a single complex rather a family of complexes. The content of bound BPS solitons on an interface given by path $\hat{\wp}$ in parameter space $\CP$ depends on a choice of ``chambers" (see Figure \ref{fig:paths}).  Transformation of the BPS content across the walls separating chambers is widely known as wall-crossing phenomenon. So we just propose a choice of a path defining the supersymmetric interface crossing a sequence of chambers in such a way that the Landau-Ginzburg complex and Khovanov complex become isomorphic rather than just only quasi-isomorphic. A physical meaning of such path choice is to mimic isotypical decomposition of interfaces since the $\CR$-interface is supposed to be a categorification of $R$-matrix that becomes diagonal in the isotypical basis.

This choice may seem artificial since it is not presented by a certain path $\hat \wp$ rather by different contributions from different paths. This can be done if one considers an equivalence class of complexes with respect to the wall-crossing morphisms. Despite this artificial flavor the supercharge $Q$ acts locally in link diagram intersections by gluing or cutting cycles analogously to the differential in the Khovanov homology. Consequently we conclude that the $\fs\fu_2$ Landau-Ginzburg cohomology also satisfies the skein exact triangle sequence:
\be
\ldots \to {\bf LGCoh}\left[\begin{array}{c}
	\begin{tikzpicture}[scale=0.5]
	\draw [ultra thick] (-0.5,-0.5) to [out= 45, in =135] (0.5,-0.5) (-0.5,0.5) to [out=315, in=225] (0.5,0.5);
	\end{tikzpicture}
\end{array}\right] \to
{\bf LGCoh}\left[\begin{array}{c}
	\begin{tikzpicture}[scale=0.5,xscale=-1]
	\draw[ultra thick] (-0.5,-0.5) -- (0.5,0.5);
	\draw[ultra thick] (0.5,-0.5) -- (0.1,-0.1);
	\draw[ultra thick] (-0.1,0.1) -- (-0.5,0.5);
	\end{tikzpicture}
\end{array}\right]\to
{\bf LGCoh}\left[\begin{array}{c}
	\begin{tikzpicture}[scale=0.5]
	\draw [ultra thick] (-0.5,-0.5) to [out= 45, in =315] (-0.5,0.5) (0.5,-0.5) to [out=135, in=225] (0.5,0.5);
	\end{tikzpicture}
\end{array}\right]\to
t\;{\bf LGCoh}\left[\begin{array}{c}
	\begin{tikzpicture}[scale=0.5]
	\draw [ultra thick] (-0.5,-0.5) to [out= 45, in =135] (0.5,-0.5) (-0.5,0.5) to [out=315, in=225] (0.5,0.5);
	\end{tikzpicture}
\end{array}\right] \to\ldots
\ee

\section{Landau-Ginzburg $\fs\fu_n$ Link Cohomology}\label{sec:su_n}
Here we exploit the same Landau-Ginzburg model on the monopole moduli space, however it is natural to consider magentic monopoles in the $SU(n)$ Yang-Mills-Higgs theory on $\IR^3$. In this case ``types'' of monopoles are labelled by positive roots fo $\fs\fu_n$. The monopole moduli spaces are graded by $n-1$ non-negative monopole numbers $m_k$ we gather into a vector $\vec m$.

The superpotential for what one could call a naive monopole model analog, or quasi-classical monopole model, when monopole centers are well-separated, see \eqref{q-c_mono}, is given in \cite{Braverman}. Here we use a low energy effective description with a $\fs\fu_n$ Yang-Yang superpotential that in terms of integrals \eqref{intgrl} gives a free field formulation for $SU(n)$ WZW-model with $N$ punctures correspondingly \cite{GMOMS}:
\be\label{YY_sl_n}
\begin{split}
	W=\sum\lm_{a=1}^N \sum\lm_{i=1}^{n-1}\sum\lm_{s=1}^{m_i} k_{a,i}\log(z_a-w_{i,s})-2\sum\lm_{i=1}^{n-1}\sum\lm_{1\leq r\neq s\leq m_i}\log(w_{i,s}-w_{i,r})+\\
	+\sum\lm_{1\leq i\neq j\leq n-1}\sum\lm_{r=1}^{m_i} \sum\lm_{s=1}^{m_j}\log(w_{i,r}-w_{j,s})+\sum\lm_{i=1}^{n-1}\sum\lm_{s=1}^{m_i} c_i w_{i,s}
\end{split}
\ee
Here $k_{i,a}$ is a vector of hieghest weight of representation $R_a$ inserted in puncture $z_a$.

Let us stress again a peculiarity of the transition from the monopole model to the Yang-Yang model we mentioned in the beginning of Section \ref{sec:su_2}. During this transition new singularities are created on the target space: this makes wall-crossing formulas in their initial formulation inapplicable due to purely flavour solitons and eventually breaks the link invariance. Therefore despite we work with the Yang-Yang superpotential \eqref{YY_sl_n} in actual calculations we imply that the actual model has a target space given by the universal cover of the monopole moduli space with a superpotential without singularities.

\subsection{Classification of vacua}
If the punctures are well-separated $|c_i(z_a-z_b)|\gg 1$ we again have a  decomposition of vacua:
\be
\IV(\{z_a\};\{\vec k_a\}; \vec m)=\bigoplus\lm_{\sum\lm_{a=1}^{N}\vec\mu_a=\vec m}\IV(z_1;\vec k_1;\vec \mu_1)\times\ldots\times \IV(z_n;\vec k_n;\vec \mu_n)
\ee
And it is enough to study only vacua classification of a single puncture model. It is not surprising that we have the following statement:
\be
	\IV(z;\vec k;\vec m)\cong v_{\vec k - \sum\lm_{\vec{\alpha}_i\in\Delta^+}m_i\vec \alpha_i}
\ee
Where $v_{\lambda}$ is an invariant subspace of weight $\lambda$ and genrically,
\be
\bigoplus\lm_{\vec m}\IV(z;\vec k;\vec m)\cong V_{\vec k}
\ee
where $V_{\lambda}$ is irreducible $\fs\fu_n$-representation module of the highest weight $\lambda$. This statement follows naturally from the relation between chiral operator basis and stable basis (see \cite{Maulik-Okunkov,Aganagic:2016jmx}). 

Again we will restrict ourselves to links colored by the fundamental representation, however we will have to color strands by both fundamental $\Box$ and anti-fundamental $\bar{\Box}$ representations. Both representations are $n$ dimensional of heighest weights ${\vec k}_{\Box}=(1,0,0,\ldots)$ and ${\vec k}_{\bar\Box}=(\ldots,0,0,1)$ correspondingly, those are spanned by $n$ 1-dimensional vacuum solutions:
\be
\begin{split}
\Box:\quad \IV\left(z;\Box; (0,0,0,\ldots)\right),\; \IV\left(z;\Box; (1,0,0,\ldots)\right),\; \IV\left(z;\Box; (1,1,0,\ldots)\right),\ldots\\
\bar \Box:\quad \ldots,\; \IV\left(z;\bar\Box; (\ldots,0,1,1)\right),\;\IV\left(z;\bar\Box; (\ldots,0,0,1)\right),\; \IV\left(z;\bar\Box; (\ldots,0,0,0)\right)
\end{split}
\ee
We will denote them appearing in this order as $i$ and $\bar i$ correspondingly. It is simple to define what these solutions are, especially if one assumes some ordering, say all $c_i$ are real and satisfy $c_1\gg c_2\gg \ldots \gg c_{n-1}$:
\be
\begin{split}
	j:\quad w_{k,1}=-\sum\lm_{i=1}^k \left(\sum\lm_{s=i}^{j-1}c_s\right)^{-1}\sim -c_k^{-1},\quad k=1,\ldots, j-1\\
	\bar j:\quad w_{k,1}=-\sum\lm_{i=k}^{n-1}\left(\sum\lm_{s=n-j+1}^{i}c_s\right)^{-1}\sim -(n-k)c_{n-j+1}^{-1}, \quad k=n-j+1,\ldots,n-1
\end{split}
\ee

And in a way analogous to $\fs\fu_2$ case one arrives to the following
\begin{theorem}
There exists a basis in the chiral ring of the $\fs\fu_n$ Yang-Yang-Landau-Ginzburg model, s.t. the $\nabla$-connection \eqref{connection} has the following form of the Knizhnik-Zamolodchikov connection:
\be
\nabla=d-\zeta^{-1}\sum\lm_a dz_a\left[\sum\lm_{b\neq a} \frac{\eta^{ij}\;T_i^{(a)}\otimes T_j^{(b)}}{z_a-z_b}+\sum\lm_i c_i m_i^{(a)}\right]
\ee
\end{theorem}
Here $\eta$ is the Killing form on the algebra $\fs\fu_n$.

\subsection{Braiding interfaces}

We will consider the following types of braiding interfaces (the interface evolution axis $x$ flows in the upward direction in all diagrams, see Figure \ref{fig:tangle}):
$$
\begin{array}{c}
	\begin{tikzpicture}
	\begin{scope}[scale=0.8]
	\draw[ultra thick, ->] (-0.5,-0.5) -- (0.5,0.5); 
	\draw[ultra thick, ->] (-0.1,0.1) -- (-0.5,0.5);
	\draw[ultra thick] (0.5,-0.5) -- (0.1,-0.1);
	\end{scope}
	\end{tikzpicture}
\end{array}\qquad 
\begin{array}{c}
\begin{tikzpicture}
\begin{scope}[scale=0.8, xscale=-1]
\draw[ultra thick, ->] (-0.5,-0.5) -- (0.5,0.5); 
\draw[ultra thick, ->] (-0.1,0.1) -- (-0.5,0.5);
\draw[ultra thick] (0.5,-0.5) -- (0.1,-0.1);
\end{scope}
\end{tikzpicture}
\end{array}\qquad
%
%
$$
The corresponding strand is marked by the fundamental representation if the arrow on the strand flows in the same direction as the interface ``time" $x$, and by the anti-fundamental one otherwise. Obviously, there are six more possible orientations, however we will not discuss them here since they are related to mentioned two by ``tri-dent-like" moves.

It is simple to notice that if both strands are marked by the  fundamental representation there is a very simple admissible vacua classification. Admissible filling vectors and vacua are:
\be\label{decomp}
\begin{split}
\IV\left(\{z_a,z_b \};\{\Box,\Box \};(\underbrace{2,\ldots,2}_{i\;{\rm times}},\underbrace{1,\ldots,1}_{j-i\;{\rm times}},0,\ldots,0)\right)=(i\otimes j)\oplus (j\otimes i),\quad {\rm for}\; i<j\\
\IV\left(\{z_a,z_b \};\{\Box,\Box \};(\underbrace{2,\ldots,2}_{i\;{\rm times}},0,\ldots,0)\right)=i\otimes i
\end{split}
\ee
Therefore the spectral curve factorizes to polynomials of order two and one according to two possible cases described above:
\be
\prod\lm_i\left(p-\frac{t_i}{x}\right)\prod\lm_{j>i}\left(p^2-\left(\frac{q_{ij}}{x}+\frac{s_{ij}}{x^2}\right)p+\tilde c_{ij}\right)=0,\quad \lambda=p\; dx
\ee
it is not really crucial what are numbers $t_i$, $q_{ij}$ and $s_{ij}$, a relevant information about this curve is discriminants of polynomials for $i\otimes j$ pairs: they read $D_{ij}=x^{-2}+{\tilde c}_{ij}^2$, where effective parameter ${\tilde c}_{ij}={\vec c}\; ({\vec m}_i-{\vec m}_j)$. Therefore the behavior of $i\otimes j$ vacuum is analogous to the behavior of $+-$ vacuum in $\fs\fu_2$ case discussed in Section \ref{sec:su_2}, and similarly for $\bar i\otimes \bar j$ vacua. Therefore without re-derivation we just adopt the results of Section \ref{sec:su_2} and construct following expressions for interfaces:
\begin{subequations}
\be
\CE\left(\begin{array}{c}
	\begin{tikzpicture}
	\begin{scope}[scale=0.7]
	\draw[ultra thick, ->] (-0.5,-0.5) -- (0.5,0.5); 
	\draw[ultra thick, ->] (-0.1,0.1) -- (-0.5,0.5);
	\draw[ultra thick] (0.5,-0.5) -- (0.1,-0.1);
	\end{scope}
	\end{tikzpicture}
\end{array}\right)= \bigoplus\lm_{i,j} q^{-\delta_{ij}}\begin{array}{c}
\begin{tikzpicture}
\begin{scope}[scale=0.8]
\node[above] at (0,2) {$i$};
\node[above] at (1,2) {$j$};
\node[below] at (0,0) {$j$};
\node[below] at (1,0) {$i$};
\draw[ultra thick, purple] (0,0) -- (0,0.5) to[out=90,in=210] (0.5,1) to[out=30,in=270] (1,1.5) -- (1,2) (1,0) -- (1,0.5) to[out=90,in=330] (0.6,0.9) (0.4,1.1) to[out=150,in=270] (0,1.5) -- (0,2);
\end{scope}
\end{tikzpicture}
\end{array}\bigoplus\lm_{\substack{i,j\\ j>i}} q t^{-1}\begin{array}{c}
\begin{tikzpicture}
\begin{scope}[scale=0.8]
\node[above] at (0,2) {$j$};
\node[above] at (1,2) {$i$};
\node[below] at (0,0) {$j$};
\node[below] at (1,0) {$i$};
\draw[->] (0.6,0.5) -- (1,0.5) (0,0.5) -- (0.6,0.5);
\node[below] at (0.5,0.5) {$12$};
\draw[ultra thick, purple] (0,0) -- (0,0.5) to[out=90,in=210] (0.5,1) to[out=30,in=270] (1,1.5) -- (1,2) (1,0) -- (1,0.5) to[out=90,in=330] (0.6,0.9) (0.4,1.1) to[out=150,in=270] (0,1.5) -- (0,2);
\end{scope}
\end{tikzpicture}
\end{array}\bigoplus\lm_{\substack{i,j\\ j>i}} q^{-1}\begin{array}{c}
\begin{tikzpicture}
\begin{scope}[scale=0.8]
\node[above] at (0,2) {$j$};
\node[above] at (1,2) {$i$};
\node[below] at (0,0) {$j$};
\node[below] at (1,0) {$i$};
\draw[->] (0.6,0.5) -- (1,0.5) (0,0.5) -- (0.6,0.5);
\node[below] at (0.5,0.5) {$12'$};
\draw[ultra thick, purple] (0,0) -- (0,0.5) to[out=90,in=210] (0.5,1) to[out=30,in=270] (1,1.5) -- (1,2) (1,0) -- (1,0.5) to[out=90,in=330] (0.6,0.9) (0.4,1.1) to[out=150,in=270] (0,1.5) -- (0,2);
\end{scope}
\end{tikzpicture}
\end{array}\\
\CE\left(\begin{array}{c}
	\begin{tikzpicture}
	\begin{scope}[scale=0.7,xscale=-1]
	\draw[ultra thick, ->] (-0.5,-0.5) -- (0.5,0.5); 
	\draw[ultra thick, ->] (-0.1,0.1) -- (-0.5,0.5);
	\draw[ultra thick] (0.5,-0.5) -- (0.1,-0.1);
	\end{scope}
	\end{tikzpicture}
\end{array}\right)= \bigoplus\lm_{i,j}q^{\delta_{ij}}\begin{array}{c}
\begin{tikzpicture}
\begin{scope}[scale=0.8, xscale=-1]
\node[above] at (0,2) {$j$};
\node[above] at (1,2) {$i$};
\node[below] at (0,0) {$i$};
\node[below] at (1,0) {$j$};
\draw[ultra thick, purple] (0,0) -- (0,0.5) to[out=90,in=210] (0.5,1) to[out=30,in=270] (1,1.5) -- (1,2) (1,0) -- (1,0.5) to[out=90,in=330] (0.6,0.9) (0.4,1.1) to[out=150,in=270] (0,1.5) -- (0,2);
\end{scope}
\end{tikzpicture}
\end{array}\bigoplus\lm_{\substack{i,j\\ j>i}} q^{-1} t\begin{array}{c}
\begin{tikzpicture}
\begin{scope}[scale=0.8, xscale=-1]
\node[above] at (0,2) {$j$};
\node[above] at (1,2) {$i$};
\node[below] at (0,0) {$j$};
\node[below] at (1,0) {$i$};
\draw[->] (0,1.5) -- (0.4,1.5) (1,1.5) -- (0.4,1.5);
\node[above] at (0.5,1.5) {$21$};
\draw[ultra thick, purple] (0,0) -- (0,0.5) to[out=90,in=210] (0.5,1) to[out=30,in=270] (1,1.5) -- (1,2) (1,0) -- (1,0.5) to[out=90,in=330] (0.6,0.9) (0.4,1.1) to[out=150,in=270] (0,1.5) -- (0,2);
\end{scope}
\end{tikzpicture}
\end{array}\bigoplus\lm_{\substack{i,j\\ j>i}} q\begin{array}{c}
\begin{tikzpicture}
\begin{scope}[scale=0.8, xscale=-1]
\node[above] at (0,2) {$j$};
\node[above] at (1,2) {$i$};
\node[below] at (0,0) {$j$};
\node[below] at (1,0) {$i$};
\draw[->] (0,1.5) -- (0.4,1.5) (1,1.5) -- (0.4,1.5);
\node[above] at (0.5,1.5) {$21'$};
\draw[ultra thick, purple] (0,0) -- (0,0.5) to[out=90,in=210] (0.5,1) to[out=30,in=270] (1,1.5) -- (1,2) (1,0) -- (1,0.5) to[out=90,in=330] (0.6,0.9) (0.4,1.1) to[out=150,in=270] (0,1.5) -- (0,2);
\end{scope}
\end{tikzpicture}
\end{array}
\ee
\end{subequations}
Here solitons are marked by detours analogous to ones depicted in Figure \ref{fig:sc_su_2}. Sheets $1$ and $2$ correspond to vacua $j\otimes i$ and $i\otimes j$.

\subsection{Fusion/defusion interfaces}

A natural place to search for construction of fusion/defusion interfaces is again isotypical decomposition. The product of a representation and its conjugate always contain a trivial representation:
$$
R\otimes \bar R=\emptyset\oplus\ldots
$$
The trivial representation in this setting is treated as a wave function of a constant $x$ slice without any punctures (see Figure \ref{fig:tangle}). And the Clebsh-Gordan coefficients between $R$, $\bar R$ and $\emptyset$ give expressions for the Euler characteristics of the fusion/defusion interfaces. As we have seen in the previous section the isotypical decomposition can be invoked in the construction of ``half"-$\CR$-interfaces, therefore it is natural to guess the structure of fusion/defusion interfaces by considering $\CR$-interfaces for $\Box\otimes\bar \Box$ and $\bar \Box\otimes \Box$ vacua.

We are interested in the vacuum set
\be
\IV\left(\{z_a,z_b\};\vec{k}_{\Box}+\vec{k}_{\bar{\Box}} ;(1,\ldots,1)\right)=\bigoplus\lm_{i=1}^n (i\otimes\bar i)
\ee
since it has a zero weight ${\vec k}_{\Box}+{\vec k}_{\bar \Box}-\sum\lm_{\vec{\alpha}_i\in\Delta^+}m_i\vec \alpha_i=0$ as an empty space should have. However this vacuum set does not split and the resulting spectral curve has a rather complicated structure. For example, for $\fs\fu_3$, one has:
\be
p^3+\left(-2 c_1-c_2-\frac{3}{x}\right)p^2+\left(\frac{4 c_1}{x}+\frac{2 c_2}{x}+c_1^2+c_2 c_1 \right)p+\left(-\frac{c_1^2}{x}-\frac{c_2 c_1}{x}\right)=0,\quad \lambda=p\; dx
\ee
We depict a generic structure of $\fs\fu_n$ curve for this vacuum in Figure \ref{fig:sc_su_n}. We have denoted vacua accordingly $j=(j \otimes \bar j)$ and $\bar j=(\bar j\otimes j)$ at $x\to +\I\infty$. We construct fusion/defusion interfaces analogously to \cite[Section 4.3]{GM} as half-twists. However, it is simpler to define different interfaces in different scale orderings: $c_1\gg c_2\gg \ldots \gg c_{n-1}$, or $c_1\ll c_2\ll \ldots \ll c_{n-1}$. So we use both implying that one can change the ordering homotopically, or assume that parameters $c_i$ are slightly $x$-dependent.
\begin{figure}[h]
\begin{center}
$
\begin{array}{c|c}
c_1\gg c_2\gg \ldots \gg c_{n-1} & c_1\ll c_2\ll \ldots \ll c_{n-1}\\
\hline
&\\
\begin{array}{c}
\begin{tikzpicture}
\begin{scope}[scale=0.6]
\draw[orange, ultra thick, domain=-0.72:0.72, samples=200]  plot ({1.51*(-3/4 + cos(\x r))*(1+0.05*sin(50*\x r))}, {1.51*(sin(\x r))*(1+0.05*sin(50*\x r))});
\end{scope}
\begin{scope}[scale=1.2]
\draw[orange, ultra thick, domain=-0.72:0.72, samples=200]  plot ({1.51*(-3/4 + cos(\x r))*(1+0.05*sin(50*\x r))}, {1.51*(sin(\x r))*(1+0.05*sin(50*\x r))});
\end{scope}
\draw[thick] (-2,1.2) -- (0,1.2) to[out=300,in=30] (0,0) (0,1.2) to[out=60,in=180] (2,1.8);
\draw[thick] (-2,0.6) -- (0,0.6) to[out=300,in=60] (0,0) (0,0.6) to[out=60,in=180] (2,1.2);
\draw[thick] (0.19,0.84) to[out=330,in=180] (2,0.6);
\filldraw (0,1.5) circle (0.05) (0,1.7) circle (0.05) (0,1.9) circle (0.05);
\begin{scope}[yscale=-1]
\draw[thick] (-2,1.2) -- (0,1.2) to[out=300,in=30] (0,0) (0,1.2) to[out=60,in=180] (2,1.8);
\draw[thick] (-2,0.6) -- (0,0.6) to[out=300,in=60] (0,0) (0,0.6) to[out=60,in=180] (2,1.2);
\draw[thick] (0.19,0.84) to[out=330,in=180] (2,0.6);
\filldraw (0,1.5) circle (0.05) (0,1.7) circle (0.05) (0,1.9) circle (0.05);
\end{scope}
\begin{scope}[shift={(0,0.6)}]
\draw[purple, ultra thick] (-0.1,-0.1) -- (0.1,0.1) (0.1,-0.1) -- (-0.1,0.1);
\end{scope}
\begin{scope}[shift={(0,1.2)}]
\draw[purple, ultra thick] (-0.1,-0.1) -- (0.1,0.1) (0.1,-0.1) -- (-0.1,0.1);
\end{scope}
\begin{scope}[shift={(0,-0.6)}]
\draw[purple, ultra thick] (-0.1,-0.1) -- (0.1,0.1) (0.1,-0.1) -- (-0.1,0.1);
\end{scope}
\begin{scope}[shift={(0,-1.2)}]
\draw[purple, ultra thick] (-0.1,-0.1) -- (0.1,0.1) (0.1,-0.1) -- (-0.1,0.1);
\end{scope}
\draw[fill=white] (0,0) circle (0.1);
\node[left] at (-2,1.2) {$(3,2)$};
\node[left] at (-2,0.6) {$(2,1)$};
\node[left] at (-2,-1.2) {$(3,2)'$};
\node[left] at (-2,-0.6) {$(2,1)'$};
\node[right] at (2,-0.6) {$(1,n)$};
\draw[thick, ->] (-1,1.4) to[out=240,in=120] (-1,0.4);
\begin{scope}[shift={(-1,1.6)}]
\begin{scope}[scale=0.5]
\draw[ultra thick, ->] (-0.5,0) to[out=90,in=180] (0,1) to[out=0,in=90] (0.5,0);
\end{scope}
\end{scope}
\draw[thick, ->] (1,-0.4) to[out=300,in=60] (1,-2);
\begin{scope}[shift={(1,-2.2)}]
\begin{scope}[scale=0.5, yscale=-1]
\draw[ultra thick, ->] (-0.5,0) to[out=90,in=180] (0,1) to[out=0,in=90] (0.5,0);
\end{scope}
\end{scope}
\end{tikzpicture}
\end{array}
&
\begin{array}{c}
\begin{tikzpicture}
\begin{scope}[scale=-1]
\begin{scope}[scale=0.6]
\draw[orange, ultra thick, domain=-0.72:0.72, samples=200]  plot ({1.51*(-3/4 + cos(\x r))*(1+0.05*sin(50*\x r))}, {1.51*(sin(\x r))*(1+0.05*sin(50*\x r))});
\end{scope}
\begin{scope}[scale=1.2]
\draw[orange, ultra thick, domain=-0.72:0.72, samples=200]  plot ({1.51*(-3/4 + cos(\x r))*(1+0.05*sin(50*\x r))}, {1.51*(sin(\x r))*(1+0.05*sin(50*\x r))});
\end{scope}
\draw[thick] (-2,1.2) -- (0,1.2) to[out=300,in=30] (0,0) (0,1.2) to[out=60,in=180] (2,1.8);
\draw[thick] (-2,0.6) -- (0,0.6) to[out=300,in=60] (0,0) (0,0.6) to[out=60,in=180] (2,1.2);
\draw[thick] (0.19,0.84) to[out=330,in=180] (2,0.6);
\filldraw (0,1.5) circle (0.05) (0,1.7) circle (0.05) (0,1.9) circle (0.05);
\begin{scope}[yscale=-1]
\draw[thick] (-2,1.2) -- (0,1.2) to[out=300,in=30] (0,0) (0,1.2) to[out=60,in=180] (2,1.8);
\draw[thick] (-2,0.6) -- (0,0.6) to[out=300,in=60] (0,0) (0,0.6) to[out=60,in=180] (2,1.2);
\draw[thick] (0.19,0.84) to[out=330,in=180] (2,0.6);
\filldraw (0,1.5) circle (0.05) (0,1.7) circle (0.05) (0,1.9) circle (0.05);
\end{scope}
\begin{scope}[shift={(0,0.6)}]
\draw[purple, ultra thick] (-0.1,-0.1) -- (0.1,0.1) (0.1,-0.1) -- (-0.1,0.1);
\end{scope}
\begin{scope}[shift={(0,1.2)}]
\draw[purple, ultra thick] (-0.1,-0.1) -- (0.1,0.1) (0.1,-0.1) -- (-0.1,0.1);
\end{scope}
\begin{scope}[shift={(0,-0.6)}]
\draw[purple, ultra thick] (-0.1,-0.1) -- (0.1,0.1) (0.1,-0.1) -- (-0.1,0.1);
\end{scope}
\begin{scope}[shift={(0,-1.2)}]
\draw[purple, ultra thick] (-0.1,-0.1) -- (0.1,0.1) (0.1,-0.1) -- (-0.1,0.1);
\end{scope}
\draw[fill=white] (0,0) circle (0.1);
\node[right] at (-2,1.2) {$(\overline{n-2},\overline{n-1})$};
\node[right] at (-2,0.6) {$(\overline{n-1},\overline{n})$};
\node[right] at (-2,-1.2) {$(\overline{n-2},\overline{n-1})'$};
\node[right] at (-2,-0.6) {$(\overline{n-1},\overline{n})'$};
\node[left] at (2,-0.6) {$(\bar n,\bar 1)$};
\draw[thick, ->] (-1,1.4) to[out=240,in=120] (-1,0.4);
\begin{scope}[shift={(-1,2.1)}]
\begin{scope}[scale=0.5, yscale=-1]
\draw[ultra thick, ->] (-0.5,0) to[out=90,in=180] (0,1) to[out=0,in=90] (0.5,0);
\end{scope}
\end{scope}
\draw[thick, ->] (1,-0.4) to[out=300,in=60] (1,-2);
\begin{scope}[shift={(1,-2.7)}]
\begin{scope}[scale=0.5]
\draw[ultra thick, ->] (-0.5,0) to[out=90,in=180] (0,1) to[out=0,in=90] (0.5,0);
\end{scope}
\end{scope}
\end{scope}
\end{tikzpicture}
\end{array}
\end{array}
$
\end{center}
\caption{Spectral curve for two two punctures in $\fs\fu_n$, $\Box\otimes\bar{\Box}$.}\label{fig:sc_su_n}
\end{figure}

Or, diagramatically, we have:
\begin{subequations}
\be
	\CE\left(\begin{array}{c}
		\begin{tikzpicture}
		\begin{scope}[scale=0.8]
		\draw[ultra thick] (0,0) to[out=270,in=180] (0.6,-0.5);
		\draw[ultra thick,<-] (0.6,-0.5) to[out=0,in=270] (1.2,0);
		\end{scope}
		\end{tikzpicture}
	\end{array}\right)=\begin{array}{c}
		\begin{tikzpicture}
		\begin{scope}[scale=0.8]
		\draw[ultra thick, purple] (-0.2,0) -- (1.2,0) (0,0) -- (0,1) (1,0) -- (1,1);
		\node[below] at (0,0) {$n$}; \node[below] at (1,0) {$\bar n$};
		\node[above] at (0,1) {$n$}; \node[above] at (1,1) {$\bar n$};
		\end{scope}
		\end{tikzpicture}
	\end{array}\oplus \ldots\oplus(t q^{-1})^{n-1}\begin{array}{c}
		\begin{tikzpicture}
		\begin{scope}[scale=0.8]
		\draw[ultra thick, purple] (-0.2,0) -- (1.2,0) (0,0) -- (0,1) (1,0) -- (1,1);
		\node[below] at (0,0) {$n$}; \node[below] at (1,0) {$\bar n$};
		\node[above] at (0,1) {$1$}; \node[above] at (1,1) {$\bar 1$};
		\draw (0,0.25) -- (1,0.25) (0,0.75) -- (1,0.75);
		\draw [->] (0,0.25) -- (0.6,0.25);
		\draw [->] (0,0.75) -- (0.6,0.75);
		\draw[dashed] (0,0.5) -- (1,0.5);
		\draw[dashed,->] (0,0.5) -- (0.6,0.5);
		\end{scope}
		\end{tikzpicture}
	\end{array},\\
	\quad \CE\left(\begin{array}{c}
		\begin{tikzpicture}
		\begin{scope}[scale=0.8, yscale=-1]
		\draw[ultra thick] (0,0) to[out=270,in=180] (0.6,-0.5);
		\draw[ultra thick,<-] (0.6,-0.5) to[out=0,in=270] (1.2,0);
		\end{scope}
		\end{tikzpicture}
	\end{array}\right)=\begin{array}{c}
		\begin{tikzpicture}
		\begin{scope}[scale=0.8]
		\draw[ultra thick, purple] (-0.2,1) -- (1.2,1) (0,0) -- (0,1) (1,0) -- (1,1);
		\node[below] at (0,0) {$\bar n$}; \node[below] at (1,0) {$n$};
		\node[above] at (0,1) {$\bar n$}; \node[above] at (1,1) {$n$};
		\end{scope}
		\end{tikzpicture}
	\end{array}\oplus \ldots\oplus( q t^{-1})^{n-1}\begin{array}{c}
		\begin{tikzpicture}
		\begin{scope}[scale=0.8]
		\draw[ultra thick, purple] (-0.2,1) -- (1.2,1) (0,0) -- (0,1) (1,0) -- (1,1);
		\node[below] at (0,0) {$\bar 1$}; \node[below] at (1,0) {$1$};
		\node[above] at (0,1) {$\bar n$}; \node[above] at (1,1) {$n$};
		\draw (0,0.25) -- (1,0.25) (0,0.75) -- (1,0.75);
		\draw [->] (0,0.25) -- (0.6,0.25);
		\draw [->] (0,0.75) -- (0.6,0.75);
		\draw[dashed] (0,0.5) -- (1,0.5);
		\draw[dashed,->] (0,0.5) -- (0.6,0.5);
		\end{scope}
		\end{tikzpicture}
	\end{array}
\ee
\be
	\CE\left(\begin{array}{c}
		\begin{tikzpicture}
		\begin{scope}[scale=0.8]
		\draw[ultra thick,->] (0,0) to[out=270,in=180] (0.6,-0.5);
		\draw[ultra thick] (0.6,-0.5) to[out=0,in=270] (1.2,0);
		\end{scope}
		\end{tikzpicture}
	\end{array}\right)=\begin{array}{c}
		\begin{tikzpicture}
		\begin{scope}[scale=0.8]
		\draw[ultra thick, purple] (-0.2,0) -- (1.2,0) (0,0) -- (0,1) (1,0) -- (1,1);
		\node[below] at (0,0) {$\bar 1$}; \node[below] at (1,0) {$1$};
		\node[above] at (0,1) {$\bar 1$}; \node[above] at (1,1) {$1$};
		\end{scope}
		\end{tikzpicture}
	\end{array}\oplus \ldots\oplus(t q^{-1})^{n-1}\begin{array}{c}
		\begin{tikzpicture}
		\begin{scope}[scale=0.8]
		\draw[ultra thick, purple] (-0.2,0) -- (1.2,0) (0,0) -- (0,1) (1,0) -- (1,1);
		\node[below] at (0,0) {$\bar 1$}; \node[below] at (1,0) {$1$};
		\node[above] at (0,1) {$\bar n$}; \node[above] at (1,1) {$n$};
		\draw (0,0.25) -- (1,0.25) (0,0.75) -- (1,0.75);
		\draw [->] (0,0.25) -- (0.6,0.25);
		\draw [->] (0,0.75) -- (0.6,0.75);
		\draw[dashed] (0,0.5) -- (1,0.5);
		\draw[dashed,->] (0,0.5) -- (0.6,0.5);
		\end{scope}
		\end{tikzpicture}
	\end{array},
\ee
\be
	\CE\left(\begin{array}{c}
		\begin{tikzpicture}
		\begin{scope}[scale=0.8, yscale=-1]
		\draw[ultra thick,->] (0,0) to[out=270,in=180] (0.6,-0.5);
		\draw[ultra thick] (0.6,-0.5) to[out=0,in=270] (1.2,0);
		\end{scope}
		\end{tikzpicture}
	\end{array}\right)=\begin{array}{c}
		\begin{tikzpicture}
		\begin{scope}[scale=0.8]
		\draw[ultra thick, purple] (-0.2,1) -- (1.2,1) (0,0) -- (0,1) (1,0) -- (1,1);
		\node[below] at (0,0) {$1$}; \node[below] at (1,0) {$\bar 1$};
		\node[above] at (0,1) {$1$}; \node[above] at (1,1) {$\bar 1$};
		\end{scope}
		\end{tikzpicture}
	\end{array}\oplus \ldots\oplus( q t^{-1})^{n-1}\begin{array}{c}
		\begin{tikzpicture}
		\begin{scope}[scale=0.8]
		\draw[ultra thick, purple] (-0.2,1) -- (1.2,1) (0,0) -- (0,1) (1,0) -- (1,1);
		\node[below] at (0,0) {$n$}; \node[below] at (1,0) {$\bar n$};
		\node[above] at (0,1) {$1$}; \node[above] at (1,1) {$\bar 1$};
		\draw (0,0.25) -- (1,0.25) (0,0.75) -- (1,0.75);
		\draw [->] (0,0.25) -- (0.6,0.25);
		\draw [->] (0,0.75) -- (0.6,0.75);
		\draw[dashed] (0,0.5) -- (1,0.5);
		\draw[dashed,->] (0,0.5) -- (0.6,0.5);
		\end{scope}
		\end{tikzpicture}
	\end{array}
\ee
\end{subequations}
The thick horizontal line in all these diagrams is understood as a corresponding D-brane boundary condition. We denoted multiple solitons by a dashed horizontal line, each such soliton represents a migration of a single $w_i$ field. This is an incorrect interpretation since in counterclockwise cap and cup interfaces all the field $w_i$ migrate simultaneously, they appear from intersections with descendant WKB lines $(j,n)$ and $(\bar j,\bar n)$. However for simplicity we depict and assume that those solitons are a combination of consequent soliton jumps, like in clockwise cup/cap interfaces. This simultaneous/consequent representation switches to an opposite for counterclockwise/clocwise orientation if one chooses the opposite sign of $\zeta$ for example.

\subsection{Link invariant}
The remaining $\CR$-interfaces can be constructed by suitable coupling to the fusion/defusion interfaces through tri-dent-like moves:
$$
\begin{array}{c}
\begin{tikzpicture}
\begin{scope}[scale=0.8]
\draw[ultra thick, ->] (0,0) -- (1,1);
\draw[ultra thick, ->] (1,0) -- (0.6,0.4) (0.4,0.6) -- (0,1);
\draw[ultra thick, ->] (1.5,1) -- (1.5,0) to[out=270,in=315] (1,0); 
\end{scope}
\end{tikzpicture}
\end{array}=\begin{array}{c}
\begin{tikzpicture}
\begin{scope}[scale=0.8,xscale=-1]
\draw[ultra thick, ->] (0,0) -- (1,1);
\draw[ultra thick, ->] (1,0) -- (0.6,0.4) (0.4,0.6) -- (0,1);
\draw[ultra thick, ->] (1.5,1) -- (1.5,0) to[out=270,in=315] (1,0); 
\end{scope}
\end{tikzpicture}
\end{array}
$$
Having all the $\CR$-interfaces and fusion/defusion interfaces one can define corresponding complex by simply gluing all the elementary interfaces together. The differential is given by the $Q$-action \eqref{null-web}. The resulting $\fs\fu_n$ Landau-Ginzburg cohomology for link $L$:
\be
{}_n {\bf LGCoh}(L):=H^*\left(\CE(L),Q\right)
\ee
is expected to be an invariant of link $L$. To check this one should check invariance of this construction under three Reidemeister moves and two extra conditions we call ``Reidemeister 0" moves ensuring invariance for different interface evolution $x$-axis direction choices (see Figure \ref{fig:tangle}). ``Reidemeister 0" moves include hump deletion/creation move and trident-like move depicted in diagrams \eqref{hump} and \eqref{trident} correspondingly.
Invariance under all these moves they can be easily checked along the lines of \cite[Section 6]{GM}. Let us remind that discrepancy between Yang-Yang and monopole superpotentials becomes essential in this situation.

Similarly, we can argue invariance under Reidemeister II and III using similarity between a pair of fundamental representations in $\fs\fu_n$ and $\fs\fu_2$ cases. Reidemeister moves II and III work there as they work in the Khovanov construction, therefore they work here as well.

We will not go into detailed check of all the moves, let us restrict ourselves just to Reidemeister I move as an example.

Reidemeister I move implies an isomorphism of complexes of links different by a curl up to an overall shift of $\bf P$- and $\bf F$-degrees due to the framing anomaly:
$$
\begin{array}{c}
\begin{tikzpicture}
\begin{scope}[scale=0.6]
\draw[ultra thick, ->] (0,0) -- (0,2);
\end{scope}
\end{tikzpicture}
\end{array}=q^{\#} t^{\#}
\begin{array}{c}
\begin{tikzpicture}
\begin{scope}[scale=0.6]
\draw[ultra thick, ->] (0,0) to[out=90,in=315] (-0.4,0.9) (-0.6,1.1) to (-1,1.5) to[out=135,in=225] (-1,0.5) to (-0.5,1) to[out=45,in=270] (0,2);
\end{scope}
\end{tikzpicture}
\end{array}
$$

Let us consider a complex appearing in the LHS:
\be
\begin{split}
\CE\left[\begin{array}{c}
	\begin{tikzpicture}
	\begin{scope}[scale=0.6]
	\draw[ultra thick, ->] (0,0) to[out=90,in=315] (-0.5,1) to (-1,1.5) to[out=135,in=225] (-1,0.5) to (-0.6,0.9) (-0.4,1.1) to[out=45,in=270] (0,2);
	\end{scope}
	\end{tikzpicture}
\end{array}\right]=\bigoplus\lm_{i<n}\frac{t^{n-1}}{q^n}
\begin{array}{c}
	\begin{tikzpicture}
	\begin{scope}[scale=0.8]
	\draw[ultra thick, purple] (-0.25,0) -- (1.25,0) (1,0) -- (1,1) (-0.25,4.5) -- (1.25,4.5) (1,4.5) -- (1,3.5)  (0,0) -- (0,2) (0,2.5) -- (0,4.5);
	\draw (0,0.25) -- (1,0.25) (0,0.75) -- (1,0.75);
	\draw [->] (0,0.25) -- (0.6,0.25);
	\draw [->] (0,0.75) -- (0.6,0.75);
	\draw[dashed] (0,0.5) -- (1,0.5);
	\draw[dashed,->] (0,0.5) -- (0.6,0.5);
	\begin{scope}[shift={(1,1.5)}]
	\draw[ultra thick, purple] (0,0) -- (0,0.5) to[out=90,in=210] (0.5,1) to[out=30,in=270] (1,1.5) (1,0) -- (1,0.5) to[out=90,in=330] (0.6,0.9) (0.4,1.1) to[out=150,in=270] (0,1.5);
	\end{scope}
	\node[below] at (0,0) {$\bar 1$}; \node[below] at (1,0) {$1$};
	\node[above] at (0,4.5) {$\bar n$}; \node[above] at (1,4.5) {$n$};
	\node at (1,1.25) {$n$}; \node at (2,1.25) {$i$};
	\node at (1,3.25) {$n$}; \node at (2,3.25) {$i$};
	\node at (0,2.25) {$\bar n$};
	\begin{scope}[shift={(1,1.5)}]
	\draw[->] (0.6,0.5) -- (1,0.5) (0,0.5) -- (0.6,0.5);
	\node[below] at (0.5,0.5) {$12'$};
	\end{scope}
	\end{scope}
	\end{tikzpicture}
\end{array}
\oplus\bigoplus\lm_{i<n} \frac{t^{2i-n-1}}{q^{2i-n}}\left(
\begin{array}{c}
\begin{tikzpicture}
\begin{scope}[scale=0.8]
\draw[ultra thick, purple] (-0.25,0) -- (1.25,0) (1,0) -- (1,1) (-0.25,4.5) -- (1.25,4.5) (1,4.5) -- (1,3.5)  (0,0) -- (0,2) (0,2.5) -- (0,4.5);
\draw (0,0.25) -- (1,0.25) (0,0.75) -- (1,0.75);
\draw [->] (0,0.25) -- (0.6,0.25);
\draw [->] (0,0.75) -- (0.6,0.75);
\draw[dashed] (0,0.5) -- (1,0.5);
\draw[dashed,->] (0,0.5) -- (0.6,0.5);
\begin{scope}[shift={(0,3.5)}]
\draw (0,0.25) -- (1,0.25) (0,0.75) -- (1,0.75);
\draw [->] (0,0.25) -- (0.6,0.25);
\draw [->] (0,0.75) -- (0.6,0.75);
\draw[dashed] (0,0.5) -- (1,0.5);
\draw[dashed,->] (0,0.5) -- (0.6,0.5);
\end{scope}
\begin{scope}[shift={(1,1.5)}]
\draw[ultra thick, purple] (0,0) -- (0,0.5) to[out=90,in=210] (0.5,1) to[out=30,in=270] (1,1.5) (1,0) -- (1,0.5) to[out=90,in=330] (0.6,0.9) (0.4,1.1) to[out=150,in=270] (0,1.5);
\end{scope}
\node[below] at (0,0) {$\bar 1$}; \node[below] at (1,0) {$1$};
\node[above] at (0,4.5) {$\bar n$}; \node[above] at (1,4.5) {$n$};
\node at (1,1.25) {$i$}; \node at (2,1.25) {$i$};
\node at (1,3.25) {$i$}; \node at (2,3.25) {$i$};
\node at (0,2.25) {$\bar i$};
	\node[right] at (1,4) {$\fd_3$};
\end{scope}
\end{tikzpicture}
\end{array}
\oplus
t\begin{array}{c}
	\begin{tikzpicture}
	\begin{scope}[scale=0.8]
	\draw[ultra thick, purple] (-0.25,0) -- (1.25,0) (1,0) -- (1,1) (-0.25,4.5) -- (1.25,4.5) (1,4.5) -- (1,3.5)  (0,0) -- (0,2) (0,2.5) -- (0,4.5);
	\draw (0,0.25) -- (1,0.25) (0,0.75) -- (1,0.75);
	\draw [->] (0,0.25) -- (0.6,0.25);
	\draw [->] (0,0.75) -- (0.6,0.75);
	\draw[dashed] (0,0.5) -- (1,0.5);
	\draw[dashed,->] (0,0.5) -- (0.6,0.5);
	\begin{scope}[shift={(0,3.5)}]
	\draw (0,0.25) -- (1,0.25) (0,0.75) -- (1,0.75);
	\draw [->] (0,0.25) -- (0.6,0.25);
	\draw [->] (0,0.75) -- (0.6,0.75);
	\draw[dashed] (0,0.5) -- (1,0.5);
	\draw[dashed,->] (0,0.5) -- (0.6,0.5);
	\end{scope}
	\begin{scope}[shift={(1,1.5)}]
	\draw[ultra thick, purple] (0,0) -- (0,0.5) to[out=90,in=210] (0.5,1) to[out=30,in=270] (1,1.5) (1,0) -- (1,0.5) to[out=90,in=330] (0.6,0.9) (0.4,1.1) to[out=150,in=270] (0,1.5);
	\end{scope}
	\node[below] at (0,0) {$\bar 1$}; \node[below] at (1,0) {$1$};
	\node[above] at (0,4.5) {$\bar n$}; \node[above] at (1,4.5) {$n$};
	\node at (1,1.25) {$i+1$}; \node at (2,1.25) {$i$};
	\node at (1,3.25) {$i+1$}; \node at (2,3.25) {$i$};
	\node at (0,2.25) {$\overline{i+1}$};
	\begin{scope}[shift={(1,1.5)}]
	\draw[->] (0.6,0.5) -- (1,0.5) (0,0.5) -- (0.6,0.5);
	\node[below] at (0.5,0.5) {$12$};
	\end{scope}
	\node[right] at (1,0.5) {$\fd_1$};
	\node[right] at (2,2) {$\fd_2$};
	\end{scope}
	\end{tikzpicture}
\end{array}\right)\oplus\\
\oplus\bigoplus\lm_{\substack{i,j\\ i<j<n}} \left(\frac{t}{q}\right)^{2j-2-n}\left(
\begin{array}{c}
	\begin{tikzpicture}
	\begin{scope}[scale=0.8]
	\draw[ultra thick, purple] (-0.25,0) -- (1.25,0) (1,0) -- (1,1) (-0.25,4.5) -- (1.25,4.5) (1,4.5) -- (1,3.5)  (0,0) -- (0,2) (0,2.5) -- (0,4.5);
	\draw (0,0.25) -- (1,0.25) (0,0.75) -- (1,0.75);
	\draw [->] (0,0.25) -- (0.6,0.25);
	\draw [->] (0,0.75) -- (0.6,0.75);
	\draw[dashed] (0,0.5) -- (1,0.5);
	\draw[dashed,->] (0,0.5) -- (0.6,0.5);
	\begin{scope}[shift={(0,3.5)}]
	\draw (0,0.25) -- (1,0.25) (0,0.75) -- (1,0.75);
	\draw [->] (0,0.25) -- (0.6,0.25);
	\draw [->] (0,0.75) -- (0.6,0.75);
	\draw[dashed] (0,0.5) -- (1,0.5);
	\draw[dashed,->] (0,0.5) -- (0.6,0.5);
	\end{scope}
	\begin{scope}[shift={(1,1.5)}]
	\draw[ultra thick, purple] (0,0) -- (0,0.5) to[out=90,in=210] (0.5,1) to[out=30,in=270] (1,1.5) (1,0) -- (1,0.5) to[out=90,in=330] (0.6,0.9) (0.4,1.1) to[out=150,in=270] (0,1.5);
	\end{scope}
	\node[below] at (0,0) {$\bar 1$}; \node[below] at (1,0) {$1$};
	\node[above] at (0,4.5) {$\bar n$}; \node[above] at (1,4.5) {$n$};
	\node at (1,1.25) {$j$}; \node at (2,1.25) {$i$};
	\node at (1,3.25) {$j$}; \node at (2,3.25) {$i$};
	\node at (0,2.25) {$\bar j$};
	\begin{scope}[shift={(1,1.5)}]
	\draw[->] (0.6,0.5) -- (1,0.5) (0,0.5) -- (0.6,0.5);
	\node[below] at (0.5,0.5) {$12$};
	\end{scope}
	\node[right] at (2,2) {$\fd_4$};
	\node[right] at (1,4) {$\fd_7$};
	\end{scope}
	\end{tikzpicture}
\end{array}
\oplus
t\begin{array}{c}
	\begin{tikzpicture}
	\begin{scope}[scale=0.8]
	\draw[ultra thick, purple] (-0.25,0) -- (1.25,0) (1,0) -- (1,1) (-0.25,4.5) -- (1.25,4.5) (1,4.5) -- (1,3.5)  (0,0) -- (0,2) (0,2.5) -- (0,4.5);
	\draw (0,0.25) -- (1,0.25) (0,0.75) -- (1,0.75);
	\draw [->] (0,0.25) -- (0.6,0.25);
	\draw [->] (0,0.75) -- (0.6,0.75);
	\draw[dashed] (0,0.5) -- (1,0.5);
	\draw[dashed,->] (0,0.5) -- (0.6,0.5);
	\begin{scope}[shift={(0,3.5)}]
	\draw (0,0.25) -- (1,0.25) (0,0.75) -- (1,0.75);
	\draw [->] (0,0.25) -- (0.6,0.25);
	\draw [->] (0,0.75) -- (0.6,0.75);
	\draw[dashed] (0,0.5) -- (1,0.5);
	\draw[dashed,->] (0,0.5) -- (0.6,0.5);
	\end{scope}
	\begin{scope}[shift={(1,1.5)}]
	\draw[ultra thick, purple] (0,0) -- (0,0.5) to[out=90,in=210] (0.5,1) to[out=30,in=270] (1,1.5) (1,0) -- (1,0.5) to[out=90,in=330] (0.6,0.9) (0.4,1.1) to[out=150,in=270] (0,1.5);
	\end{scope}
	\node[below] at (0,0) {$\bar 1$}; \node[below] at (1,0) {$1$};
	\node[above] at (0,4.5) {$\bar n$}; \node[above] at (1,4.5) {$n$};
	\node at (1,1.25) {$j+1$}; \node at (2,1.25) {$i$};
	\node at (1,3.25) {$j+1$}; \node at (2,3.25) {$i$};
	\node at (0,2.25) {$\overline{j+1}$};
	\begin{scope}[shift={(1,1.5)}]
	\draw[->] (0.6,0.5) -- (1,0.5) (0,0.5) -- (0.6,0.5);
	\node[below] at (0.5,0.5) {$12'$};
	\end{scope}
	\node[right] at (1,0.5) {$\fd_5$};
	\node[right] at (2,2) {$\fd_6$};
	\end{scope}
	\end{tikzpicture}
\end{array}\right)\oplus
\frac{t^{n-1}}{q^n}
\begin{array}{c}
	\begin{tikzpicture}
	\begin{scope}[scale=0.8]
	\draw[ultra thick, purple] (-0.25,0) -- (1.25,0) (1,0) -- (1,1) (-0.25,4.5) -- (1.25,4.5) (1,4.5) -- (1,3.5)  (0,0) -- (0,2) (0,2.5) -- (0,4.5);
	\draw (0,0.25) -- (1,0.25) (0,0.75) -- (1,0.75);
	\draw [->] (0,0.25) -- (0.6,0.25);
	\draw [->] (0,0.75) -- (0.6,0.75);
	\draw[dashed] (0,0.5) -- (1,0.5);
	\draw[dashed,->] (0,0.5) -- (0.6,0.5);
	\begin{scope}[shift={(1,1.5)}]
	\draw[ultra thick, purple] (0,0) -- (0,0.5) to[out=90,in=210] (0.5,1) to[out=30,in=270] (1,1.5) (1,0) -- (1,0.5) to[out=90,in=330] (0.6,0.9) (0.4,1.1) to[out=150,in=270] (0,1.5);
	\end{scope}
	\node[below] at (0,0) {$\bar 1$}; \node[below] at (1,0) {$1$};
	\node[above] at (0,4.5) {$\bar n$}; \node[above] at (1,4.5) {$n$};
	\node at (1,1.25) {$n$}; \node at (2,1.25) {$n$};
	\node at (1,3.25) {$n$}; \node at (2,3.25) {$n$};
	\node at (0,2.25) {$\bar n$};
	\end{scope}
	\end{tikzpicture}
\end{array}
\end{split}\label{su_n_RI}
\ee

Terms in brackets cancel due to corresponding null-webs we can depict schematically as:
\be
\begin{array}{c}
\begin{tikzpicture}
\draw[thick] (0,1) -- (1,-0.5) (0.5,-1) -- (1,-0.5) (2,-1) -- (1,-0.5);
\draw[ultra thick, ->] (0,0) -- (2,0);\node[right] at (2,0) {$\wp$};
\begin{scope}[shift={(0,1)}]
\draw[ultra thick, purple] (-0.1,-0.1) -- (0.1,0.1) (0.1,-0.1) -- (-0.1,0.1);
\end{scope}
\begin{scope}[shift={(0.5,-1)}]
\draw[ultra thick, purple] (-0.1,-0.1) -- (0.1,0.1) (0.1,-0.1) -- (-0.1,0.1);
\end{scope}
\begin{scope}[shift={(2,-1)}]
\draw[ultra thick, purple] (-0.1,-0.1) -- (0.1,0.1) (0.1,-0.1) -- (-0.1,0.1);
\end{scope}
\node[left] at (0.4,-1) {$\fd_2$};
\node[right] at (2.1,-1) {$\fd_3$};
\node[right] at (0.1,1) {$\fd_1$};
\end{tikzpicture}
\end{array}\qquad
\begin{array}{c}
\begin{tikzpicture}
\draw[thick] (0,1) -- (1,-0.5) (0.3,-1) -- (1,-0.5) (2,-1) -- (1,-0.5) (1,-0.5) -- (1.7,0);
\draw[ultra thick, ->] (0,0.5) -- (2,0.5);\node[right] at (2,0.5) {$\wp$};
\begin{scope}[shift={(0,1)}]
\draw[ultra thick, purple] (-0.1,-0.1) -- (0.1,0.1) (0.1,-0.1) -- (-0.1,0.1);
\end{scope}
\begin{scope}[shift={(0.3,-1)}]
\draw[ultra thick, purple] (-0.1,-0.1) -- (0.1,0.1) (0.1,-0.1) -- (-0.1,0.1);
\end{scope}
\begin{scope}[shift={(2,-1)}]
\draw[ultra thick, purple] (-0.1,-0.1) -- (0.1,0.1) (0.1,-0.1) -- (-0.1,0.1);
\end{scope}
\begin{scope}[shift={(1.7,0)}]
\draw[ultra thick, purple] (-0.1,-0.1) -- (0.1,0.1) (0.1,-0.1) -- (-0.1,0.1);
\end{scope}
\node[right] at (2.1,-1) {$\fd_7$};
\node[right] at (1.8,0) {$\fd_6$};
\node[left] at (0.2,-1) {$\fd_4$};
\node[right] at (0.1,1) {$\fd_5$};
\end{tikzpicture}
\end{array}
\ee
And remaining terms in \eqref{su_n_RI} represent exactly one strand identity interface. This can be easily seen in the explicit detour construction or by an implant argument like in \cite{GM}.

Actual calculation of link invariants in this setup is still rather involved. We consider an example of such calculation for the Hopf link in $\fs\fu_3$ in Appendix \ref{sec:Hopf}. As one sees the resulting cohmology is isomorphic to Khovanov-Rozansky homology in this example \cite{Gukov_Hopf} and Poincar\'e polynomials are related by a simple change of variables:
\be
P_{\rm KR}(q,t)=P_{\rm LG}(qt,t)
\ee

\section{On Khovanov-Rozansky Link Homology}\label{sec:compare}
\subsection{Brief review of Khovanov-Rozansky homology}

Here we follow mostly recent review \cite{guide}.

Generically, $R$-matrix when both representations are fundamental has a simple decomposition (up to an overall scaling):
\be
R=1-q P_{\tiny\arraycolsep=0pt\begin{array}{c}
		\begin{tikzpicture}
		\begin{scope}[scale=0.15]
		\draw (0,0) -- (0,1) -- (2,1) -- (2,0) -- cycle (1,0) -- (1,1);
		\end{scope}
		\end{tikzpicture}
\end{array}},\quad R^{-1}=1-q^{-1} P_{\tiny\arraycolsep=0pt\begin{array}{c}
\begin{tikzpicture}
\begin{scope}[scale=0.15]
\draw (0,0) -- (0,1) -- (2,1) -- (2,0) -- cycle (1,0) -- (1,1);
\end{scope}
\end{tikzpicture}
\end{array}}
\ee
where $P_{\tiny\arraycolsep=0pt\begin{array}{c}
	\begin{tikzpicture}
	\begin{scope}[scale=0.15]
	\draw (0,0) -- (0,1) -- (2,1) -- (2,0) -- cycle (1,0) -- (1,1);
	\end{scope}
	\end{tikzpicture}
	\end{array}}$ is a projector to the first antisymmetric representation. This projector becomes a projector to a trivial representation in the case of $\fs\fu_2$ and is non-trivial for the generic case of $\fs\fu_n$. So in the resolution of an intersection a new type of diagrams appears. It contains a tri-valent vertices for two incoming or outgoing edges and special edge corresponding to the anti-symmetric representation. So one has the following complex decomposition:
\be
\begin{split}
	E\left[
	\begin{array}{c}
		\begin{tikzpicture}[scale=0.7]
		\draw[ultra thick, ->] (-0.5,-0.5) -- (0.5,0.5);
		\draw[ultra thick] (0.5,-0.5) -- (0.1,-0.1);
		\draw[ultra thick, ->] (-0.1,0.1) -- (-0.5,0.5);
		\end{tikzpicture}
	\end{array}\right]=[q^{-1}t^{-1}] \; E\left[
	\begin{array}{c}
		\begin{tikzpicture}[scale=0.7]
		\draw [ultra thick,->] (-0.5,-0.5) to [out= 45, in =315] (-0.5,0.5); \draw [ultra thick,->] (0.5,-0.5) to [out=135, in=225] (0.5,0.5);
		\end{tikzpicture}
	\end{array}\right]\oplus E\left[
	\begin{array}{c}
		\begin{tikzpicture}[scale=0.7]
		\draw [ultra thick] (-0.5,-0.5) -- (0,-0.2) -- (0.5,-0.5);
		\draw [ultra thick, ->] (0,0.2) -- (-0.5,0.5); 
		\draw [ultra thick, ->] (0,0.2) -- (0.5,0.5);
		\draw[fill=white] (0.05,0.2) -- (0.05,-0.2) -- (-0.05,-0.2) -- (-0.05,0.2) -- cycle;
		\end{tikzpicture}
	\end{array}
	\right]\\
	E\left[
	\begin{array}{c}
		\begin{tikzpicture}[scale=0.7, xscale=-1]
		\draw[ultra thick, ->] (-0.5,-0.5) -- (0.5,0.5);
		\draw[ultra thick] (0.5,-0.5) -- (0.1,-0.1);
		\draw[ultra thick, ->] (-0.1,0.1) -- (-0.5,0.5);
		\end{tikzpicture}
	\end{array}\right]=[qt] \; E\left[
	\begin{array}{c}
		\begin{tikzpicture}[scale=0.7]
		\draw [ultra thick,->] (-0.5,-0.5) to [out= 45, in =315] (-0.5,0.5); \draw [ultra thick,->] (0.5,-0.5) to [out=135, in=225] (0.5,0.5);
		\end{tikzpicture}
	\end{array}\right]\oplus E\left[
	\begin{array}{c}
		\begin{tikzpicture}[scale=0.7]
		\draw [ultra thick] (-0.5,-0.5) -- (0,-0.2) -- (0.5,-0.5);
		\draw [ultra thick, ->] (0,0.2) -- (-0.5,0.5); 
		\draw [ultra thick, ->] (0,0.2) -- (0.5,0.5);
		\draw[fill=white] (0.05,0.2) -- (0.05,-0.2) -- (-0.05,-0.2) -- (-0.05,0.2) -- cycle;
		\end{tikzpicture}
	\end{array}
	\right]
\end{split}
\ee
Each resolved diagram forms a vertex of the resolution hypercube and it is represented by a tri-valent graph. Khovanov-Rozansky homology theory associates to each graph a $q$-graded vector space \cite{KR}. Technically this space is spanned by elements of a polynomial ring modulo certain ideals so that the elements of the diagram correspond to coloring of the resulting diagram by vectors of corresponding representations. So, for example, one associates variable $x_i$ to each oriented edge of the resolved diagram with an ideal $x_i^{n+1}=0$ to ensure that the fundamental rep is $n$-dimensional.

The differential acts by local morphisms $d=\sum\lm_{{\rm all}\; \chi} d_{\chi}$ in intersections $\chi$ mapping between two vector spaces corresponding to two resolutions. However, a simple explicit construction of this differential like in the $SU(2)$ case is still missing, one should use a rather intricate matrix factorization theory instead. Similarly to $\fs\fu_2$ case one would like to construct some analog of TQFT in terms of invariance of thin film embeddings. In this case one uses two types of films, for example see Figure \ref{fig:foam} for a film corresponding to a differential morphism. In the literature compositions of these thin films are referred to as (pre-)foams and studies of relations between morphisms are referred to as foam calculus \cite{Chun:2015gda,Lauda,Khovanov_sl_3,Mackaay}. In particular, one uses foam calculus to show local differential anti-commutativity
$$
\left[d_{\chi},d_{\chi'}\right]=0
$$
similarly to TQFT relations \eqref{commutatitivity}.

\begin{figure}
\begin{center}
\begin{tikzpicture}
\draw[ultra thick] (-1.6,0.4) -- (-0.5,0) -- (-2.4,-0.4) (-0.5,0) -- (0.5,0);
\draw[ultra thick,->] (0.5,0) -- (1.6,-0.4);
\draw[ultra thick,->] (0.5,0) -- (2.4,0.4);
\draw[fill=blue, opacity=0.2] (-1.6,0.4) -- (-0.5,0) to[out=90,in=180] (0,1) to[out=0,in=90] (0.5,0) -- (2.4,0.4) -- (2.4,2.4) to[out=190,in=0] (0.2,2.2) to[out=180,in=345] (-1.6,2.4) -- cycle;
\draw[fill=red, opacity=0.5, thick] (-0.5,0) to[out=90,in=180] (0,1) to[out=0,in=90] (0.5,0) -- cycle;
\draw[fill=blue, opacity=0.2] (-2.4,-0.4) -- (-0.5,0) to[out=90,in=180] (0,1) to[out=0,in=90] (0.5,0) -- (1.6,-0.4) -- (1.6,1.6) to[out=165,in=0] (-0.2,1.8) to[out=180,in=10] (-2.4,1.6) -- cycle;
\draw[ultra thick, ->] (-2.4,1.6) to[out=10,in=180] (-0.2,1.8) to[out=0,in=165] (1.6,1.6);
\draw[ultra thick, ->] (-1.6,2.4) to[out=345,in=180] (0.2,2.2) to[out=0,in=190] (2.4,2.4);
\end{tikzpicture}\qquad \qquad \begin{tikzpicture}
\begin{scope}[scale=-1]
\draw[ultra thick, <-] (-2.4,1.6) to[out=10,in=180] (-0.2,1.8) to[out=0,in=165] (1.6,1.6);
\draw[ultra thick, <-] (-1.6,2.4) to[out=345,in=180] (0.2,2.2) to[out=0,in=190] (2.4,2.4);
\draw[fill=blue, opacity=0.2] (-1.6,0.4) -- (-0.5,0) to[out=90,in=180] (0,1) to[out=0,in=90] (0.5,0) -- (2.4,0.4) -- (2.4,2.4) to[out=190,in=0] (0.2,2.2) to[out=180,in=345] (-1.6,2.4) -- cycle;
\draw[fill=red, opacity=0.5, thick] (-0.5,0) to[out=90,in=180] (0,1) to[out=0,in=90] (0.5,0) -- cycle;
\draw[fill=blue, opacity=0.2] (-2.4,-0.4) -- (-0.5,0) to[out=90,in=180] (0,1) to[out=0,in=90] (0.5,0) -- (1.6,-0.4) -- (1.6,1.6) to[out=165,in=0] (-0.2,1.8) to[out=180,in=10] (-2.4,1.6) -- cycle;
\draw[ultra thick, <-] (-1.6,0.4) -- (-0.5,0);
\draw[ultra thick, <-] (-2.4,-0.4) -- (-0.5,0);
\draw[ultra thick] (-0.5,0) -- (0.5,0);
\draw[ultra thick] (0.5,0) -- (1.6,-0.4);
\draw[ultra thick] (0.5,0) -- (2.4,0.4);
\end{scope}
\end{tikzpicture}
\end{center}
\caption{Simple  prefoams for a crossing resolution}\label{fig:foam}
\end{figure}

The corresponding homology exists for link $L$ \cite{KR}
\be
{}_n {\bf KRHom}(L)=H_*(E(L),d)
\ee

Due to anti-commutativity of local differentials $d_{\chi}$ the Khovanov-Rozansky homology satisfies tautologically (see Appendix \ref{sec:taut}) skein exact triangle (each full cycle in this triangle gives a homological degree shift):
\be
\begin{array}{c}
\begin{tikzpicture}
\node(A) {${}_n {\bf KRHom}\left[\begin{array}{c}
	\begin{tikzpicture}[scale=0.7]
	\draw[ultra thick, ->] (-0.5,-0.5) -- (0.5,0.5);
	\draw[ultra thick] (0.5,-0.5) -- (0.1,-0.1);
	\draw[ultra thick, ->] (-0.1,0.1) -- (-0.5,0.5);
	\end{tikzpicture}
\end{array}\right]$};
\node(B) at (-4.5,0) {${}_n {\bf KRHom}\left[\begin{array}{c}
	\begin{tikzpicture}[scale=0.7]
	\draw [ultra thick,->] (-0.5,-0.5) to [out= 45, in =315] (-0.5,0.5); \draw [ultra thick,->] (0.5,-0.5) to [out=135, in=225] (0.5,0.5);
	\end{tikzpicture}
	\end{array}\right]$};
\node(C) at (4.5,0) {${}_n {\bf KRHom}\left[\begin{array}{c}
	\begin{tikzpicture}[scale=0.7]
	\draw [ultra thick] (-0.5,-0.5) -- (0,-0.2) -- (0.5,-0.5);
	\draw [ultra thick, ->] (0,0.2) -- (-0.5,0.5); 
	\draw [ultra thick, ->] (0,0.2) -- (0.5,0.5);
	\draw[fill=white] (0.05,0.2) -- (0.05,-0.2) -- (-0.05,-0.2) -- (-0.05,0.2) -- cycle;
	\end{tikzpicture}
	\end{array}\right]$};
\path (B) edge[->] (A) (A) edge[->] (C) (C) edge[bend left, ->] node[above] {$t$} (B);
\end{tikzpicture}
\end{array}
\ee
\be
\begin{array}{c}
	\begin{tikzpicture}
	\node(A) {${}_n {\bf KRHom}\left[\begin{array}{c}
		\begin{tikzpicture}[scale=0.7,xscale=-1]
		\draw[ultra thick, ->] (-0.5,-0.5) -- (0.5,0.5);
		\draw[ultra thick] (0.5,-0.5) -- (0.1,-0.1);
		\draw[ultra thick, ->] (-0.1,0.1) -- (-0.5,0.5);
		\end{tikzpicture}
		\end{array}\right]$};
	\node(B) at (-4.5,0) {${}_n {\bf KRHom}\left[\begin{array}{c}
		\begin{tikzpicture}[scale=0.7]
		\draw [ultra thick,->] (-0.5,-0.5) to [out= 45, in =315] (-0.5,0.5); \draw [ultra thick,->] (0.5,-0.5) to [out=135, in=225] (0.5,0.5);
		\end{tikzpicture}
		\end{array}\right]$};
	\node(C) at (4.5,0) {${}_n {\bf KRHom}\left[\begin{array}{c}
		\begin{tikzpicture}[scale=0.7]
		\draw [ultra thick] (-0.5,-0.5) -- (0,-0.2) -- (0.5,-0.5);
		\draw [ultra thick, ->] (0,0.2) -- (-0.5,0.5); 
		\draw [ultra thick, ->] (0,0.2) -- (0.5,0.5);
		\draw[fill=white] (0.05,0.2) -- (0.05,-0.2) -- (-0.05,-0.2) -- (-0.05,0.2) -- cycle;
		\end{tikzpicture}
		\end{array}\right]$};
	\path (B) edge[<-] (A) (A) edge[<-] (C) (C) edge[bend left, <-] node[above] {$t$} (B);
	\end{tikzpicture}
\end{array}
\ee

\subsection{Exact skein triangle in Landau-Ginzburg $\fs\fu_n$ cohomology}
In this case we will not argue the whole equivalence of homologies due to complexity of Khovanov-Rozansky homology formulation. Similarly rules for the supercharge $Q$ in the generic $\fs\fu_n$ case are rather involved. In particular, there will be non-trivial instantons  changing sheet pairs in decomposition \eqref{decomp}. So we will only emphasize a similarity between Landau-Ginzburg cohomology and Khovanov-Rozansky homology lying at the top. It follows from the isotypical interface construction.

Using the relation between spectral covers in vacua $\Box\otimes\Box$ in $\fs\fu_2$ and $\fs\fu_n$ (see Section \ref{sec:su_2}) it is not complicated to construct analogous isotypical interface in the $\fs\fu_n$ case, however in the isotypical decomposition instead of trivial representation appears the first antisymmetric one (that is trivial in the $\fs\fu_2$ case). So after shifting the fermion degree ${\bf F}\to {\bf F}+{\bf P}$, or, equivalently, changing variables $q\to q t$ one gets:
\be
\begin{split}
\tilde\CE\left[
\begin{array}{c}
	\begin{tikzpicture}[scale=0.7]
	\draw[ultra thick, ->] (-0.5,-0.5) -- (0.5,0.5);
	\draw[ultra thick] (0.5,-0.5) -- (0.1,-0.1);
	\draw[ultra thick, ->] (-0.1,0.1) -- (-0.5,0.5);
	\end{tikzpicture}
\end{array}\right]=[q^{-1}t^{-1}] \; \tilde\CE\left[
\begin{array}{c}
	\begin{tikzpicture}[scale=0.7]
	\draw [ultra thick,->] (-0.5,-0.5) to [out= 45, in =315] (-0.5,0.5); \draw [ultra thick,->] (0.5,-0.5) to [out=135, in=225] (0.5,0.5);
	\end{tikzpicture}
\end{array}\right]\oplus \tilde\CE\left[
\begin{array}{c}
	\begin{tikzpicture}[scale=0.7]
	\draw [ultra thick] (-0.5,-0.5) -- (0,-0.2) -- (0.5,-0.5);
	\draw [ultra thick, ->] (0,0.2) -- (-0.5,0.5); 
	\draw [ultra thick, ->] (0,0.2) -- (0.5,0.5);
	\draw[fill=white] (0.05,0.2) -- (0.05,-0.2) -- (-0.05,-0.2) -- (-0.05,0.2) -- cycle;
	\end{tikzpicture}
\end{array}
\right]\\
\tilde\CE\left[
\begin{array}{c}
	\begin{tikzpicture}[scale=0.7, xscale=-1]
	\draw[ultra thick, ->] (-0.5,-0.5) -- (0.5,0.5);
	\draw[ultra thick] (0.5,-0.5) -- (0.1,-0.1);
	\draw[ultra thick, ->] (-0.1,0.1) -- (-0.5,0.5);
	\end{tikzpicture}
\end{array}\right]=[qt] \; \tilde\CE\left[
\begin{array}{c}
	\begin{tikzpicture}[scale=0.7]
	\draw [ultra thick,->] (-0.5,-0.5) to [out= 45, in =315] (-0.5,0.5); \draw [ultra thick,->] (0.5,-0.5) to [out=135, in=225] (0.5,0.5);
	\end{tikzpicture}
\end{array}\right]\oplus \tilde\CE\left[
\begin{array}{c}
	\begin{tikzpicture}[scale=0.7]
	\draw [ultra thick] (-0.5,-0.5) -- (0,-0.2) -- (0.5,-0.5);
	\draw [ultra thick, ->] (0,0.2) -- (-0.5,0.5); 
	\draw [ultra thick, ->] (0,0.2) -- (0.5,0.5);
	\draw[fill=white] (0.05,0.2) -- (0.05,-0.2) -- (-0.05,-0.2) -- (-0.05,0.2) -- cycle;
	\end{tikzpicture}
\end{array}
\right]
\end{split}
\ee
Where 
\be
\begin{split}
\begin{array}{c}
	\begin{tikzpicture}[scale=0.7]
	\draw [ultra thick] (-0.5,-0.5) -- (0,-0.2) -- (0.5,-0.5);
	\draw [ultra thick, ->] (0,0.2) -- (-0.5,0.5); 
	\draw [ultra thick, ->] (0,0.2) -- (0.5,0.5);
	\draw[fill=white] (0.05,0.2) -- (0.05,-0.2) -- (-0.05,-0.2) -- (-0.05,0.2) -- cycle;
	\end{tikzpicture}
\end{array}=\bigoplus\lm_{\substack{i,j\\ i\neq j}}
\begin{array}{c}
	\begin{tikzpicture}[scale=0.7]
	\draw [ultra thick] (-0.5,-0.5) -- (0,-0.2) -- (0.5,-0.5);
	\draw [ultra thick, ->] (0,0.2) -- (-0.5,0.5); 
	\draw [ultra thick, ->] (0,0.2) -- (0.5,0.5);
	\draw[fill=white] (0.05,0.2) -- (0.05,-0.2) -- (-0.05,-0.2) -- (-0.05,0.2) -- cycle;
	\node[below] at (-0.5,-0.5) {$j$};
	\node[below] at (0.5,-0.5) {$i$};
	\node[above] at (-0.5,0.5) {$i$};
	\node[above] at (0.5,0.5) {$j$};
	\end{tikzpicture}
\end{array}\oplus \bigoplus\lm_{\substack{i,j\\ j>i}}\left[
q\begin{array}{c}
\begin{tikzpicture}[scale=0.7]
\draw [ultra thick] (-0.5,-0.5) -- (0,-0.2) -- (0.5,-0.5);
\draw [ultra thick, ->] (0,0.2) -- (-0.5,0.5); 
\draw [ultra thick, ->] (0,0.2) -- (0.5,0.5);
\draw[fill=white] (0.05,0.2) -- (0.05,-0.2) -- (-0.05,-0.2) -- (-0.05,0.2) -- cycle;
\node[below] at (-0.5,-0.5) {$j$};
\node[below] at (0.5,-0.5) {$i$};
\node[above] at (-0.5,0.5) {$j$};
\node[above] at (0.5,0.5) {$i$};
\end{tikzpicture}
\end{array}\oplus q^{-1}\begin{array}{c}
\begin{tikzpicture}[scale=0.7]
\draw [ultra thick] (-0.5,-0.5) -- (0,-0.2) -- (0.5,-0.5);
\draw [ultra thick, ->] (0,0.2) -- (-0.5,0.5); 
\draw [ultra thick, ->] (0,0.2) -- (0.5,0.5);
\draw[fill=white] (0.05,0.2) -- (0.05,-0.2) -- (-0.05,-0.2) -- (-0.05,0.2) -- cycle;
\node[below] at (-0.5,-0.5) {$i$};
\node[below] at (0.5,-0.5) {$j$};
\node[above] at (-0.5,0.5) {$i$};
\node[above] at (0.5,0.5) {$j$};
\end{tikzpicture}
\end{array}\right]=\\
=\bigoplus\lm_{j>i}\left(
\begin{array}{c}
	\begin{tikzpicture}
	\begin{scope}[scale=0.8]
	\draw[purple,ultra thick] (0,0) -- (0,2) (1,0) -- (1,2);
	\draw[->] (0.6,1.5) -- (1,1.5) (0,1.5) -- (0.6,1.5);
	\node[below] at (0.5,1.5) {$\fd_4$};
	\node[below] at (0,0) {$j$}; \node[below] at (1,0) {$i$};
	\node[above] at (0,2) {$i$}; \node[above] at (1,2) {$j$};
	\end{scope}
	\end{tikzpicture}
\end{array}\oplus 
\begin{array}{c}
	\begin{tikzpicture}
	\begin{scope}[scale=0.8]
	\draw[purple,ultra thick] (0,0) -- (0,2) (1,0) -- (1,2);
	\draw[->] (0.4,0.5) -- (0,0.5) (1,0.5) -- (0.4,0.5);
	\node[above] at (0.5,0.5) {$\fd_3$};
	\node[below] at (0,0) {$i$}; \node[below] at (1,0) {$j$};
	\node[above] at (0,2) {$j$}; \node[above] at (1,2) {$i$};
	\end{scope}
	\end{tikzpicture}
\end{array}\oplus q
\begin{array}{c}
	\begin{tikzpicture}
	\begin{scope}[scale=0.8]
	\draw[purple,ultra thick] (0,0) -- (0,2) (1,0) -- (1,2);
	\draw[->] (0.6,0.5) -- (1,0.5) (0,0.5) -- (0.6,0.5);
	\draw[->] (0.4,1) -- (0,1) (1,1) -- (0.4,1); 
	\node[above] at (0.5,1) {$\fd_6$};
	\node[below] at (0.5,0.5) {$\fd_1$};
	\node[below] at (0,0) {$j$}; \node[below] at (1,0) {$i$};
	\node[above] at (0,2) {$j$}; \node[above] at (1,2) {$i$};
	\end{scope}
	\end{tikzpicture}
\end{array}
\oplus q^{-1}
\begin{array}{c}
	\begin{tikzpicture}
	\begin{scope}[scale=0.8]
	\draw[purple,ultra thick] (0,0) -- (0,2) (1,0) -- (1,2);
	\draw[->] (0.4,0.5) -- (0,0.5) (1,0.5) -- (0.4,0.5);
	\draw[->] (0.6,1) -- (1,1) (0,1) -- (0.6,1); 
	\node[above] at (0.5,1) {$\fd_5$};
	\node[below] at (0.5,0.5) {$\fd_3$};
	\node[below] at (0,0) {$i$}; \node[below] at (1,0) {$j$};
	\node[above] at (0,2) {$i$}; \node[above] at (1,2) {$j$};
	\end{scope}
	\end{tikzpicture}
\end{array}
\right)
\end{split}\label{P}
\ee
and the corresponding Euler characteristic of this interface is projector $P_{\tiny\arraycolsep=0pt\begin{array}{c}
	\begin{tikzpicture}
	\begin{scope}[scale=0.15]
	\draw (0,0) -- (0,1) -- (2,1) -- (2,0) -- cycle (1,0) -- (1,1);
	\end{scope}
	\end{tikzpicture}
	\end{array}}$. Here solitons in the interface correspond to detours depicted in diagram \eqref{isotyp}.

For fusion/defusion interfaces we have:
\begin{subequations}
	\be
	\tilde\CE\left(\begin{array}{c}
		\begin{tikzpicture}
		\begin{scope}[scale=0.8]
		\draw[ultra thick] (0,0) to[out=270,in=180] (0.6,-0.5);
		\draw[ultra thick,<-] (0.6,-0.5) to[out=0,in=270] (1.2,0);
		\end{scope}
		\end{tikzpicture}
	\end{array}\right)=\begin{array}{c}
	\begin{tikzpicture}
	\begin{scope}[scale=0.8]
	\draw[ultra thick, purple] (-0.2,0) -- (1.2,0) (0,0) -- (0,1) (1,0) -- (1,1);
	\node[below] at (0,0) {$n$}; \node[below] at (1,0) {$\bar n$};
	\node[above] at (0,1) {$n$}; \node[above] at (1,1) {$\bar n$};
	\end{scope}
	\end{tikzpicture}
\end{array}\oplus \ldots\oplus q^{-(n-1)}\begin{array}{c}
\begin{tikzpicture}
\begin{scope}[scale=0.8]
\draw[ultra thick, purple] (-0.2,0) -- (1.2,0) (0,0) -- (0,1) (1,0) -- (1,1);
\node[below] at (0,0) {$n$}; \node[below] at (1,0) {$\bar n$};
\node[above] at (0,1) {$1$}; \node[above] at (1,1) {$\bar 1$};
\draw (0,0.25) -- (1,0.25) (0,0.75) -- (1,0.75);
\draw [->] (0,0.25) -- (0.6,0.25);
\draw [->] (0,0.75) -- (0.6,0.75);
\draw[dashed] (0,0.5) -- (1,0.5);
\draw[dashed,->] (0,0.5) -- (0.6,0.5);
\end{scope}
\end{tikzpicture}
\end{array}
\ee
\be
\tilde\CE\left(\begin{array}{c}
	\begin{tikzpicture}
	\begin{scope}[scale=0.8, yscale=-1]
	\draw[ultra thick] (0,0) to[out=270,in=180] (0.6,-0.5);
	\draw[ultra thick,<-] (0.6,-0.5) to[out=0,in=270] (1.2,0);
	\end{scope}
	\end{tikzpicture}
\end{array}\right)=\begin{array}{c}
\begin{tikzpicture}
\begin{scope}[scale=0.8]
\draw[ultra thick, purple] (-0.2,1) -- (1.2,1) (0,0) -- (0,1) (1,0) -- (1,1);
\node[below] at (0,0) {$\bar n$}; \node[below] at (1,0) {$n$};
\node[above] at (0,1) {$\bar n$}; \node[above] at (1,1) {$n$};
\end{scope}
\end{tikzpicture}
\end{array}\oplus \ldots\oplus q^{n-1}\begin{array}{c}
\begin{tikzpicture}
\begin{scope}[scale=0.8]
\draw[ultra thick, purple] (-0.2,1) -- (1.2,1) (0,0) -- (0,1) (1,0) -- (1,1);
\node[below] at (0,0) {$\bar 1$}; \node[below] at (1,0) {$1$};
\node[above] at (0,1) {$\bar n$}; \node[above] at (1,1) {$n$};
\draw (0,0.25) -- (1,0.25) (0,0.75) -- (1,0.75);
\draw [->] (0,0.25) -- (0.6,0.25);
\draw [->] (0,0.75) -- (0.6,0.75);
\draw[dashed] (0,0.5) -- (1,0.5);
\draw[dashed,->] (0,0.5) -- (0.6,0.5);
\end{scope}
\end{tikzpicture}
\end{array}
\ee
\be
\tilde\CE\left(\begin{array}{c}
	\begin{tikzpicture}
	\begin{scope}[scale=0.8]
	\draw[ultra thick,->] (0,0) to[out=270,in=180] (0.6,-0.5);
	\draw[ultra thick] (0.6,-0.5) to[out=0,in=270] (1.2,0);
	\end{scope}
	\end{tikzpicture}
\end{array}\right)=\begin{array}{c}
\begin{tikzpicture}
\begin{scope}[scale=0.8]
\draw[ultra thick, purple] (-0.2,0) -- (1.2,0) (0,0) -- (0,1) (1,0) -- (1,1);
\node[below] at (0,0) {$\bar 1$}; \node[below] at (1,0) {$1$};
\node[above] at (0,1) {$\bar 1$}; \node[above] at (1,1) {$1$};
\end{scope}
\end{tikzpicture}
\end{array}\oplus \ldots\oplus q^{-(n-1)}\begin{array}{c}
\begin{tikzpicture}
\begin{scope}[scale=0.8]
\draw[ultra thick, purple] (-0.2,0) -- (1.2,0) (0,0) -- (0,1) (1,0) -- (1,1);
\node[below] at (0,0) {$\bar 1$}; \node[below] at (1,0) {$1$};
\node[above] at (0,1) {$\bar n$}; \node[above] at (1,1) {$n$};
\draw (0,0.25) -- (1,0.25) (0,0.75) -- (1,0.75);
\draw [->] (0,0.25) -- (0.6,0.25);
\draw [->] (0,0.75) -- (0.6,0.75);
\draw[dashed] (0,0.5) -- (1,0.5);
\draw[dashed,->] (0,0.5) -- (0.6,0.5);
\end{scope}
\end{tikzpicture}
\end{array},
\ee
\be
\tilde\CE\left(\begin{array}{c}
	\begin{tikzpicture}
	\begin{scope}[scale=0.8, yscale=-1]
	\draw[ultra thick,->] (0,0) to[out=270,in=180] (0.6,-0.5);
	\draw[ultra thick] (0.6,-0.5) to[out=0,in=270] (1.2,0);
	\end{scope}
	\end{tikzpicture}
\end{array}\right)=\begin{array}{c}
\begin{tikzpicture}
\begin{scope}[scale=0.8]
\draw[ultra thick, purple] (-0.2,1) -- (1.2,1) (0,0) -- (0,1) (1,0) -- (1,1);
\node[below] at (0,0) {$1$}; \node[below] at (1,0) {$\bar 1$};
\node[above] at (0,1) {$1$}; \node[above] at (1,1) {$\bar 1$};
\end{scope}
\end{tikzpicture}
\end{array}\oplus \ldots\oplus q ^{n-1}\begin{array}{c}
\begin{tikzpicture}
\begin{scope}[scale=0.8]
\draw[ultra thick, purple] (-0.2,1) -- (1.2,1) (0,0) -- (0,1) (1,0) -- (1,1);
\node[below] at (0,0) {$n$}; \node[below] at (1,0) {$\bar n$};
\node[above] at (0,1) {$1$}; \node[above] at (1,1) {$\bar 1$};
\draw (0,0.25) -- (1,0.25) (0,0.75) -- (1,0.75);
\draw [->] (0,0.25) -- (0.6,0.25);
\draw [->] (0,0.75) -- (0.6,0.75);
\draw[dashed] (0,0.5) -- (1,0.5);
\draw[dashed,->] (0,0.5) -- (0.6,0.5);
\end{scope}
\end{tikzpicture}
\end{array}
\ee
\end{subequations}
Apparently, the cohomology of the unknot is properly normalized:
\be
{}_n{\bf LGCoh}\left(\begin{array}{c}
	\begin{tikzpicture}
	\draw[ultra thick] (0,0) circle (0.3);
	\end{tikzpicture}
\end{array} \right)=\bigoplus\lm_{j=0}^{n-1}q^{2j-n+1}\IZ
\ee

The next step in this construction is to provide operators $Q_{\chi}$ acting in each intersection resolution $\chi$ locally. An important property of locality is that $Q_{\chi}$ and $Q_{\chi'}$ anti-commute for different intersections $\chi$ and $\chi'$. In this case we are able to say that 
\be
Q=\sum\lm_{\chi} Q_{\chi}
\ee
And a tautological consequence of this construction (see Appendix \ref{sec:taut}) is an exact triangle for LG cohomology:
\be
\begin{array}{c}
	\begin{tikzpicture}
	\node(A) {${}_n {\bf LGCoh}\left[\begin{array}{c}
		\begin{tikzpicture}[scale=0.7]
		\draw[ultra thick, ->] (-0.5,-0.5) -- (0.5,0.5);
		\draw[ultra thick] (0.5,-0.5) -- (0.1,-0.1);
		\draw[ultra thick, ->] (-0.1,0.1) -- (-0.5,0.5);
		\end{tikzpicture}
		\end{array}\right]$};
	\node(B) at (-4.5,0) {${}_n {\bf LGCoh}\left[\begin{array}{c}
		\begin{tikzpicture}[scale=0.7]
		\draw [ultra thick,->] (-0.5,-0.5) to [out= 45, in =315] (-0.5,0.5); \draw [ultra thick,->] (0.5,-0.5) to [out=135, in=225] (0.5,0.5);
		\end{tikzpicture}
		\end{array}\right]$};
	\node(C) at (4.5,0) {${}_n {\bf LGCoh}\left[\begin{array}{c}
		\begin{tikzpicture}[scale=0.7]
		\draw [ultra thick] (-0.5,-0.5) -- (0,-0.2) -- (0.5,-0.5);
		\draw [ultra thick, ->] (0,0.2) -- (-0.5,0.5); 
		\draw [ultra thick, ->] (0,0.2) -- (0.5,0.5);
		\draw[fill=white] (0.05,0.2) -- (0.05,-0.2) -- (-0.05,-0.2) -- (-0.05,0.2) -- cycle;
		\end{tikzpicture}
		\end{array}\right]$};
	\path (B) edge[->] (A) (A) edge[->] (C) (C) edge[bend left, ->] node[above] {$t$} (B);
	\end{tikzpicture}
\end{array}
\ee
\be
\begin{array}{c}
	\begin{tikzpicture}
	\node(A) {${}_n {\bf LGCoh}\left[\begin{array}{c}
		\begin{tikzpicture}[scale=0.7,xscale=-1]
		\draw[ultra thick, ->] (-0.5,-0.5) -- (0.5,0.5);
		\draw[ultra thick] (0.5,-0.5) -- (0.1,-0.1);
		\draw[ultra thick, ->] (-0.1,0.1) -- (-0.5,0.5);
		\end{tikzpicture}
		\end{array}\right]$};
	\node(B) at (-4.5,0) {${}_n {\bf LGCoh}\left[\begin{array}{c}
		\begin{tikzpicture}[scale=0.7]
		\draw [ultra thick,->] (-0.5,-0.5) to [out= 45, in =315] (-0.5,0.5); \draw [ultra thick,->] (0.5,-0.5) to [out=135, in=225] (0.5,0.5);
		\end{tikzpicture}
		\end{array}\right]$};
	\node(C) at (4.5,0) {${}_n {\bf LGCoh}\left[\begin{array}{c}
		\begin{tikzpicture}[scale=0.7]
		\draw [ultra thick] (-0.5,-0.5) -- (0,-0.2) -- (0.5,-0.5);
		\draw [ultra thick, ->] (0,0.2) -- (-0.5,0.5); 
		\draw [ultra thick, ->] (0,0.2) -- (0.5,0.5);
		\draw[fill=white] (0.05,0.2) -- (0.05,-0.2) -- (-0.05,-0.2) -- (-0.05,0.2) -- cycle;
		\end{tikzpicture}
		\end{array}\right]$};
	\path (B) edge[<-] (A) (A) edge[<-] (C) (C) edge[bend left, <-] node[above] {$t$} (B);
	\end{tikzpicture}
\end{array}
\ee

However let us avoid  constructing supercharge action explicitly, instead we propose implicit arguments for local supercharges' anti-commutativity.
The fact that local operators $Q_{\chi}$ exist is rather obvious. There is always a map between the last element of \eqref{P} and identity interface corresponding to contraction of solitons $\fd_3$ and $\fd_5$. Surely, $Q_{\chi}$ contains much more non-trivial elements. They appear due to collision of solitons bound to the intersection and result into emission of different soliton showers that spread across the link interacting with other solitons bound to other intersections and fusion/defusion interfaces. 

Let us argue that all the local supercharges $Q_{\chi}$ anti-commute. Consider two simple neighboring intersections and apply Reidemeister II move: 
$$
\begin{array}{c}
\begin{tikzpicture}
\begin{scope}[rotate=-45]
\draw[ultra thick, red,  ->] (0,-1) -- (0,1);
\draw[ultra thick, blue, ->] (0.5,-1) -- (0.5,1);
\draw[ultra thick, ->] (1,0) -- (0.6,0) (0.4,0) -- (0.1,0) (-0.1,0) -- (-0.5,0);
\end{scope}
\end{tikzpicture}
\end{array}=
\begin{array}{c}
\begin{tikzpicture}
\begin{scope}[rotate=-45]
\draw[ultra thick, red, ->] (0,-1) to[out=90,in=225] (0.25,-0.5) to[out=45,in=270] (0.5,0) to[out=90,in=315] (0.25,0.5) to[out=135,in=270] (0,1);
\draw[ultra thick, blue, ->] (0.5,-1) to[out=90,in=315] (0.3,-0.55) (0.2,-0.45) to[out=135,in=270] (0,0) to[out=90,in=225] (0.2,0.45) (0.3,0.55) to[out=45,in=270] (0.5,1);
\draw[ultra thick, ->] (1,0) -- (0.6,0) (0.4,0) -- (0.1,0) (-0.1,0) -- (-0.5,0);
\end{scope}
\end{tikzpicture}
\end{array}=
\begin{array}{c}
\begin{tikzpicture}
\begin{scope}[rotate=-45]
\draw[ultra thick, red,  ->] (0,-1) -- (0,-0.5) (0.5,-0.5) -- (0.5,0.5) (0,0.5) --  (0,1);
\draw[ultra thick, blue, ->] (0.5,-1) -- (0.5,-0.5) (0,-0.5) -- (0,0.5) (0.5,0.5) --  (0.5,1);
\draw[ultra thick, ->] (1,0) -- (0.6,0) (0.4,0) -- (0.1,0) (-0.1,0) -- (-0.5,0);
\end{scope}
\end{tikzpicture}
\end{array}
$$
Two new intersections appear, there we apply expansion of the $\CR$-interfaces and cancel extra contributions by the fake pair argument. The result is the same link, however the $\CR$-interfaces and corresponding local supercharge actions are permuted. This implies that there for any local pair of actions $Q_1$ and $Q_2$ there is another pair $\tilde Q_1$ and $\tilde Q_2$ applied in the opposite order so that $Q_2Q_1=-\tilde Q_1\tilde Q_2$. 

Other types of two neighbor intersections can be dealt with by auxiliary isotopy moves, then one uses the exchange argument proposed above:
$$
\begin{array}{c}
\begin{tikzpicture}
\begin{scope}[rotate=-45]
\draw[ultra thick, ->] (0,-1) -- (0,1);
\draw[ultra thick, <-] (0.5,-1) -- (0.5,1);
\draw[ultra thick, ->] (1,0) -- (0.6,0) (0.4,0) -- (0.1,0) (-0.1,0) -- (-0.5,0);
\end{scope}
\end{tikzpicture}
\end{array}= \begin{array}{c}
\begin{tikzpicture}
\begin{scope}[rotate=-45]
\draw[ultra thick, ->] (0,-1) -- (0,1);
\draw[ultra thick, ->] (1.5,1) to[out=270,in=45] (1,0) to[out=225,in=0] (0.7,-0.5) to[out=180,in=270] (0.5,0) to[out=90,in=180] (0.7,0.5) to[out=0,in=135] (0.95,0.05) (1.05,-0.05) to[out=315,in=90] (1.5,-1);
\draw[ultra thick, ->] (2,0) -- (1.1,0) (0.9,0) -- (0.6,0) (0.4,0) -- (0.1,0) (-0.1,0) -- (-0.5,0);
\end{scope}
\end{tikzpicture}
\end{array},\quad
\begin{array}{c}
\begin{tikzpicture}
\begin{scope}[rotate=-45]
\draw[ultra thick, ->] (0,-1) -- (0,1);
\draw[ultra thick, ->] (0.5,-1) -- (0.5,-0.1) (0.5,0.1) -- (0.5,1);
\draw[ultra thick, ->] (1,0) -- (0.1,0) (-0.1,0) -- (-0.5,0);
\end{scope}
\end{tikzpicture}
\end{array}= \begin{array}{c}
\begin{tikzpicture}
\begin{scope}[rotate=-45]
\draw[ultra thick, ->] (0,-1) -- (0,1);
\draw[ultra thick, ->] (2,0) -- (1.1,0) (0.9,0) -- (0.6,0) (0.4,0) -- (0.1,0) (-0.1,0) -- (-0.5,0);
\draw[ultra thick, ->] (1.5,-1) -- (1.5,-0.1) (1.5,0.1) to[out=90,in=0] (1.25,0.5)  to[out=180,in=90] (1,0) to[out=270,in=0] (0.75,-0.5) to[out=180,in=270] (0.5,0) -- (0.5,1);
\end{scope}
\end{tikzpicture}
\end{array}
$$

Apparently, one can extend these arguments to arbitrary pair of intersections, so all the local supercharges anti-commute.

\appendix

\section{Vacua in $\fs\fu_2$  LG model}\label{sec:vac_su_2}
In \cite{GM} vacua were associated to the fundamental representation of $\fs\fu_2$, in particular, for $k_a=1$ two solutions denoted by $(+)$ and $(-)$ (spin up -- spin down) were described. The field theory prescribes to associate to a puncture with a parameter $k$ an irreducible representation of dimension $k+1$, here we check this idea. In particular, we show that a system of vacuum equations for a single puncture:
\be\label{App_Eqs}
\frac{k}{w_i}-\sum\lm_{j\neq i}\frac{2}{w_i-w_j}+c=0,\quad \forall i=1,\ldots,n
\ee
has a single solution modulo permutation group $S_n$ for each $n$ in the interval $0\leq n\leq k$, and no solutions otherwise.

We define a resolvent $\rho(z)$ for this system:
\be
\rho(z):=\sum\lm_{i=1}^n \frac{1}{z-w_i^*}
\ee
where $w_i^*$ is a solution of \eqref{App_Eqs}. Obviously, $\rho(z)$ is an invariant of the permutation group $S_n$ acting on $w_i$, therefore to count equivalence classes $\{w_i^* \}/S_n$ we can count all distinct resolvent solutions.

We multiply each $i^{\rm th}$ equation by $(z-w_i^*)^{-1}$, sum over index $i$ and massage the result to get:
\be\label{App_resolv}
\rho(z)^2+\rho'(z)-\left(c+\frac{k}{z}\right)\rho(z)+\frac{c n}{z}=0
\ee
The resolvent expansion we are searching for reads:
\be\label{App_BC}
\rho(z)=\frac{n}{z}+\sum\lm_{s=1}^{\infty} \frac{T_s}{z^{s+1}}
\ee
where
\be
T_s=\sum\lm_i (w_i^*)^s
\ee
For example,
\be
T_1=\frac{-k n+n^2-n}{c},\quad T_2=\frac{n \left(k^2-3 k n+3 k+2 n^2-4 n+2\right)}{c^2},\quad \ldots
\ee
First of all notice there is a unique solution of \eqref{App_resolv} with boundary conditions \eqref{App_BC}, therefore for each $n$ equivalence class of solutions  $\{w_i^* \}/S_n$ is a single point.

Furthermore, even if we have a solution for $\rho(z)$ this solution may be incompatible with a system \eqref{App_Eqs} if $w_i^*=0$ for some $i$ or $w_i^*=w_j^*$ for some pair $(i,j)$.

It is simple to reformulate this incompatibility condition as a specific locus. Consider the following characteristic polynomial:
\be
P(x):=\prod\lm_{i=1}^n (x-w_i^*)=\sum\lm_{j=0}^n c_j x^j
\ee
The locus $w_i^*=w_j^*$ for some pair $(i,j)$ is defined by the discriminant ${\bf D}_x(P)=0$ and the locus $w_i^*=0$ for some $i$ is defined by $c_0=0$. In this way we construct the full singular locus as a union of mentioned loci:
\be
\Delta:=c_0 {\bf D}_x (P)
\ee
Eventually, using the famous relation between the momenta $T_s$ and the characteristic polynomial:
\be
P(x)=\left[x^ne^{-\sum\lm_{s\geq 1} \frac{T_s}{s x^s}}\right]_+=\chi_{\wedge^n\Box}(kT_k)
\ee
where by $[\star]_+$ we imply a truncation to non-negative powers of $x$, and $\chi$ is the Schur polynomial (see a nice review in \cite{characters}). We derive several first expressions for $\Delta$ for various $n$:
\be
\begin{split}
	\Delta(n=1,k)=\frac{k}{c},\quad \Delta(n=2,k)=-\frac{4 k (k-1)^2}{c^4},\quad \Delta(n=3,k)=\frac{108 k(k-1)^2(k-2)^3}{c^9},\ldots 
\end{split}
\ee
Thus we guess eventually the generic formula:
\be
\Delta(n,k)=(-1)^{g(n)}\frac{\prod\lm_{m=1}^n m^m}{c^{n^2}}\prod\lm_{m=0}^{n-1}(k-m)^{m+1}
\ee
where $g(n)$ is some integer valued function of $n$.

Equations \eqref{App_Eqs} have a single solution modulo permutation group if and only if $\Delta(n,k)\neq 0$, so we have one solution for all values of $n$  in the interval $0\leq n\leq k$.

\section{Examples of spectral analysis for SQM instantons}
\subsection{Airy model}\label{sec:Airy}
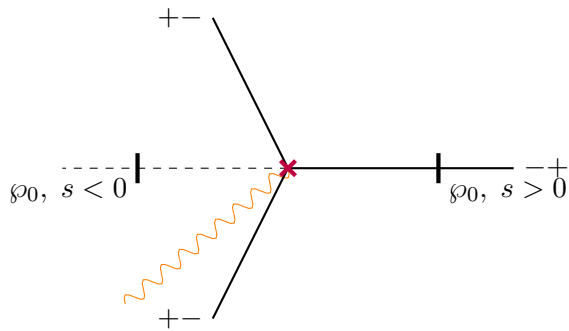
\begin{figure}[h!]
	\begin{center}
		\begin{tikzpicture}
		\begin{scope}[rotate=220]
		\draw [orange, domain=0:0.94] plot (\x, {0.1*sin(20*\x r)});
		\begin{scope}[shift={(0.94,0)}]
		\draw [orange, domain=0:0.94] plot (\x, {0.1*sin(20*\x r)});
		\end{scope}
		\begin{scope}[shift={(1.88,0)}]
		\draw [orange, domain=0:0.94] plot (\x, {0.1*sin(20*\x r)});
		\end{scope}
		\end{scope}
		\draw[thick] (0,0) -- (3,0) (0,0) -- (-1,2) (0,0) -- (-1,-2);
		\draw[dashed] (0,0) -- (-3,0);
		\begin{scope}[shift={(0,0)}]
		\draw[ultra thick, purple] (-0.1,-0.1) -- (0.1,0.1) (0.1,-0.1) -- (-0.1,0.1);
		\end{scope}
		\node[right] at (3,0) {$-+$};
		\node[left] at (-1,2) {$+-$};
		\node[left] at (-1,-2) {$+-$};
		\draw[ultra thick] (2,-0.2) -- (2,0.2);
		\draw[ultra thick] (-2,-0.2) -- (-2,0.2);
		\node[below right] at (2,0) {$\wp_0,\; s>0$};
		\node[below left] at (-2,0) {$\wp_0,\; s<0$};
		\end{tikzpicture}
	\end{center}
	\caption{Spectral curve for the Airy model.} \label{fig:Airy}
\end{figure}

Here we consider a simple illustrating example of the Airy model. Consider the following cubic height function on a real line parameterized by a coordinate $x$:
\be
h=\frac{x^3}{3}-s x
\ee
If $s>0$ this theory has two real vacua $x_{\pm}=\pm s^{\frac{1}{2}}$ and an instanton solution connecting them reads:
\be
x(\tau)=s^{\frac{1}{2}}\tanh \left(s^{\frac{1}{2}}\tau\right)
\ee
Let us rederive the same result using spectral analysis approach. First we promote fields and parameters to complex ones $s\to z$, $x\to Z$, $h\to H$:
\be
H=\frac{Z^3}{3}-z Z
\ee
Now function $H$ can be considered as a superpotential. The chiral ring is spanned by ${\bf 1}$, and ${\bf Z}$ with the following ring relations:
\be
{\bf 1}\cdot {\bf 1}= {\bf 1},\quad {\bf 1}\cdot {\bf Z}= {\bf Z}, \quad  {\bf Z}\cdot {\bf Z}=z {\bf 1}
\ee
The action of the derivative $\p_z$ acts as a multiplication by ${\bf Z}$ in the chiral ring. The resulting connection coincides with the Airy connection:
\be
\nabla=d-\zeta^{-1}dz\left(\begin{array}{cc}
	0 & -1\\
	-z & 0 \\
\end{array}\right)
\ee
Its flat sections are spanned by Airy functions. The spectral curve is very simple:
\be
p^2-z=0,\quad \lambda =p dz
\ee
The WKB web $\CW_1$ for $\zeta=1$  have the form depicted in Figure \ref{fig:Airy}. Now we construct $\wp_0$ around the point $s$ and see that for $s\geq 0$ there is an
instanton flowing from vacuum $-s^{\frac{1}{2}}$  to vacuum $s^{\frac{1}{2}}$, and there is no instanton otherwise.

\subsection{Airy${}^2$ model}\label{sec:Airy-2}
As an illustrative example of $Q$-matrix model sign calculation we consider Airy${}^2$ model, just two non-interacting Airy models:
\be
h=\left(\frac{x_1^3}{3}-s_1 x_1\right)+\left(\frac{x_2^3}{3}-s_2 x_2\right)
\ee
The complex in this case is just a tensor product of complexes:
\be\label{diag1}
\begin{array}{c}
	\xymatrix{
		& 0\ar[d] & 0\ar[d] & \\	
		0\ar[r] & \IZ(-s^{\frac{1}{2}},-s^{\frac{1}{2}})_1 \ar[r]^{Q_1}\ar[d]_{Q_2}& \IZ(+s^{\frac{1}{2}},-s^{\frac{1}{2}})_2 \ar[r]\ar[d]^{Q_2}& 0\\
		0 \ar[r]& \IZ(-s^{\frac{1}{2}},+s^{\frac{1}{2}})_3\ar[r]^{Q_1}\ar[d] & \IZ(+s^{\frac{1}{2}},+s^{\frac{1}{2}})_4\ar[r]\ar[d] & 0\\
		& 0 & 0 & \\
	}
\end{array}
\ee
And the supercharge $Q=Q_1+Q_2$. The condition $Q^2=0$ implies that supercharges $Q_1$ and $Q_2$ anti-commute, in interms of matrix elements this statement reads: 
\be
\frac{\langle ++|Q_2|+-\rangle\langle+-|Q_1|--\rangle}{\langle ++|Q_1|-+\rangle\langle-+|Q_2|--\rangle}=-1
\ee
The corresponding spectral curve is highly degenerate, so we resolve it a little bit:
\begin{center}
	\begin{tikzpicture}
	\draw[thick] (-1,1) -- (2,1) (1,0.5) -- (-2,0.5) (-1,0) -- (2,0) (1,-0.5) -- (-2,-0.5);
	\begin{scope}[shift={(-1,1)}]
	\draw[ultra thick, purple] (-0.1,-0.1) -- (0.1,0.1) (0.1,-0.1) -- (-0.1,0.1);
	\end{scope}
	\begin{scope}[shift={(1,0.5)}]
	\draw[ultra thick, purple] (-0.1,-0.1) -- (0.1,0.1) (0.1,-0.1) -- (-0.1,0.1);
	\end{scope}
	\begin{scope}[shift={(-1,0)}]
	\draw[ultra thick, purple] (-0.1,-0.1) -- (0.1,0.1) (0.1,-0.1) -- (-0.1,0.1);
	\end{scope}
	\begin{scope}[shift={(1,-0.5)}]
	\draw[ultra thick, purple] (-0.1,-0.1) -- (0.1,0.1) (0.1,-0.1) -- (-0.1,0.1);
	\end{scope}
	\draw (-1.3,1) to[out=90,in=90] (-0.7,1) -- (-0.7,0) to[out=270,in=270] (-1.3,0) -- (-1.3,1); 
	\draw (0.7,0.5) to[out=90,in=90] (1.3,0.5) -- (1.3,-0.5) to[out=270,in=270] (0.7,-0.5) -- (0.7,0.5);
	\draw[ultra thick] (0,-0.7) -- (0,1.2);
	\node[above] at (0,1.2) {$\wp_0$};
	\node[above left] at (-1.3,1) {$\theta_1$};
	\node[below right] at (1.3,-0.5) {$\theta_2$};
	\node[left] at (-2,-0.5) {$13$};
	\node[left] at (-2,0.5) {$24$};
	\node[right] at (2,0) {$12$};
	\node[right] at (2,1) {$34$};
	\end{tikzpicture}
\end{center}
Here we used vacua marking as in subscripts in diagram \eqref{diag1}. If one takes off the resolution of the spectral curve the encircled branching points collide. Each matrix element corresponds to a detour along a WKB line:
\be
\begin{split}
	\langle+-|Q_1|--\rangle\rightsquigarrow \fd_{12}\\
	\langle++|Q_1|-+\rangle\rightsquigarrow \fd_{34}\\
	\langle-+|Q_2|--\rangle\rightsquigarrow \fd_{13}\\
	\langle++|Q_2|+-\rangle\rightsquigarrow \fd_{24}
\end{split}
\ee
Calculating corresponidng signs we get
\be
\frac{\langle ++|Q_2|+-\rangle\langle+-|Q_1|--\rangle}{\langle ++|Q_1|-+\rangle\langle-+|Q_2|--\rangle}=\frac{\theta_2\theta_1}{\theta_1\theta_2}=-1
\ee

\subsection{``Good" and ``bad" map choices: quintic height function}\label{sec:good-bad}
The simplest example of good and bad choice of map $z_a$ can be demonstrated in a quintic height function:
\be
h=\frac{x^5}{5}-\frac{1}{3}(z_1+z_2)x^3+z_1z_2 x
\ee
The vacuum equation reads:
\be
h'=(x^2-z_1)(x^2-z_2)=0
\ee
If we choose $z_1=a^2$, $z_2=-b^2$ for $a$, $b$ being real the system has four vacua $\{{\color{red}a},{\color{blue}-a},{\color{green!60!black}\I b},{\color{violet}-\I b} \}$. Only two first vacua are real, and there is an instanton connecting them just like in the Airy model example.

It is simple to construct this instanton in terms of WKB webs. For this purpose we make a choice (``good" one) of a map $z_1(x)=x,\; z_2(x)=-b^2$. Obviously, the contributing branching point is located at 0, when two first vacua collide. And one gets a detour due to the corresponding WKB line:
\be
\begin{array}{c}
\begin{tikzpicture}
\draw[->] (-2,0) -- (2,0); \node[right] at (2,0) {${\rm Re}\; x$};
\draw[red, thick] (0,0) to[out=10,in=180] (1,0.1) -- (2,0.1);
\draw[blue, thick] (0,0) to[out=350,in=180] (1,-0.1) -- (2,-0.1);
\begin{scope}[shift={(0,0)}]
\draw[ultra thick, purple] (-0.1,-0.1) -- (0.1,0.1) (0.1,-0.1) -- (-0.1,0.1);
\end{scope}
\draw[ultra thick] (1.5,-0.2) -- (1.5,0.2); \node[above] at (1.5,0.2) {$\wp$};
\node[below right] at (1.5,-0.1) {$a^2$};
\node[below] at (0,-0.1) {$0$};
\end{tikzpicture}
\end{array}
\ee
Here we have splitted projections of the WKB line and marked them with the corresponding color code.

However one can choose a different map: $z_1(x)=a^2,\; z_2(x)=x$. This is a ``bad" choice of a map since there is no value of parameter $x$, even complex one, when two first vacua collide, instead there are two values: $x=0$ when two last vacua collide, and $x=a^2$ when the first and last pair of vacua collide. The corresponding instanton is constructed as a descendant combination of WKB lines from these two sources:
\be
\begin{array}{c}
\begin{tikzpicture}
\draw[->] (-2,0) -- (2.5,0);
\draw[thick, green!60!black] (0,0) to[out=10,in=180] (1,0.1) to[out=0,in=190] (2,0.2);
\draw[thick, violet] (0,0) to[out=350,in=180] (1,-0.1) to[out=0,in=170] (2,-0.2);
\draw[red, thick] (2,0.2) to[out=170,in=0] (1,0.3) -- (-2,0.3);
\draw[blue, thick] (2,-0.2) to[out=190,in=0] (1,-0.3) -- (-2,-0.3);
\begin{scope}[shift={(0,0)}]
\draw[ultra thick, purple] (-0.1,-0.1) -- (0.1,0.1) (0.1,-0.1) -- (-0.1,0.1);
\end{scope}
\begin{scope}[shift={(2,0.2)}]
\draw[ultra thick, purple] (-0.1,-0.1) -- (0.1,0.1) (0.1,-0.1) -- (-0.1,0.1);
\end{scope}
\begin{scope}[shift={(2,-0.2)}]
\draw[ultra thick, purple] (-0.1,-0.1) -- (0.1,0.1) (0.1,-0.1) -- (-0.1,0.1);
\end{scope}
\draw[ultra thick] (-1.5,-0.4) -- (-1.5,0.4); \node[above] at (-1.5,0.4) {$\wp$};
\node[right] at (2.5,0) {${\rm Re}\; x$};
\node[below] at (0,-0.3) {$0$};
\node[below] at (2,-0.3) {$a^2$};
\node[below] at (-1.5,-0.4) {$-b^2$};
\end{tikzpicture}
\end{array}
\ee

\section{Hopf link LG cohomology for $\fs\fu_3$}\label{sec:Hopf}

Generically, calculation of link cohomologies becomes quite involved for more complicated links and higher $n$.
In this section as the first non-trivial but still simple example we construct explicitly cohomology ${}_3 {\bf LGCoh}({\rm Hopf})$ for the Hopf link and compare it to the analogous Khovanov-Rozansky homology.

For the link complex we have:
\be\label{LG_su3_Hopf}
\CE\left(\begin{array}{c}
	\begin{tikzpicture}
	\draw[ultra thick] ([shift=(70:0.5)]0,0) arc (70:410:0.5);
	\draw[ultra thick] ([shift=(250:0.5)]0.5,0) arc (250:360:0.5);
	\draw[ultra thick] ([shift=(0:0.5)]0.5,0) arc (0:230:0.5);
	\draw[thick,->] ([shift=(350:0.5)]0,0) arc (350:360:0.5);
	\draw[thick,<-] ([shift=(180:0.5)]0.5,0) arc (180:190:0.5);
	\end{tikzpicture}
\end{array}\right)=\begin{array}{c}
	\begin{tikzpicture}
	\begin{scope}[scale=0.8]
	\draw[ultra thick, purple] (-0.25,0) -- (1.25,0) (1.75,0) -- (3.25,0) (1,0) -- (1,1) (2,0) -- (2,1) (-0.25,6.5) -- (1.25,6.5) (1.75,6.5) -- (3.25,6.5) (1,6.5) -- (1,5.5) (2,6.5) -- (2,5.5) (0,0) -- (0,3) (0,3.5) -- (0,6.5) (3,0) -- (3,3) (3,3.5) -- (3,6.5);
	\begin{scope}[shift={(1,1.5)}]
	\draw[ultra thick, purple] (0,0) -- (0,0.5) to[out=90,in=210] (0.5,1) to[out=30,in=270] (1,1.5) (1,0) -- (1,0.5) to[out=90,in=330] (0.6,0.9) (0.4,1.1) to[out=150,in=270] (0,1.5);
	\end{scope}
	\begin{scope}[shift={(1,3.5)}]
	\draw[ultra thick, purple] (0,0) -- (0,0.5) to[out=90,in=210] (0.5,1) to[out=30,in=270] (1,1.5) (1,0) -- (1,0.5) to[out=90,in=330] (0.6,0.9) (0.4,1.1) to[out=150,in=270] (0,1.5);
	\end{scope}
	\node[below] at (0,0) {$\bar 1$}; \node[below] at (1,0) {$1$}; \node[below] at (2,0) {$3$}; \node[below] at (3,0) {$\bar 3$};
	\node[above] at (0,6.5) {$\bar 3$}; \node[above] at (1,6.5) {$3$}; \node[above] at (2,6.5) {$1$}; \node[above] at (3,6.5) {$\bar 1$};
	\node at (1,1.25) {$3$}; \node at (2,1.25) {$1$};
	\node at (0,3.25) {$\bar 3$}; \node at (1,3.25) {$3$}; \node at (2,3.25) {$1$}; \node at (3,3.25) {$\bar 1$};
	\node at (1,5.25) {$3$}; \node at (2,5.25) {$1$};
	\draw[->] (0.6,0.25) -- (1,0.25) (0,0.25) -- (0.6,0.25);
	\draw[->] (0.6,0.75) -- (1,0.75) (0,0.75) -- (0.6,0.75);
	\begin{scope}[shift={(2,0)}]
	\draw[->] (0.6,0.25) -- (1,0.25) (0,0.25) -- (0.6,0.25);
	\draw[->] (0.6,0.75) -- (1,0.75) (0,0.75) -- (0.6,0.75);
	\end{scope}
	\begin{scope}[shift={(1,1.5)}]
	\draw[->] (0.6,0.5) -- (1,0.5) (0,0.5) -- (0.6,0.5);
	\node[below] at (0.5,0.5) {$12'$};
	\end{scope}
	\begin{scope}[shift={(1,3.5)}]
	\draw[->] (0.6,0.5) -- (1,0.5) (0,0.5) -- (0.6,0.5);
	\node[below] at (0.5,0.5) {$12'$};
	\end{scope}
	\end{scope}
	\end{tikzpicture}
\end{array}\oplus q^{10} t^{-8} \begin{array}{c}
	\begin{tikzpicture}
	\begin{scope}[scale=0.8]
	\draw[ultra thick, purple] (-0.25,0) -- (1.25,0) (1.75,0) -- (3.25,0) (1,0) -- (1,1) (2,0) -- (2,1) (-0.25,6.5) -- (1.25,6.5) (1.75,6.5) -- (3.25,6.5) (1,6.5) -- (1,5.5) (2,6.5) -- (2,5.5) (0,0) -- (0,3) (0,3.5) -- (0,6.5) (3,0) -- (3,3) (3,3.5) -- (3,6.5);
	\begin{scope}[shift={(1,1.5)}]
	\draw[ultra thick, purple] (0,0) -- (0,0.5) to[out=90,in=210] (0.5,1) to[out=30,in=270] (1,1.5) (1,0) -- (1,0.5) to[out=90,in=330] (0.6,0.9) (0.4,1.1) to[out=150,in=270] (0,1.5);
	\end{scope}
	\begin{scope}[shift={(1,3.5)}]
	\draw[ultra thick, purple] (0,0) -- (0,0.5) to[out=90,in=210] (0.5,1) to[out=30,in=270] (1,1.5) (1,0) -- (1,0.5) to[out=90,in=330] (0.6,0.9) (0.4,1.1) to[out=150,in=270] (0,1.5);
	\end{scope}
	\node[below] at (0,0) {$\bar 1$}; \node[below] at (1,0) {$1$}; \node[below] at (2,0) {$3$}; \node[below] at (3,0) {$\bar 3$};
	\node[above] at (0,6.5) {$\bar 3$}; \node[above] at (1,6.5) {$3$}; \node[above] at (2,6.5) {$1$}; \node[above] at (3,6.5) {$\bar 1$};
	\node at (1,1.25) {$1$}; \node at (2,1.25) {$3$};
	\node at (0,3.25) {$\bar 1$}; \node at (1,3.25) {$3$}; \node at (2,3.25) {$1$}; \node at (3,3.25) {$\bar 3$};
	\node at (1,5.25) {$1$}; \node at (2,5.25) {$3$};
	\begin{scope}[shift={(0,5.5)}]
	\draw[->] (0.6,0.25) -- (1,0.25) (0,0.25) -- (0.6,0.25);
	\draw[->] (0.6,0.75) -- (1,0.75) (0,0.75) -- (0.6,0.75);
	\end{scope}
	\begin{scope}[shift={(2,5.5)}]
	\draw[->] (0.6,0.25) -- (1,0.25) (0,0.25) -- (0.6,0.25);
	\draw[->] (0.6,0.75) -- (1,0.75) (0,0.75) -- (0.6,0.75);
	\end{scope}
	\end{scope}
	\end{tikzpicture}
\end{array}
\oplus \ee
\be\nn
\oplus q^6 t^{-4}\left(
\begin{array}{c}
	\begin{tikzpicture}
	\begin{scope}[scale=0.8]
	\draw[ultra thick, purple] (-0.25,0) -- (1.25,0) (1.75,0) -- (3.25,0) (1,0) -- (1,1) (2,0) -- (2,1) (-0.25,6.5) -- (1.25,6.5) (1.75,6.5) -- (3.25,6.5) (1,6.5) -- (1,5.5) (2,6.5) -- (2,5.5) (0,0) -- (0,3) (0,3.5) -- (0,6.5) (3,0) -- (3,3) (3,3.5) -- (3,6.5);
	\begin{scope}[shift={(1,1.5)}]
	\draw[ultra thick, purple] (0,0) -- (0,0.5) to[out=90,in=210] (0.5,1) to[out=30,in=270] (1,1.5) (1,0) -- (1,0.5) to[out=90,in=330] (0.6,0.9) (0.4,1.1) to[out=150,in=270] (0,1.5);
	\end{scope}
	\begin{scope}[shift={(1,3.5)}]
	\draw[ultra thick, purple] (0,0) -- (0,0.5) to[out=90,in=210] (0.5,1) to[out=30,in=270] (1,1.5) (1,0) -- (1,0.5) to[out=90,in=330] (0.6,0.9) (0.4,1.1) to[out=150,in=270] (0,1.5);
	\end{scope}
	\node[below] at (0,0) {$\bar 1$}; \node[below] at (1,0) {$1$}; \node[below] at (2,0) {$3$}; \node[below] at (3,0) {$\bar 3$};
	\node[above] at (0,6.5) {$\bar 3$}; \node[above] at (1,6.5) {$3$}; \node[above] at (2,6.5) {$1$}; \node[above] at (3,6.5) {$\bar 1$};
	\node at (1,1.25) {$2$}; \node at (2,1.25) {$1$};
	\node at (0,3.25) {$\bar 2$}; \node at (1,3.25) {$2$}; \node at (2,3.25) {$1$}; \node at (3,3.25) {$\bar 1$};
	\node at (1,5.25) {$2$}; \node at (2,5.25) {$1$};
	\draw[->] (0.6,0.5) -- (1,0.5) (0,0.5) -- (0.6,0.5);
	\begin{scope}[shift={(2,0)}]
	\draw[->] (0.6,0.25) -- (1,0.25) (0,0.25) -- (0.6,0.25);
	\draw[->] (0.6,0.75) -- (1,0.75) (0,0.75) -- (0.6,0.75);
	\end{scope}
	\begin{scope}[shift={(0,5.5)}]
	\draw[->] (0.6,0.5) -- (1,0.5) (0,0.5) -- (0.6,0.5);
	\end{scope}
	\begin{scope}[shift={(1,1.5)}]
	\draw[->] (0.6,0.5) -- (1,0.5) (0,0.5) -- (0.6,0.5);
	\node[below] at (0.5,0.5) {$12$};
	\end{scope}
	\begin{scope}[shift={(1,3.5)}]
	\draw[->] (0.6,0.5) -- (1,0.5) (0,0.5) -- (0.6,0.5);
	\node[below] at (0.5,0.5) {$12$};
	\end{scope}
	\end{scope}
	\end{tikzpicture}
\end{array}\oplus
\begin{array}{c}
	\begin{tikzpicture}
	\begin{scope}[scale=0.8]
	\draw[ultra thick, purple] (-0.25,0) -- (1.25,0) (1.75,0) -- (3.25,0) (1,0) -- (1,1) (2,0) -- (2,1) (-0.25,6.5) -- (1.25,6.5) (1.75,6.5) -- (3.25,6.5) (1,6.5) -- (1,5.5) (2,6.5) -- (2,5.5) (0,0) -- (0,3) (0,3.5) -- (0,6.5) (3,0) -- (3,3) (3,3.5) -- (3,6.5);
	\begin{scope}[shift={(1,1.5)}]
	\draw[ultra thick, purple] (0,0) -- (0,0.5) to[out=90,in=210] (0.5,1) to[out=30,in=270] (1,1.5) (1,0) -- (1,0.5) to[out=90,in=330] (0.6,0.9) (0.4,1.1) to[out=150,in=270] (0,1.5);
	\end{scope}
	\begin{scope}[shift={(1,3.5)}]
	\draw[ultra thick, purple] (0,0) -- (0,0.5) to[out=90,in=210] (0.5,1) to[out=30,in=270] (1,1.5) (1,0) -- (1,0.5) to[out=90,in=330] (0.6,0.9) (0.4,1.1) to[out=150,in=270] (0,1.5);
	\end{scope}
	\node[below] at (0,0) {$\bar 1$}; \node[below] at (1,0) {$1$}; \node[below] at (2,0) {$3$}; \node[below] at (3,0) {$\bar 3$};
	\node[above] at (0,6.5) {$\bar 3$}; \node[above] at (1,6.5) {$3$}; \node[above] at (2,6.5) {$1$}; \node[above] at (3,6.5) {$\bar 1$};
	\node at (1,1.25) {$3$}; \node at (2,1.25) {$2$};
	\node at (0,3.25) {$\bar 3$}; \node at (1,3.25) {$3$}; \node at (2,3.25) {$2$}; \node at (3,3.25) {$\bar 2$};
	\node at (1,5.25) {$3$}; \node at (2,5.25) {$2$};
	\begin{scope}[shift={(2,0)}]
	\draw[->] (0.6,0.5) -- (1,0.5) (0,0.5) -- (0.6,0.5);
	\end{scope}
	\begin{scope}[shift={(0,0)}]
	\draw[->] (0.6,0.25) -- (1,0.25) (0,0.25) -- (0.6,0.25);
	\draw[->] (0.6,0.75) -- (1,0.75) (0,0.75) -- (0.6,0.75);
	\end{scope}
	\begin{scope}[shift={(2,5.5)}]
	\draw[->] (0.6,0.5) -- (1,0.5) (0,0.5) -- (0.6,0.5);
	\end{scope}
	\begin{scope}[shift={(1,1.5)}]
	\draw[->] (0.6,0.5) -- (1,0.5) (0,0.5) -- (0.6,0.5);
	\node[below] at (0.5,0.5) {$12$};
	\end{scope}
	\begin{scope}[shift={(1,3.5)}]
	\draw[->] (0.6,0.5) -- (1,0.5) (0,0.5) -- (0.6,0.5);
	\node[below] at (0.5,0.5) {$12$};
	\end{scope}
	\end{scope}
	\end{tikzpicture}
\end{array}
\right)\oplus q^8 t^{-6}\left(
\begin{array}{c}
	\begin{tikzpicture}
	\begin{scope}[scale=0.8]
	\draw[ultra thick, purple] (-0.25,0) -- (1.25,0) (1.75,0) -- (3.25,0) (1,0) -- (1,1) (2,0) -- (2,1) (-0.25,6.5) -- (1.25,6.5) (1.75,6.5) -- (3.25,6.5) (1,6.5) -- (1,5.5) (2,6.5) -- (2,5.5) (0,0) -- (0,3) (0,3.5) -- (0,6.5) (3,0) -- (3,3) (3,3.5) -- (3,6.5);
	\begin{scope}[shift={(1,1.5)}]
	\draw[ultra thick, purple] (0,0) -- (0,0.5) to[out=90,in=210] (0.5,1) to[out=30,in=270] (1,1.5) (1,0) -- (1,0.5) to[out=90,in=330] (0.6,0.9) (0.4,1.1) to[out=150,in=270] (0,1.5);
	\end{scope}
	\begin{scope}[shift={(1,3.5)}]
	\draw[ultra thick, purple] (0,0) -- (0,0.5) to[out=90,in=210] (0.5,1) to[out=30,in=270] (1,1.5) (1,0) -- (1,0.5) to[out=90,in=330] (0.6,0.9) (0.4,1.1) to[out=150,in=270] (0,1.5);
	\end{scope}
	\node[below] at (0,0) {$\bar 1$}; \node[below] at (1,0) {$1$}; \node[below] at (2,0) {$3$}; \node[below] at (3,0) {$\bar 3$};
	\node[above] at (0,6.5) {$\bar 3$}; \node[above] at (1,6.5) {$3$}; \node[above] at (2,6.5) {$1$}; \node[above] at (3,6.5) {$\bar 1$};
	\node at (1,1.25) {$1$}; \node at (2,1.25) {$2$};
	\node at (0,3.25) {$\bar 1$}; \node at (1,3.25) {$2$}; \node at (2,3.25) {$1$}; \node at (3,3.25) {$\bar 2$};
	\node at (1,5.25) {$1$}; \node at (2,5.25) {$2$};
	\begin{scope}[shift={(2,0)}]
	\draw[->] (0.6,0.5) -- (1,0.5) (0,0.5) -- (0.6,0.5);
	\end{scope}
	\begin{scope}[shift={(0,5.5)}]
	\draw[->] (0.6,0.25) -- (1,0.25) (0,0.25) -- (0.6,0.25);
	\draw[->] (0.6,0.75) -- (1,0.75) (0,0.75) -- (0.6,0.75);
	\end{scope}
	\begin{scope}[shift={(2,5.5)}]
	\draw[->] (0.6,0.5) -- (1,0.5) (0,0.5) -- (0.6,0.5);
	\end{scope}
	\end{scope}
	\end{tikzpicture}
\end{array}\oplus
\begin{array}{c}
	\begin{tikzpicture}
	\begin{scope}[scale=0.8]
	\draw[ultra thick, purple] (-0.25,0) -- (1.25,0) (1.75,0) -- (3.25,0) (1,0) -- (1,1) (2,0) -- (2,1) (-0.25,6.5) -- (1.25,6.5) (1.75,6.5) -- (3.25,6.5) (1,6.5) -- (1,5.5) (2,6.5) -- (2,5.5) (0,0) -- (0,3) (0,3.5) -- (0,6.5) (3,0) -- (3,3) (3,3.5) -- (3,6.5);
	\begin{scope}[shift={(1,1.5)}]
	\draw[ultra thick, purple] (0,0) -- (0,0.5) to[out=90,in=210] (0.5,1) to[out=30,in=270] (1,1.5) (1,0) -- (1,0.5) to[out=90,in=330] (0.6,0.9) (0.4,1.1) to[out=150,in=270] (0,1.5);
	\end{scope}
	\begin{scope}[shift={(1,3.5)}]
	\draw[ultra thick, purple] (0,0) -- (0,0.5) to[out=90,in=210] (0.5,1) to[out=30,in=270] (1,1.5) (1,0) -- (1,0.5) to[out=90,in=330] (0.6,0.9) (0.4,1.1) to[out=150,in=270] (0,1.5);
	\end{scope}
	\node[below] at (0,0) {$\bar 1$}; \node[below] at (1,0) {$1$}; \node[below] at (2,0) {$3$}; \node[below] at (3,0) {$\bar 3$};
	\node[above] at (0,6.5) {$\bar 3$}; \node[above] at (1,6.5) {$3$}; \node[above] at (2,6.5) {$1$}; \node[above] at (3,6.5) {$\bar 1$};
	\node at (1,1.25) {$2$}; \node at (2,1.25) {$3$};
	\node at (0,3.25) {$\bar 2$}; \node at (1,3.25) {$3$}; \node at (2,3.25) {$2$}; \node at (3,3.25) {$\bar 3$};
	\node at (1,5.25) {$2$}; \node at (2,5.25) {$3$};
	\begin{scope}[shift={(0,5.5)}]
	\draw[->] (0.6,0.5) -- (1,0.5) (0,0.5) -- (0.6,0.5);
	\end{scope}
	\begin{scope}[shift={(2,5.5)}]
	\draw[->] (0.6,0.25) -- (1,0.25) (0,0.25) -- (0.6,0.25);
	\draw[->] (0.6,0.75) -- (1,0.75) (0,0.75) -- (0.6,0.75);
	\end{scope}
	\begin{scope}[shift={(0,0)}]
	\draw[->] (0.6,0.5) -- (1,0.5) (0,0.5) -- (0.6,0.5);
	\end{scope}
	\end{scope}
	\end{tikzpicture}
\end{array}
\right)\oplus
\ee
\be\nn
\oplus q^2t^{-2}\left[ \left(
\begin{array}{c}
	\begin{tikzpicture}
	\begin{scope}[scale=0.8]
	\draw[ultra thick, purple] (-0.25,0) -- (1.25,0) (1.75,0) -- (3.25,0) (1,0) -- (1,1) (2,0) -- (2,1) (-0.25,6.5) -- (1.25,6.5) (1.75,6.5) -- (3.25,6.5) (1,6.5) -- (1,5.5) (2,6.5) -- (2,5.5) (0,0) -- (0,3) (0,3.5) -- (0,6.5) (3,0) -- (3,3) (3,3.5) -- (3,6.5);
	\begin{scope}[shift={(1,1.5)}]
	\draw[ultra thick, purple] (0,0) -- (0,0.5) to[out=90,in=210] (0.5,1) to[out=30,in=270] (1,1.5) (1,0) -- (1,0.5) to[out=90,in=330] (0.6,0.9) (0.4,1.1) to[out=150,in=270] (0,1.5);
	\end{scope}
	\begin{scope}[shift={(1,3.5)}]
	\draw[ultra thick, purple] (0,0) -- (0,0.5) to[out=90,in=210] (0.5,1) to[out=30,in=270] (1,1.5) (1,0) -- (1,0.5) to[out=90,in=330] (0.6,0.9) (0.4,1.1) to[out=150,in=270] (0,1.5);
	\end{scope}
	\node[below] at (0,0) {$\bar 1$}; \node[below] at (1,0) {$1$}; \node[below] at (2,0) {$3$}; \node[below] at (3,0) {$\bar 3$};
	\node[above] at (0,6.5) {$\bar 3$}; \node[above] at (1,6.5) {$3$}; \node[above] at (2,6.5) {$1$}; \node[above] at (3,6.5) {$\bar 1$};
	\node at (1,1.25) {$2$}; \node at (2,1.25) {$1$};
	\node at (0,3.25) {$\bar 2$}; \node at (1,3.25) {$2$}; \node at (2,3.25) {$1$}; \node at (3,3.25) {$\bar 1$};
	\node at (1,5.25) {$2$}; \node at (2,5.25) {$1$};
	\draw[->] (0.6,0.5) -- (1,0.5) (0,0.5) -- (0.6,0.5);
	\begin{scope}[shift={(0,5.5)}]
	\draw[->] (0.6,0.5) -- (1,0.5) (0,0.5) -- (0.6,0.5);
	\node[below] at (0.5,0.5) {$\fd_1$};
	\end{scope}
	\begin{scope}[shift={(2,0)}]
	\draw[->] (0.6,0.25) -- (1,0.25) (0,0.25) -- (0.6,0.25);
	\draw[->] (0.6,0.75) -- (1,0.75) (0,0.75) -- (0.6,0.75);
	\end{scope}
	\begin{scope}[shift={(1,1.5)}]
	\draw[->] (0.6,0.5) -- (1,0.5) (0,0.5) -- (0.6,0.5);
	\node[below] at (0.5,0.5) {$12'$};
	\node[right] at (1,0.5) {$\fd_2$};
	\end{scope}
	\begin{scope}[shift={(1,3.5)}]
	\draw[->] (0.6,0.5) -- (1,0.5) (0,0.5) -- (0.6,0.5);
	\node[below] at (0.5,0.5) {$12'$};
	\node[right] at (1,0.5) {$\fd_3$};
	\end{scope}
	\end{scope}
	\end{tikzpicture}
\end{array}\oplus \begin{array}{c}
	\begin{tikzpicture}
	\begin{scope}[scale=0.8]
	\draw[ultra thick, purple] (-0.25,0) -- (1.25,0) (1.75,0) -- (3.25,0) (1,0) -- (1,1) (2,0) -- (2,1) (-0.25,6.5) -- (1.25,6.5) (1.75,6.5) -- (3.25,6.5) (1,6.5) -- (1,5.5) (2,6.5) -- (2,5.5) (0,0) -- (0,3) (0,3.5) -- (0,6.5) (3,0) -- (3,3) (3,3.5) -- (3,6.5);
	\begin{scope}[shift={(1,1.5)}]
	\draw[ultra thick, purple] (0,0) -- (0,0.5) to[out=90,in=210] (0.5,1) to[out=30,in=270] (1,1.5) (1,0) -- (1,0.5) to[out=90,in=330] (0.6,0.9) (0.4,1.1) to[out=150,in=270] (0,1.5);
	\end{scope}
	\begin{scope}[shift={(1,3.5)}]
	\draw[ultra thick, purple] (0,0) -- (0,0.5) to[out=90,in=210] (0.5,1) to[out=30,in=270] (1,1.5) (1,0) -- (1,0.5) to[out=90,in=330] (0.6,0.9) (0.4,1.1) to[out=150,in=270] (0,1.5);
	\end{scope}
	\node[below] at (0,0) {$\bar 1$}; \node[below] at (1,0) {$1$}; \node[below] at (2,0) {$3$}; \node[below] at (3,0) {$\bar 3$};
	\node[above] at (0,6.5) {$\bar 3$}; \node[above] at (1,6.5) {$3$}; \node[above] at (2,6.5) {$1$}; \node[above] at (3,6.5) {$\bar 1$};
	\node at (1,1.25) {$3$}; \node at (2,1.25) {$2$};
	\node at (0,3.25) {$\bar 3$}; \node at (1,3.25) {$3$}; \node at (2,3.25) {$2$}; \node at (3,3.25) {$\bar 2$};
	\node at (1,5.25) {$3$}; \node at (2,5.25) {$2$};
	\draw[->] (0.6,0.25) -- (1,0.25) (0,0.25) -- (0.6,0.25);
	\draw[->] (0.6,0.75) -- (1,0.75) (0,0.75) -- (0.6,0.75);
	\begin{scope}[shift={(2,0)}]
	\draw[->] (0.6,0.5) -- (1,0.5) (0,0.5) -- (0.6,0.5);
	\end{scope}
	\begin{scope}[shift={(2,5.5)}]
	\draw[->] (0.6,0.5) -- (1,0.5) (0,0.5) -- (0.6,0.5);
	\node[below] at (0.5,0.5) {$\fd_4$};
	\end{scope}
	\begin{scope}[shift={(1,1.5)}]
	\draw[->] (0.6,0.5) -- (1,0.5) (0,0.5) -- (0.6,0.5);
	\node[below] at (0.5,0.5) {$12'$};
	\node[right] at (1,0.5) {$\fd_5$};
	\end{scope}
	\begin{scope}[shift={(1,3.5)}]
	\draw[->] (0.6,0.5) -- (1,0.5) (0,0.5) -- (0.6,0.5);
	\node[below] at (0.5,0.5) {$12'$};
	\node[right] at (1,0.5) {$\fd_6$};
	\end{scope}
	\end{scope}
	\end{tikzpicture}
\end{array}
\right)\oplus t\left(\begin{array}{c}
	\begin{tikzpicture}
	\begin{scope}[scale=0.8]
	\draw[ultra thick, purple] (-0.25,0) -- (1.25,0) (1.75,0) -- (3.25,0) (1,0) -- (1,1) (2,0) -- (2,1) (-0.25,6.5) -- (1.25,6.5) (1.75,6.5) -- (3.25,6.5) (1,6.5) -- (1,5.5) (2,6.5) -- (2,5.5) (0,0) -- (0,3) (0,3.5) -- (0,6.5) (3,0) -- (3,3) (3,3.5) -- (3,6.5);
	\begin{scope}[shift={(1,1.5)}]
	\draw[ultra thick, purple] (0,0) -- (0,0.5) to[out=90,in=210] (0.5,1) to[out=30,in=270] (1,1.5) (1,0) -- (1,0.5) to[out=90,in=330] (0.6,0.9) (0.4,1.1) to[out=150,in=270] (0,1.5);
	\end{scope}
	\begin{scope}[shift={(1,3.5)}]
	\draw[ultra thick, purple] (0,0) -- (0,0.5) to[out=90,in=210] (0.5,1) to[out=30,in=270] (1,1.5) (1,0) -- (1,0.5) to[out=90,in=330] (0.6,0.9) (0.4,1.1) to[out=150,in=270] (0,1.5);
	\end{scope}
	\node[below] at (0,0) {$\bar 1$}; \node[below] at (1,0) {$1$}; \node[below] at (2,0) {$3$}; \node[below] at (3,0) {$\bar 3$};
	\node[above] at (0,6.5) {$\bar 3$}; \node[above] at (1,6.5) {$3$}; \node[above] at (2,6.5) {$1$}; \node[above] at (3,6.5) {$\bar 1$};
	\node at (1,1.25) {$3$}; \node at (2,1.25) {$1$};
	\node at (0,3.25) {$\bar 3$}; \node at (1,3.25) {$3$}; \node at (2,3.25) {$1$}; \node at (3,3.25) {$\bar 1$};
	\node at (1,5.25) {$3$}; \node at (2,5.25) {$1$};
	\draw[->] (0.6,0.25) -- (1,0.25) (0,0.25) -- (0.6,0.25);
	\draw[->] (0.6,0.75) -- (1,0.75) (0,0.75) -- (0.6,0.75);
	\node[above] at (0.5,0.75) {$\fd_7$};
	\begin{scope}[shift={(2,0)}]
	\draw[->] (0.6,0.25) -- (1,0.25) (0,0.25) -- (0.6,0.25);
	\draw[->] (0.6,0.75) -- (1,0.75) (0,0.75) -- (0.6,0.75);
	\node[above] at (0.5,0.75) {$\fd_8$};
	\end{scope}
	\begin{scope}[shift={(1,1.5)}]
	\draw[->] (0.6,0.5) -- (1,0.5) (0,0.5) -- (0.6,0.5);
	\node[below] at (0.5,0.5) {$12'$};
	\node[right] at (1,0.5) {$\fd_9$};
	\end{scope}
	\begin{scope}[shift={(1,3.5)}]
	\draw[->] (0.6,0.5) -- (1,0.5) (0,0.5) -- (0.6,0.5);
	\node[below] at (0.5,0.5) {$12$};
	\node[right] at (1,0.5) {$\fd_{10}$};
	\end{scope}
	\end{scope}
	\end{tikzpicture}
\end{array}\oplus \begin{array}{c}
	\begin{tikzpicture}
	\begin{scope}[scale=0.8]
	\draw[ultra thick, purple] (-0.25,0) -- (1.25,0) (1.75,0) -- (3.25,0) (1,0) -- (1,1) (2,0) -- (2,1) (-0.25,6.5) -- (1.25,6.5) (1.75,6.5) -- (3.25,6.5) (1,6.5) -- (1,5.5) (2,6.5) -- (2,5.5) (0,0) -- (0,3) (0,3.5) -- (0,6.5) (3,0) -- (3,3) (3,3.5) -- (3,6.5);
	\begin{scope}[shift={(1,1.5)}]
	\draw[ultra thick, purple] (0,0) -- (0,0.5) to[out=90,in=210] (0.5,1) to[out=30,in=270] (1,1.5) (1,0) -- (1,0.5) to[out=90,in=330] (0.6,0.9) (0.4,1.1) to[out=150,in=270] (0,1.5);
	\end{scope}
	\begin{scope}[shift={(1,3.5)}]
	\draw[ultra thick, purple] (0,0) -- (0,0.5) to[out=90,in=210] (0.5,1) to[out=30,in=270] (1,1.5) (1,0) -- (1,0.5) to[out=90,in=330] (0.6,0.9) (0.4,1.1) to[out=150,in=270] (0,1.5);
	\end{scope}
	\node[below] at (0,0) {$\bar 1$}; \node[below] at (1,0) {$1$}; \node[below] at (2,0) {$3$}; \node[below] at (3,0) {$\bar 3$};
	\node[above] at (0,6.5) {$\bar 3$}; \node[above] at (1,6.5) {$3$}; \node[above] at (2,6.5) {$1$}; \node[above] at (3,6.5) {$\bar 1$};
	\node at (1,1.25) {$3$}; \node at (2,1.25) {$1$};
	\node at (0,3.25) {$\bar 3$}; \node at (1,3.25) {$3$}; \node at (2,3.25) {$1$}; \node at (3,3.25) {$\bar 1$};
	\node at (1,5.25) {$3$}; \node at (2,5.25) {$1$};
	\draw[->] (0.6,0.25) -- (1,0.25) (0,0.25) -- (0.6,0.25);
	\draw[->] (0.6,0.75) -- (1,0.75) (0,0.75) -- (0.6,0.75);
	\node[above] at (0.5,0.75) {$\fd_{11}$};
	\begin{scope}[shift={(2,0)}]
	\draw[->] (0.6,0.25) -- (1,0.25) (0,0.25) -- (0.6,0.25);
	\draw[->] (0.6,0.75) -- (1,0.75) (0,0.75) -- (0.6,0.75);
	\node[above] at (0.5,0.75) {$\fd_{12}$};
	\end{scope}
	\begin{scope}[shift={(1,1.5)}]
	\draw[->] (0.6,0.5) -- (1,0.5) (0,0.5) -- (0.6,0.5);
	\node[below] at (0.5,0.5) {$12$};
	\node[right] at (1,0.5) {$\fd_{13}$};
	\end{scope}
	\begin{scope}[shift={(1,3.5)}]
	\draw[->] (0.6,0.5) -- (1,0.5) (0,0.5) -- (0.6,0.5);
	\node[below] at (0.5,0.5) {$12'$};
	\node[right] at (1,0.5) {$\fd_{14}$};
	\end{scope}
	\end{scope}
	\end{tikzpicture}
\end{array}\right)\right.
\ee
\be\nn
\left. \oplus t^2 \begin{array}{c}
	\begin{tikzpicture}
	\begin{scope}[scale=0.8]
	\draw[ultra thick, purple] (-0.25,0) -- (1.25,0) (1.75,0) -- (3.25,0) (1,0) -- (1,1) (2,0) -- (2,1) (-0.25,6.5) -- (1.25,6.5) (1.75,6.5) -- (3.25,6.5) (1,6.5) -- (1,5.5) (2,6.5) -- (2,5.5) (0,0) -- (0,3) (0,3.5) -- (0,6.5) (3,0) -- (3,3) (3,3.5) -- (3,6.5);
	\begin{scope}[shift={(1,1.5)}]
	\draw[ultra thick, purple] (0,0) -- (0,0.5) to[out=90,in=210] (0.5,1) to[out=30,in=270] (1,1.5) (1,0) -- (1,0.5) to[out=90,in=330] (0.6,0.9) (0.4,1.1) to[out=150,in=270] (0,1.5);
	\end{scope}
	\begin{scope}[shift={(1,3.5)}]
	\draw[ultra thick, purple] (0,0) -- (0,0.5) to[out=90,in=210] (0.5,1) to[out=30,in=270] (1,1.5) (1,0) -- (1,0.5) to[out=90,in=330] (0.6,0.9) (0.4,1.1) to[out=150,in=270] (0,1.5);
	\end{scope}
	\node[below] at (0,0) {$\bar 1$}; \node[below] at (1,0) {$1$}; \node[below] at (2,0) {$3$}; \node[below] at (3,0) {$\bar 3$};
	\node[above] at (0,6.5) {$\bar 3$}; \node[above] at (1,6.5) {$3$}; \node[above] at (2,6.5) {$1$}; \node[above] at (3,6.5) {$\bar 1$};
	\node at (1,1.25) {$3$}; \node at (2,1.25) {$1$};
	\node at (0,3.25) {$\bar 3$}; \node at (1,3.25) {$1$}; \node at (2,3.25) {$3$}; \node at (3,3.25) {$\bar 1$};
	\node at (1,5.25) {$3$}; \node at (2,5.25) {$1$};
	\draw[->] (0.6,0.25) -- (1,0.25) (0,0.25) -- (0.6,0.25);
	\draw[->] (0.6,0.75) -- (1,0.75) (0,0.75) -- (0.6,0.75);
	\begin{scope}[shift={(2,0)}]
	\draw[->] (0.6,0.25) -- (1,0.25) (0,0.25) -- (0.6,0.25);
	\draw[->] (0.6,0.75) -- (1,0.75) (0,0.75) -- (0.6,0.75);
	\end{scope}
	\end{scope}
	\end{tikzpicture}
\end{array}\right]\oplus q^4t^{-4}\left[
\left(\begin{array}{c}
	\begin{tikzpicture}
	\begin{scope}[scale=0.8]
	\draw[ultra thick, purple] (-0.25,0) -- (1.25,0) (1.75,0) -- (3.25,0) (1,0) -- (1,1) (2,0) -- (2,1) (-0.25,6.5) -- (1.25,6.5) (1.75,6.5) -- (3.25,6.5) (1,6.5) -- (1,5.5) (2,6.5) -- (2,5.5) (0,0) -- (0,3) (0,3.5) -- (0,6.5) (3,0) -- (3,3) (3,3.5) -- (3,6.5);
	\begin{scope}[shift={(1,1.5)}]
	\draw[ultra thick, purple] (0,0) -- (0,0.5) to[out=90,in=210] (0.5,1) to[out=30,in=270] (1,1.5) (1,0) -- (1,0.5) to[out=90,in=330] (0.6,0.9) (0.4,1.1) to[out=150,in=270] (0,1.5);
	\end{scope}
	\begin{scope}[shift={(1,3.5)}]
	\draw[ultra thick, purple] (0,0) -- (0,0.5) to[out=90,in=210] (0.5,1) to[out=30,in=270] (1,1.5) (1,0) -- (1,0.5) to[out=90,in=330] (0.6,0.9) (0.4,1.1) to[out=150,in=270] (0,1.5);
	\end{scope}
	\node[below] at (0,0) {$\bar 1$}; \node[below] at (1,0) {$1$}; \node[below] at (2,0) {$3$}; \node[below] at (3,0) {$\bar 3$};
	\node[above] at (0,6.5) {$\bar 3$}; \node[above] at (1,6.5) {$3$}; \node[above] at (2,6.5) {$1$}; \node[above] at (3,6.5) {$\bar 1$};
	\node at (1,1.25) {$1$}; \node at (2,1.25) {$1$};
	\node at (0,3.25) {$\bar 1$}; \node at (1,3.25) {$1$}; \node at (2,3.25) {$1$}; \node at (3,3.25) {$\bar 1$};
	\node at (1,5.25) {$1$}; \node at (2,5.25) {$1$};
	\begin{scope}[shift={(0,5.5)}]
	\draw[->] (0.6,0.25) -- (1,0.25) (0,0.25) -- (0.6,0.25);
	\draw[->] (0.6,0.75) -- (1,0.75) (0,0.75) -- (0.6,0.75);
	\end{scope}
	\begin{scope}[shift={(2,0)}]
	\draw[->] (0.6,0.25) -- (1,0.25) (0,0.25) -- (0.6,0.25);
	\draw[->] (0.6,0.75) -- (1,0.75) (0,0.75) -- (0.6,0.75);
	\end{scope}
	\end{scope}
	\end{tikzpicture}
\end{array}\oplus \begin{array}{c}
	\begin{tikzpicture}
	\begin{scope}[scale=0.8]
	\draw[ultra thick, purple] (-0.25,0) -- (1.25,0) (1.75,0) -- (3.25,0) (1,0) -- (1,1) (2,0) -- (2,1) (-0.25,6.5) -- (1.25,6.5) (1.75,6.5) -- (3.25,6.5) (1,6.5) -- (1,5.5) (2,6.5) -- (2,5.5) (0,0) -- (0,3) (0,3.5) -- (0,6.5) (3,0) -- (3,3) (3,3.5) -- (3,6.5);
	\begin{scope}[shift={(1,1.5)}]
	\draw[ultra thick, purple] (0,0) -- (0,0.5) to[out=90,in=210] (0.5,1) to[out=30,in=270] (1,1.5) (1,0) -- (1,0.5) to[out=90,in=330] (0.6,0.9) (0.4,1.1) to[out=150,in=270] (0,1.5);
	\end{scope}
	\begin{scope}[shift={(1,3.5)}]
	\draw[ultra thick, purple] (0,0) -- (0,0.5) to[out=90,in=210] (0.5,1) to[out=30,in=270] (1,1.5) (1,0) -- (1,0.5) to[out=90,in=330] (0.6,0.9) (0.4,1.1) to[out=150,in=270] (0,1.5);
	\end{scope}
	\node[below] at (0,0) {$\bar 1$}; \node[below] at (1,0) {$1$}; \node[below] at (2,0) {$3$}; \node[below] at (3,0) {$\bar 3$};
	\node[above] at (0,6.5) {$\bar 3$}; \node[above] at (1,6.5) {$3$}; \node[above] at (2,6.5) {$1$}; \node[above] at (3,6.5) {$\bar 1$};
	\node at (1,1.25) {$2$}; \node at (2,1.25) {$2$};
	\node at (0,3.25) {$\bar 2$}; \node at (1,3.25) {$2$}; \node at (2,3.25) {$2$}; \node at (3,3.25) {$\bar 2$};
	\node at (1,5.25) {$2$}; \node at (2,5.25) {$2$};
	\begin{scope}[shift={(0,0)}]
	\draw[->] (0.6,0.5) -- (1,0.5) (0,0.5) -- (0.6,0.5);
	\end{scope}
	\begin{scope}[shift={(0,5.5)}]
	\draw[->] (0.6,0.5) -- (1,0.5) (0,0.5) -- (0.6,0.5);
	\end{scope}
	\begin{scope}[shift={(2,0)}]
	\draw[->] (0.6,0.5) -- (1,0.5) (0,0.5) -- (0.6,0.5);
	\end{scope}
	\begin{scope}[shift={(2,5.5)}]
	\draw[->] (0.6,0.5) -- (1,0.5) (0,0.5) -- (0.6,0.5);
	\end{scope}
	\end{scope}
	\end{tikzpicture}
\end{array}\oplus
\begin{array}{c}
	\begin{tikzpicture}
	\begin{scope}[scale=0.8]
	\draw[ultra thick, purple] (-0.25,0) -- (1.25,0) (1.75,0) -- (3.25,0) (1,0) -- (1,1) (2,0) -- (2,1) (-0.25,6.5) -- (1.25,6.5) (1.75,6.5) -- (3.25,6.5) (1,6.5) -- (1,5.5) (2,6.5) -- (2,5.5) (0,0) -- (0,3) (0,3.5) -- (0,6.5) (3,0) -- (3,3) (3,3.5) -- (3,6.5);
	\begin{scope}[shift={(1,1.5)}]
	\draw[ultra thick, purple] (0,0) -- (0,0.5) to[out=90,in=210] (0.5,1) to[out=30,in=270] (1,1.5) (1,0) -- (1,0.5) to[out=90,in=330] (0.6,0.9) (0.4,1.1) to[out=150,in=270] (0,1.5);
	\end{scope}
	\begin{scope}[shift={(1,3.5)}]
	\draw[ultra thick, purple] (0,0) -- (0,0.5) to[out=90,in=210] (0.5,1) to[out=30,in=270] (1,1.5) (1,0) -- (1,0.5) to[out=90,in=330] (0.6,0.9) (0.4,1.1) to[out=150,in=270] (0,1.5);
	\end{scope}
	\node[below] at (0,0) {$\bar 1$}; \node[below] at (1,0) {$1$}; \node[below] at (2,0) {$3$}; \node[below] at (3,0) {$\bar 3$};
	\node[above] at (0,6.5) {$\bar 3$}; \node[above] at (1,6.5) {$3$}; \node[above] at (2,6.5) {$1$}; \node[above] at (3,6.5) {$\bar 1$};
	\node at (1,1.25) {$3$}; \node at (2,1.25) {$3$};
	\node at (0,3.25) {$\bar 3$}; \node at (1,3.25) {$3$}; \node at (2,3.25) {$3$}; \node at (3,3.25) {$\bar 3$};
	\node at (1,5.25) {$3$}; \node at (2,5.25) {$3$};
	\begin{scope}[shift={(0,0)}]
	\draw[->] (0.6,0.25) -- (1,0.25) (0,0.25) -- (0.6,0.25);
	\draw[->] (0.6,0.75) -- (1,0.75) (0,0.75) -- (0.6,0.75);
	\end{scope}
	\begin{scope}[shift={(2,5.5)}]
	\draw[->] (0.6,0.25) -- (1,0.25) (0,0.25) -- (0.6,0.25);
	\draw[->] (0.6,0.75) -- (1,0.75) (0,0.75) -- (0.6,0.75);
	\end{scope}
	\end{scope}
	\end{tikzpicture}
\end{array}\right)\oplus
\right.
\ee
\be\nn
\oplus t\left(
\begin{array}{c}
	\begin{tikzpicture}
	\begin{scope}[scale=0.8]
	\draw[ultra thick, purple] (-0.25,0) -- (1.25,0) (1.75,0) -- (3.25,0) (1,0) -- (1,1) (2,0) -- (2,1) (-0.25,6.5) -- (1.25,6.5) (1.75,6.5) -- (3.25,6.5) (1,6.5) -- (1,5.5) (2,6.5) -- (2,5.5) (0,0) -- (0,3) (0,3.5) -- (0,6.5) (3,0) -- (3,3) (3,3.5) -- (3,6.5);
	\begin{scope}[shift={(1,1.5)}]
	\draw[ultra thick, purple] (0,0) -- (0,0.5) to[out=90,in=210] (0.5,1) to[out=30,in=270] (1,1.5) (1,0) -- (1,0.5) to[out=90,in=330] (0.6,0.9) (0.4,1.1) to[out=150,in=270] (0,1.5);
	\end{scope}
	\begin{scope}[shift={(1,3.5)}]
	\draw[ultra thick, purple] (0,0) -- (0,0.5) to[out=90,in=210] (0.5,1) to[out=30,in=270] (1,1.5) (1,0) -- (1,0.5) to[out=90,in=330] (0.6,0.9) (0.4,1.1) to[out=150,in=270] (0,1.5);
	\end{scope}
	\node[below] at (0,0) {$\bar 1$}; \node[below] at (1,0) {$1$}; \node[below] at (2,0) {$3$}; \node[below] at (3,0) {$\bar 3$};
	\node[above] at (0,6.5) {$\bar 3$}; \node[above] at (1,6.5) {$3$}; \node[above] at (2,6.5) {$1$}; \node[above] at (3,6.5) {$\bar 1$};
	\node at (1,1.25) {$2$}; \node at (2,1.25) {$1$};
	\node at (0,3.25) {$\bar 2$}; \node at (1,3.25) {$2$}; \node at (2,3.25) {$1$}; \node at (3,3.25) {$\bar 1$};
	\node at (1,5.25) {$2$}; \node at (2,5.25) {$1$};
	\draw[->] (0.6,0.5) -- (1,0.5) (0,0.5) -- (0.6,0.5);
	\begin{scope}[shift={(2,0)}]
	\draw[->] (0.6,0.25) -- (1,0.25) (0,0.25) -- (0.6,0.25);
	\draw[->] (0.6,0.75) -- (1,0.75) (0,0.75) -- (0.6,0.75);
	\end{scope}
	\begin{scope}[shift={(0,5.5)}]
	\draw[->] (0.6,0.5) -- (1,0.5) (0,0.5) -- (0.6,0.5);
	\end{scope}
	\begin{scope}[shift={(1,1.5)}]
	\draw[->] (0.6,0.5) -- (1,0.5) (0,0.5) -- (0.6,0.5);
	\node[below] at (0.5,0.5) {$12'$};
	\end{scope}
	\begin{scope}[shift={(1,3.5)}]
	\draw[->] (0.6,0.5) -- (1,0.5) (0,0.5) -- (0.6,0.5);
	\node[below] at (0.5,0.5) {$12$};
	\end{scope}
	\end{scope}
	\end{tikzpicture}
\end{array}\oplus
\begin{array}{c}
	\begin{tikzpicture}
	\begin{scope}[scale=0.8]
	\draw[ultra thick, purple] (-0.25,0) -- (1.25,0) (1.75,0) -- (3.25,0) (1,0) -- (1,1) (2,0) -- (2,1) (-0.25,6.5) -- (1.25,6.5) (1.75,6.5) -- (3.25,6.5) (1,6.5) -- (1,5.5) (2,6.5) -- (2,5.5) (0,0) -- (0,3) (0,3.5) -- (0,6.5) (3,0) -- (3,3) (3,3.5) -- (3,6.5);
	\begin{scope}[shift={(1,1.5)}]
	\draw[ultra thick, purple] (0,0) -- (0,0.5) to[out=90,in=210] (0.5,1) to[out=30,in=270] (1,1.5) (1,0) -- (1,0.5) to[out=90,in=330] (0.6,0.9) (0.4,1.1) to[out=150,in=270] (0,1.5);
	\end{scope}
	\begin{scope}[shift={(1,3.5)}]
	\draw[ultra thick, purple] (0,0) -- (0,0.5) to[out=90,in=210] (0.5,1) to[out=30,in=270] (1,1.5) (1,0) -- (1,0.5) to[out=90,in=330] (0.6,0.9) (0.4,1.1) to[out=150,in=270] (0,1.5);
	\end{scope}
	\node[below] at (0,0) {$\bar 1$}; \node[below] at (1,0) {$1$}; \node[below] at (2,0) {$3$}; \node[below] at (3,0) {$\bar 3$};
	\node[above] at (0,6.5) {$\bar 3$}; \node[above] at (1,6.5) {$3$}; \node[above] at (2,6.5) {$1$}; \node[above] at (3,6.5) {$\bar 1$};
	\node at (1,1.25) {$3$}; \node at (2,1.25) {$2$};
	\node at (0,3.25) {$\bar 3$}; \node at (1,3.25) {$3$}; \node at (2,3.25) {$2$}; \node at (3,3.25) {$\bar 2$};
	\node at (1,5.25) {$3$}; \node at (2,5.25) {$2$};
	\begin{scope}[shift={(2,0)}]
	\draw[->] (0.6,0.5) -- (1,0.5) (0,0.5) -- (0.6,0.5);
	\end{scope}
	\begin{scope}[shift={(0,0)}]
	\draw[->] (0.6,0.25) -- (1,0.25) (0,0.25) -- (0.6,0.25);
	\draw[->] (0.6,0.75) -- (1,0.75) (0,0.75) -- (0.6,0.75);
	\end{scope}
	\begin{scope}[shift={(2,5.5)}]
	\draw[->] (0.6,0.5) -- (1,0.5) (0,0.5) -- (0.6,0.5);
	\end{scope}
	\begin{scope}[shift={(1,1.5)}]
	\draw[->] (0.6,0.5) -- (1,0.5) (0,0.5) -- (0.6,0.5);
	\node[below] at (0.5,0.5) {$12'$};
	\end{scope}
	\begin{scope}[shift={(1,3.5)}]
	\draw[->] (0.6,0.5) -- (1,0.5) (0,0.5) -- (0.6,0.5);
	\node[below] at (0.5,0.5) {$12$};
	\end{scope}
	\end{scope}
	\end{tikzpicture}
\end{array}\oplus
\begin{array}{c}
	\begin{tikzpicture}
	\begin{scope}[scale=0.8]
	\draw[ultra thick, purple] (-0.25,0) -- (1.25,0) (1.75,0) -- (3.25,0) (1,0) -- (1,1) (2,0) -- (2,1) (-0.25,6.5) -- (1.25,6.5) (1.75,6.5) -- (3.25,6.5) (1,6.5) -- (1,5.5) (2,6.5) -- (2,5.5) (0,0) -- (0,3) (0,3.5) -- (0,6.5) (3,0) -- (3,3) (3,3.5) -- (3,6.5);
	\begin{scope}[shift={(1,1.5)}]
	\draw[ultra thick, purple] (0,0) -- (0,0.5) to[out=90,in=210] (0.5,1) to[out=30,in=270] (1,1.5) (1,0) -- (1,0.5) to[out=90,in=330] (0.6,0.9) (0.4,1.1) to[out=150,in=270] (0,1.5);
	\end{scope}
	\begin{scope}[shift={(1,3.5)}]
	\draw[ultra thick, purple] (0,0) -- (0,0.5) to[out=90,in=210] (0.5,1) to[out=30,in=270] (1,1.5) (1,0) -- (1,0.5) to[out=90,in=330] (0.6,0.9) (0.4,1.1) to[out=150,in=270] (0,1.5);
	\end{scope}
	\node[below] at (0,0) {$\bar 1$}; \node[below] at (1,0) {$1$}; \node[below] at (2,0) {$3$}; \node[below] at (3,0) {$\bar 3$};
	\node[above] at (0,6.5) {$\bar 3$}; \node[above] at (1,6.5) {$3$}; \node[above] at (2,6.5) {$1$}; \node[above] at (3,6.5) {$\bar 1$};
	\node at (1,1.25) {$2$}; \node at (2,1.25) {$1$};
	\node at (0,3.25) {$\bar 2$}; \node at (1,3.25) {$2$}; \node at (2,3.25) {$1$}; \node at (3,3.25) {$\bar 1$};
	\node at (1,5.25) {$2$}; \node at (2,5.25) {$1$};
	\draw[->] (0.6,0.5) -- (1,0.5) (0,0.5) -- (0.6,0.5);
	\begin{scope}[shift={(2,0)}]
	\draw[->] (0.6,0.25) -- (1,0.25) (0,0.25) -- (0.6,0.25);
	\draw[->] (0.6,0.75) -- (1,0.75) (0,0.75) -- (0.6,0.75);
	\end{scope}
	\begin{scope}[shift={(0,5.5)}]
	\draw[->] (0.6,0.5) -- (1,0.5) (0,0.5) -- (0.6,0.5);
	\end{scope}
	\begin{scope}[shift={(1,1.5)}]
	\draw[->] (0.6,0.5) -- (1,0.5) (0,0.5) -- (0.6,0.5);
	\node[below] at (0.5,0.5) {$12$};
	\end{scope}
	\begin{scope}[shift={(1,3.5)}]
	\draw[->] (0.6,0.5) -- (1,0.5) (0,0.5) -- (0.6,0.5);
	\node[below] at (0.5,0.5) {$12'$};
	\end{scope}
	\end{scope}
	\end{tikzpicture}
\end{array}\oplus
\begin{array}{c}
	\begin{tikzpicture}
	\begin{scope}[scale=0.8]
	\draw[ultra thick, purple] (-0.25,0) -- (1.25,0) (1.75,0) -- (3.25,0) (1,0) -- (1,1) (2,0) -- (2,1) (-0.25,6.5) -- (1.25,6.5) (1.75,6.5) -- (3.25,6.5) (1,6.5) -- (1,5.5) (2,6.5) -- (2,5.5) (0,0) -- (0,3) (0,3.5) -- (0,6.5) (3,0) -- (3,3) (3,3.5) -- (3,6.5);
	\begin{scope}[shift={(1,1.5)}]
	\draw[ultra thick, purple] (0,0) -- (0,0.5) to[out=90,in=210] (0.5,1) to[out=30,in=270] (1,1.5) (1,0) -- (1,0.5) to[out=90,in=330] (0.6,0.9) (0.4,1.1) to[out=150,in=270] (0,1.5);
	\end{scope}
	\begin{scope}[shift={(1,3.5)}]
	\draw[ultra thick, purple] (0,0) -- (0,0.5) to[out=90,in=210] (0.5,1) to[out=30,in=270] (1,1.5) (1,0) -- (1,0.5) to[out=90,in=330] (0.6,0.9) (0.4,1.1) to[out=150,in=270] (0,1.5);
	\end{scope}
	\node[below] at (0,0) {$\bar 1$}; \node[below] at (1,0) {$1$}; \node[below] at (2,0) {$3$}; \node[below] at (3,0) {$\bar 3$};
	\node[above] at (0,6.5) {$\bar 3$}; \node[above] at (1,6.5) {$3$}; \node[above] at (2,6.5) {$1$}; \node[above] at (3,6.5) {$\bar 1$};
	\node at (1,1.25) {$3$}; \node at (2,1.25) {$2$};
	\node at (0,3.25) {$\bar 3$}; \node at (1,3.25) {$3$}; \node at (2,3.25) {$2$}; \node at (3,3.25) {$\bar 2$};
	\node at (1,5.25) {$3$}; \node at (2,5.25) {$2$};
	\begin{scope}[shift={(2,0)}]
	\draw[->] (0.6,0.5) -- (1,0.5) (0,0.5) -- (0.6,0.5);
	\end{scope}
	\begin{scope}[shift={(0,0)}]
	\draw[->] (0.6,0.25) -- (1,0.25) (0,0.25) -- (0.6,0.25);
	\draw[->] (0.6,0.75) -- (1,0.75) (0,0.75) -- (0.6,0.75);
	\end{scope}
	\begin{scope}[shift={(2,5.5)}]
	\draw[->] (0.6,0.5) -- (1,0.5) (0,0.5) -- (0.6,0.5);
	\end{scope}
	\begin{scope}[shift={(1,1.5)}]
	\draw[->] (0.6,0.5) -- (1,0.5) (0,0.5) -- (0.6,0.5);
	\node[below] at (0.5,0.5) {$12$};
	\end{scope}
	\begin{scope}[shift={(1,3.5)}]
	\draw[->] (0.6,0.5) -- (1,0.5) (0,0.5) -- (0.6,0.5);
	\node[below] at (0.5,0.5) {$12'$};
	\end{scope}
	\end{scope}
	\end{tikzpicture}
\end{array}
\right)\oplus
\ee
\be\nn
\oplus \left.t^2\left(
\begin{array}{c}
	\begin{tikzpicture}
	\begin{scope}[scale=0.8]
	\draw[ultra thick, purple] (-0.25,0) -- (1.25,0) (1.75,0) -- (3.25,0) (1,0) -- (1,1) (2,0) -- (2,1) (-0.25,6.5) -- (1.25,6.5) (1.75,6.5) -- (3.25,6.5) (1,6.5) -- (1,5.5) (2,6.5) -- (2,5.5) (0,0) -- (0,3) (0,3.5) -- (0,6.5) (3,0) -- (3,3) (3,3.5) -- (3,6.5);
	\begin{scope}[shift={(1,1.5)}]
	\draw[ultra thick, purple] (0,0) -- (0,0.5) to[out=90,in=210] (0.5,1) to[out=30,in=270] (1,1.5) (1,0) -- (1,0.5) to[out=90,in=330] (0.6,0.9) (0.4,1.1) to[out=150,in=270] (0,1.5);
	\end{scope}
	\begin{scope}[shift={(1,3.5)}]
	\draw[ultra thick, purple] (0,0) -- (0,0.5) to[out=90,in=210] (0.5,1) to[out=30,in=270] (1,1.5) (1,0) -- (1,0.5) to[out=90,in=330] (0.6,0.9) (0.4,1.1) to[out=150,in=270] (0,1.5);
	\end{scope}
	\node[below] at (0,0) {$\bar 1$}; \node[below] at (1,0) {$1$}; \node[below] at (2,0) {$3$}; \node[below] at (3,0) {$\bar 3$};
	\node[above] at (0,6.5) {$\bar 3$}; \node[above] at (1,6.5) {$3$}; \node[above] at (2,6.5) {$1$}; \node[above] at (3,6.5) {$\bar 1$};
	\node at (1,1.25) {$2$}; \node at (2,1.25) {$1$};
	\node at (0,3.25) {$\bar 2$}; \node at (1,3.25) {$1$}; \node at (2,3.25) {$2$}; \node at (3,3.25) {$\bar 1$};
	\node at (1,5.25) {$2$}; \node at (2,5.25) {$1$};
	\draw[->] (0.6,0.5) -- (1,0.5) (0,0.5) -- (0.6,0.5);
	\begin{scope}[shift={(2,0)}]
	\draw[->] (0.6,0.25) -- (1,0.25) (0,0.25) -- (0.6,0.25);
	\draw[->] (0.6,0.75) -- (1,0.75) (0,0.75) -- (0.6,0.75);
	\end{scope}
	\begin{scope}[shift={(0,5.5)}]
	\draw[->] (0.6,0.5) -- (1,0.5) (0,0.5) -- (0.6,0.5);
	\end{scope}
	\end{scope}
	\end{tikzpicture}
\end{array}\oplus
\begin{array}{c}
	\begin{tikzpicture}
	\begin{scope}[scale=0.8]
	\draw[ultra thick, purple] (-0.25,0) -- (1.25,0) (1.75,0) -- (3.25,0) (1,0) -- (1,1) (2,0) -- (2,1) (-0.25,6.5) -- (1.25,6.5) (1.75,6.5) -- (3.25,6.5) (1,6.5) -- (1,5.5) (2,6.5) -- (2,5.5) (0,0) -- (0,3) (0,3.5) -- (0,6.5) (3,0) -- (3,3) (3,3.5) -- (3,6.5);
	\begin{scope}[shift={(1,1.5)}]
	\draw[ultra thick, purple] (0,0) -- (0,0.5) to[out=90,in=210] (0.5,1) to[out=30,in=270] (1,1.5) (1,0) -- (1,0.5) to[out=90,in=330] (0.6,0.9) (0.4,1.1) to[out=150,in=270] (0,1.5);
	\end{scope}
	\begin{scope}[shift={(1,3.5)}]
	\draw[ultra thick, purple] (0,0) -- (0,0.5) to[out=90,in=210] (0.5,1) to[out=30,in=270] (1,1.5) (1,0) -- (1,0.5) to[out=90,in=330] (0.6,0.9) (0.4,1.1) to[out=150,in=270] (0,1.5);
	\end{scope}
	\node[below] at (0,0) {$\bar 1$}; \node[below] at (1,0) {$1$}; \node[below] at (2,0) {$3$}; \node[below] at (3,0) {$\bar 3$};
	\node[above] at (0,6.5) {$\bar 3$}; \node[above] at (1,6.5) {$3$}; \node[above] at (2,6.5) {$1$}; \node[above] at (3,6.5) {$\bar 1$};
	\node at (1,1.25) {$3$}; \node at (2,1.25) {$2$};
	\node at (0,3.25) {$\bar 3$}; \node at (1,3.25) {$2$}; \node at (2,3.25) {$3$}; \node at (3,3.25) {$\bar 2$};
	\node at (1,5.25) {$3$}; \node at (2,5.25) {$2$};
	\begin{scope}[shift={(2,0)}]
	\draw[->] (0.6,0.5) -- (1,0.5) (0,0.5) -- (0.6,0.5);
	\end{scope}
	\begin{scope}[shift={(0,0)}]
	\draw[->] (0.6,0.25) -- (1,0.25) (0,0.25) -- (0.6,0.25);
	\draw[->] (0.6,0.75) -- (1,0.75) (0,0.75) -- (0.6,0.75);
	\end{scope}
	\begin{scope}[shift={(2,5.5)}]
	\draw[->] (0.6,0.5) -- (1,0.5) (0,0.5) -- (0.6,0.5);
	\end{scope}
	\end{scope}
	\end{tikzpicture}
\end{array}\oplus \begin{array}{c}
	\begin{tikzpicture}
	\begin{scope}[scale=0.8]
	\draw[ultra thick, purple] (-0.25,0) -- (1.25,0) (1.75,0) -- (3.25,0) (1,0) -- (1,1) (2,0) -- (2,1) (-0.25,6.5) -- (1.25,6.5) (1.75,6.5) -- (3.25,6.5) (1,6.5) -- (1,5.5) (2,6.5) -- (2,5.5) (0,0) -- (0,3) (0,3.5) -- (0,6.5) (3,0) -- (3,3) (3,3.5) -- (3,6.5);
	\begin{scope}[shift={(1,1.5)}]
	\draw[ultra thick, purple] (0,0) -- (0,0.5) to[out=90,in=210] (0.5,1) to[out=30,in=270] (1,1.5) (1,0) -- (1,0.5) to[out=90,in=330] (0.6,0.9) (0.4,1.1) to[out=150,in=270] (0,1.5);
	\end{scope}
	\begin{scope}[shift={(1,3.5)}]
	\draw[ultra thick, purple] (0,0) -- (0,0.5) to[out=90,in=210] (0.5,1) to[out=30,in=270] (1,1.5) (1,0) -- (1,0.5) to[out=90,in=330] (0.6,0.9) (0.4,1.1) to[out=150,in=270] (0,1.5);
	\end{scope}
	\node[below] at (0,0) {$\bar 1$}; \node[below] at (1,0) {$1$}; \node[below] at (2,0) {$3$}; \node[below] at (3,0) {$\bar 3$};
	\node[above] at (0,6.5) {$\bar 3$}; \node[above] at (1,6.5) {$3$}; \node[above] at (2,6.5) {$1$}; \node[above] at (3,6.5) {$\bar 1$};
	\node at (1,1.25) {$3$}; \node at (2,1.25) {$1$};
	\node at (0,3.25) {$\bar 3$}; \node at (1,3.25) {$3$}; \node at (2,3.25) {$1$}; \node at (3,3.25) {$\bar 1$};
	\node at (1,5.25) {$3$}; \node at (2,5.25) {$1$};
	\draw[->] (0.6,0.25) -- (1,0.25) (0,0.25) -- (0.6,0.25);
	\draw[->] (0.6,0.75) -- (1,0.75) (0,0.75) -- (0.6,0.75);
	\begin{scope}[shift={(2,0)}]
	\draw[->] (0.6,0.25) -- (1,0.25) (0,0.25) -- (0.6,0.25);
	\draw[->] (0.6,0.75) -- (1,0.75) (0,0.75) -- (0.6,0.75);
	\end{scope}
	\begin{scope}[shift={(1,1.5)}]
	\draw[->] (0.6,0.5) -- (1,0.5) (0,0.5) -- (0.6,0.5);
	\node[below] at (0.5,0.5) {$12$};
	\end{scope}
	\begin{scope}[shift={(1,3.5)}]
	\draw[->] (0.6,0.5) -- (1,0.5) (0,0.5) -- (0.6,0.5);
	\node[below] at (0.5,0.5) {$12$};
	\end{scope}
	\end{scope}
	\end{tikzpicture}
\end{array}
\right)\right]
\ee

Supercharge $Q$ does not change $\bf P$-degree, therefore components of this complex of definite $\bf P$-degree are honest subcomplexes:
\be
\CE=\bigoplus\lm_{\alpha=0}^5 q^{2\alpha}\CE_{2\alpha}
\ee
Subcomplexes of $\bf P$-degrees $0$, $6$, $8$, $10$ are rather simple, supercharges act there by 0. However complexes of $\bf P$-degrees $2$ and $4$ are non-trivial.

Let us consider them consequently: first we consider $\CE_2$. We label all the generators of the complex by $\Psi_i$, $i=1,\ldots,5$ in the order as they appear in expansion \eqref{LG_su3_Hopf}, then we have:
\be
\CE_2=t^{-2}\left(0\to \IZ[0] \Psi_1\oplus\IZ[0]\Psi_2 \mathop{\longrightarrow}^{Q_1} \IZ[1] \Psi_3\oplus \IZ[1]\Psi_4 \mathop{\longrightarrow}^{Q_2} \IZ[2] \Psi_5\to 0\right)
\ee 
We will draw all the null-webs as in a matrix form as they saturate corresponding matrix elements:
\be
Q_1=\left(
\begin{array}{c|c}
	(1)\begin{array}{c}
		\begin{tikzpicture}
		\begin{scope}[scale=0.65,xscale=-1]
		\draw[thick] (0,0) -- (4,4) (1,2) -- (1,1) -- (2,0) (2,3) -- (2,2) -- (3,1);
		\begin{scope}[shift={(0,0)}]
		\draw[purple, ultra thick] (-0.1,-0.1) -- (0.1,0.1) (0.1,-0.1) -- (-0.1,0.1);
		\end{scope}
		\begin{scope}[shift={(1,2)}]
		\draw[purple, ultra thick] (-0.1,-0.1) -- (0.1,0.1) (0.1,-0.1) -- (-0.1,0.1);
		\end{scope}
		\begin{scope}[shift={(2,0)}]
		\draw[purple, ultra thick] (-0.1,-0.1) -- (0.1,0.1) (0.1,-0.1) -- (-0.1,0.1);
		\end{scope}
		\begin{scope}[shift={(2,3)}]
		\draw[purple, ultra thick] (-0.1,-0.1) -- (0.1,0.1) (0.1,-0.1) -- (-0.1,0.1);
		\end{scope}
		\begin{scope}[shift={(3,1)}]
		\draw[purple, ultra thick] (-0.1,-0.1) -- (0.1,0.1) (0.1,-0.1) -- (-0.1,0.1);
		\end{scope}
		\begin{scope}[shift={(4,4)}]
		\draw[purple, ultra thick] (-0.1,-0.1) -- (0.1,0.1) (0.1,-0.1) -- (-0.1,0.1);
		\end{scope}
		\draw[ultra thick, <-] (0,3.5) -- (4.5,3.5); \node[right] at (0,3.5) {$\wp$};
		\node[right] at (0,0) {$\fd_1$}; \node[left] at (2,0) {$\fd_3$};
		\node[right] at (1,2) {$\fd_{10}$}; \node[right] at (2,3) {$\fd_9$};
		\node[left] at (3,1) {$\fd_2$}; \node[above] at (4,4) {$\fd_7$};
		\end{scope}
		\end{tikzpicture}
	\end{array} &
	(1)\begin{array}{c}
		\begin{tikzpicture}
		\begin{scope}[scale=0.65,xscale=-1]
		\draw[thick] (0,0) -- (4,4) (1,2) -- (1,1) -- (2,0) (2,3) -- (2,2) -- (3,1);
		\begin{scope}[shift={(0,0)}]
		\draw[purple, ultra thick] (-0.1,-0.1) -- (0.1,0.1) (0.1,-0.1) -- (-0.1,0.1);
		\end{scope}
		\begin{scope}[shift={(1,2)}]
		\draw[purple, ultra thick] (-0.1,-0.1) -- (0.1,0.1) (0.1,-0.1) -- (-0.1,0.1);
		\end{scope}
		\begin{scope}[shift={(2,0)}]
		\draw[purple, ultra thick] (-0.1,-0.1) -- (0.1,0.1) (0.1,-0.1) -- (-0.1,0.1);
		\end{scope}
		\begin{scope}[shift={(2,3)}]
		\draw[purple, ultra thick] (-0.1,-0.1) -- (0.1,0.1) (0.1,-0.1) -- (-0.1,0.1);
		\end{scope}
		\begin{scope}[shift={(3,1)}]
		\draw[purple, ultra thick] (-0.1,-0.1) -- (0.1,0.1) (0.1,-0.1) -- (-0.1,0.1);
		\end{scope}
		\begin{scope}[shift={(4,4)}]
		\draw[purple, ultra thick] (-0.1,-0.1) -- (0.1,0.1) (0.1,-0.1) -- (-0.1,0.1);
		\end{scope}
		\draw[ultra thick, <-] (0,3.5) -- (4.5,3.5); \node[right] at (0,3.5) {$\wp$};
		\node[right] at (0,0) {$\fd_1$}; \node[left] at (2,0) {$\fd_3$};
		\node[right] at (1,2) {$\fd_{14}$}; \node[right] at (2,3) {$\fd_{13}$};
		\node[left] at (3,1) {$\fd_2$}; \node[above] at (4,4) {$\fd_{11}$};
		\end{scope}
		\end{tikzpicture}
	\end{array}\\
	\hline
	(1)\begin{array}{c}
		\begin{tikzpicture}
		\begin{scope}[scale=0.65,xscale=-1]
		\draw[thick] (0,0) -- (4,4) (1,2) -- (1,1) -- (2,0) (2,3) -- (2,2) -- (3,1);
		\begin{scope}[shift={(0,0)}]
		\draw[purple, ultra thick] (-0.1,-0.1) -- (0.1,0.1) (0.1,-0.1) -- (-0.1,0.1);
		\end{scope}
		\begin{scope}[shift={(1,2)}]
		\draw[purple, ultra thick] (-0.1,-0.1) -- (0.1,0.1) (0.1,-0.1) -- (-0.1,0.1);
		\end{scope}
		\begin{scope}[shift={(2,0)}]
		\draw[purple, ultra thick] (-0.1,-0.1) -- (0.1,0.1) (0.1,-0.1) -- (-0.1,0.1);
		\end{scope}
		\begin{scope}[shift={(2,3)}]
		\draw[purple, ultra thick] (-0.1,-0.1) -- (0.1,0.1) (0.1,-0.1) -- (-0.1,0.1);
		\end{scope}
		\begin{scope}[shift={(3,1)}]
		\draw[purple, ultra thick] (-0.1,-0.1) -- (0.1,0.1) (0.1,-0.1) -- (-0.1,0.1);
		\end{scope}
		\begin{scope}[shift={(4,4)}]
		\draw[purple, ultra thick] (-0.1,-0.1) -- (0.1,0.1) (0.1,-0.1) -- (-0.1,0.1);
		\end{scope}
		\draw[ultra thick, <-] (0,3.5) -- (4.5,3.5); \node[right] at (0,3.5) {$\wp$};
		\node[right] at (0,0) {$\fd_4$}; \node[left] at (2,0) {$\fd_6$};
		\node[right] at (1,2) {$\fd_{10}$}; \node[right] at (2,3) {$\fd_9$};
		\node[left] at (3,1) {$\fd_5$}; \node[above] at (4,4) {$\fd_8$};
		\end{scope}
		\end{tikzpicture}
	\end{array}&
	(1)\begin{array}{c}
		\begin{tikzpicture}
		\begin{scope}[scale=0.65,xscale=-1]
		\draw[thick] (0,0) -- (4,4) (1,2) -- (1,1) -- (2,0) (2,3) -- (2,2) -- (3,1);
		\begin{scope}[shift={(0,0)}]
		\draw[purple, ultra thick] (-0.1,-0.1) -- (0.1,0.1) (0.1,-0.1) -- (-0.1,0.1);
		\end{scope}
		\begin{scope}[shift={(1,2)}]
		\draw[purple, ultra thick] (-0.1,-0.1) -- (0.1,0.1) (0.1,-0.1) -- (-0.1,0.1);
		\end{scope}
		\begin{scope}[shift={(2,0)}]
		\draw[purple, ultra thick] (-0.1,-0.1) -- (0.1,0.1) (0.1,-0.1) -- (-0.1,0.1);
		\end{scope}
		\begin{scope}[shift={(2,3)}]
		\draw[purple, ultra thick] (-0.1,-0.1) -- (0.1,0.1) (0.1,-0.1) -- (-0.1,0.1);
		\end{scope}
		\begin{scope}[shift={(3,1)}]
		\draw[purple, ultra thick] (-0.1,-0.1) -- (0.1,0.1) (0.1,-0.1) -- (-0.1,0.1);
		\end{scope}
		\begin{scope}[shift={(4,4)}]
		\draw[purple, ultra thick] (-0.1,-0.1) -- (0.1,0.1) (0.1,-0.1) -- (-0.1,0.1);
		\end{scope}
		\draw[ultra thick, <-] (0,3.5) -- (4.5,3.5); \node[right] at (0,3.5) {$\wp$};
		\node[right] at (0,0) {$\fd_4$}; \node[left] at (2,0) {$\fd_6$};
		\node[right] at (1,2) {$\fd_{14}$}; \node[right] at (2,3) {$\fd_{13}$};
		\node[left] at (3,1) {$\fd_5$}; \node[above] at (4,4) {$\fd_{12}$};
		\end{scope}
		\end{tikzpicture}
	\end{array}
\end{array}\right)
\ee
Here  we denoted corresponding solitons as in expansion \eqref{LG_su3_Hopf}.

The second differential is simply contraction of soliton-anti-soliton pairs:
\be
Q_2=\left(\begin{array}{c|c}
	(1)\begin{array}{c}
		\begin{tikzpicture}
		\draw[thick, ->] (0,0.8) -- (0,1.5) (0,0) -- (0,0.8);
		\draw[thick, ->] (0.2,0.8) -- (0.2,1.5) (0.2,0) -- (0.2,0.8);
		\begin{scope}[shift={(0,0)}]
		\draw[purple, ultra thick] (-0.1,-0.1) -- (0.1,0.1) (0.1,-0.1) -- (-0.1,0.1);
		\end{scope}
		\begin{scope}[shift={(0.2,0)}]
		\draw[purple, ultra thick] (-0.1,-0.1) -- (0.1,0.1) (0.1,-0.1) -- (-0.1,0.1);
		\end{scope}
		\draw (-0.1,0.2) -- (0.3,0.2) to[out=0,in=0] (0.3,-0.2) -- (-0.1,-0.2) to[out=180,in=180] (-0.1,0.2);
		\node[left] at (0,1.5) {$\fd_9$};
		\node[right] at (0.2,1.5) {$\fd_{10}$};
		\draw[ultra thick, ->] (-0.5,1) -- (0.7,1); \node[right] at (0.7,1) {$\wp$};
		\end{tikzpicture}
	\end{array} & (-1) \begin{array}{c}
		\begin{tikzpicture}
		\draw[thick, ->] (0,0.8) -- (0,1.5) (0,0) -- (0,0.8);
		\draw[thick, ->] (0.2,0.8) -- (0.2,1.5) (0.2,0) -- (0.2,0.8);
		\begin{scope}[shift={(0,0)}]
		\draw[purple, ultra thick] (-0.1,-0.1) -- (0.1,0.1) (0.1,-0.1) -- (-0.1,0.1);
		\end{scope}
		\begin{scope}[shift={(0.2,0)}]
		\draw[purple, ultra thick] (-0.1,-0.1) -- (0.1,0.1) (0.1,-0.1) -- (-0.1,0.1);
		\end{scope}
		\draw (-0.1,0.2) -- (0.3,0.2) to[out=0,in=0] (0.3,-0.2) -- (-0.1,-0.2) to[out=180,in=180] (-0.1,0.2);
		\node[left] at (0,1.5) {$\fd_{13}$};
		\node[right] at (0.2,1.5) {$\fd_{14}$};
		\draw[ultra thick, ->] (-0.5,1) -- (0.7,1); \node[right] at (0.7,1) {$\wp$};
		\end{tikzpicture}
	\end{array}
\end{array}\right)
\ee
And we have put $-1$ sign to one of the elements by analogy with the Airy${}^2$ model discussed in Appendix \ref{sec:Airy-2}, so that $Q_2 Q_1=0$. I this case we derive the following cohomology:
\be
H^*(\CE_2,Q)=\IZ[-2](\Psi_1-\Psi_2)
\ee

Generators of the next subcomplex $\CE_4$ we also denote by $\Psi_i$, $i=1,\ldots,10$ as they appear in \eqref{LG_su3_Hopf}, then we have:
\be
\begin{split}
	\CE_4=t^{-4}\left(0 \to \IZ[0]\Psi_1 \oplus \IZ[0]\Psi_2\oplus \IZ[0]\Psi_3\mathop{\longrightarrow}^{Q_1} \IZ[1]\Psi_4\oplus \IZ[1]\Psi_5 \oplus\right. \\ \left. \oplus  \IZ[1]\Psi_6 \oplus \IZ[1]\Psi_7\mathop{\longrightarrow}^{Q_2}\IZ[2]\Psi_8\oplus \IZ[2]\Psi_9\oplus \IZ[2]\Psi_{10}\to 0\right)
\end{split}
\ee
We will not draw all the null webs corresponding to the matrix elements of the supercharge: they can be constructed in analogy to the previous case by moving solitons from fusion interfaces to defusion interfaces and contracting soliton -- anti-soliton pairs:
\be
Q_1=\left(\begin{array}{ccc}
	1 & 1 & 0\\
	0 & 1 & 1\\
	1 & 1 & 0\\
	0 & 1 & 1\\
\end{array}\right),\quad
Q_2=\left(
\begin{array}{cccc}
	1 & 0 & -1 & 0\\
	0 & 1 & 0 & -1 \\
	1 & 1 & -1 & -1\\
\end{array}
\right)
\ee
The resulting cohomology reads:
\be
H^*(\CE_4,Q)=\IZ[-4](\Psi_1-\Psi_2+\Psi_3)\oplus \IZ[-2] (\Psi_8+\Psi_9)
\ee

The resulting Poincar\'e polynomial reads:
\be
P_{\rm LG}\left( q,t \left|\begin{array}{c}
	\begin{tikzpicture}
	\draw[ultra thick] ([shift=(70:0.5)]0,0) arc (70:410:0.5);
	\draw[ultra thick] ([shift=(250:0.5)]0.5,0) arc (250:360:0.5);
	\draw[ultra thick] ([shift=(0:0.5)]0.5,0) arc (0:230:0.5);
	\draw[thick,->] ([shift=(350:0.5)]0,0) arc (350:360:0.5);
	\draw[thick,<-] ([shift=(180:0.5)]0.5,0) arc (180:190:0.5);
	\end{tikzpicture}
\end{array}\right.\right)=1+q^2 t^{-2}+q^4(t^{-4}+t^{-2})+2 q^6 t^{-4}+2 q^8 t^{-6}+q^{10} t^{-8}
\ee
It should be compared to Khovanov-Rozansky polynomial (see \cite{Gukov_Hopf} for value $N=3$):
\be
P_{\rm KR}\left( q,t\right)=P_{\rm LG}\left( qt,t\right)=1+q^2+q^4(1+t^{2})+2 q^6 t^{2}+2 q^8 t^{2}+q^{10} t^{2}
\ee

\section{On exact skein triangle tautology}\label{sec:taut}

The statement that the Khovanov (Khovanov-Rozansky) homological construction satisfies the skein exact triangle relation is somewhat tautological since necessary elements are just built in the construction. However this property is not so obvious for other link homology constructions. So let us review what elements of the construction are crucial for the skein exact triangle. We exploit the following
\begin{theorem}
  (\cite[Lemma 7.1]{skein}). Suppose there is a family of complexes $(C_i,d_i)$ for $i\in\IZ$ and chain maps
  $$
  f_i:\quad C_i\longrightarrow C_{i+1}
  $$
  Suppose that the composite chain-map $f_{i+1}f_i$ is chain-homotopic to 0, i.e. there are
  $$
  g_i:\quad  C_i\longrightarrow C_{i+2}
  $$
  such that
  $$
  f_{i+1}f_i+d_{i+2}g_i+g_id_i=0
  $$
  Moreover suppose maps 
  $$
  g_{i+1}f_i+f_{i+2}g_i: \quad C_i\longrightarrow C_{i+3}
  $$
  induce an isomorphism in homology. Then the induced maps in homology:
  $$
  (f_i)_*:\quad H^*(C_i,d_i)\longrightarrow H^*(C_{i+1},d_{i+1})
  $$
  form an exact sequence.
\end{theorem}

Schematically, we can depict this system of maps in the following way:
$$
\begin{array}{c}
\begin{tikzpicture}
\node (A) at (0,0) {$\ldots$};
\node (B) at (2,0) {$C_{i}$};
\node (C) at (4,0) {$C_{i+1}$};
\node (D) at (6,0) {$C_{i+2}$};
\node (E) at (8,0) {$C_{i+3}$};
\node (F) at (10,0) {$\ldots$};
\path (B) edge[->, loop above] node[above] {$d_i$} (B) (C) edge[->, loop above] node[above] {$d_{i+1}$} (C) (D) edge[->, loop above] node[above] {$d_{i+2}$} (D) (E) edge[->, loop above] node[above] {$d_{i+3}$} (E) (A) edge[->] node[above] {$f_{i-1}$} (B) (B) edge[->] node[above] {$f_{i}$} (C) (C) edge[->] node[above] {$f_{i+1}$} (D) (D) edge[->] node[above] {$f_{i+2}$} (E) (E) edge[->] node[above] {$f_{i+3}$} (F) (B) edge[->, bend right] node [below] {$g_{i}$} (D) (C) edge[->, bend right] node [below] {$g_{i+1}$} (E);
\end{tikzpicture}
\end{array}
$$

It is not complicated to construct all these maps for Khovanov-Rozansky complex explicitly. Let us denote complexes corresponding to two resolutions of intersection $\chi$ as
$$
C_{\chi}^{(-)}=E\left[
\begin{array}{c}
\begin{tikzpicture}[scale=0.7]
\draw [ultra thick] (-0.5,-0.5) -- (0,-0.2) -- (0.5,-0.5);
\draw [ultra thick, ->] (0,0.2) -- (-0.5,0.5); 
\draw [ultra thick, ->] (0,0.2) -- (0.5,0.5);
\draw[fill=white] (0.05,0.2) -- (0.05,-0.2) -- (-0.05,-0.2) -- (-0.05,0.2) -- cycle;
\end{tikzpicture}
\end{array}
\right], \quad
C_\chi^{(+)}=q^{-1} t^{-1}\; E\left[
\begin{array}{c}
\begin{tikzpicture}[scale=0.7]
\draw [ultra thick,->] (-0.5,-0.5) to [out= 45, in =315] (-0.5,0.5); \draw [ultra thick,->] (0.5,-0.5) to [out=135, in=225] (0.5,0.5);
\end{tikzpicture}
\end{array}\right],\quad C_{\chi}^{(0)}=E\left[
\begin{array}{c}
\begin{tikzpicture}[scale=0.7]
\draw[ultra thick, ->] (-0.5,-0.5) -- (0.5,0.5);
\draw[ultra thick] (0.5,-0.5) -- (0.1,-0.1);
\draw[ultra thick, ->] (-0.1,0.1) -- (-0.5,0.5);
\end{tikzpicture}
\end{array}\right]
$$
So we have 
$$
C_\chi^{(0)}=C_{\chi}^{(+)}\oplus C_{\chi}^{(-)}
$$
Thus we construct the complexes:
$$
C_i=C^{(-)}_{\chi},\quad C_{i+1}=C^{(0)}_{\chi},\quad C_{i+2}=C^{(+)}_{\chi},\quad C_{i+3}=t C_i
$$
and maps:
$$
d_i=d',\quad d_{i+1}=d,\quad d_{i+2}=d',\quad d_{i+3}=td_i
$$
$$
f_i=P_-,\quad f_{i+1}=P_+,\quad f_{i+2}=d_{\chi},\quad f_{i+3}=tf_i
$$
$$
g_i=0,\quad g_{i+1}=-tP_-,\quad g_{i+2}=tP_+,\quad g_{i+3}=tg_i
$$
where $d'=\sum\lm_{\chi'\neq \chi}d_{\chi'}$ and $P_a$ is a projector to a subspace marked by $a$. These maps satisfy all the necessary conditions \emph{provided} that $d_{\chi}$ and $d_{\chi'}$ \emph{anti-commute} for all pairs of intersections $\chi$ and $\chi'$.

\end{document}